\documentclass[journal]{IEEEtran}

\usepackage{times}
\usepackage{epsfig}
\usepackage{graphicx}
\usepackage{amsmath}
\usepackage{bm}
\usepackage{amssymb}
\usepackage{mathtools, cuted}
\usepackage{mdframed}
\usepackage[]{subfigure}
\usepackage{multirow}
\usepackage{colortbl}
\usepackage{mathrsfs,amsmath}
\usepackage{dsfont}
\usepackage[normalem]{ulem}
\usepackage[table]{xcolor}
\newcommand{\argmin}{\arg\!\min}
\newcommand{\argmax}{\arg\!\max}
\usepackage{boldline}
\usepackage{algorithm2e}
\usepackage{array}
\usepackage{booktabs}
\usepackage{makecell}

\usepackage{setspace}

\newcommand{\specialcell}[2][c]{\begin{tabular}[#1]{@{}c@{}}#2\end{tabular}}
\makeatletter
\newcommand{\thickhline}{%
	\noalign {\ifnum 0=`}\fi \hrule height 5pt
	\futurelet \reserved@a \@xhline
}
\newcolumntype{"}{@{\hskip\tabcolsep\vrule width 5pt\hskip\tabcolsep}}
\makeatother

\newif\ifdebugon
\debugonfalse

\definecolor{MyDarkBlue}{rgb}{0,0.08,0.5}
\definecolor{MyDarkRed}{rgb}{0.7,0.02,0.02}
\definecolor{MyDarkGreen}{rgb}{0.02,0.30,0.02}
\definecolor{MyDarkOrange}{rgb}{0.40,0.2,0.02}
\definecolor{MyRed}{rgb}{1.0,0.0,0.0}

\ifdebugon
\newcommand{\katie}[1]{\textcolor{MyDarkRed}{[Katie: #1]}}
\else
\newcommand{\katie}[1]{}
\fi

\ifdebugon
\newcommand{\michael}[1]{\textcolor{MyDarkBlue}{[Michael: #1]}}
\else
\newcommand{\michael}[1]{}
\fi

\ifdebugon
\newcommand{\delete}[1]{\textred{\sout{#1}}}
\else
\newcommand{\delete}[1]{}
\fi

\ifdebugon

\else

\fi

\makeatletter
\newcommand{\Spvek}[2][r]{%
	\gdef\@VORNE{1}
	\left[\hskip-\arraycolsep%
	\begin{array}{#1}\vekSp@lten{#2}\end{array}%
	\hskip-\arraycolsep\right]}

\def\vekSp@lten#1{\xvekSp@lten#1;vekL@stLine;}
\def\vekL@stLine{vekL@stLine}
\def\xvekSp@lten#1;{\def\temp{#1}%
	\ifx\temp\vekL@stLine
	\else
	\ifnum\@VORNE=1\gdef\@VORNE{0}
	\else\@arraycr\fi%
	#1%
	\expandafter\xvekSp@lten
	\fi}
\makeatother

\newcommand{\xpos}{\varrho}
\newcommand{\ypos}{\delta}
\newcommand{\FTmtx}{\bm{F}}
\newcommand{\vecFTmtx}{\mathbf{f} }
\newcommand{\vis}{\Gamma}
\newcommand{\bLambda}{\bm{\Lambda} }
\newcommand{\bR}{\bm{R} }
\newcommand{\evolve}{\bm{A} }
\newcommand{\bQ}{\bm{Q} }
\newcommand{\bmu}{\bm{\mu} }
\newcommand{\nmeas}{K}
\newcommand{\ntime}{N}
\newcommand{\meas}{ \bm{y}}
\newcommand{\im}{ \bm{x}}
\newcommand{\npix}{M}
\newcommand{\ntimes}{N}
\newcommand{\ntele}{P}

\usepackage[pagebackref=true,breaklinks=true,letterpaper=true,colorlinks,bookmarks=false]{hyperref}

\begin{document}

\title{Reconstructing Video from Interferometric \\ Measurements of Time-Varying Sources}

\author{Katherine L.\ Bouman$^{1,2}$, 
	Michael D.\ Johnson$^2$, 
	Adrian V.\ Dalca$^{1,3}$,
	Andrew A.\ Chael$^2$,  \\
	Freek Roelofs$^{4}$, 
	Sheperd S.\ Doeleman$^2$,
	William T.\ Freeman$^{1,5}$\\ \vspace{0.1in}
    {\small
	$^1${Massachusetts Institute of Technology, CSAIL}
	$^2${Harvard-Smithsonian Center for Astrophysics} \\
    $^3${Massachusetts General Hospital, HMS}
    $^4${Radboud University}
    $^5${Google Research} \vspace{-0.3in}
    }
}

\maketitle

\begin{abstract}	
	Very long baseline interferometry (VLBI) makes it possible to recover images of astronomical sources with extremely high angular resolution. 
	Most recently, the Event Horizon Telescope (EHT) has extended VLBI to short millimeter wavelengths with a goal of achieving angular resolution sufficient for imaging the event horizons of nearby supermassive black holes.
	VLBI provides measurements related to the underlying source image through a sparse set spatial frequencies. 
	An image can then be recovered from these measurements by making assumptions about the underlying image. 
	One of the most important assumptions made by conventional imaging methods is that over the course of a night's observation the image is static. 
	However, for quickly evolving sources, such as the galactic center's supermassive black hole (SgrA*) targeted by the EHT, this assumption is violated and these conventional imaging approaches fail.  
	In this work we propose a new way to model VLBI measurements that allows us to recover both the appearance and dynamics of an evolving source by reconstructing a video rather than a static image.  
By modeling VLBI measurements using a Gaussian Markov Model, we are able to propagate information across observations in time to reconstruct a video,
while simultaneously learning about the dynamics of the source's emission region.
	We demonstrate our proposed Expectation-Maximization (EM) algorithm, StarWarps, on realistic synthetic observations of black holes, and show how it substantially improves results compared to conventional imaging algorithms. Additionally, we demonstrate StarWarps on real VLBI data of the M87 Jet from the VLBA. 
\end{abstract}

\begin{figure*}[h!]
	\vspace{-.1in}
	\begin{center}
		\begin{tabular}{ c  c  c  c  c  }
			\large{\textsf{17 GST / 0 Hrs}}  &   \large{\textsf{20 GST / 3 Hrs}}   &\large{\textsf{23 GST / 6 Hrs}} &\large{\textsf{2 GST / 9 Hrs}}  &\large{\textsf{6 GST / 13 Hrs}}    \\
			{{\includegraphics[width=.17\linewidth]{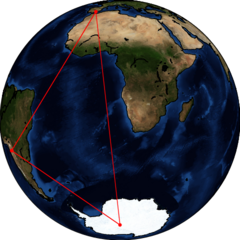}} }  & {{\includegraphics[width=.17\linewidth]{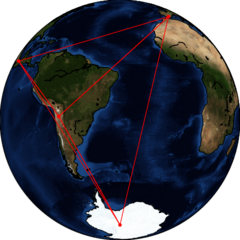}} } &
			{\includegraphics[width=.17\linewidth]{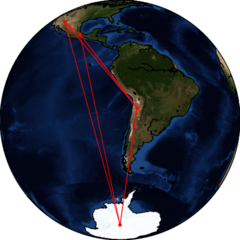}} 
			&
			{\includegraphics[width=.17\linewidth]{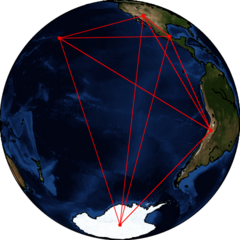}} &
			{\includegraphics[width=.17\linewidth]{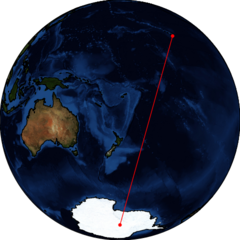}} 
			
			\\
			\Large{$\Downarrow$}  &   \Large{$\Downarrow$}  & \Large{$\Downarrow$} & \Large{$\Downarrow$}& \Large{$\Downarrow$}     \\
			& \vspace{-.1in}
			\\
			{{\includegraphics[width=.17\linewidth]{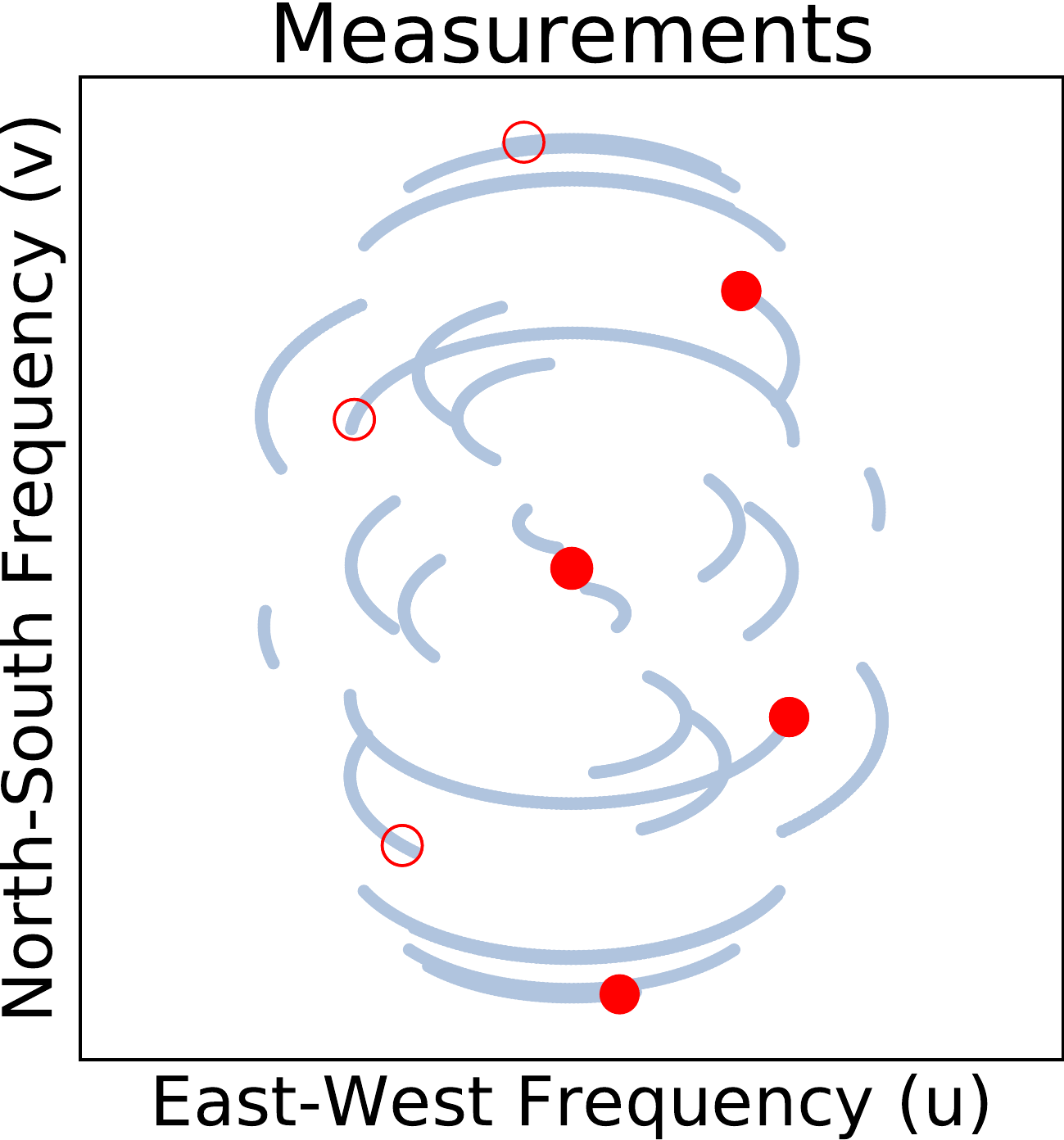}} } &
			{{\includegraphics[width=.17\linewidth]{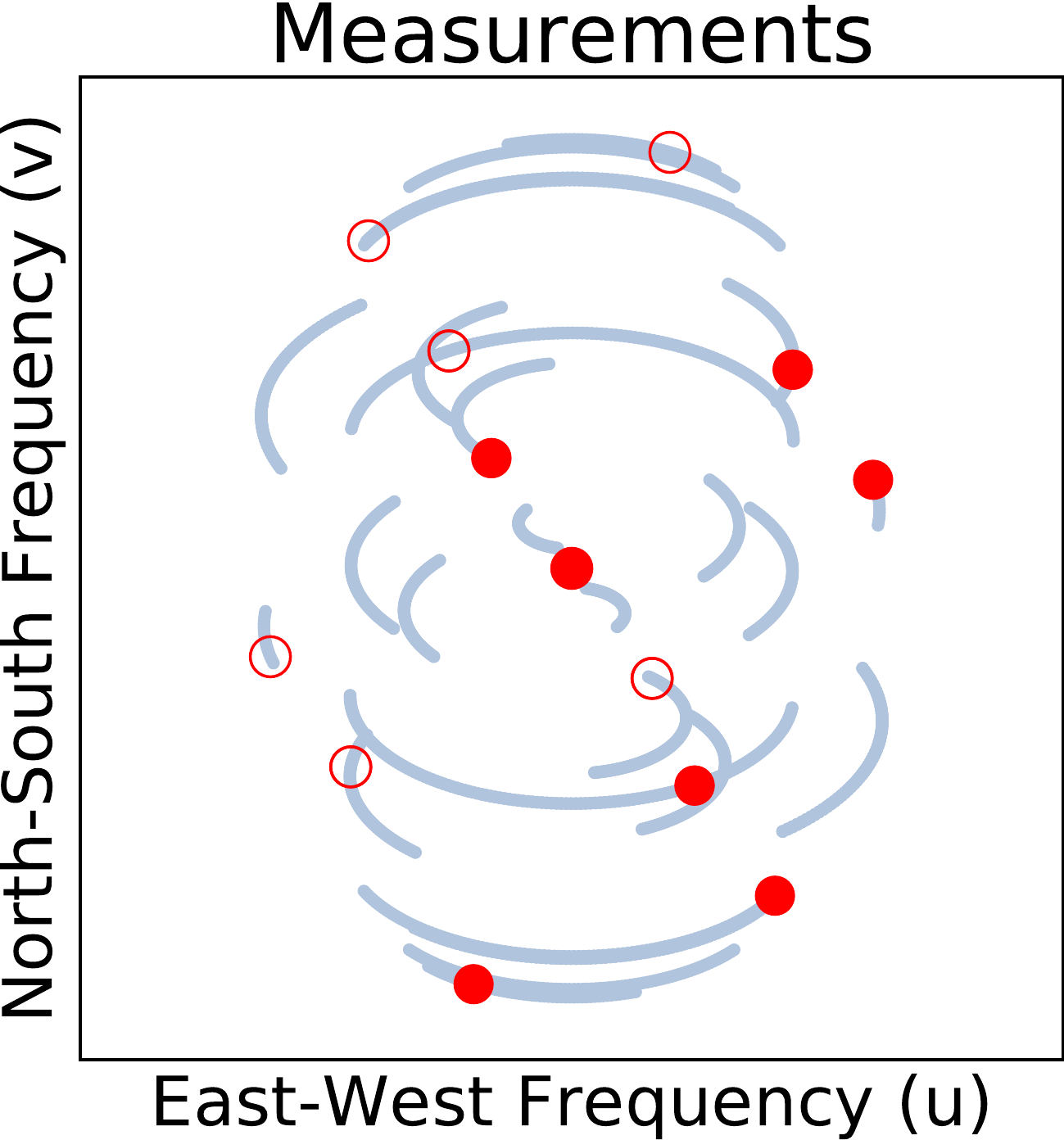}} } &
			{\includegraphics[width=.17\linewidth]{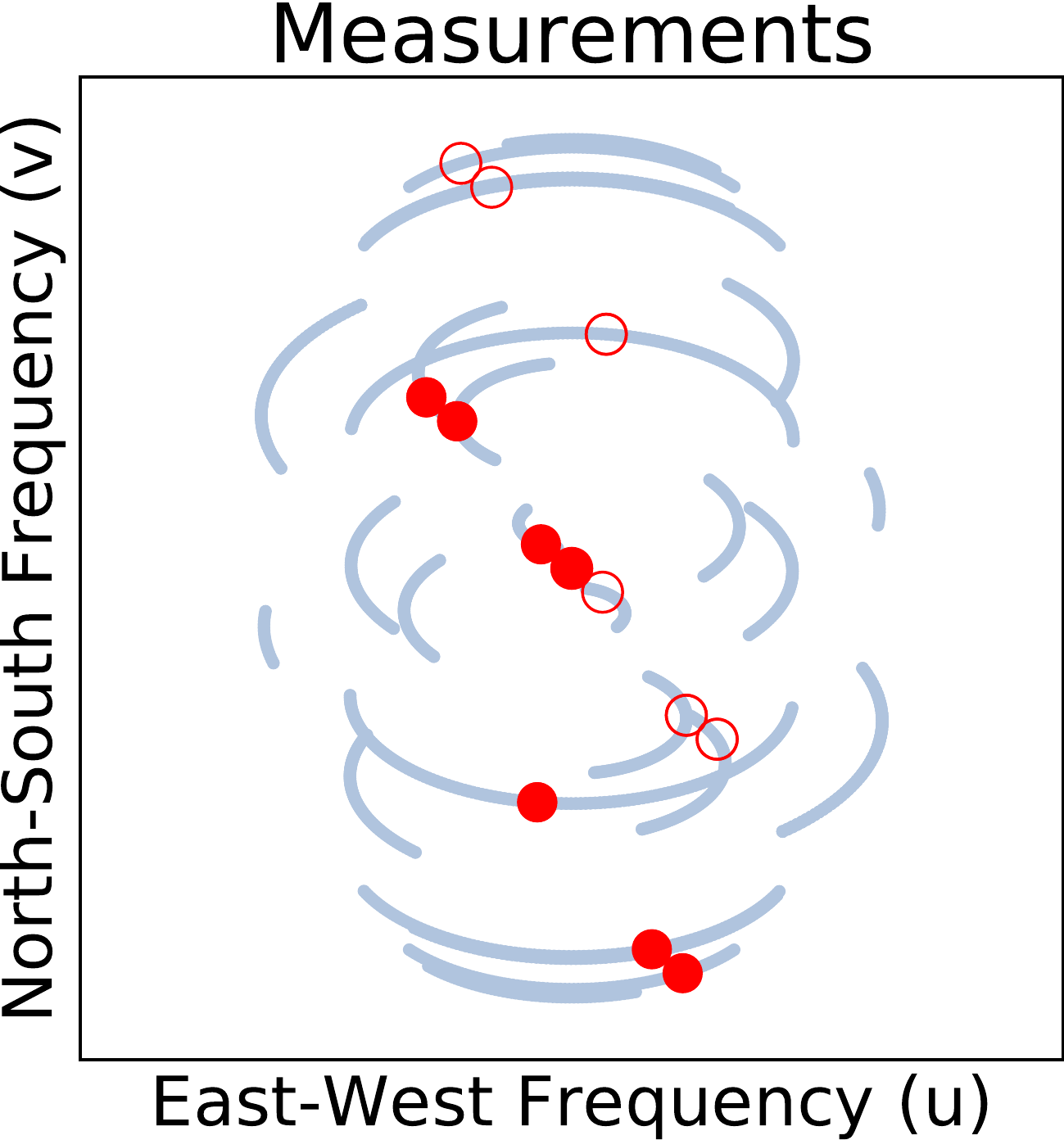}} 
			&
			{\includegraphics[width=.17\linewidth]{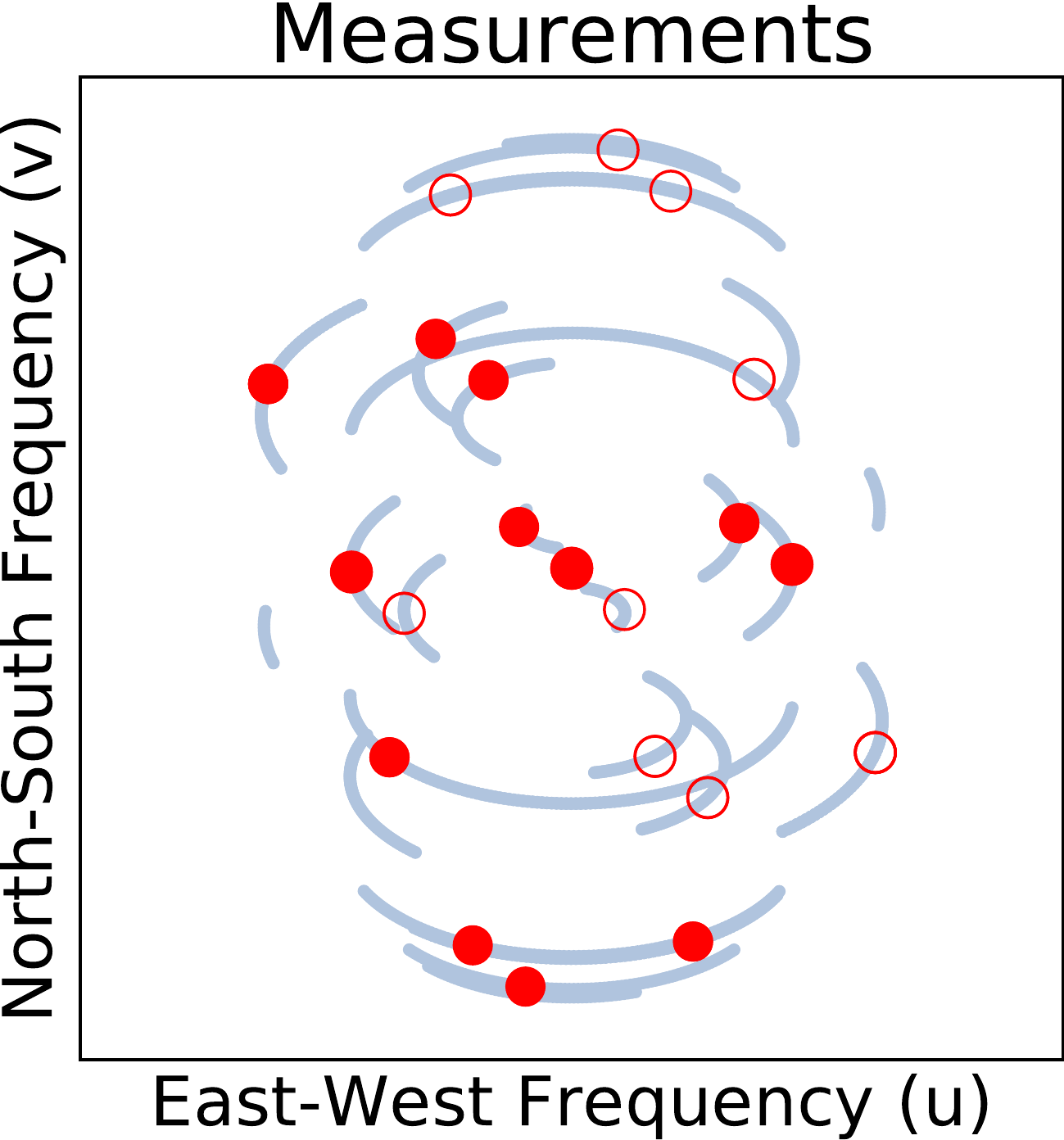}} &
			{\includegraphics[width=.17\linewidth]{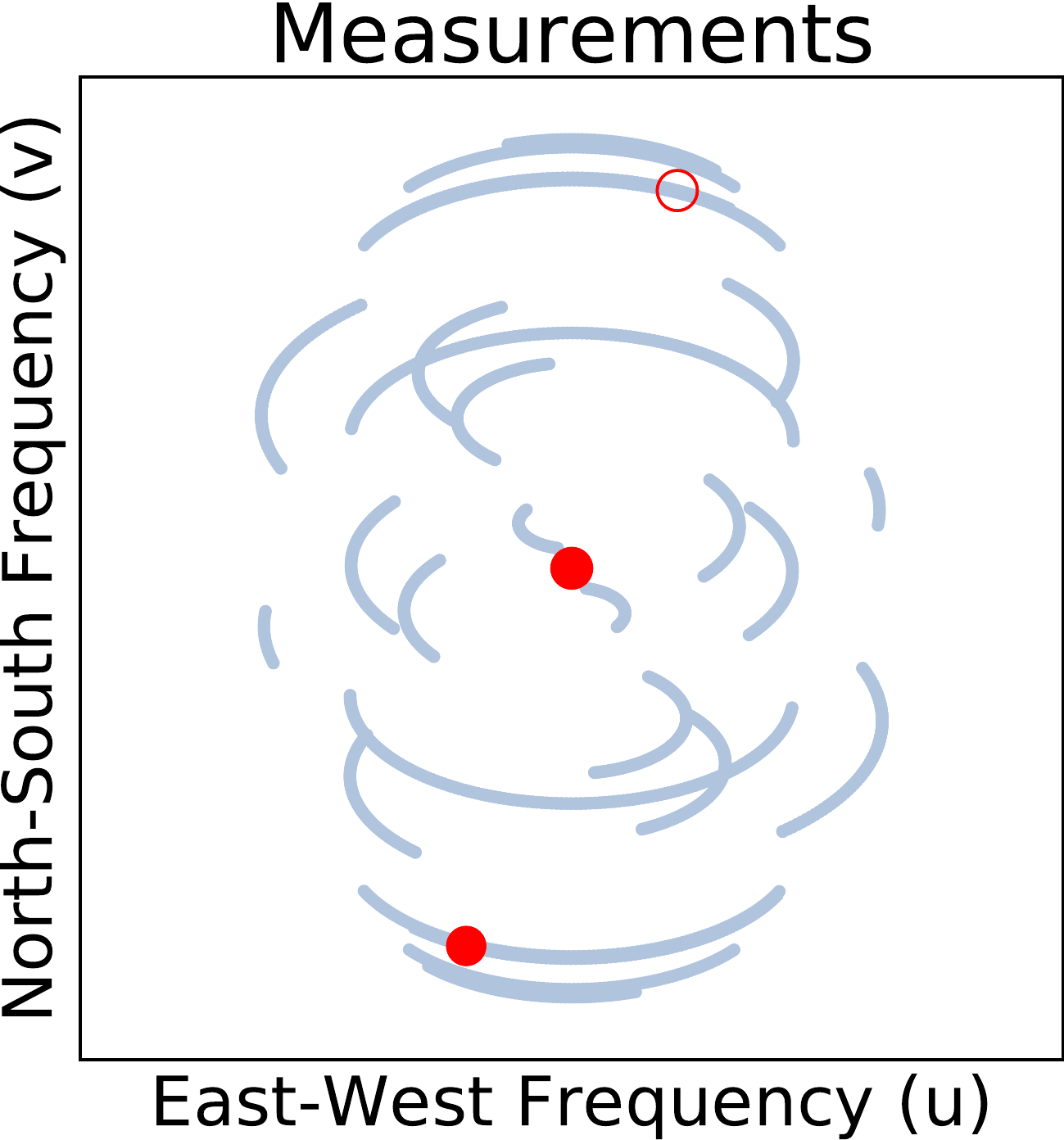}} 
			\\
		\end{tabular}
		\caption{{\bf Earth Rotation Synthesis:} For every two telescopes in the interferometric array, we obtain a single measurement (visibility) related to the underlying source image's 2D spatial frequency. This frequency is related to the baseline between the telescopes in the direction perpendicular to the observing source. 
			It is prohibitively difficult to reconstruct a faithful image from just these small number of measurements. 
			However, as the Earth rotates, the projected baselines change, and we observe new measurements related to different regions of the 2D frequency plane. 
			As time progresses (specified by the Greenwich Sidereal Time (GST)), the projected baselines change and the red dots on the frequency coverage plot (bottom row), which indicate the current measurements, change position. 
			Assuming the source is static, this amounts to carving out elliptical paths in the frequency plane that are all related to the same image. The set of all frequencies sampled over an observation is shown by the transparent blue lines. However, when a source evolves over the course of a night's observation, each instantaneous set of measurements (shown in red on each frequency plot) is only related to a single image.
			As light emitted from the source is real-valued we obtain two measurements on opposite sides of the frequency plane -- each independent set of measurements displayed as either the open or closed red circles. The Earth diagrams are shown from the point of view of an observer at Sgr A*. Note that although the telescope array is not changing, the number of visibilities at a given time changes depending on how many sites can still see Sgr A*. 
		} 
		\label{fig:earthrotationsythesis}
	\end{center}
	\vspace{-.1in}
\end{figure*}

\vspace{-.1in}
\section{Introduction}

Imaging distant astronomical sources with high resolving power at radio wavelengths would require single-dish telescopes with prohibitively large diameters due to the inverse relationship between angular resolution and telescope diameter. Very long baseline interferometry (VLBI) is a technique that alleviates the need for building an impossibly large single-dish telescope by simultaneously observing a common source from an array of telescopes distributed around the Earth. This technique makes it possible to emulate samples from a single-dish telescope with a diameter equal to the maximum distance between telescopes in the array, at the expense of having to handle missing data~\cite{TMS}. Thus far, VLBI has been primarily used to image sources that are static on the time scale of a day's observation. In this work, we extend the technique's applicability to imaging time-varying sources by reconstructing a video of the source. 

VLBI measurements place a sparse set of constraints on
the 
spatial frequencies of the underlying source image. In particular, each pair of telescopes provides information about a single 2D spatial frequency. This frequency is related to the baseline vector connecting the two telescope sites, projected orthogonal to the direction of the target source~\cite{TMS}. Thus, at a single time, for an array with $\ntele$ telescopes, at most ${\ntele (\ntele - 1)}/{2}$ spatial frequencies are measured. 
For example, an array of 6 telescopes would yield only 15 measurements. 
However, as the Earth rotates, 
the baseline vector connecting each pair of telescopes changes. This allows the array to sample additional spatial frequencies along elliptical paths in the frequency domain~\cite{TMS}. Refer to Figure~\ref{fig:earthrotationsythesis}. Combining the different measurements taken as the Earth rotates is referred to as {\it Earth Rotation Synthesis}.
Earth rotation synthesis 
is essential for building up enough measurements to constrain image reconstruction.

The task of reconstructing an image from these sparse constraints is highly ill-posed and relies heavily on assumptions made about the underlying image~\cite{hogbom1974aperture, taylor1999synthesis, rusenimaging}. 
If a source is static, the VLBI measurements -- taken over time as the Earth rotates -- all correspond to the same underlying image. 
Under a static source assumption, 
recently developed VLBI image reconstruction techniques have been demonstrated on small telescope arrays~\cite{bouman2016computational, andrew, kazu, Fish_2016_Imaging, akiyama2017superresolution}.
However, for an evolving source, measurements are no longer sampled from the same image, and these reconstruction algorithms quickly break down.


Although most 
astronomical sources are static over the time scale of a night's observation, 
some notable
sources have detectable structural changes on much shorter timescales. 
For instance,  the Galactic Center supermassive black hole, Sagittarius A* (SgrA*), has an estimated mass of only four-million solar masses~\cite{Ghez_2008}. 
This implies that SgrA* is quickly evolving, with an innermost stable circular orbit of just 4 to 30 minutes, depending on the spin of the black hole~\cite{orbitalperiod}. 
Previous observations have shown that SgrA* varies dramatically over a night's observation on the scale of its predicted event horizon, in both total-intensity and polarization~\cite{Fish_2011, Johnson_2015}. 

SgrA* is a prime target for the Event Horizon Telescope (EHT) -- an international project whose goal is to take the first image of the immediate environment around a black hole~\cite{Doeleman_2009}.
Realizing this goal would not only substantiate the existence of a black holes' event horizon, but also aid in studying general relativity in the strong field regime~\cite{Doeleman_2008, Johannsen_2010}.
Unfortunately, the amount of variation predicted for SgrA* suggests that conventional VLBI imaging techniques will be inappropriate for observations taken by the EHT~\cite{rusenimaging,orbitalperiod}. 
Thus, in this work we present a new imaging algorithm for time-varying sources that models the VLBI observations as being from a Gaussian Markov Model. 
Our dynamic imaging algorithm allows for an evolving emission region by simultaneously reconstructing both images and motion trajectories - essentially reconstructing a video rather than a static image.

In Section~\ref{sec:meas} we review the basics of interferometric imaging and the data traditionally used to constrain image reconstruction. In Section~\ref{sec:setup} we summarize standard approaches to VLBI imaging and define a generalized expression for data consistency. In Section~\ref{sec:static} we explain how static imaging can be performed under a multivariate Gaussian prior. Highlighting the approach's benefits first in the case of a static source becomes helpful when discussing our proposed solution to the more complex dynamic imaging problem, StarWarps, in Section~\ref{sec:dynamic_model}, and when deriving our proposed EM inference algorithm in Section~\ref{sec:dynamic_inference}. Results can be seen in Section~\ref{sec:results}, followed by concluding remarks in Section~\ref{sec:conclusion}.


\section{VLBI Measurements and Data Products}
\label{sec:meas}

\begin{figure*}[h!]
	\vspace{-.3in}
	\begin{center}
		\begin{tabular}{  c  c   }
			\large{\textsf{Visibility Amplitude}}   &\large{\textsf{Closure Phase}}      \\ 
			\small{\textsf{ALMA-SPT Baseline}}   &\small{\textsf{ALMA-SPT-PV Triangle}}      \\ 
			\includegraphics[width=.47\linewidth]{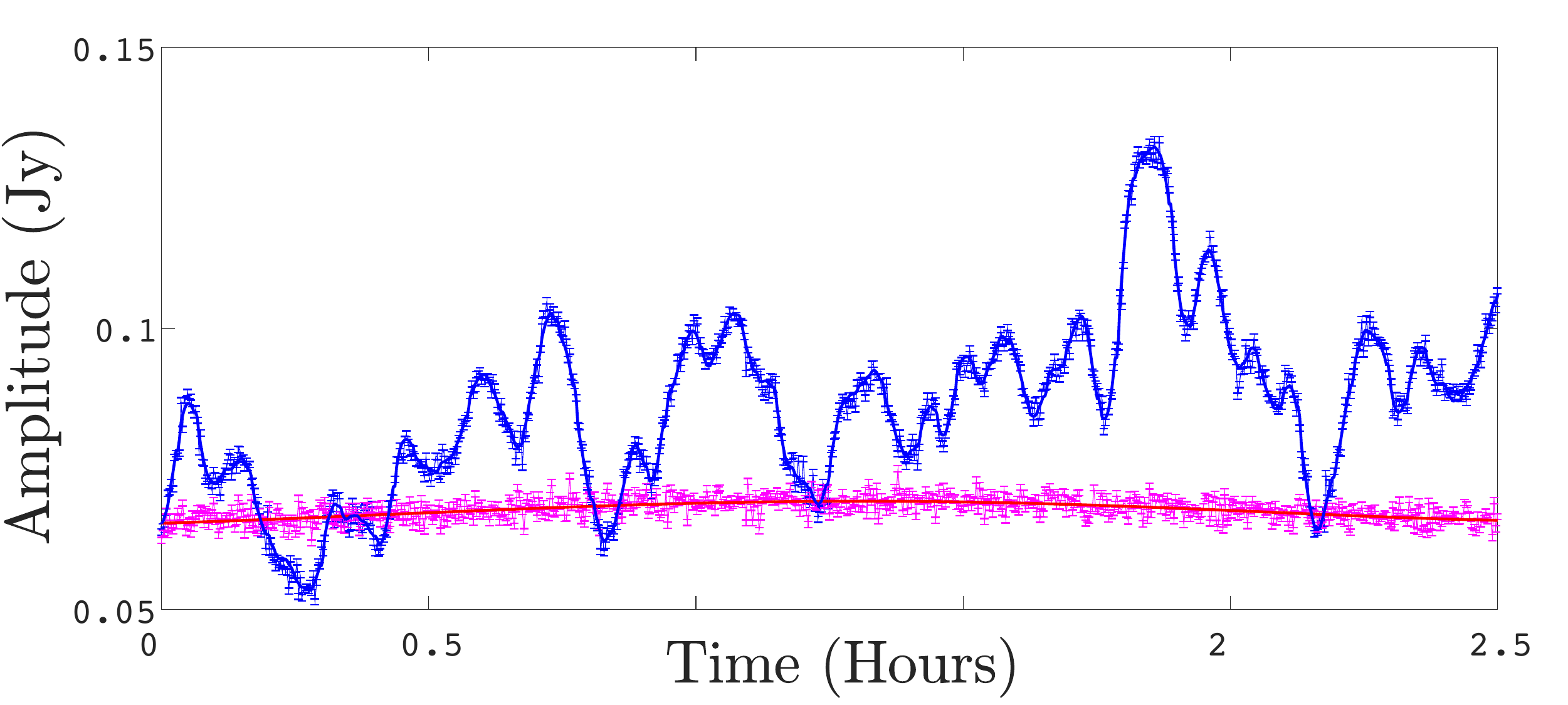} &
			\includegraphics[width=.47\linewidth]{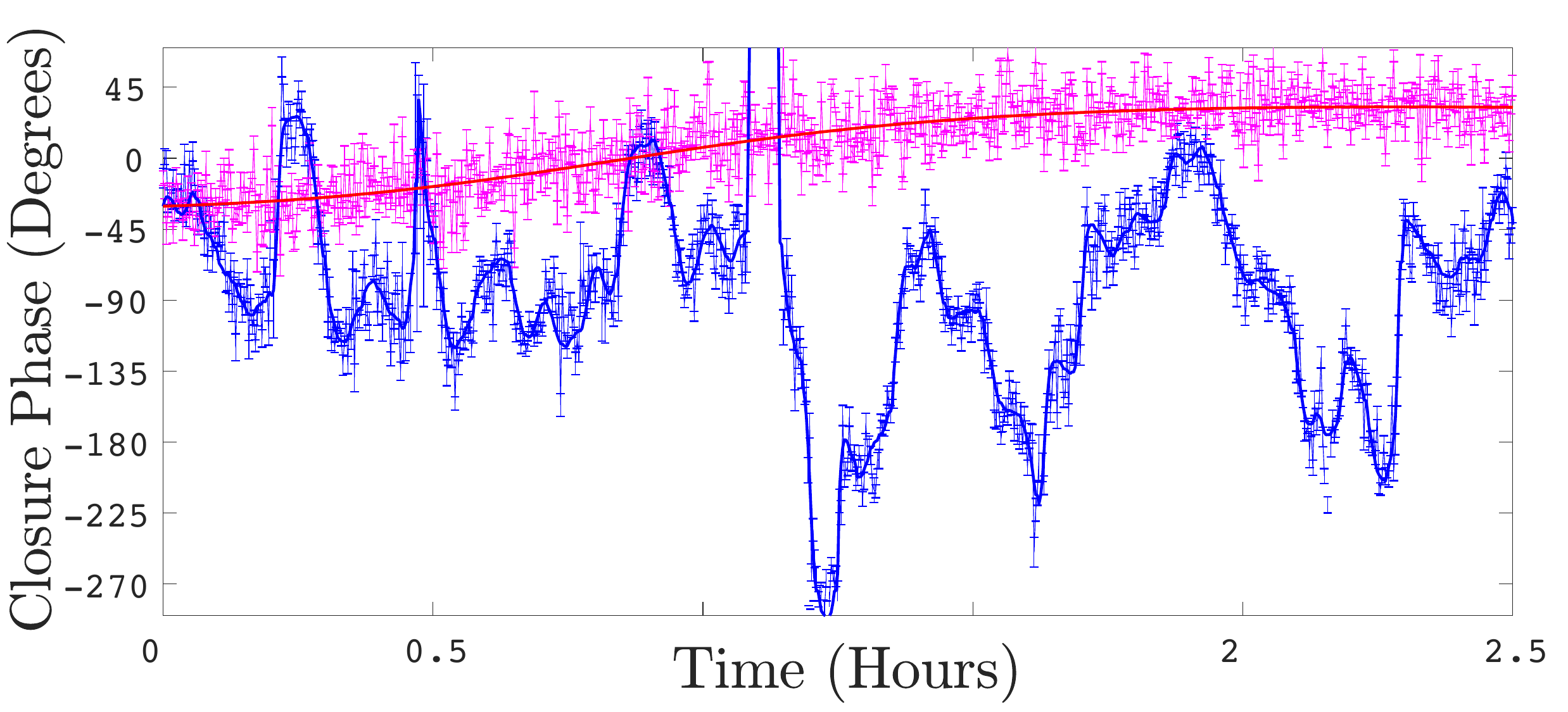} 
			\\
		\end{tabular}
		\caption{{\bf Simulated data under a static vs. varying source:} Contrasting of data observed from a static emission region (magenta) to that of a varying emission region (blue) over the course of 2.5 hours. Although both sequences start with the same image, the visibility amplitude and closure phase both begin to deviate from the static image very quickly. The ideal observation for the static and time-varying source is shown by the solid red and blue lines, respectively. We also show sample measurements with their respective error bars in the same colors. This data is simulated at a sampling interval of 11 seconds using the EHT2017 array from the frames in Video 3 presented in Section~\ref{sec:results}. This figure shows 1 of the ${\nmeas \choose 2}$ possible visibility amplitudes and 1 of ${\nmeas \choose 3}$ possible closure phases (phase of the bispectrum). }
		\label{fig:amplandphase}
	\end{center}
	\vspace{-.2in}
\end{figure*}

A single-dish telescope is diffraction limited, with an angular resolution dictated by the ratio of wavelength to dish diameter~\cite{TMS}. This governing law limits the highest achievable angular resolution of traditional single-dish telescopes. For example, past observations and simulations suggest that the emission surrounding SgrA* subtends approximately $2.5 \times 10^{-10}$ radians, or $50$ $\mu$-arcseconds~\cite{Doeleman_2008}. Imaging this emission region at $1.3$ mm wavelength would require an impossibly large single-dish telescope with a $13,000$ km diameter. However, simultaneously collecting data from an array of telescopes, called an interferometer, allows us to overcome the single-dish diffraction limit, and create a virtual telescope as large as the maximum distance between telescopes in the array. When these telescopes are distributed globally, using separate clocks and recording systems, this technique is referred to as Very Long Baseline Interferometry (VLBI)~\cite{TMS}.

An interferometer consists of $\ntele$ telescopes simultaneously observing and recording time-stamped data-streams of light traveling from a common source~\cite{TMS}.
From these measurements a number of different {\it data products} can be computed. 
Depending on the imaging technique and the quality of the measurements, different data products may be used. In this section we review a number of common data products used in VLBI imaging, and throughout this paper. 



\vspace{-.2in}
\subsection{Visibilities }
\label{sec:vis}

The time-averaged cross-correlation of the recorded scalar electric fields at two telescopes, called a {\it visibility}, provides information about the spatial structure of the observed source\footnote{Although the particular length of time-averaging varies per experiment, for the EHT data is averaged over sub-second intervals before additional coherent averaging is done to increase SNR (on an order of 5-100 seconds).}. Formally, the van Cittert-Zernike theorem states that a visibility, $\vis(I, u,v)$ is related to the ideal source image $I(\xpos, \ypos)$ through a Fourier transform: 
\begin{align}
\vis(I, u,v) \approx \int_{\xpos} \int_{\ypos} e^{-i2 \pi (u\xpos + v \ypos) } I(\xpos, \ypos) d\xpos d\ypos
\label{eq:vcz}
\end{align}
\noindent{where $(\xpos,\ypos)$ is the angular sky coordinate in radians, and $(u,v)$ is the dimensionless baseline vector between two telescopes, measured in wavelengths and orthogonal to the source direction~\cite{TMS}. Note that each visibility is a complex value with both an amplitude and a phase. As each visibility is calculated from a pair of telescopes, for a $\ntele$ telescope array we can obtain up to $\frac{\ntele (\ntele - 1)}{2}$ visibility measurements at a single instant in time. The image spatial frequency sampled by a visibility is related to the distance between the telescopes; telescopes that are father apart inform us about higher spatial frequencies of the underlying image.}
For radio wavelengths, the thermal noise appearing on perfectly calibrated visibilities is 
circularly Gaussian in the real-imaginary plane\footnote{In practice, the thermal noise is calculated from first principles using measured SEFD values at each telescope (see Eq~\ref{eq:thermal}) or estimated empirically. }~\cite{taylor1999synthesis}.




\vspace{-.1in}
\subsection{Bispectrum and Closure Phases }
\label{sec:bispec}


The van Cittert-Zernike theorem (Equation~\ref{eq:vcz}) is derived under the assumption that light is moving through a vacuum. 
However, in reality, inhomogeneity in the Earth's atmosphere causes the light to move at different velocities towards each telescope. At short observing wavelengths, this propagation delay has a large effect on the phase of measurements, and renders the absolute visibility phase unusable for image reconstruction~\cite{monnier2013radio}. Although absolute phase measurements cannot be used, a property called phase closure allows us to still obtain some information from these phases. 

Consider three telescopes, denoted by $i, j$, and $k$, observing the same source. From each pair of telescopes we compute a visibility: $\vis_{i,j}, \vis_{j,k}, \vis_{k,i}$. Rapidly changing atmospheric propagation delays, resulting in phase errors of $\phi_i$ and $\phi_j$ to telescopes $i$ and $j$ respectively, will cause error in the visibility's phase: 
\begin{equation} \vis_{i,j}^{\mbox{\tiny{meas}}} = e^{i(\phi_i - \phi_j)}\vis_{i,j}^{\mbox{\tiny{ideal}}}. 
\end{equation}
\noindent{However, by multiplying the visibilities from 3 different telescopes in a closed loop, we obtain a term that is invariant to the atmosphere (Equation~\ref{eq:phaseclosure})~\cite{jennison}.   }
\begin{align}  
\notag \vis^{\mbox{\tiny{meas}}}_{i,j}\vis^{\mbox{\tiny{meas}}}_{j,k}\vis^{\mbox{\tiny{meas}}}_{k,i} &= e^{i(\phi_i-\phi_j)}\vis^{\mbox{\tiny{ideal}}}_{i,j}e^{i(\phi_j-\phi_k)}\vis^{\mbox{\tiny{ideal}}}_{j,k}e^{i(\phi_k-\phi_i)}\vis^{\mbox{\tiny{ideal}}}_{k,i} \\
&=\vis^{\mbox{\tiny{ideal}}}_{i,j}\vis^{\mbox{\tiny{ideal}}}_{j,k}\vis^{\mbox{\tiny{ideal}}}_{k,i}. 
\label{eq:phaseclosure}
 \end{align}

We refer to the multiplication of these three complex visibilities in a closed loop 
as the {\it bispectrum}, and the phase of the bispectrum 
as the {\it closure phase}.
Although these data products are invariant to atmospheric noise, they come at the cost of having a fewer number of independent data points - only $\frac{(\ntele-1)(\ntele-2)}{2}$ compared to  $\frac{\ntele(\ntele-1)}{2}$ independent visibilities. Additionally, when using these data products, information related to the absolute location of the source is lost. 

As each bispectrum (Eq.~\ref{eq:phaseclosure}) is the multiplication of three visibilities with Gaussian noise, its noise is not Gaussian. 
Although the noise cannot be fully expressed simply with a covariance matrix, we follow~\cite{bouman2016computational} and approximate the bispectrum's distribution as Gaussian with a best-fit covariance matrix.
Refer to the supplemental material for a derivation of this approximation. 




\vspace{-.1in}
\subsection{Visibility Amplitudes }
\label{sec:visamp}




Bispectrum measurements not only entangle the phase, but also the amplitudes of the three corresponding spatial frequency components. 
Although atmospheric inhomogeneity causes substantial phase errors in each complex visibility, with careful calibration the amplitude of the visibility can be well estimated. Thus, we have found that during image reconstruction it is quite helpful to additionally constrain the visibility amplitude in addition to the bispectrum or closure phase. Empirically we have found this to be especially important when reconstructing an image with a large field of view. 
As thermal noise is isotropic in the real-imaginary plane, a perfectly gain-calibrated visibility amplitude has the same standard deviation of noise as a perfectly calibrated complex visibility \cite{TMS}. 

\vspace{-.1in}
\section{Previous Imaging Approaches}
\label{sec:setup}

VLBI image reconstruction has similarities with other spectral image reconstruction problems, such as Synthetic Aperture Radar (SAR), Magnetic Resonance Imaging (MRI), Computed Tomography (CT), and seismic interferometry~\cite{bracewell2004fourier, lustig2007sparse, 1456966,demanet}. 
However, VLBI image reconstruction faces a number of unique challenges.
For instance, SAR, MRI, and CT are generally not plagued by large corruption of the signal's phase, as is the case due to atmospheric differences in mm/sub-mm VLBI. Additionally, the frequency coverage obtained in VLBI is dictated by where there are existing telescopes, and often contains large gaps of unsampled space.  
Accounting for these differences is crucial.


Traditional interferometric imaging methods, such as CLEAN, are not inherently capable of handing the atmosphere's corruption of the visibility phases in mm/sub-mm wavelengths, even for static sources~\cite{hogbom1974aperture, taylor1999synthesis}. Thus, they frequently have trouble imaging with sparse, short-wavelength interferometric arrays like the EHT. 
However, Bayesian-style methods have made it possible to handle sparse and heterogeneous arrays in the mm/sub-mm regime, and can often even achieve accurate super-resolved images in the case of a static source~\cite{Narayan_Nityananda_1986, skilling1990quantified,  andrew}.


In order to reconstruct an image from the observed VLBI data, $\meas$, 
these methods first approximate the continuous image $I(\xpos, \ypos)$ as a square $M \times M$ array of values, $\im$, that represents the source image's flux over a specified field of view (FOV)\footnote{The FOV for a time-varying source can be estimated from the variation in visibilities obtained over simultanoulsy observed frequency bands. }. Using this representation, the imaging methods aim to solve 
\begin{align}
\hat{\im} = \argmin_{\im} \left[ \chi(\im,\meas) - \gamma \mathcal{R}(\im) \right]
\label{eq:setup}
\end{align}
\noindent{where $\chi(\im,\meas)$ indicates how inconsistent the image, $\im$, is with the observed data, $\meas$, and $\mathcal{R}(\im)$ expresses how likely we are to have observed the image $\im$. 
	These two terms often have different preferences for the ``best" image, 
	and fight against each other in selecting $\hat{\im}$.
	Their relative impact in this minimization process is specified with the hyperparameter $\gamma$. Equation~\ref{eq:setup} can be interpreted probabilistically when 
	$- \chi(\im,\meas) = \log p(\meas|\im) $, $ \mathcal{R}(\im) = \log p(\im)$, and $\gamma=1$. In this case, the minimizing cost function is a log-posterior 
	with maximum a posteriori (MAP) estimate $\hat{\im}$, and can be solved using a Bayesian inference method. 
	While many methods do not have a probabilistic interpretation, their formulation leads to a similar optimization.
}

Multiple algorithms have taken this Bayesian-style approach to VLBI imaging. These algorithms often define a similar data inconsistency measure, $\chi(\im, \meas)$ (refer to Section~\ref{sec:dataproducts}), but vary in what characterizes a ``good'' image, $\mathcal{R}(\im)$. Maximum entropy, total variation, sparsity, and patch-based priors have all been used to construct $\mathcal{R}(\im)$, and have been demonstrated in imaging both optical and radio interferometry data taken with sparse, heterogeneous arrays of static sources~\cite{bouman2016computational, andrew, Fish_2016_Imaging, baron2010novel, buscher1994direct, Suksmono, lofar}. 

VLBI data taken from an evolving source are significantly different from that of a static source~\cite{flaring, freek}. 
For example, Figure~\ref{fig:amplandphase} shows a simulated visibility amplitude and closure phase expected for a source evolving over time, and compares it to the data products expected if the same source were static.
As the data can no longer be explained by a single image with similar structure, most static imaging methods break down on this data and are unable to produce accurate results.

Recent work has attempted 
to recover a single average image by first normalizing the data and then smoothing it within a characteristic time-frame~\cite{freek}\footnote{In the case of no atmospheric error the visibilities are smoothed. In the case of atmospheric error the bispectrum and visibility amplitudes are smoothed.}. However, as this method was designed to be applied to multi-epoch data, it is often unable to recover an improved reconstruction for a single day's observation when there is significant source variability. (see Section~\ref{sec:results}). 

In~\cite{Johnson_dynamical} we simultaneously developed an alternative method for time-variability imaging that also reconstructs a video of the underlying emission region from sparse interferometric measurements. 
In that work, we develop a more flexible framework with fewer model constraints. However, this modeling choice leads to a much more difficult optimization problem that is prone to local minima, and often requires ad hoc methods to achieve satisfactory convergence.
This frequently leads to a solution inconsistent with the true structure of the source images when the data is especially sparse or noisy. Thus, the strengths and weaknesses of this alternative method are complementary to those of the approach we develop here. Similarly, in~\cite{anderson1979optimal} a Kalman Filter based approach was proposed for video generation from sparse interferometric measurements. This early work has striking similarities to parts of our own presented approach; however, the method was not demonstrated on interferometric data -- real or synthetic -- and does not address many of the limiting characteristics of EHT-quality data. 

The problem of reconstructing video of a moving subject from sparse frequency samples appears in the MRI literature and is often referred to as Dynamic MRI (dMRI)~\cite{tsao2012mri, liang1994efficient, fessler, chun2012spatial}. Recent dMRI approaches reconstruct time-evolving MRI images using a variety of sophisticated regularization techniques, such as non-linear manifold~\cite{nakarmi2017kernel} and sparsity models~\cite{chen2008prior, biswas}. Although this problem has similarities to VLBI, there are substantial differences that make it not possible to blindly apply approaches developed for dMRI to VLBI. As discussed in Section~\ref{sec:dataproducts}, due to atmospheric inhomogeneity, VLBI measurements for short wavelength observations lose absolute phase information. Not only does this mean it is not possible to simply use a linear DTFT relationship to recover the underlying image from the data, but additionally the absolute position of the image on the sky is lost. Therefore, the common strawman approach of using a sliding-window to share data across time results in a series of frames with no spatial consistency, leading to a flickering video. 

\vspace{-.1in}
\subsection{Data Consistency}
\label{sec:dataproducts}

Methods often constrain the image reconstruction using a different set of data products. However, most can be generalized through a common definition of $\chi(\im, \meas)$. This generalization will become helpful in defining models in Sections~\ref{sec:static} and~\ref{sec:dynamic_model}.


Let $\meas$ be a $\nmeas$-dimensional real vector of, possibly heterogeneous, data products. 
We define a function $f(\im)$ to return the expected value of $\meas$ if $\im$ were the true underlying image. This measurement function, $f(\im)$, is composed of a set of sub-functions, $g_k(\im)$, that each simulate an ideally observed data product: 
\begin{align}
f(\im) = \Spvek{g_1(\im); g_2(\im); ...; g_{\nmeas}(\im)} : \Re^{ M^2 \times 1} \rightarrow \Re^{ \nmeas \times 1 }.
\end{align}

Although interferometric observables  (e.g. visibilities, bispectrum) are in general complex valued, we break up these complex quantities into their real and imaginary parts in the vector $\meas$.
For example, if we wish to constrain image reconstruction with complex visibilities for $P$ observing telescope sites, then $\meas_{2k} = \Re \left[ \vis(u_k,v_k) \right]$ and  $\meas_{2k+1} = \Im \left[ \vis(u_k,v_k) \right]$ for $K=\ntele(\ntele-1) /2$. In this case, 
\begin{align}
g_{2k} ( \im ) = & \Re[ \vecFTmtx(u_k,v_k) ]^T  \im\\
g_{2k+1} ( \im ) = & \Im[ \vecFTmtx(u_k,v_k) ]^T \im ,
\end{align} 
for the complex row-vector $\vecFTmtx(u_k,v_k)$ that extracts a single spatial frequency $(u_k,v_k)$ from the vectorized image $\im$. 

%

Using these functions, we define $\chi(\im,\meas)$ as
\begin{align}
\chi(\im, \meas) &\propto   \sum_{k=1}^{\nmeas} \frac{(\meas[k] - g_k(\im))^2 }{\sigma[k]^2} \\
& = (\meas - f(\im))^T \bR^{-1} (\meas - f(\im)),
\label{eq:chi2}
\end{align}
\noindent{for $\bR = \mathrm{diag} [\sigma[1]^2, .., \sigma[K]^2]$ composed of the variance of noise on each $\meas[k]$. If $\meas$ is sampled from a Gaussian distribution with mean $f(\im)$ and covariance $\bR$ then $\frac{1}{K}\chi(\im, \meas) \approx 1$. }

Depending on the data products selected for $\meas$, the measurement function, $f(\im)$, may be a linear or non-linear function of $\im$. 
Visibilities can be extracted by performing a linear matrix operation similar to a Discrete Time Fourier Transform (DTFT): $f(\im) = \FTmtx \im$~\cite{bouman2016computational}. However, in the presence of atmospheric noise, 
$\meas$ may be populated with data products that are invariant to atmospheric inhomogeneity.
The bispectrum, closure phase, and visibility amplitude data products are invariant to this error, but come at the cost of requiring a non-linear measurement function, $f(\im)$~\cite{bouman2016computational, andrew, buscher1994direct}.

\section{Static Model \& Inference}
\label{sec:static}


Before discussing our proposed approach to dynamic imaging for time-varying sources, we first review 
interferometric imaging for a static source and discuss a simple, yet instructive, approach using a multivariate Gaussian image prior. 
The intent of this section is {\it not} to present a novel and competitive static imaging method, but instead to set up the tools necessary to easily understand dynamic imaging in Sections~\ref{sec:dynamic_model} and~\ref{sec:dynamic_inference}.

We measure 
a vector of real values $\meas$ 
that are generated by observing a static source's emission region image, $I(\xpos, \ypos)$. These measurements are extremely sparse and noisy, and thus do not fully characterize the underlying image. 
For example, a simple PCA analysis on rows of the DTFT matrix $\FTmtx$ for the EHT 2017 campaign (see the uv-coverage of Fig.~\ref{fig:staticimaging}) shows that 95\% of the variance can be described using only 1624 measurement sub-functions, $g(\im)$, for a 10000 pixel image; essentially $\FTmtx$ constrains only 16\% of the unknowns. 
To solve this problem, we impose a prior distribution on $\im$ and seek a maximum a posteriori (MAP) estimate of the underlying image given these sparse observations. We adopt the model presented in~\cite{bouman2016computational} to represent $I(\xpos, \ypos)$ as vectorized coefficients, $\im$. Using this representation, we define our observation model as:
\begin{align}
\meas & \sim \mathcal{N}_{\meas}(f(\im), \bR), \\
\im & \sim \mathcal{N}_{\im}(\bmu, \bLambda),
\end{align}
\noindent{where $\mathcal{N}_{z}(m, \Sigma)$ is the multivariate normal distribution of $z$ with mean $m$ and covariance $\Sigma$. In the chosen model, both the data likelihood, $p(\meas|\im)$, and the underlying image prior, $p(\im)$, are multivariate normal distributions. The posterior probability is written in terms of these two terms: 
}
\begin{align} 
\label{eq:bayes}
p(\im|\meas) & \propto p(\meas|\im) p(\im) \\
\notag & = \mathcal{N}_{\meas}(f(\im), \bR) \mathcal{N}_{\im}(\bmu, \bLambda).
\end{align}

Unfortunately, $p(\meas|\im)$ is not truly Gaussian when $\meas$ is composed of bispectra or closure phases, as each visibility is used to compute multiple terms. However, as discussed in Section~\ref{sec:bispec}, we assume that each term of $\meas$ is independent and can be described with a Gaussian noise model. This approximation has been shown to be a good approximation in practice (see the supplemental material)~\cite{TMS,bouman2016computational}.


\begin{figure}[tb]
                       	\begin{center}
                       		\begin{tabular}{  c | c | c   }
                       			&\large{\textsf{$\bLambda$}}   &\large{\textsf{SAMPLES }} \hspace{.55in}   \\ \hline
                     
&\vspace{-.1in}& \\
                       			\multirow{1}{*}[.6in]{ \rotatebox[origin=t]{90}{\large{\textsf{a = 2}} }}
                                &
{{ \includegraphics[width=0.2\linewidth]{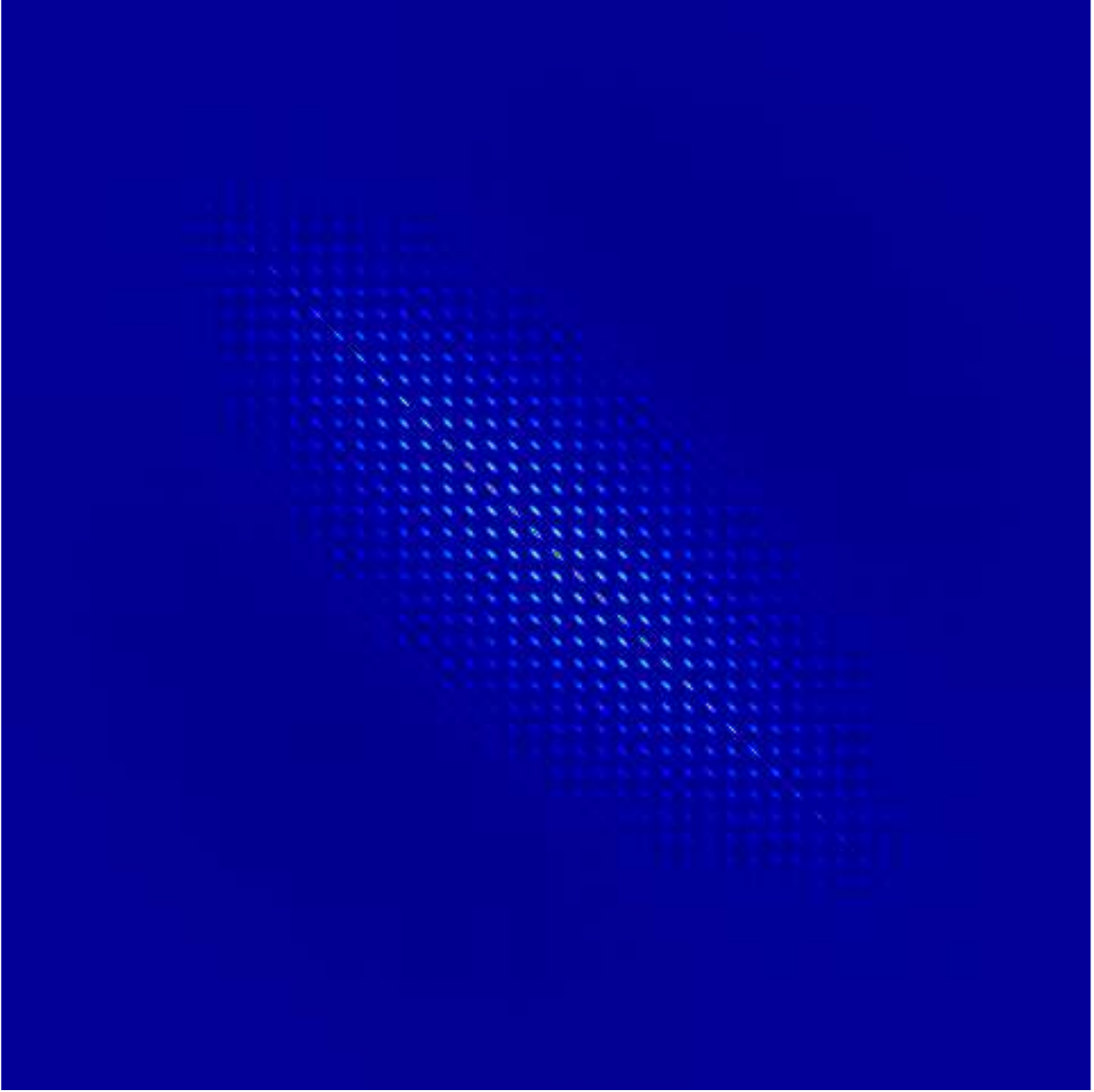}} } &
                       			{ \includegraphics[width=0.2\linewidth]{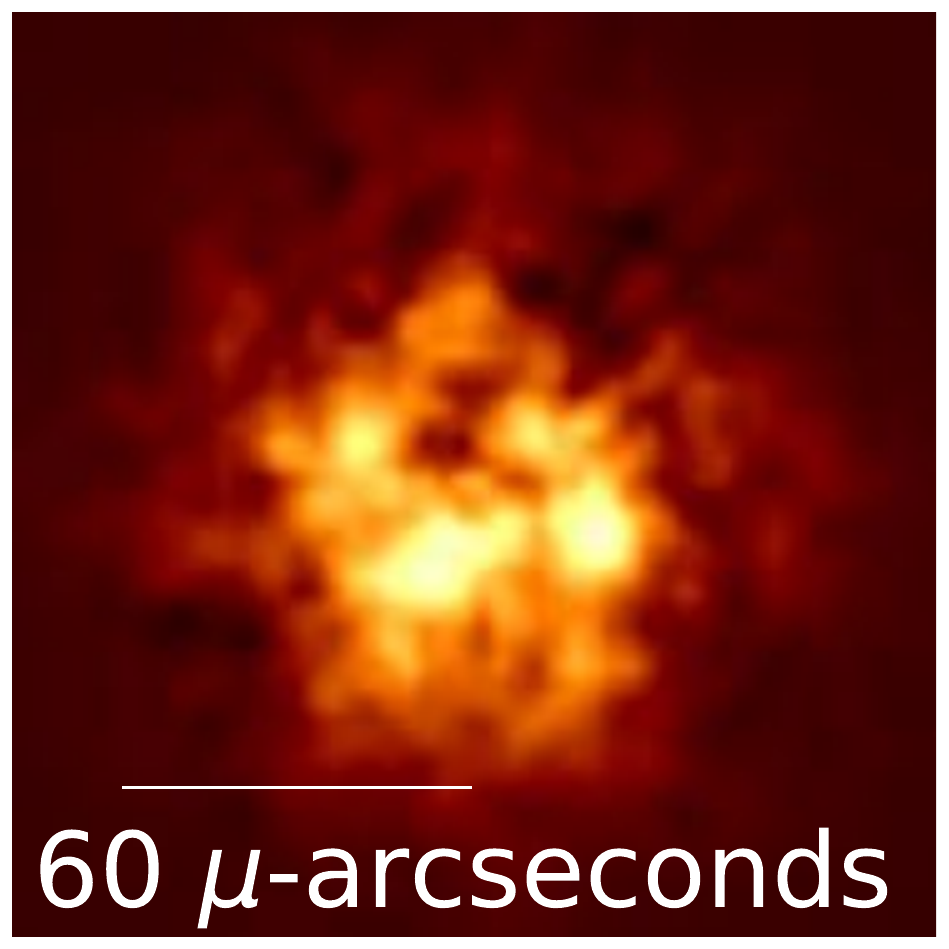}} \includegraphics[width=0.2\linewidth]{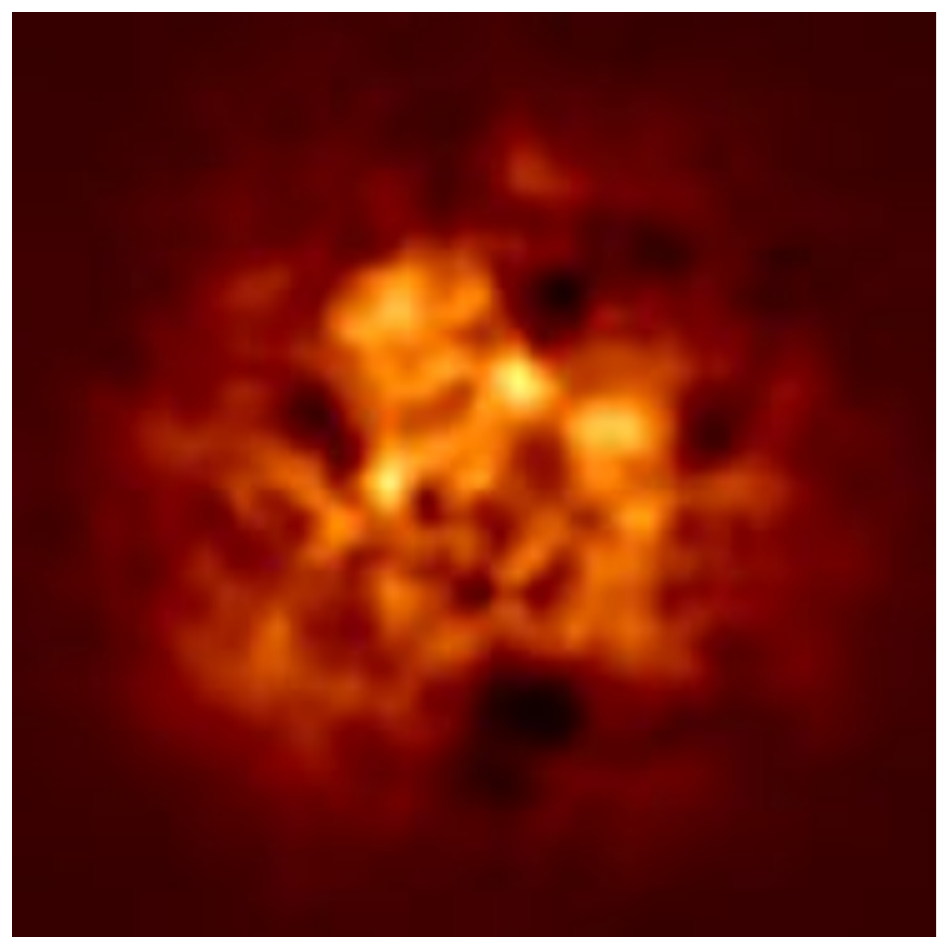} 
                       			\multirow{3}{*}[.6in]{ \includegraphics[width=0.155\linewidth]{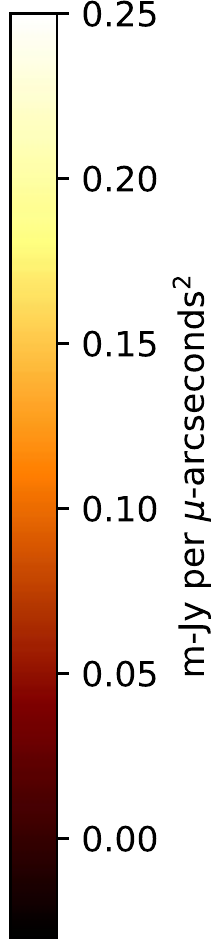} }
                                \\
                       			&\vspace{-.1in}&\\
                       			\multirow{1}{*}[.6in]{ \rotatebox[origin=t]{90}{\large{\textsf{a = 3}} }} & 
                       			{{ \includegraphics[height=0.2\linewidth]{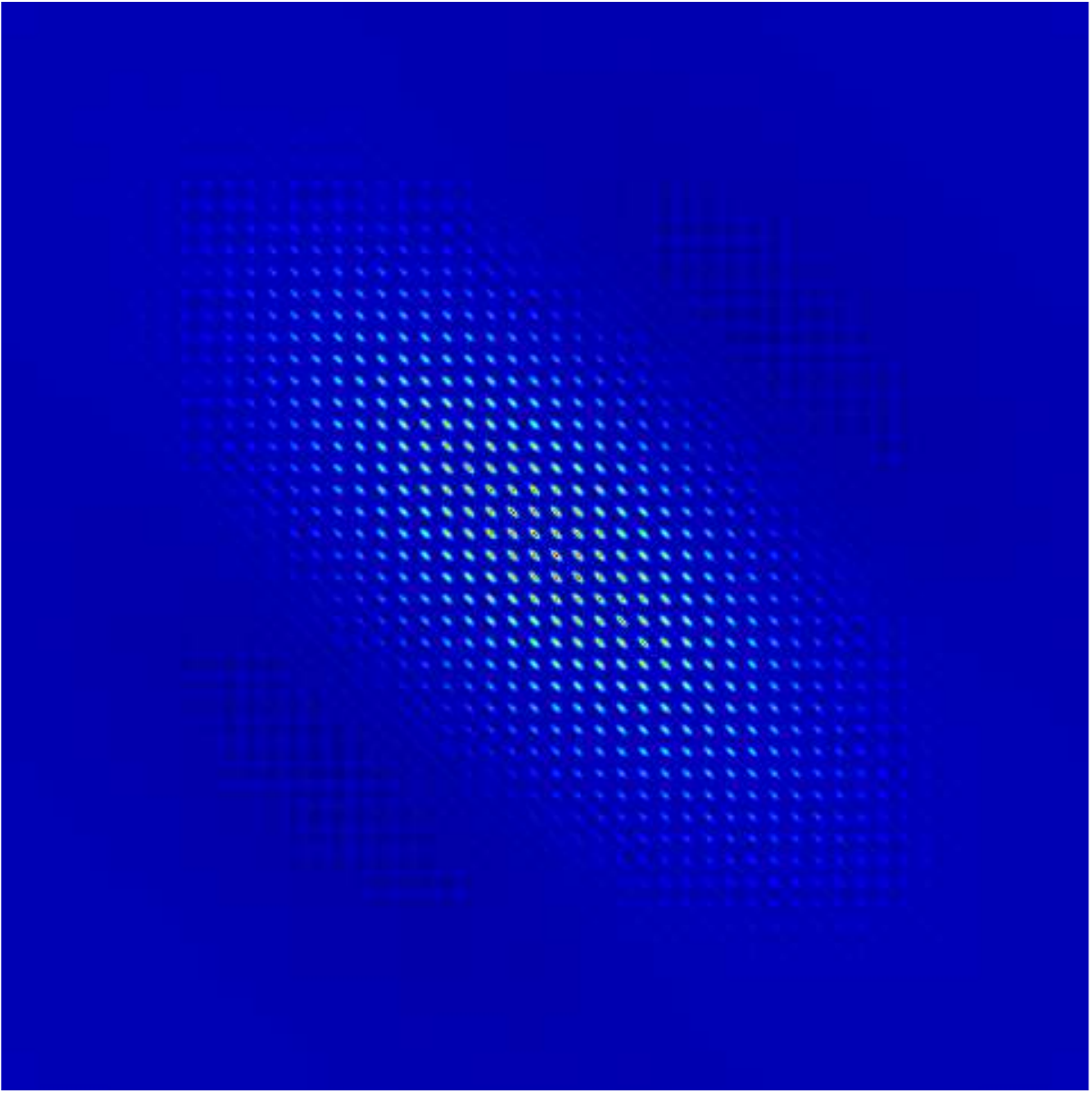}} } &
                       			{ \includegraphics[height=0.2\linewidth]{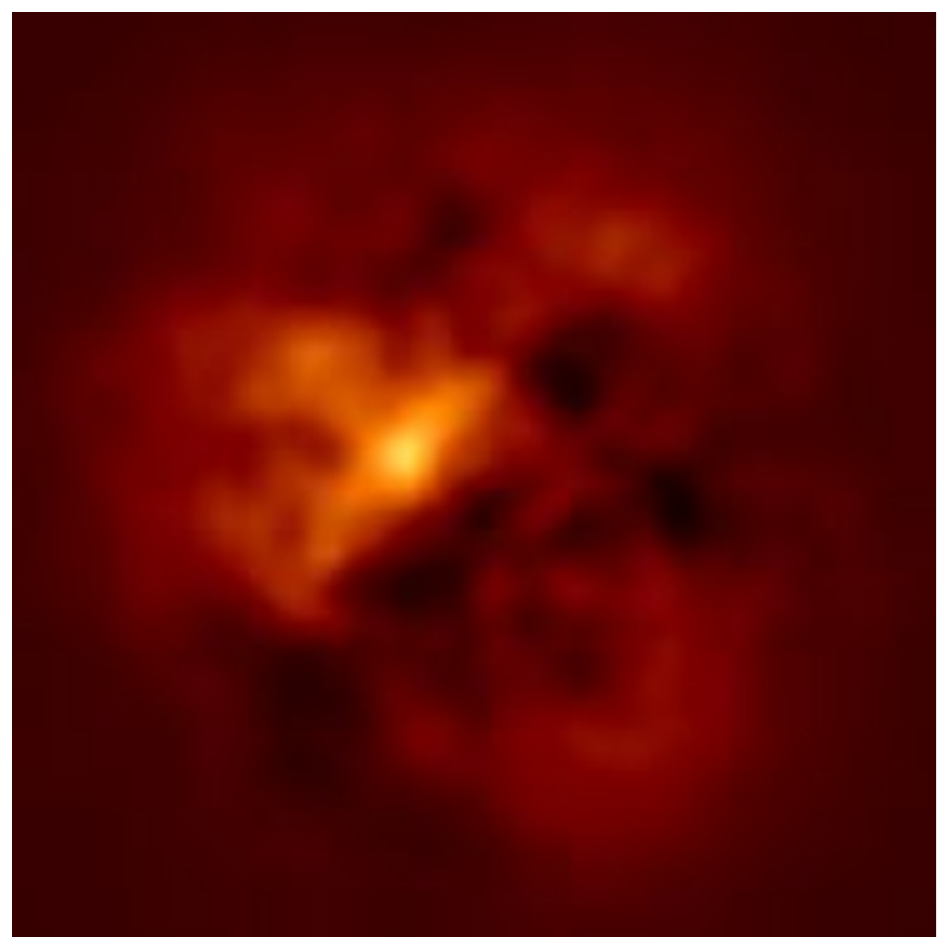}} \includegraphics[height=0.2\linewidth]{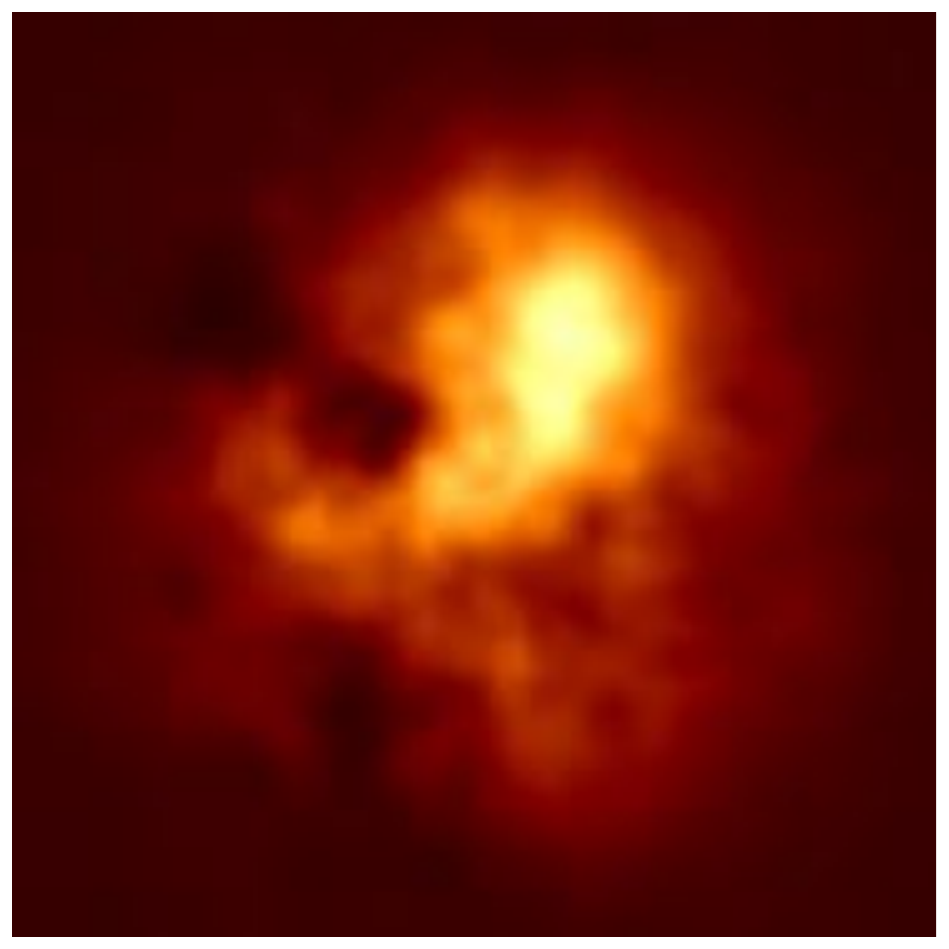}  
                       			\hspace{.65in}
                       			\\
                       			&\vspace{-.1in}& \\
                       			\multirow{1}{*}[.6in]{ \rotatebox[origin=t]{90}{\large{\textsf{a = 4}} }} & 
                       			{{ \includegraphics[height=0.2\linewidth]{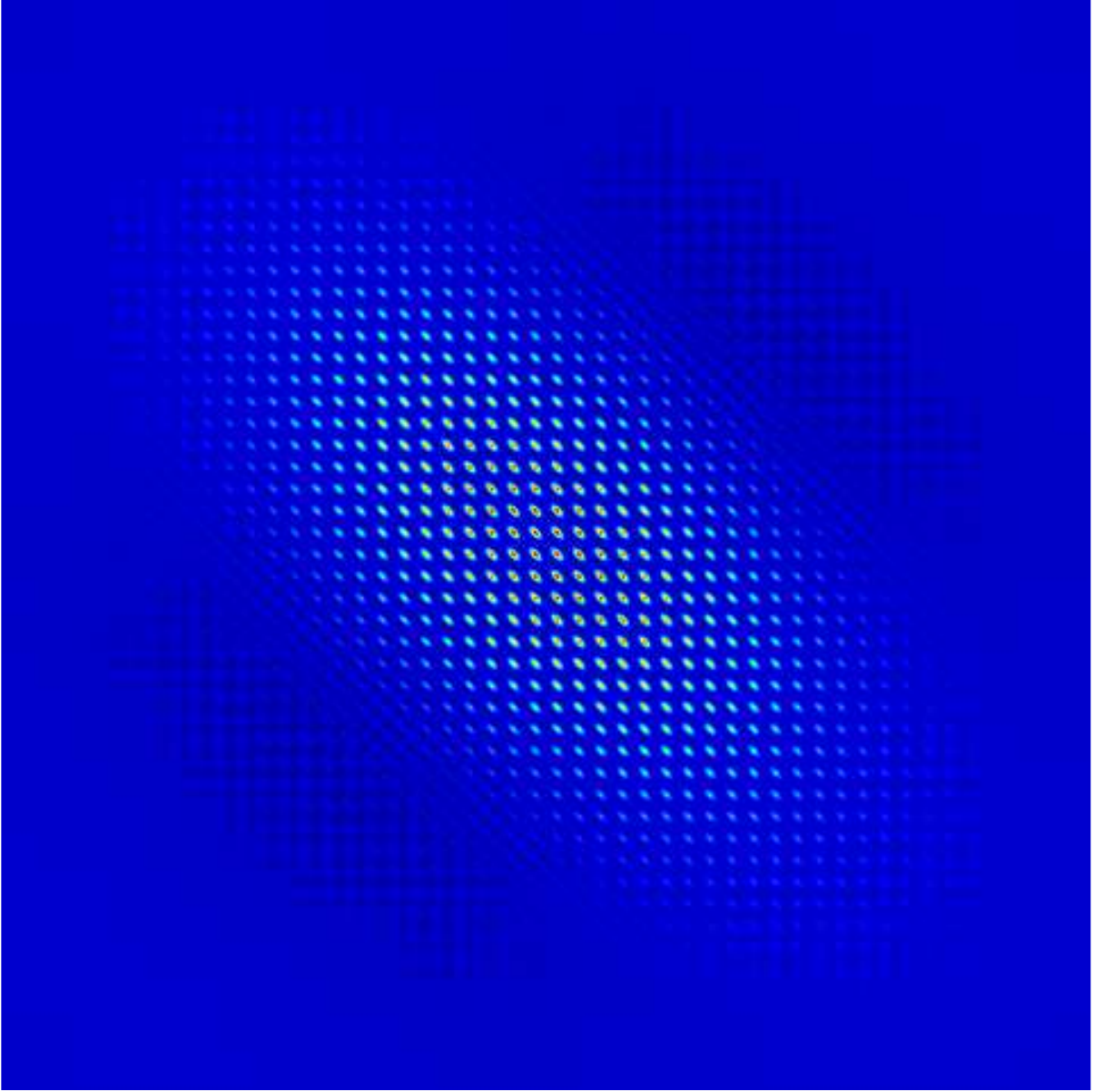}} } &
                       			{ \includegraphics[height=0.2\linewidth]{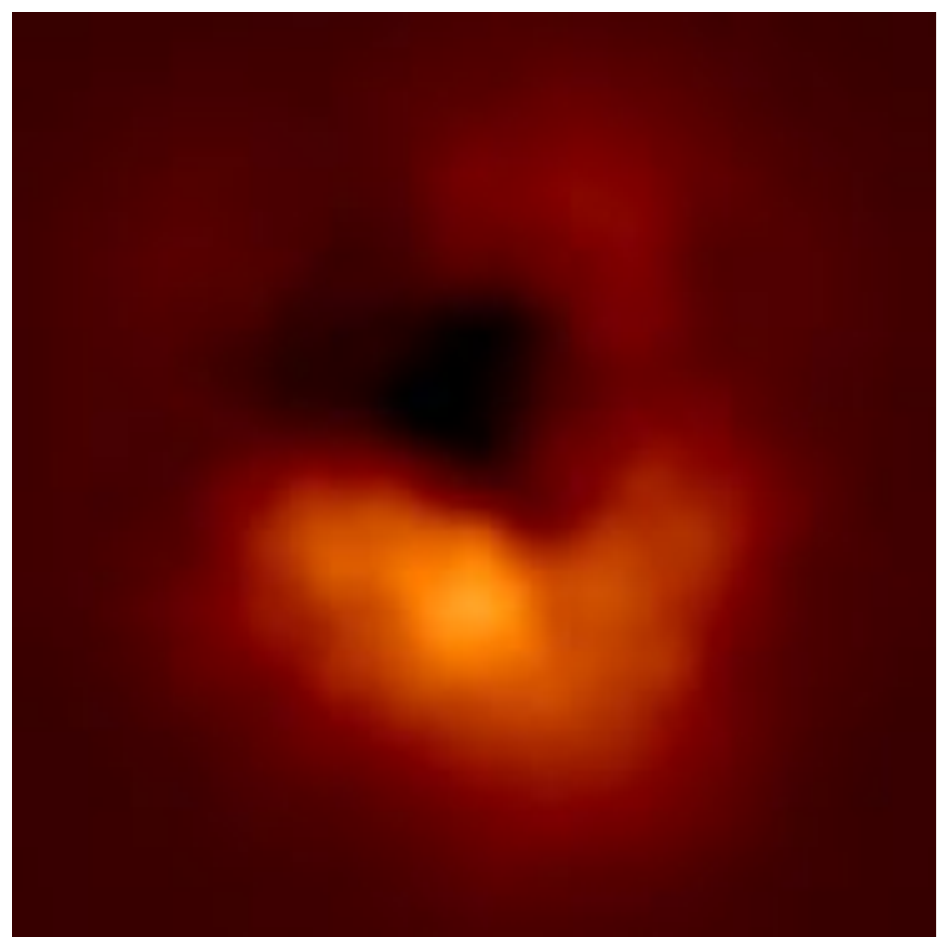}} \includegraphics[height=0.2\linewidth]{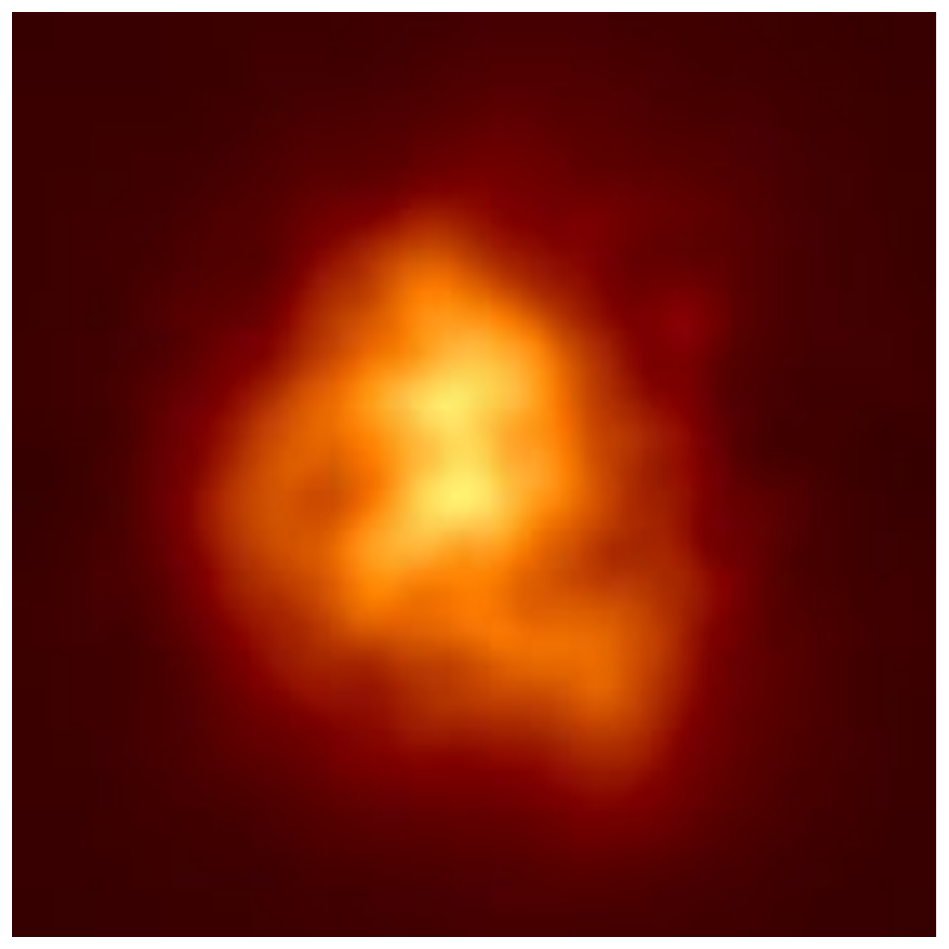}                        			
                       			\hspace{.65in}
                       			\\            	
                       		\end{tabular}
                       		\caption{\footnotesize{{\bf Gaussian Image Prior:} The covariance matrix constructed for $a=2,3,4$ along with image samples from the prior  $\mathcal{N}_x(\mu, \Lambda)$. The image samples have a field of view of 160 $\mu$-arcseconds. Notice that as $a$ increases, the sampled images appear smoother (i.e., the prior encourages smoother structure). In these examples $\mu$ is a 2D Gaussian image with standard deviation of 75 $\mu$-arcseconds. and $c=0.5$. 
                       			}}
                       			\label{fig:priorsamples}
\end{center}
\vspace{-.2in}
\end{figure}

\begin{figure*}[h!]
	\vspace{-.0in}
	\setlength{\tabcolsep}{1pt}
	\begin{center}
		\begin{tabular}{ c  c  | c  c  c c  }
			 \hspace*{-1.0cm}  
             \multirow{4}{*}[-0in]{ \rotatebox[origin=t]{0}{ {\vspace*{1in} \includegraphics[trim=0cm 0 0 -4cm,height=.43\linewidth]{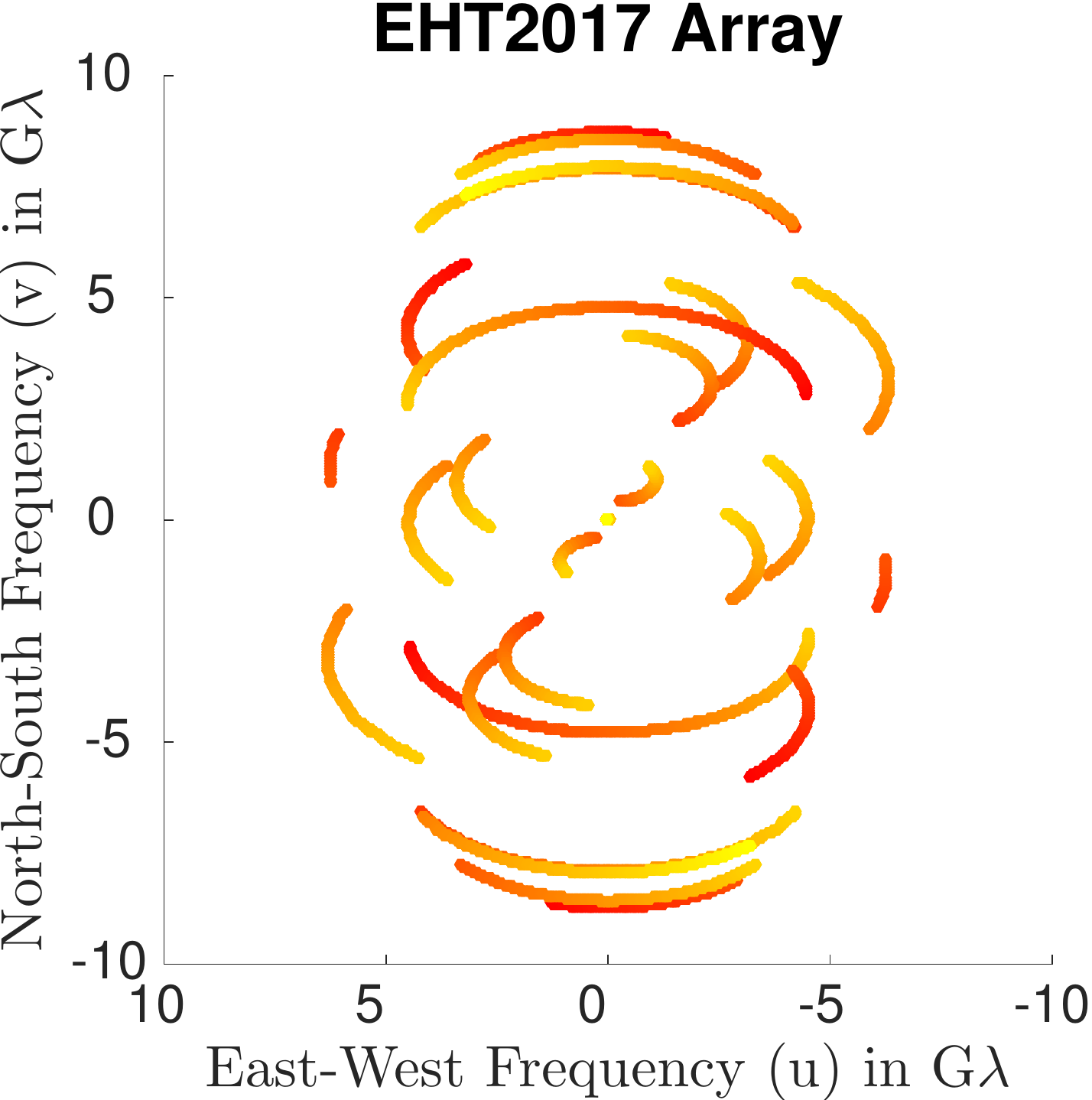}}
			 		\qquad  }}
			 &  \hspace{-0.7cm}  \large{\textsf{Truth}}   & &  \large{\textsf{a = 2}} & \large{\textsf{a = 5}}  &  \large{\textsf{a = 10}}    \\
			&  \hspace{-0.5cm} {{\includegraphics[height=.13\linewidth]{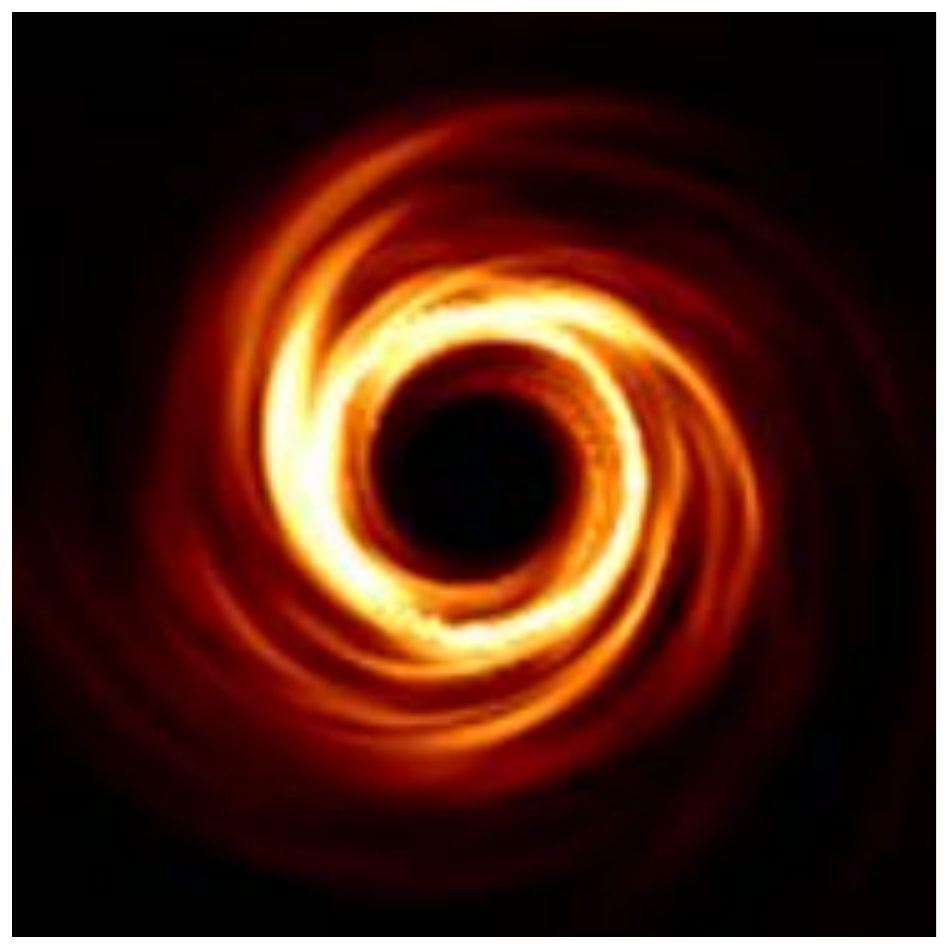}} } &
			\multirow{1}{*}[0.82in]{ \rotatebox[origin=t]{90}{ \small{\textsf{Gauss. Recon.}} }}
			&
			{\includegraphics[height=.13\linewidth]{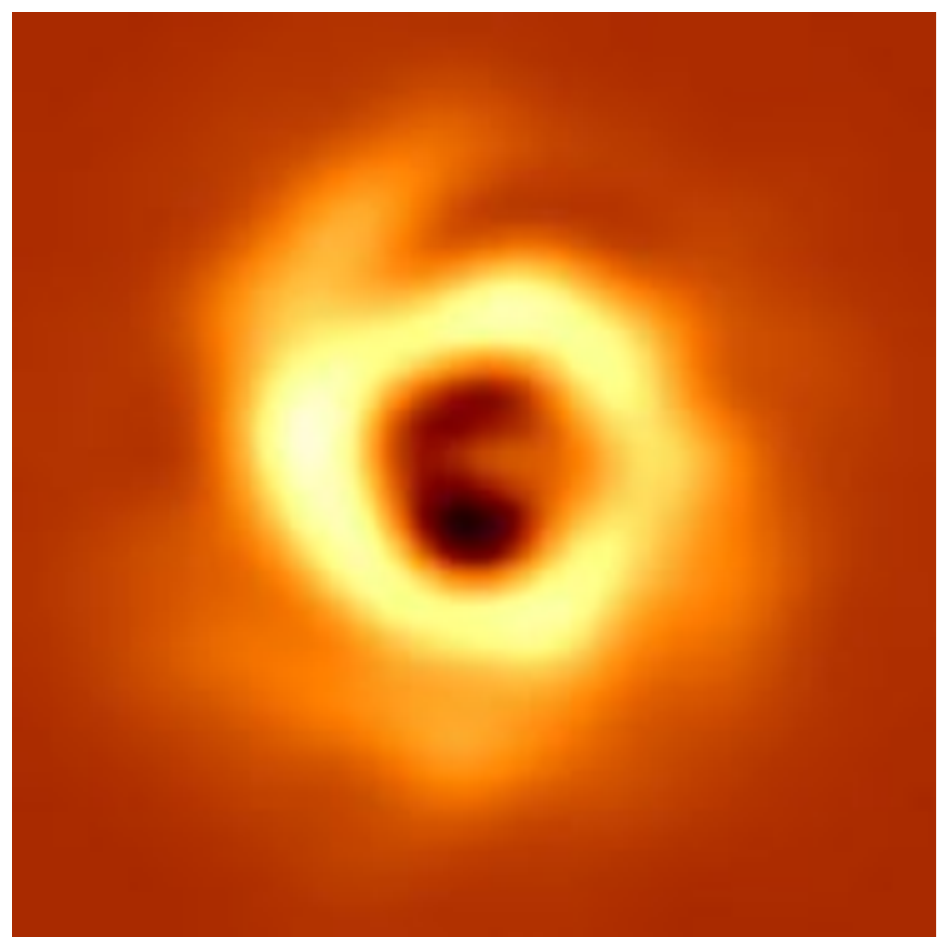}} &
			{\includegraphics[height=.13\linewidth]{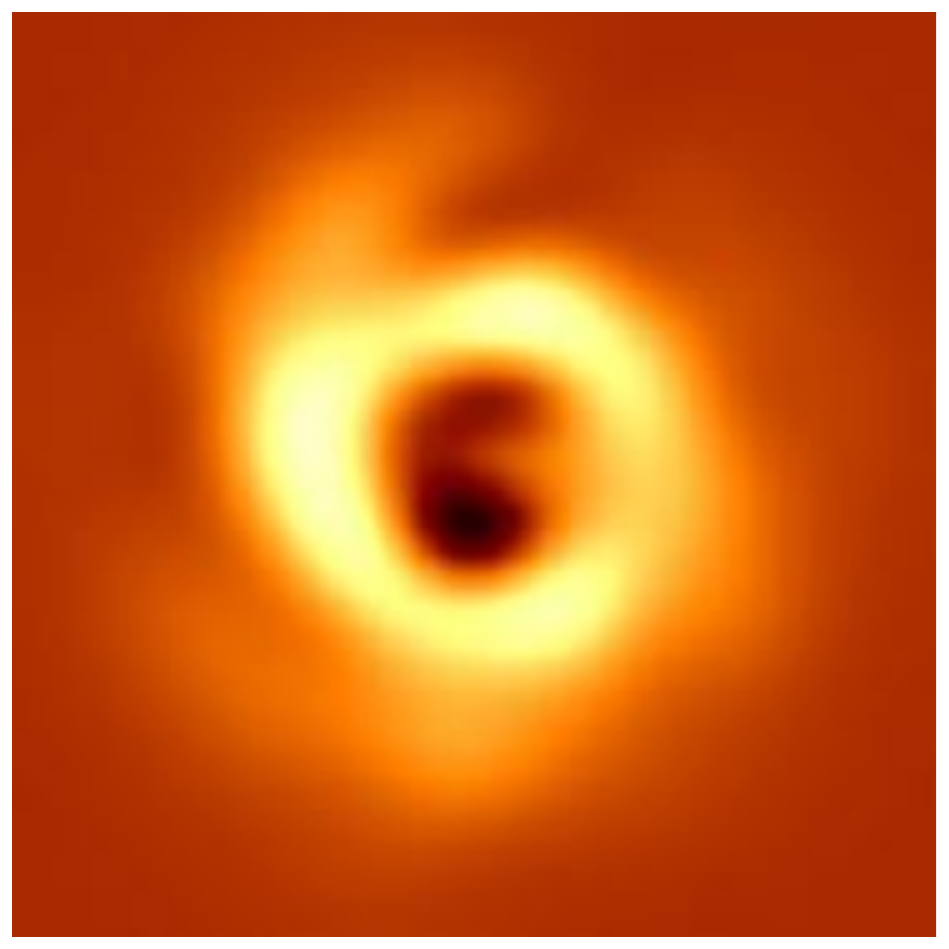}} 
			&
			{\includegraphics[height=.13\linewidth]{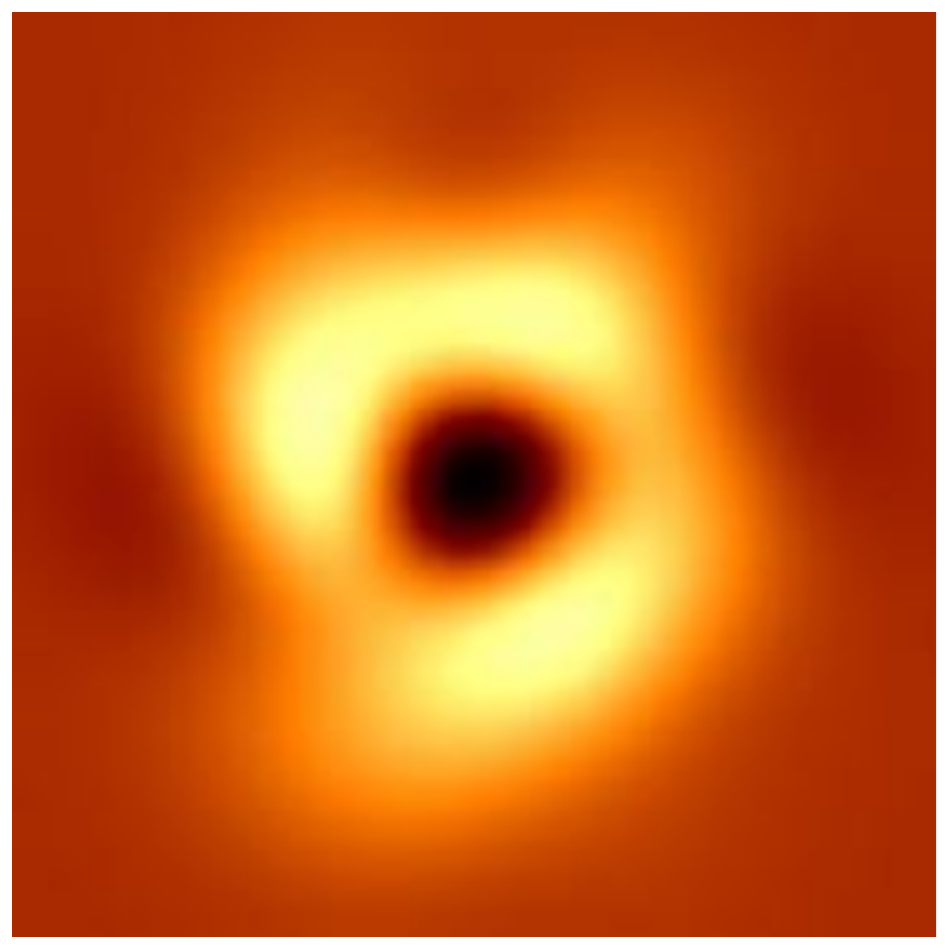}} 
			\\
			& \vspace{-.0in}  \hspace{-0.8cm} \large{\textsf{MEM \& TV}}  & &  \multicolumn{3}{c}{ \includegraphics[width=.25\linewidth]{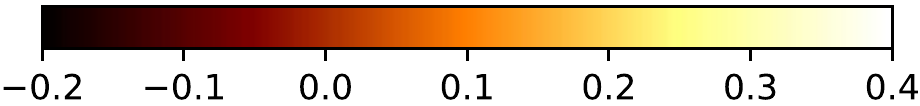} }
			\\
			&\hspace{-0.5cm} {{\includegraphics[height=.13\linewidth]{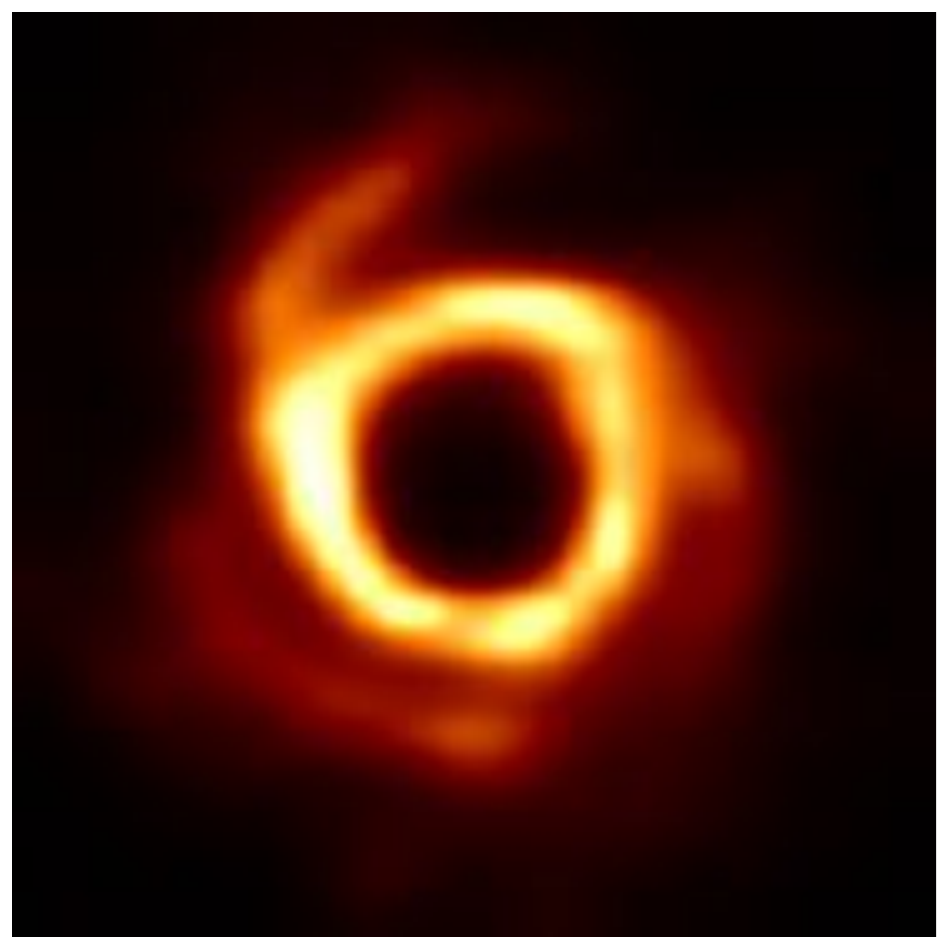}} } &
			\multirow{1}{*}[0.82in]{ \rotatebox[origin=t]{90}{ \small{\textsf{Clipped Recon.}} }}
			&
			\includegraphics[height=.13\linewidth]{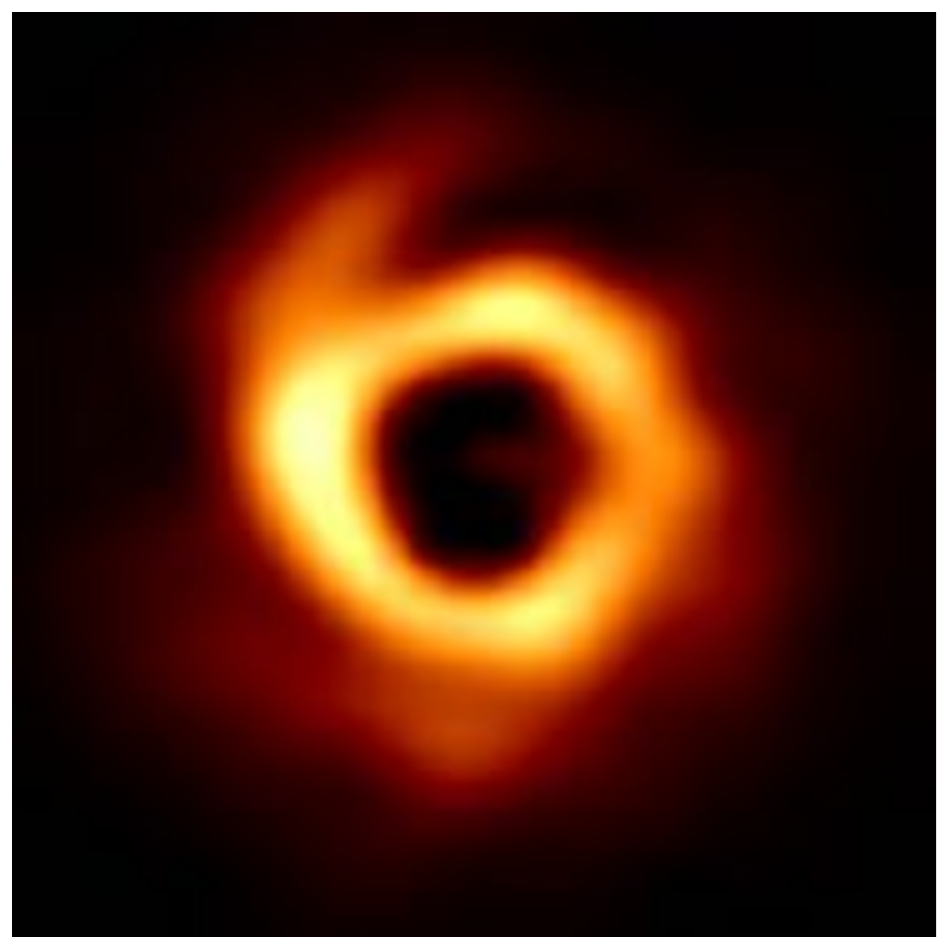} &
			\includegraphics[height=.13\linewidth]{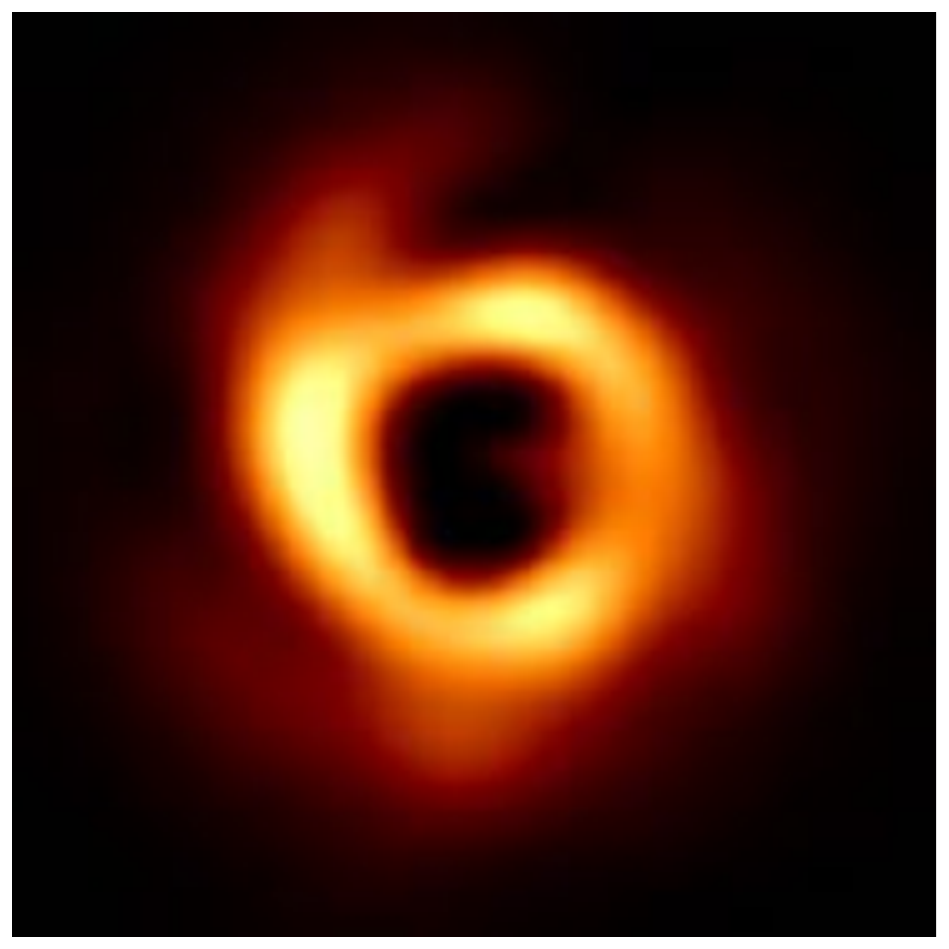} &
			\includegraphics[height=.13\linewidth]{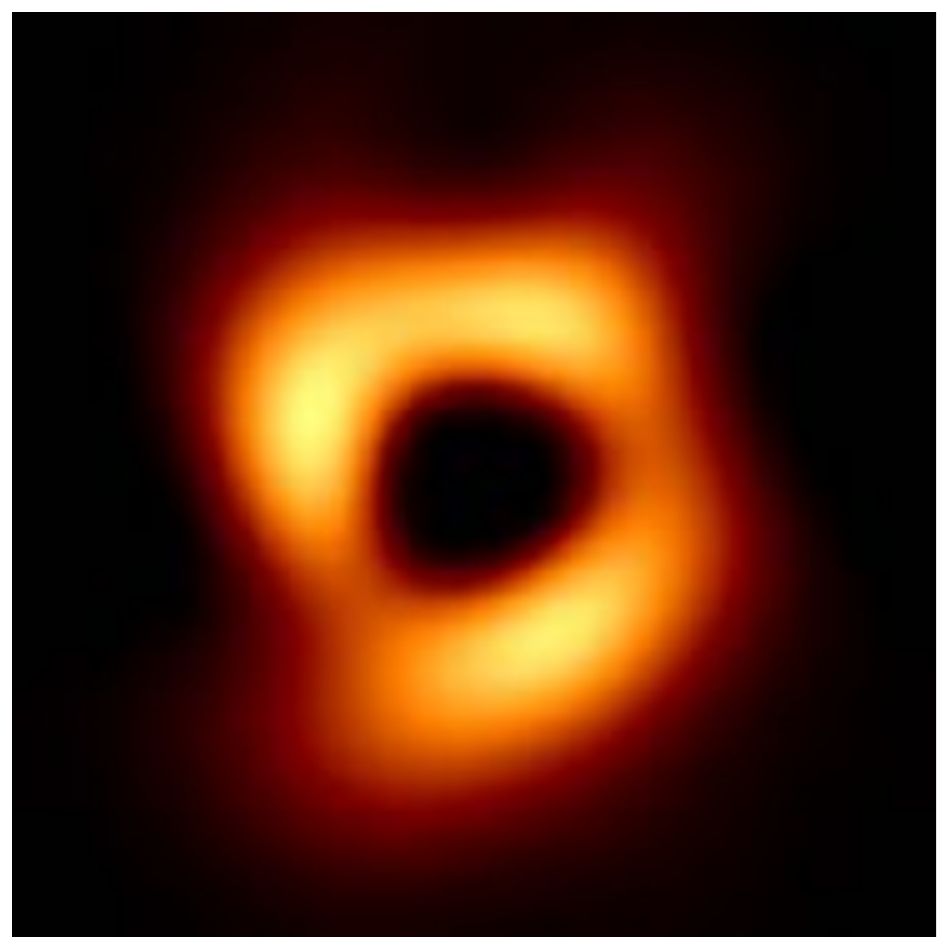} 
			\\
			& \vspace{-.0in} \hspace{-.8cm} \large{\textsf{CHIRP}}  & &  \multicolumn{3}{c}{ \includegraphics[width=.25\linewidth]{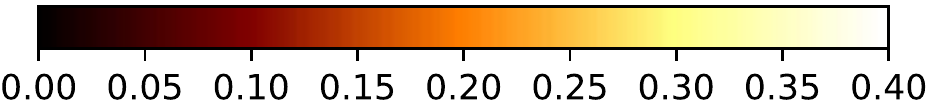} }
			\\ 
			&\hspace{-0.5cm} {{\includegraphics[width=.13\linewidth,trim=0.0cm -1.5cm 0.0cm 0.0cm]{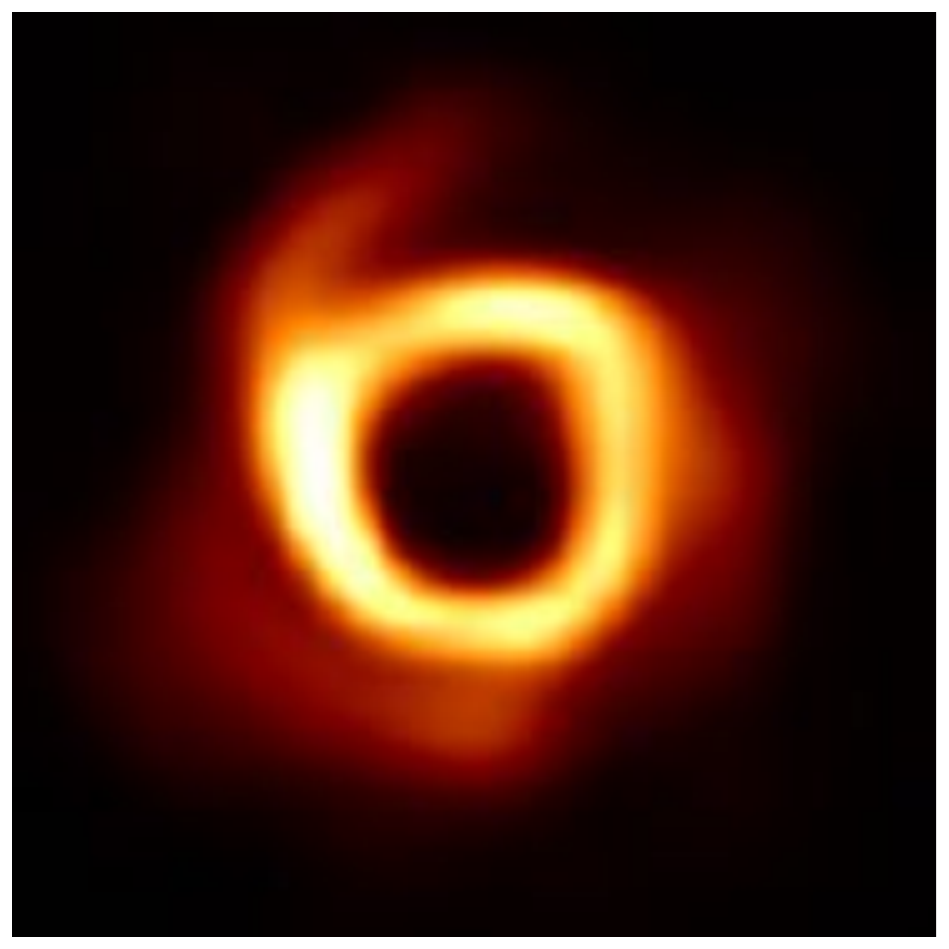}} \vspace{0.04cm} } &
			\multirow{1}{*}[1.05in]{ \rotatebox[origin=t]{90}{ \small{\textsf{Diagonal Std. Dev. }} }}
			&
			\hspace{-.1in} \includegraphics[width=.138\linewidth]{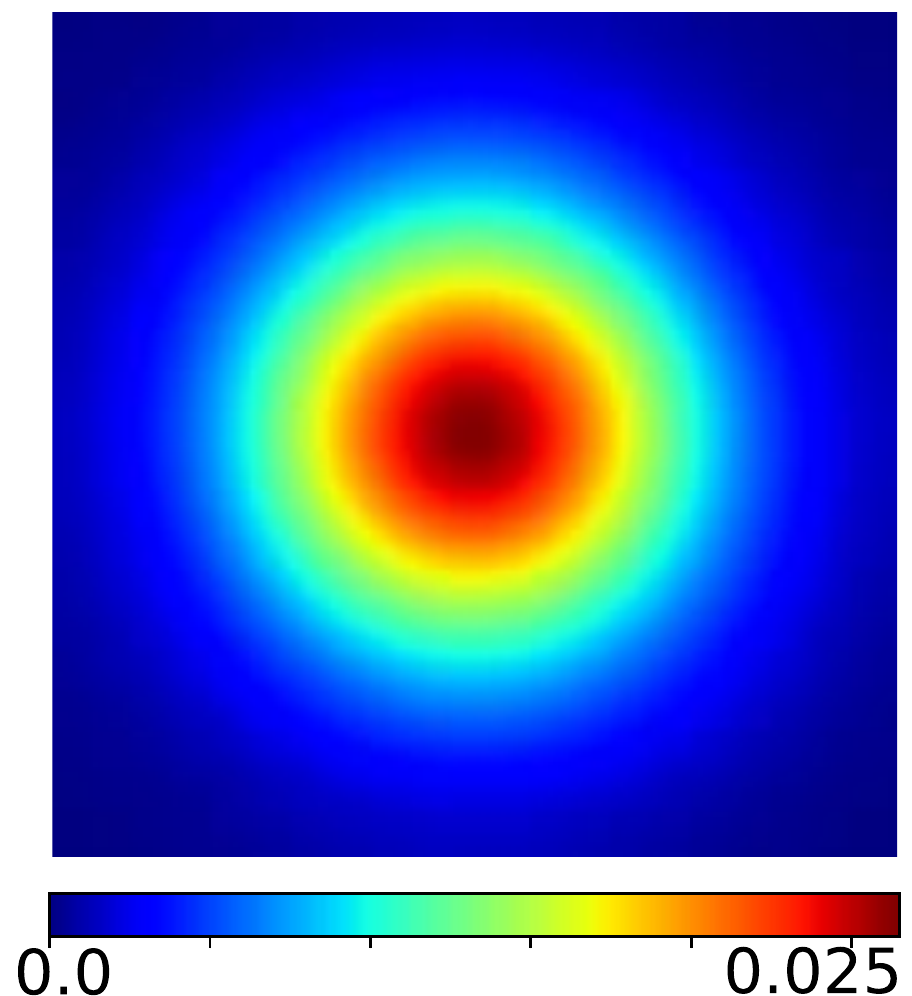} &
			\includegraphics[width=.148\linewidth]{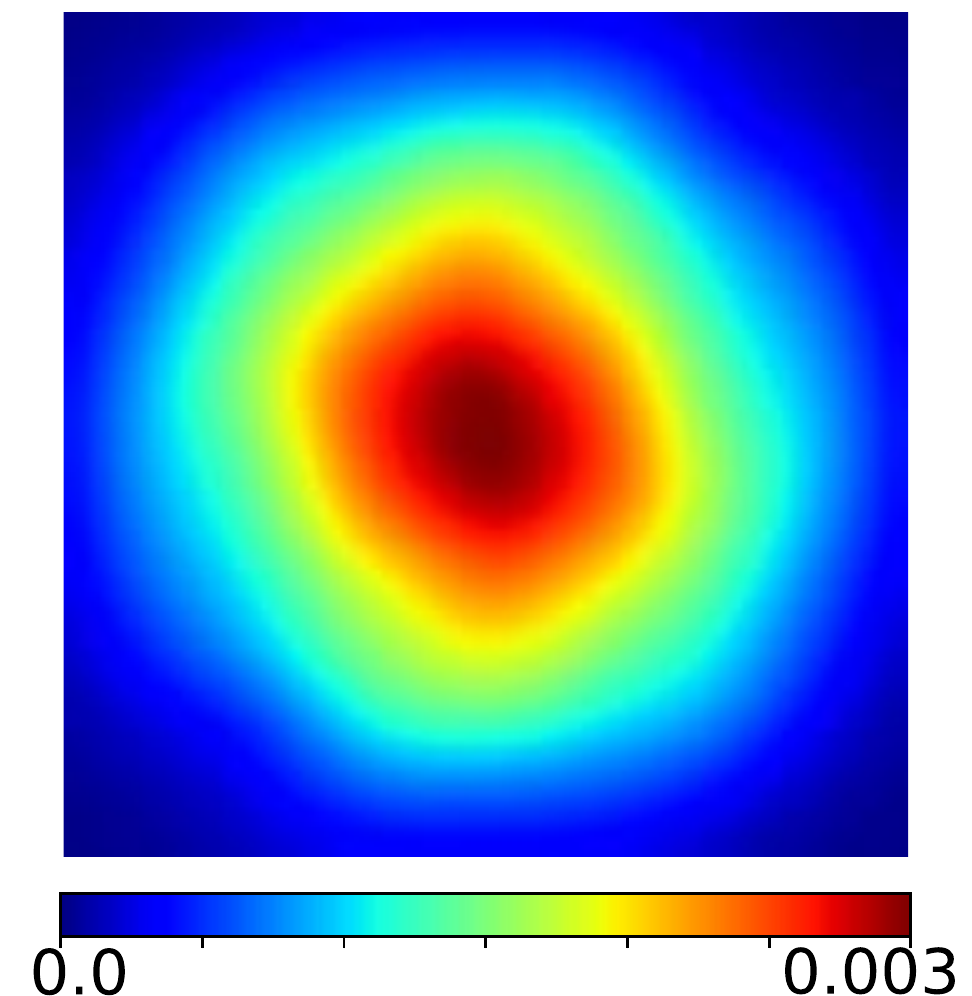} 
			&\hspace{-.1in}
			\includegraphics[height=.145\linewidth]{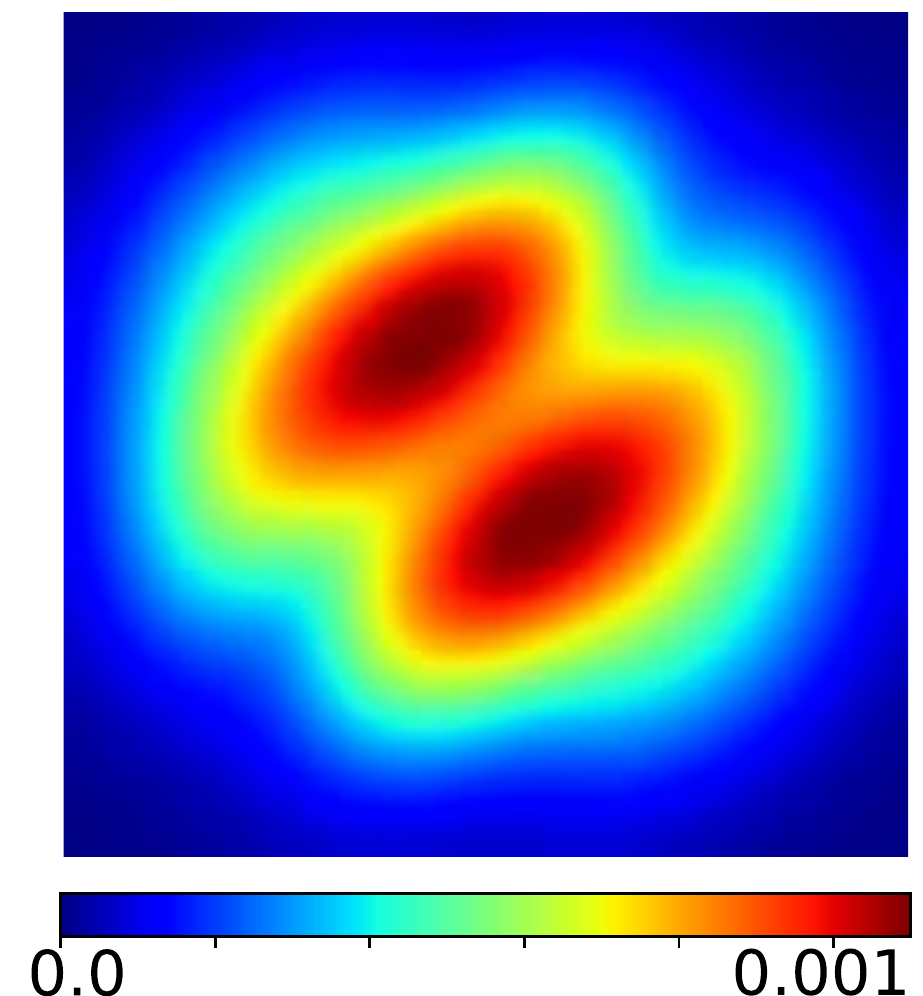} 
		\end{tabular}
		\caption{{\bf Static Imaging Comparison:} Results of static imaging using a multivariate Gaussian prior ( \textsf{a} = 2, 5, 10) compared to state-of-the-art reconstruction methods using MEM \& TV regularizers~\cite{andrew} as well as patch-based regularizers (CHIRP)~\cite{bouman2016computational}. All images are shown with a field of view of 160 $\mu$-arcseconds. Data is generated using a static image with the uv-coverage of the EHT2017 array shown on the left (see Section~\ref{sec:results}). The uv-coverage is colored by time, as indicated by the colorbar in Figure~\ref{fig:uvcov2}. Note however that in this static imaging case the time of measurements is not relevant. Although the previous algorithms (MEM \& TV and CHIRP) both produce better results, the Gaussian reconstruction is able to correctly get the broad structure of the underlying image. Since we do not impose positivity, negative values are reconstructed. However, by clipping the resulting image we can see that the result aligns well with the true static image. The Gaussian prior model also allows us to easily estimate our reconstructed image uncertainty. We visualize the diagonal entries of the posterior covariance matrix as the reshaped standard deviation image. Note that as the smoothness parameter \textsf{a} is increased, the per-pixel standard deviation becomes smaller, but the structure of the standard deviation deviates from what was specified in the prior (recall $\bLambda$ is scaled by $\bmu$, which we have specified as a 2D Gaussian in this work). For large \textsf{a} the uncertainty is shown to be primarily in the diagonal north-west to south-east direction, due to the lack of spatial frequencies sampled by the telescope array in this direction. To avoid approximations and best show the recovered posterior covariance matrices, atmospheric error has not been included in the data used to recover these images. The scaling of the colormaps is in mili-Jansky per squared $\mu$-arcsecond. } 
		\label{fig:staticimaging}
	\end{center}
	\vspace{-.2in}
\end{figure*}

\vspace{-.2in}
\subsection{Multivariate Gaussian Image Prior}
\label{sec:gauss_prior}

A prior distribution on $\im$ constrains the space of possible solutions during inference, and can be defined in a variety of ways.
For instance, maximum entropy, sparsity, and patch priors have been all used previously for VLBI imaging~\cite{andrew,kazu,bouman2016computational, rusenimaging}.
In this work we instead choose to define the underlying image, $\im$, as being a sample from the distribution $\mathcal{N}_{\im}(\bmu, \bLambda)$. This choice leads to less sharp image reconstructions compared to richer priors, 
but its simplicity allows for a cleaner understanding of our solutions. This proves especially valuable in propagating uncertainties during dynamic imaging (refer to Section~\ref{sec:dynamic_inference}). 


Studies have shown that the average power spectrum of an image often falls with the inverse of spatial frequency in the form $1/(u^2 + v^2)^{a/2}$, where $a$ is a value that specifies the smoothness of the image~\cite{torralba2003statistics}. 
As the amplitude of a spatial frequency is linearly related to the image itself, this statistical property can also be enforced by specifying the covariance in a prior distribution. Specifically, 
\begin{align}
& \hspace{.5in} \bLambda'  =    \bm{W}^{*T}  \mbox{diag} \left[ \bm{b}  \right]  \bm{W}  \\ 
b[i] = & \begin{cases} 
({u[i]}^2 + {v[i]}^2)^{-a/2}  & {u[i]}^2 + {v[i]}^2 > 0 \\
\epsilon & {u[i]}^2 + {v[i]}^2  = 0 \\
\end{cases}
\end{align}
\noindent{for DFT matrix $\bm{W}$ of size $M^2 \times M^2$ 
for an $M\times M$ pixel image and a small positive value, $\epsilon$. Each row of $\bm{W}$ and $\bm{b}$ 
	corresponds to a $(u,v)$ coordinate in the 2D grid of frequencies, \{ $S \times S$ \}, for }
\begin{align}
S &= \left\{ \frac{m-M/2}{FOV} \right\}, m\in \mathbb{Z}: m \in [0,M-1],
\end{align}
where $FOV$ is the image's field of view in radians.
To specify the variance of each pixel and help encourage positivity, we modify the amplitude of the covariance by left and right multiplying by ${c \cdot \mbox{diag}[\bmu ]} $:
\begin{align}
\bLambda = c^2 \mbox{diag}[\bmu ]^T \bLambda' \hspace{0.01in} \mbox{diag}[\bmu ] 
\end{align}
\noindent{A $c$ value of 1/3 implies that 99\% of flux values sampled from $\mathcal{N}_x(\bmu, \bLambda)$ will be positive. In this work we have chosen $\mu$ to be a circular Gaussian with a standard deviation of 40-50\% of the reconstructed FOV to encourage most of the flux to stay near the center of the image and away from the edges. Figure~\ref{fig:priorsamples} shows the covariance matrix constructed for $a=2,3,4$ along with images sampled from the prior  $\mathcal{N}_x(\bmu, \bLambda)$. Notice that as $a$ increases, the sampled images are smoother. Thus, $a$ provides the ability to tune the desired smoothness of the inferred images. 
}


\subsection{Inference}
\label{sec:static_inf}


Our goal is to find the most likely image, $\im$, that describes the data products we have observed, $\meas$. A maximum a posteriori (MAP) solution is found by maximizing the log-posterior from Equation~\ref{eq:bayes}:
\begin{align}
\hat{\im}  = \argmax_{\im} & \log p(\im|\meas) \\
\notag =  \argmin_{\im}  &\left[ (f(\im)-\meas)^T \bR^{-1} (f(\im)-\meas) \right. \\
& \left. + (\im-\bmu)^T \bLambda^{-1} (\im-\bmu) \right] .
\end{align}

Note the similarities of this equation's structure to that of previous static imaging methods in Equation~\ref{eq:setup} and~\ref{eq:chi2}. 
Although the hyperparameter $\gamma$ is no longer explicit, the scaling of $\bLambda$ acts like this hyperparameter and balances influence of the measured data with influence of the prior.


\vspace{0.1in}
\subsubsection{Linear Measurements}

As explained in Section~\ref{sec:dataproducts}, $f(\im)$ is linear when $\meas$ is composed solely of calibrated complex visibilities with no 
atmospheric error.
In this case -- when $f(\im) = \FTmtx \im$ --
a closed-form solution of $\hat{\im}$ can be found 
through traditional Weiner filtering. Specifically, we can compute the most likely estimate of each $\im$ as: 
\begin{align}
\hat{\im} &=  \bmu  + \bLambda {\bf F}^T ( \bR + \FTmtx \bLambda \FTmtx^{T} )^{-1} (  \meas -  \FTmtx \bmu ) .
\label{eq:map}
\end{align}
\noindent{
	In the limit of having no prior information about the underlying image $\im$, e.g.,  $\bLambda = \lim_{\lambda \to\infty} \lambda \mathds{1}$ for identity matrix $\mathds{1}$, this MAP solution reduces to $\hat{\im} =  \FTmtx^{+} \meas$. In other words, in the absence of prior image assumptions, the noise on each measurement, $\bR$, is no longer relevant and the reconstructed image is simply obtained by inverting $\meas = \FTmtx \im$.
	This is very similar to reconstructing the ``dirty image''~\cite{taylor1999synthesis}.
	


The same solution can also be obtained by evaluating 
the posterior distribution. 
With a linear measurement function, $f(\im)$, the proposed Gaussian formulation leads to a closed-form expression for the posterior. In particular, 
\begin{align}
p(\im|\meas) & = \mathcal{N}_{\im} (\hat{\im}, \bm{C} ). 
\end{align}
for covariance matrix
\begin{align}
\bm{C} = \bLambda - \bLambda \FTmtx^T ( \bR + \FTmtx \bLambda \FTmtx^T )^{-1} \FTmtx \bLambda .
\end{align}
\noindent Estimating uncertainty with the covariance matrix is useful in understanding what regions of the reconstructed image we trust. This becomes especially helpful when propagating information in dynamical imaging, as will be demonstrated in Figure~\ref{fig:propinfo}.

\vspace{0.1in}
\subsubsection{Non-linear Measurements}
When $f(\im)$ is a non-linear function of $\im$, as is the case when the data products in $\meas$ are invariant to atmospheric noise, a closed-form solution does not exist. 
However, to solve for the optimal $\im$ we linearize $f(\im)$ to obtain an approximate solution, $\hat{\im}$. Using a first order Taylor series expansion approximation around $\tilde{\im}$, we approximate the data likelihood as
\begin{align}
p(\meas|\im) = \mathcal{N}_{\meas}(f(\im),\bR) \approx \mathcal{N}_{\meas} \left( f( \tilde{\im} ) +  \dot{\FTmtx} (\im - \tilde{\im} )  , \bR \right) , 
\end{align}
\noindent{ for $\dot{\FTmtx} = \left. \frac{df(\im)}{d \im} \right| _{\tilde{\im} }$. Using this approximation, 
	the optimal $\hat{\im}$ is}
\begin{align}
\hat{\im} &=  \bmu  + \bLambda \dot{\FTmtx}^T ( \bR + \dot{\FTmtx} \bLambda \dot{\FTmtx}^{T} )^{-1} (  y  - f(\tilde{\im}) +  \dot{\FTmtx} (\tilde{\im}-\bmu) ) .
\label{eq:approxoptimal}
\end{align}

To further improve the solution, we solve Equation~\ref{eq:approxoptimal} iteratively by updating  $\hat{\im}$ and setting $\tilde{\im} = \hat{\im} $ until convergence.
Note that in the case that $f(\im)$ is linear, $\dot{\FTmtx} = \FTmtx$ and Equation~\ref{eq:approxoptimal} reduces to Equation~\ref{eq:map}. We compare results of this reconstruction method to other state-of-the-art methods for a static source in Figure~\ref{fig:staticimaging}. Figure~\ref{fig:staticimaging} demonstrates that, although this approach does not outperform other state-of-the-art static imaging methods, reasonable results are achieved despite a simpler image regularizer and optimization procedure. This simpler approach will become useful in developing a dynamic imaging approach.

\section{Dynamic Model}
\label{sec:dynamic_model}

Earth rotation synthesis inherently assumes that the source being imaged is static over the course of an observation~\cite{taylor1999synthesis}. 
If this assumption holds, it is possible to collect more than $\ntele (\ntele-1)/2$ measurements that inform us about the underlying source through earth rotation synthesis
However, in the case of an evolving source, as is predicted to be the case for SgrA*, this assumption is violated -- measurements taken at different times throughout the observation correspond to different underlying source images. 

At each time $t=1,...,\ntime$ we measure a vector of data products $\meas_t$, that are observed
from an evolving source image, $\im_t$. Our goal is to reconstruct the $\ntime$ instantaneous images $ {\bf X} = \{\im_1, ..., \im_\ntime \}$ using the set of sparse observations $ {\bf Y} = \{\meas_1, ..., \meas_\ntime \}$. We define a dynamic imaging 
model for this observed data as potentials ($\varphi$) of an undirected tree graph (see Figure~\ref{fig:model}):
\begin{align}
\varphi_{\meas_t | \im_t} &=  \mathcal{N}_{\meas_t} ( f_t ( \im_{t} ) , \bR_t ), \\
\varphi_{\im_t} &=  \mathcal{N}_{\im_1} ( \bmu_t, \bLambda_t ), \\
\varphi_{\im_t|\im_{t-1}} &=  \mathcal{N}_{\im_t} ( \evolve \im_{t-1}, \bQ ), 
\label{eq:evolution_potential}
\end{align}
for $\bLambda_t = \mathrm{diag}[\bmu_t]^T \bLambda'\mathrm{diag}[\bmu_t]$.

Similar to the static imaging model, each set of observed data $\meas_t$ taken at time $t$ is related to the underlying instantaneous source image, $\im_t$, through the functional relationship, $f_t(\im_t)$, and $\im_t$ is encouraged to be a sample from a multivariate Gaussian distribution. However, new to this dynamic imaging model is the addition of (\ref{eq:evolution_potential}) that describes how images evolve over time. 
If we assume that there is no evolution between neighboring images in time ($\evolve=\mathds{1}, \bQ = \bm{0}$), this dynamic model reduces to that of static imaging. 
Using the Hammersley-Clifford Theorem~\cite{hammersley1971markov}, the joint distribution of this dynamic model can be written as a product of its potential functions: 
\begin{align}
p({\bf X}, {\bf Y} ; \evolve )  \propto \prod_{t=1}^{\ntime} \varphi_{\meas_t |\im_t} \prod_{t=1}^{N} \varphi_{\im_t}   \prod_{t=2}^{\ntime} \varphi_{\im_t|\im_{t-1}} .
\label{eq:likelihood}
\end{align}

\begin{figure}
	\centering
	\includegraphics[width=0.8\linewidth]{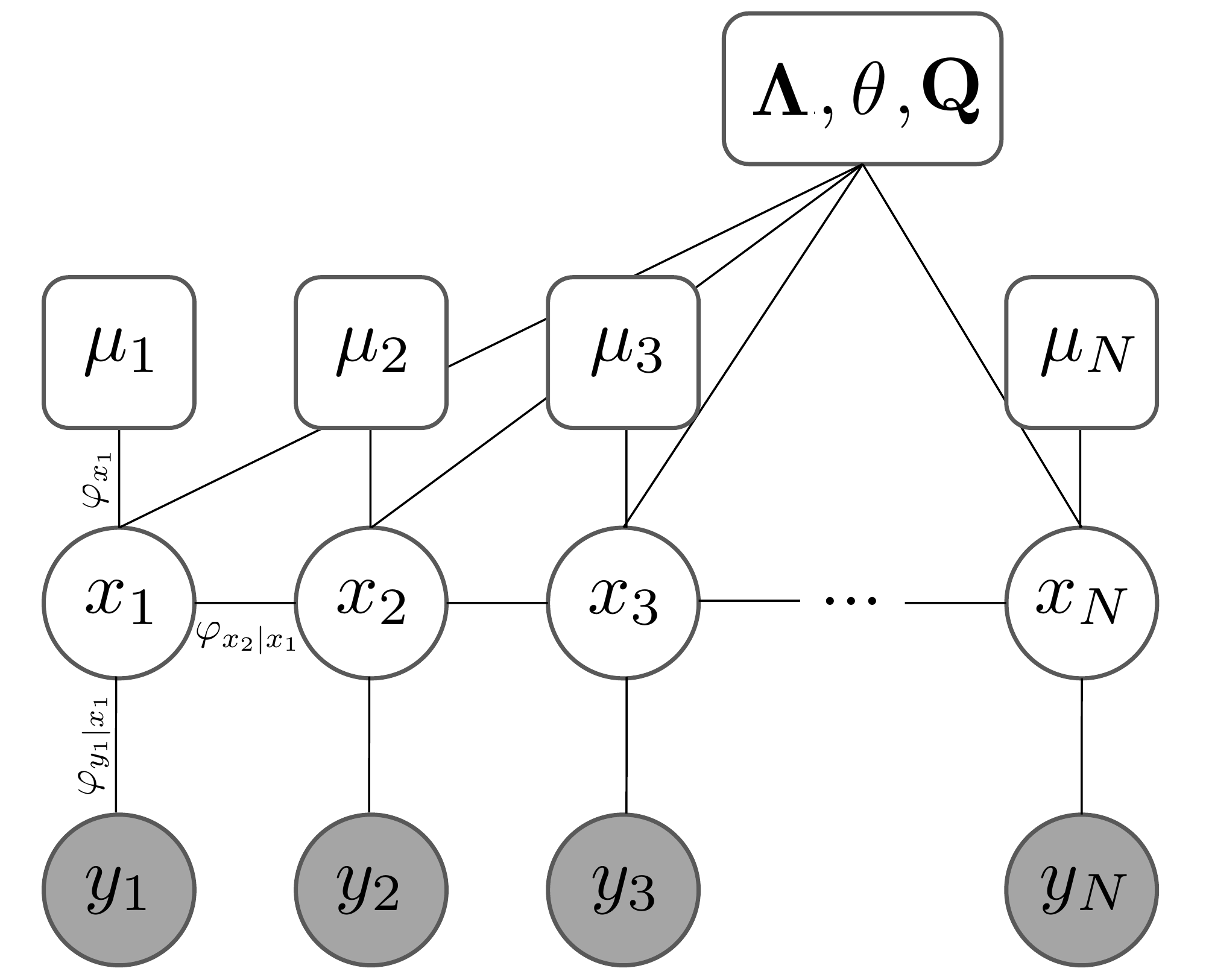}
\caption{{\bf Graphical Representation of our Dynamic Imaging Model}: At each time $t$ we observe a vector of data products $\meas_t$ corresponding to the instantaneous source image $\im_t$. 
	We assume each image $\im_t$ is related to its adjacent neighbors in time, $\im_{t-1}$ and $\im_{t+1}$, and is also related to a multivariate Gaussian distribution specified by mean $\bmu_t$ and covariance $\bLambda$. The persistent global evolution of the source images over time is specified by $\evolve$, which is further parameterized by $\theta$. Additional intensity perturbations in time are constrained by the covariance matrix $\bQ$. In this diagram, squares indicate parameters, circles are variables, and shaded circles indicate the variable is observed. }
\label{fig:model}
\vspace{-.2in}
\end{figure}

\vspace{-.2in}
\subsection{Evolution Model}
\label{sec:evolution}

Each image $\im_t$ is related to the previous image $\im_{t-1}$ through a linear relationship: $\im_t \approx \evolve \im_{t-1}$.  Matrix $\evolve$ (size $\npix^2 \times \npix^2$) 
defines the evolution of the source's emission region between time steps.
 For instance, $\evolve=\mathds{1}$ indicates that, on average, the source image does not change, whereas $\evolve=2 \mathds{1}$ indicates that the image's brightness doubles at each time step. Since the evolution matrix $\evolve$ is not time dependent, the underlying source image evolves similarly over the entire observation. However, at each time the image can deviate slightly from this persistent evolution. 
 The amount of allowed intensity deviation is expressed in the time-invariant covariance matrix $\bQ$.  

We assume that the evolution of the emission region over time is primarily described by small perturbations on top of a persistent 2D projected flow of material that preserves total flux.
We treat each source image like a 2D array of light pulses originating at locations $( \bm{\xpos}, \bm{\ypos})$. These pulses can shift around, causing motion in the image. 
As described by the Shift Theorem, the shift of a pulse by $\Delta$ will change the phase of its Fourier Transform by $2 \pi f \Delta$ for each frequency $f$. 
Thus, under small motions, we can write $\evolve$ in terms of the image's full $M^2 \times M^2$ DFT matrix $\bm{W}$ (see Section~\ref{sec:gauss_prior}), and a column-vector of $M$ pixel shifts: $\bm{s} = (\bm{s}_{ \bm{\xpos}}, \bm{s}_{\bm{\ypos}})$:
\begin{align}
\evolve = \Re \left[ \bm{W}^{*T} \hspace{0.01in}  \left(  \exp \left[ -i 2 \pi ( \bm{u} \bm{s}_{\xpos}^T + \bm{v} \bm{s}_{\ypos}^T ) \right]  \odot  \bm{W}  \right) \right] . 
\end{align}
Applying $\evolve$ to a vectorized image $\im_t$ of light pulses results in a new image, $\im_{t+1}$, where the pulses have been shifted according to $\bm{s}$ and re-interpolated on the 2D DFT grid. Note that in the case $\bm{s} = \bm{0}$ then $\evolve = \bm{W}^{*T} \bm{W} = \mathds{1}$.




The above parameterization of evolution matrix $\evolve$ in terms of $\bm{s}$ allows for independent, arbitrary shifts of each pulse of light, resulting in $2 \npix^2$ shift parameters. However, as neighboring material generally moves together, the pixel shifts should have a much lower intrinsic dimensionality. 
To address this, and simultaneously reduce the number of free parameters, we instead describe motion $\evolve$ using a low-dimensional subspace, parameterized by $\theta$. The length of $\theta$, $D$, is much smaller than the number of unconstrained shift parameters, $2 \npix^2$. 
We define a motion basis $\mathcal{M} = \Spvek{\mathcal{M}_{\xpos}, \mathcal{M}_\ypos}^T$ of size $2\npix^2 \times D+1$, and restrict the motion at every time step to be a linear function of this motion:
\begin{align}
\Spvek{\bm{\xpos}_{t+1}; \bm{\ypos}_{t+1}} = \Spvek{\bm{\xpos}_t; \bm{\ypos}_t} + \Spvek{\bm{s}_{\xpos_t}; \bm{s}_{\ypos_t}} =  \Spvek{\mathcal{M}_{\xpos_t}; \mathcal{M}_{\ypos_t}} \Spvek{1; \theta} .
\end{align}

This parameterization allows us to describe a wide variety of motion (or warp) fields. 
Generic ``smooth" warp fields can be described by using a truncated Discrete Cosine Transform (DCT) basis as $\mathcal{M}$.
However, more compressed motion bases can also be used~\cite{lowdim14, erikmiller}. 
In this work, results are shown using an affine transformation parametrized with a four-dimensional $\theta$ that captures rotation, shear, and scaling. As an affine transformation, $\theta$, acting on a pulse at location $(\xpos_t, \ypos_t)$ results in moving the pulse to location
    \begin{align}
    \Spvek{{\xpos}_{t+1}; {\ypos}_{t+1}} = \Spvek{\theta_1 \hspace{.1in}\theta_2 ; 
    	\theta_3 \hspace{.1in}\theta_4  } \Spvek{{\xpos}_{t}; {\ypos}_{t}} = \Spvek{ \xpos_t \hspace{.1in} \ypos_t \hspace{.1in}0 \hspace{.1in} 0 ; 
    	0 \hspace{.1in}0 \hspace{.1in} \xpos_t \hspace{.1in} \ypos_t } \Spvek{\theta_1; \theta_2; \theta_3; \theta_4},
    \end{align}
 in this work we define, 
   \begin{align}
  \Spvek{\mathcal{M}_{\xpos_t}; \mathcal{M}_{\ypos_t}}  =  \Spvek{\bm{0} \hspace{.1in} \bm{\xpos}_t \hspace{.1in} \bm{\ypos}_t \hspace{.1in} \bm{0} \hspace{.1in} \bm{0} ; \bm{0} \hspace{.1in}
   	\bm{0} \hspace{.1in} \bm{0} \hspace{.1in} \bm{\xpos}_t \hspace{.1in} \bm{\ypos}_t } .
   \end{align}
For example, using this motion basis with $\theta_1 = \cos \phi$, $\theta_2 = \sin \phi$, $\theta_3 = -\sin \phi$, and $\theta_4 = \cos \phi$ would specify that every time step the image is rotated by $\phi$ radians.







\section{Dynamic Imaging Inference \& Learning}
\label{sec:dynamic_inference}


We solve for the best set of $\ntime$ images $ \bm{X} $ constrained by the $\ntime$ vectors of sparse observations $\bm{Y}$. In general, we assume that $f_t(.)$, $\bR_t$, $\bmu_t$, $\bLambda_t$, $\bQ$ are known/specified model parameters. However, $\evolve$, which defines how the source evolves, is not necessarily known ahead of time. If there is reason to believe that only small perturbations exist in the source image over time, then a reasonable assumption is to set $\evolve = \mathds{1}$. However, in the case of large persistent motion this may fail to give informative results. 

We begin in Section~\ref{sec:dynamic_inference_known} by discussing how to solve for $\bm{X}$ when $\evolve$ is known. In this case, the model contains no unspecified parameters and the goal is to simply solve for the latent images. 
In Section~\ref{sec:dynamic_inference_unknown}, we forgo this assumption and no longer assume that $\evolve$ is known. In this case, we jointly solve for $\evolve$ and $\bm{X}$ by first learning $\evolve$'s parameters $\theta$ using an Expectation-Maximization (EM) algorithm before solving for the latent images, $\bm{X}$. We refer to our proposed method as StarWarps. 

\subsection{Known Evolution}
\label{sec:dynamic_inference_known}

Given all of the model parameters and observed data, our goal is to estimate the optimal set of latent images, $\bm{X}$. 
In static imaging we set up an optimization problem that allowed us to easily solve for the most likely latent image, $\im$, given the observed data, $\meas$. 
In the proposed dynamic model, a similar closed-form solution exists in the case of a linear $f(\im)$ and diagonal $\bR_t$ and $\bQ$ matrices~\cite{fessler}. However, this requires us to invert a large $\npix^4 \times \ntime^2$ non-block-diagonal matrix.   
Thus, instead of the MAP estimate, we compute the most likely instantaneous image at each time, $t$, given all of the observed data $\bm{Y}$.
In particular, we estimate the marginal distribution of each $\im_t$, $p(\im_t | \bm{Y})$, by integrating out the other latent images in time, and set $\hat{\im}_t$ equal to the mean of each distribution. 



Since we have defined our dynamic model in terms of Gaussian distributions, we can efficiently solve for $p(\im_t | \meas_1,...,\meas_\ntime)$ by marginalizing out the latent images $ \{ \im_1,...,\im_{t-1},\im_{t+1},...,\im_{\ntime} \}$ using the Elimination Algorithm~\cite{graphicalmodels}. Specifically, we derive a function proportional to the marginal distributions. This function is evaluated using a two-pass algorithm, which consists of a forward pass and a backward pass. Each pass, outlined in Algorithms~\ref{alg:forward} and~\ref{alg:backward}, propagates information using recursive updates that compute distributions proportional to 
$p(\im_t, \meas_1,...,\meas_{t-1})$ and $p( \meas_{t},...,\meas_{\ntime} | \im_t )$ 
for each $\im_t$ in the forward and backward pass, respectively. By combining these terms we obtain
\begin{align}
\label{eq:marg1}
p(\im_t|\bm{Y}) & = \mathcal{N}_{x_t}(\hat{\im}_t, \bm{C}_t ) \\
\notag & \propto \mathcal{N}_{x_t}({\bf z}^{\alpha}_{t|t-1}, {\bf P}^{\alpha}_{t|t-1})  \mathcal{N}_{x_{t}}( {\bf z}^{\beta}_{t|t} , {\bf P}^{\beta}_{t|t} ),
\end{align}
which, as shown in the supplemental material, has mean $\hat{\im}_t$ and covariance $\bf{C}_t$:
{\footnotesize
	\begin{align}
	\notag    \hat{\im}_t & = {\bf P}^{\beta}_{t|t} ( {\bf P}^{\alpha}_{t|t-1} + {\bf P}^{\beta}_{t|t})^{-1} {\bf z}^{\alpha}_{t|t-1}  +   {\bf P}^{\alpha}_{t|t-1}( {\bf P}^{\alpha}_{t|t-1} + {\bf P}^{\beta}_{t|t})^{-1} {\bf z}^{\beta}_{t|t} \\
	\bf{C}_t & =  {\bf P}^{\alpha}_{t|t-1}( {\bf P}^{\alpha}_{t|t-1} + {\bf P}^{\beta}_{t|t})^{-1} {\bf P}^{\beta}_{t|t} , 
	\label{eq:marg2}
	\end{align}
}where ${\bf z}_{t|\tau}^{\alpha}$, ${\bf P}_{t|\tau}^{\alpha}$ are the estimates of the mean and covariance of $\im_t$ using observations at time steps $1$ through $\tau$. Similarly, ${\bf z}_{t|\tau}^{\beta}$, ${\bf P}_{t|\tau}^{\beta}$ are the estimates of the mean and covariance of $\im_t$ using observations $\tau$ through $\ntimes$.

For generality, we have listed the forward and backward algorithms in terms of non-linear measurement functions, $f_t(\im_t)$ with derivative $\dot{\FTmtx}$. In this case, similar to our static model inference in Section~\ref{sec:static_inf}, we linearize the solution around $\tilde{\im}_t$ to get an approximate estimate. To improve the solution of the forward and backward terms, each step in the forward pass can be iteratively re-solved, updating $\tilde{\im}_t$ at each iteration. The values of $\tilde{\im}_t$ are then fixed for the backwards pass. Recall 
that when $f_t(\im)$ is linear in $\im$ then $f_t(\im) = \FTmtx_t \im  = \dot{\FTmtx}_t \im$, and the $\hat{\im}$ will converge to the optimal solution in a single update.

The above inference algorithm is similar to 
the forward-backward algorithm used for Gaussian Hidden Markov Models~\cite{graphicalmodels}. 
In fact, removing the $\varphi_{\im_t}$ term for $t>1$ in Equation~\ref{eq:likelihood} yields the familiar form of a Gaussian Hidden Markov Model. In this case, inference reduces to the traditional Kalman filtering and smoothing (extended Kalman filtering in the case of non-linear $f_t(\im)$)~\cite{anderson1979optimal}. Although this simpler formulation can sometimes produce acceptable results, in our typical scenario of especially sparse or noisy data keeping the additional potential terms helps to further constrain the problem, and results in better reconstructions. 

\RestyleAlgo{boxruled}
\begin{algorithm}[t]
	\caption{Forward Updates {\footnotesize $t = 1 \rightarrow 2 \rightarrow ... \rightarrow \ntime$ } \label{alg:forward} }
	
	{\bf Predict:}
	
	{ \footnotesize
		\begin{align}
		\notag {\bf z}^{\alpha}_{t|t-1} &= \evolve {\bf z}^{\alpha}_{t-1|t-1} \\
		\notag {\bf P}^{\alpha}_{t|t-1} &= \bQ  + \evolve {\bf P}^{\alpha}_{t-1|t-1} \evolve^T
		\end{align}
		\begin{align}
		\notag {\bf z}^{\alpha*}_{t|t-1} &= \bLambda_t ( \bLambda_t + {\bf P}^{\alpha}_{t|t-1} )^{-1} {\bf z}^{\alpha}_{t|t-1} + {\bf P}^{\alpha}_{t|t-1} ( \bLambda_t + {\bf P}^{\alpha}_{t|t-1} )^{-1} \bmu_t \\
		\notag {\bf P}^{\alpha*}_{t|t-1} & = \bLambda_t ( \bLambda_t + {\bf P}^{\alpha}_{t|t-1} )^{-1} {\bf P}^{\alpha}_{t|t-1} 
		\end{align}
	}
	
	{\bf Update:}
	
	{\footnotesize
		\begin{align}
		\notag \meas_\Delta &= (  \meas_t + \dot{\FTmtx} \tilde{\im}_t - f(\tilde{\im}_t) -  \dot{\FTmtx} {\bf z}^{\alpha*}_{t|t-1} ) \\
		\notag {\bf z}^{\alpha}_{t|t} &  =  {\bf z}^{\alpha*}_{t|t-1}  + {\bf P}^{\alpha*}_{t|t-1} \dot{\FTmtx}_t^T ( \bR_t +  \dot{\FTmtx}_t {\bf P}^{\alpha*}_{t|t-1} \dot{\FTmtx}_t^T )^{-1} \meas_\Delta  \\
		\notag {\bf P}^{\alpha}_{t|t} & = {\bf P}^{\alpha*}_{t|t-1} - {\bf P}^{\alpha*}_{t|t-1}\dot{\FTmtx}_t^T ( \bR_t + \dot{\FTmtx}_t {\bf P}^{\alpha*}_{t|t-1} \dot{\FTmtx}_t^T )^{-1} \dot{\FTmtx}_t {\bf P}^{\alpha*}_{t|t-1} 
		\end{align}
	}

	{\bf Initialization:}
	
	{\footnotesize
		\begin{align}
		\notag {\bf z}^{\alpha*}_{1|0} = \bmu_1 \mbox{   ,   } {\bf P}^{\alpha}_{1,0}   = \bLambda_1
		\end{align}
	}

\end{algorithm}

\RestyleAlgo{boxruled}
\begin{algorithm}[b]
	\caption{Backward Updates: {\footnotesize $t = \ntime \rightarrow \ntime-1 \rightarrow ... \rightarrow 1 \hspace{-.1in}$ }  \label{alg:backward}}
	
	{\bf Predict:}
	{\footnotesize
		\begin{align}
		\notag {\bf z}^{\beta*}_{t|t+1}  & \hspace{-.02in} = \hspace{-.02in} \bmu_t  + \bLambda_t  \evolve^T ( \bQ + {\bf P}^{\beta}_{t+1|t+1} +  \evolve \bLambda_t  \evolve^T )^{-1} (  {\bf z}^{\beta}_{t+1|t+1} \hspace{-.13in} -   \evolve \bmu_t ) \\
		\notag  {\bf P}^{\beta*}_{t|t+1} &  \hspace{-.02in}= \hspace{-.02in} \bLambda_t - \bLambda_t \evolve^T (  \bQ + {\bf P}^{\beta}_{t+1|t+1} + \evolve \bLambda_t  \evolve^T )^{-1}  \evolve \bLambda_t 
		\end{align}
	}

	{\bf Update:}
	
	{\footnotesize
		\begin{align}
		\notag \meas_\Delta &= (  \meas_t + \dot{\FTmtx} \tilde{\im_t} - f(\tilde{\im}_t) -  \dot{\FTmtx}_t {\bf z}^{\beta*}_{t|t+1} ) \\
		\notag  {\bf z}^{\beta}_{t|t} &  =  {\bf z}^{\beta*}_{t|t+1}  + {\bf P}^{\beta*}_{t|t+1} \dot{\FTmtx}_t^T ( \bR_t +  \dot{\FTmtx}_t {\bf P}^{\beta*}_{t|t+1} \dot{\FTmtx}_t^T )^{-1} \meas_\Delta \\
		\notag  {\bf P}^{\beta}_{t|t} & = {\bf P}^{\beta*}_{t|t+1} - {\bf P}^{\beta*}_{t|t+1}\dot{\FTmtx}_t^T ( \bR_t + \dot{\FTmtx}_t {\bf P}^{\beta*}_{t|t+1} \dot{\FTmtx}_t^T )^{-1} \dot{\FTmtx}_t {\bf P}^{\beta*}_{t|t+1} 
		\end{align}
	}

	{\bf Initialization:}
	
	{\footnotesize
		\begin{align}
		\notag  {\bf z}^{\beta*}_{\ntime|\ntime+1} = \bmu_{\ntime} \mbox{   ,   } {\bf P}^{\beta*}_{\ntime|\ntime+1}   = \bLambda_{\ntime}
		\end{align}
	}

\end{algorithm}

\vspace{-.1in}

\subsection{Unknown Evolution}
\label{sec:dynamic_inference_unknown}




If the evolution matrix $\evolve$ is unknown, it is not possible to simply solve for $\bm{X}$ in the way outlined in Section~\ref{sec:dynamic_inference_known}. 
Instead we choose to use an Expectation-Maximization (EM) algorithm to recover $\evolve$ (parameterized by $\theta$), and then subsequently use the procedure presented in Section~\ref{sec:dynamic_inference_known} to recover $\bm{X}$. 

The EM algorithm defines an iterative process that
solves for 
the evolution parameters $\theta$ that maximize the complete likelihood in Equation~\ref{eq:likelihood} when all of the underlying images, $\bm{X}$, are unknown (latent). 
Each iteration of EM improves the log-likelihood of the data under the defined objective function and is especially useful when the likelihood is from an exponential family, as is the case in our proposed model. In particular the EM algorithm consists of the following two iterative steps:

\begin{itemize}
	\item Expectation step (E step): Calculate the expected value of the log likelihood function (see Equation~\ref{eq:likelihood}), with respect to the conditional distribution of $\bm{X}$ given $\bm{Y}$ under the current estimate of the $\theta$ parameters, $\theta^{(i)}$: 
	\begin{align}
	Q(\theta|\theta^{(i)}) = E_{ {\bf X}| {\bf Y}, \theta^{(i)}} [ \log p ({\bf X}, {\bf Y} | \theta) ]
	\label{eq:qfunction}
	\end{align}
	\item Maximization step (M step): Find the parameter that maximizes: 
	\begin{align}
	\theta^{(i+1)} = \argmax_{\theta} Q(\theta | \theta^{(i)} ) .
	\label{eq:argmaxem}
	\end{align}
\end{itemize}

We solve for $\theta$ using gradient ascent. As $\evolve$ is a function of $\theta$, we must 
compute the derivative of $Q(\theta|\theta^{(i)})$ using the chain rule. We compute this derivative with respect to each element $j$ in $\theta$: 
\begin{align}
\frac{d}{d \theta[j] } Q(\theta|\theta^{(i)} ) &=  \sum_p \sum_q  \frac{d Q(\theta|\theta^{(i)} )  }{d \evolve[p,q]} \frac{d \evolve[p,q]}{d \theta[j] } 
\end{align}
Using the low-dimensional subspace evolution model proposed in Section~\ref{sec:evolution}, the derivative $\frac{d \evolve[p,q]}{d \theta[j] }$ can be computed as \katie{I took out taking just the real portion to make it nicer looking, should probably resolve that}
\begin{align}
\frac{d \evolve }{d \theta[j]} = -i 2 \pi \theta[j] \evolve  \left( \bm{u} \mathcal{M}_{\xpos}[:,j+1]^T  + \bm{v} \mathcal{M}_{\ypos}[:,j+1]^T  \right).
\end{align}
By expanding and taking the derivative of the log-likelihood from Equation~\ref{eq:likelihood} with respect to $\evolve$, we obtain the expression 
\begin{align}
& \hspace{-.1in}  \frac{d}{d \evolve} Q(\theta|\theta^{(i)}) 
=   -\frac{1}{2} \sum_{t=2}^{N} \left[ 2 \bQ^{-1} \evolve  E_{ \bm{X}| \bm{Y}, \theta^{(i)} } \left[ \im_{t-1}   \im_{t-1}^T\right]   \right. \\
\notag & \hspace{0.2in} \left. - \bQ^{-1} E_{ \bm{X}| \bm{Y}, \theta^{(i)} }  \left[ \im_t \im_{t-1}^T \right] -  \bQ^{-1} E_{ \bm{X}| \bm{Y}, \theta^{(i)} } \left[ \im_{t-1} \im_t^T \right]   \right].
\end{align}
By inspecting this expression we can see that the sufficient statistics we require to maximize the log-likelihood are the expected value of $\im_t \im_t^T$ and $\im_{t-1} \im_t^T$ under the distribution $p(\bm{X} | \bm{Y}; \theta^{(i)} ) $. Conveniently, these sufficient statistics can be computed using from the set of ${\bf z}$'s and $\bf{P}$'s computed in Section~\ref{sec:dynamic_inference_known}.  From the marginal distributions (Equations~\ref{eq:marg1} and~\ref{eq:marg2}) derived in Section~\ref{sec:dynamic_inference_known}, we obtain
\begin{align}
E_{ \bm{X}| \bm{Y}, \theta^{(i)} } \left[ \im_{t}   \im_{t}^T\right] = \hat{\im}_t  \hat{\im}_t ^T + {\bf C}_{t}.
\end{align}
The sufficient statistic $E_{ \bm{X}| \bm{Y}, \theta^{(i)} } \left[ \im_{t-1} \im_t^T \right]$ is a bit trickier to obtain, but can also be calculated using the same forward-backward terms, as shown in the supplemental material. Mathematically,	
\begin{align}
E_{ \bm{X}| \bm{Y}, \theta^{(i)} } \left[ \im_{t-1} \im_t^T \right] = \hat{\im}_{t-1} \hat{\im}_t^T  + \bm{\xi}_3 \bm{\xi}_1^{T -1}, 
\end{align}	
where $\bm{\xi}_1$ and $\bm{\xi}_3$ are defined according to
    \begin{align}
    & p(\im_t, \im_{t-1}|\bm{Y})  = \mathcal{N}_{\im_{t-1}}(\bm{\xi}_1\im_t + \xi_2, \bm{\xi}_3) \\
    \notag & \propto \mathcal{N}_{\im_{t-1}} ({\bf z}^{\alpha}_{t-1|t-1}, {\bf P}^{\alpha}_{t-1|t-1} ) \mathcal{N}_{\im_t} ( \evolve \im_{t-1}, \bQ )  \mathcal{N}_{\im_{t}} ({\bf z}^{\beta}_{t|t}, {\bf P}^{\beta}_{t|t} ).
    \end{align}
	
	
	

	

	To learn the parameters $\theta$, we iterate between computing sufficient statistics of $\bm{X}$ given the current estimate of parameters, $\theta^{(i)}$, 
	and solving for new parameters that maximize the updated log-likelihood, $\theta^{(i+1)}$, under those statistics. 
	Once convergence has been reached, we return the parameters $\hat{\theta}$ and the optimal set of instantaneous images under that transformation $E[\bm{X}] =  \{ \hat{\im}_t \}_{t=1}^{\ntimes} $.
	
	Note that since our EM method's maximization step requires solving a non-convex problem we likely will only find a local-maximum of $\theta$ at each step. Nonetheless, the log-likelihood is guaranteed to increase for a $\theta$ that increases $Q(\theta|\theta^{(i)})$ (Equation~\ref{eq:qfunction})~\cite{little2014statistical}. This class of algorithms, which do not necessarily find the optimal $\theta$ at each iteration, are more rigorously referred to as ``Generalized EM"~\cite{little2014statistical}. In the case of a linear $f(\im)$ this EM procedure is exact and the log-likelihood increases at every iteration. In the case of a non-linear $f(\im)$ the forward-backward algorithm in Section~\ref{sec:dynamic_inference_known} provides only an approximation of the true sufficient statistics. Nonetheless, we empirically find that, when we fix each latent image's linearization point, the log-likelihood consistently improves.

	

	

\begin{figure*}[h!]
	\begin{center}
		\hspace*{-.4in}
		\vspace*{-0.3in}
		\begin{tabular}{   c | c  c  c  c  }
			& \large{\textsf{VIDEO 1 }} &\large{\textsf{VIDEO 2}}   &\large{\textsf{VIDEO 3}} &\large{\textsf{VIDEO 4}}      \\ 
			&\vspace{-.1in}&&&\\
			& \large{\textsf{Pure Rotation}} &\large{\textsf{Rotating Hotspot}}   &\large{\textsf{Face-on Disk}} &\large{\textsf{Edge-on Disk}}      \\ \hline
			&\vspace{-.1in}&&&\\
			\multirow{1}{*}[0.9in]{ \rotatebox[origin=t]{90}{\small{\textsf{SINGLE FRAME}} }}
			&
			{{\includegraphics[height=.15\linewidth]{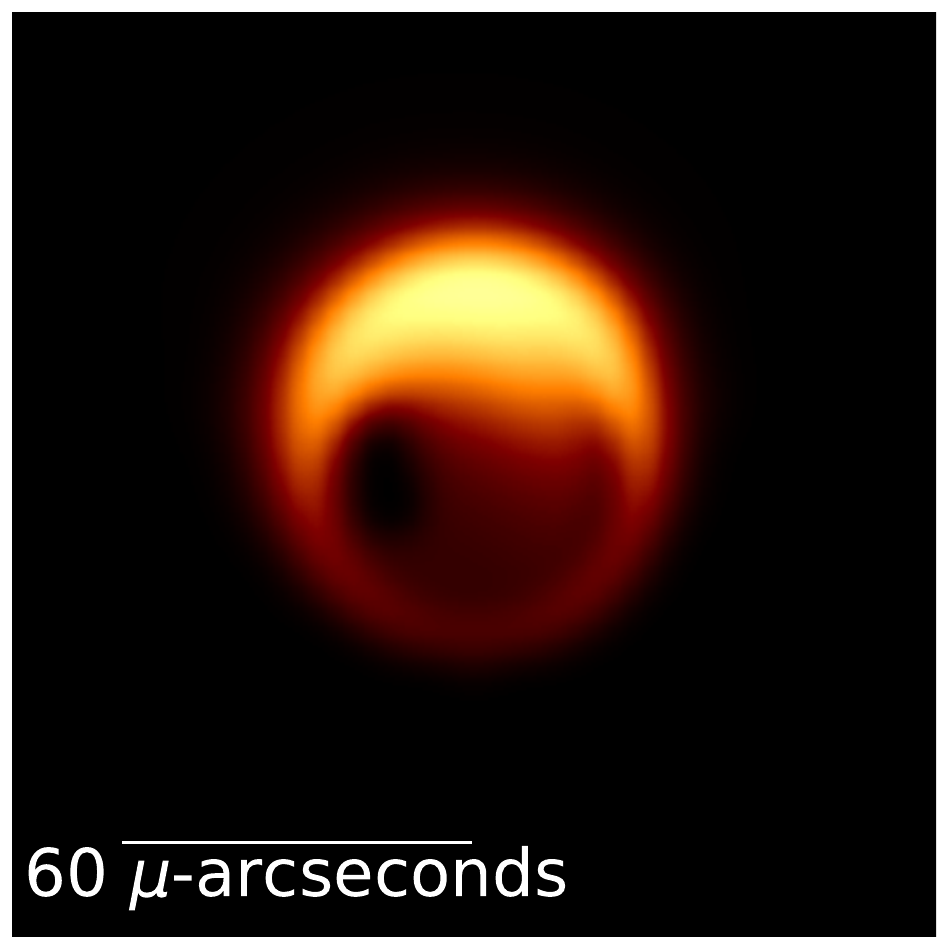}} } 
			\multirow{2}{*}[.9in]{ \includegraphics[width=0.06\linewidth]{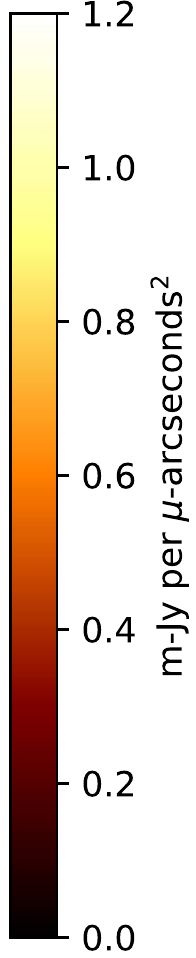} }
			&
			\includegraphics[height=.15\linewidth]{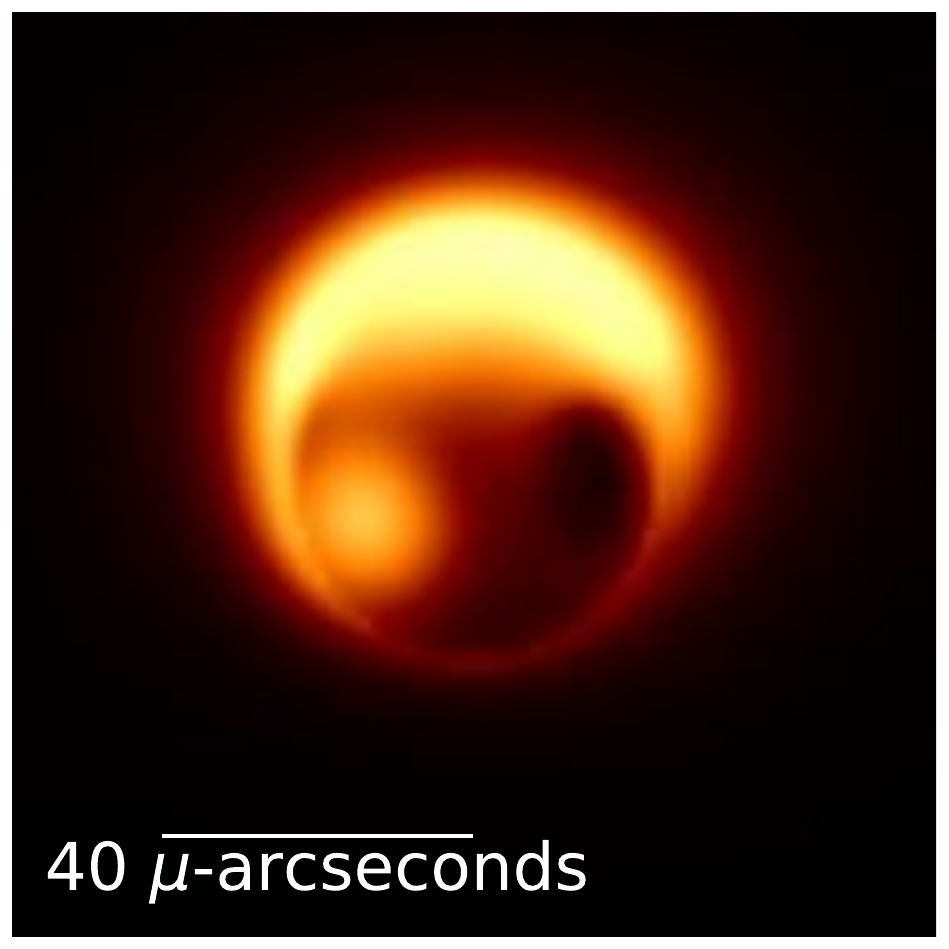} 
			\multirow{2}{*}[.9in]{ \includegraphics[width=0.06\linewidth]{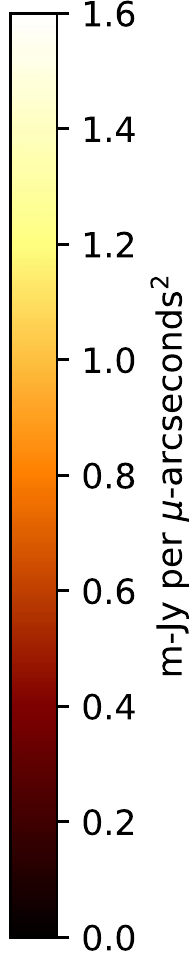} }
			&
			\includegraphics[height=.15\linewidth]{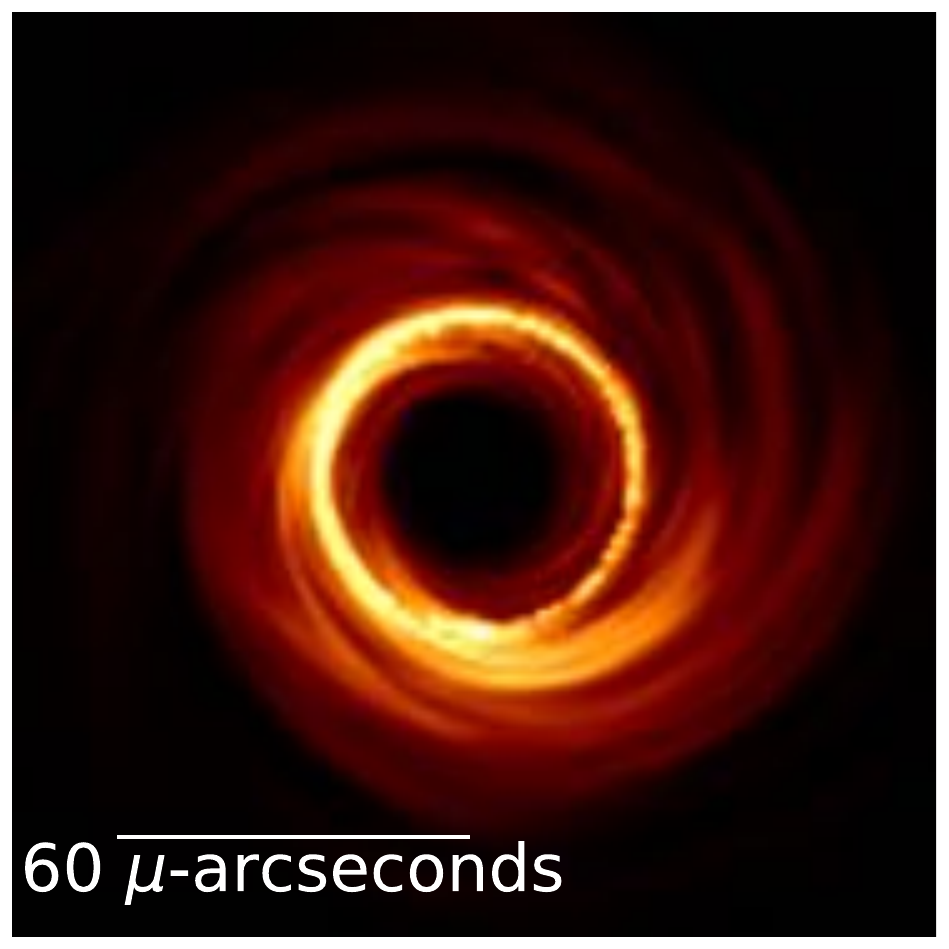} 
			\multirow{2}{*}[.9in]{ \includegraphics[width=0.0668\linewidth]{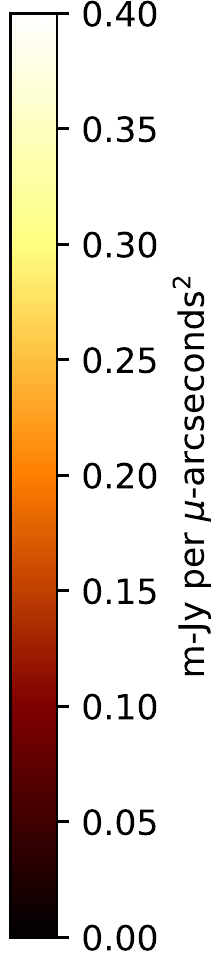} }
			&
			\includegraphics[height=.15\linewidth]{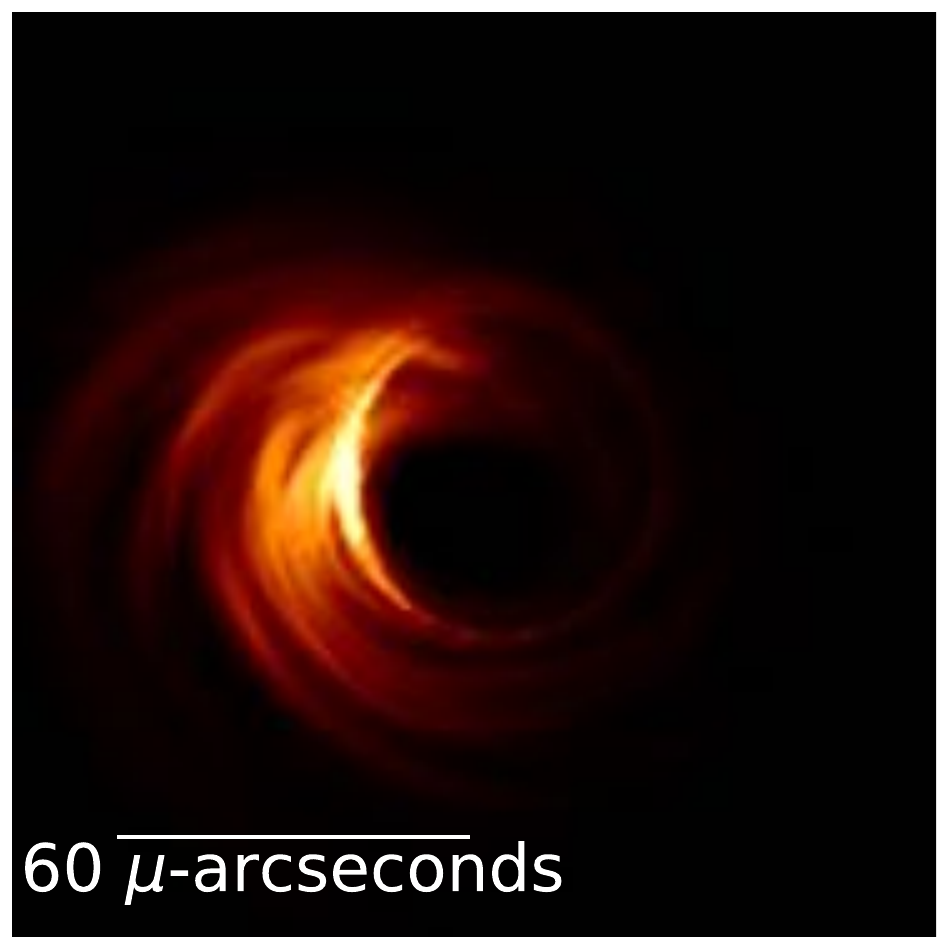} 
			\multirow{2}{*}[.9in]{ \includegraphics[width=0.06\linewidth]{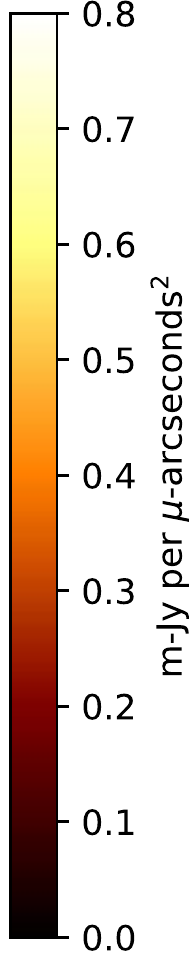} }
			\\
			\multirow{1}{*}[0.7in]{ \rotatebox[origin=t]{90}{\small{\textsf{MEAN}} }}
			&
			{{\includegraphics[height=.15\linewidth]{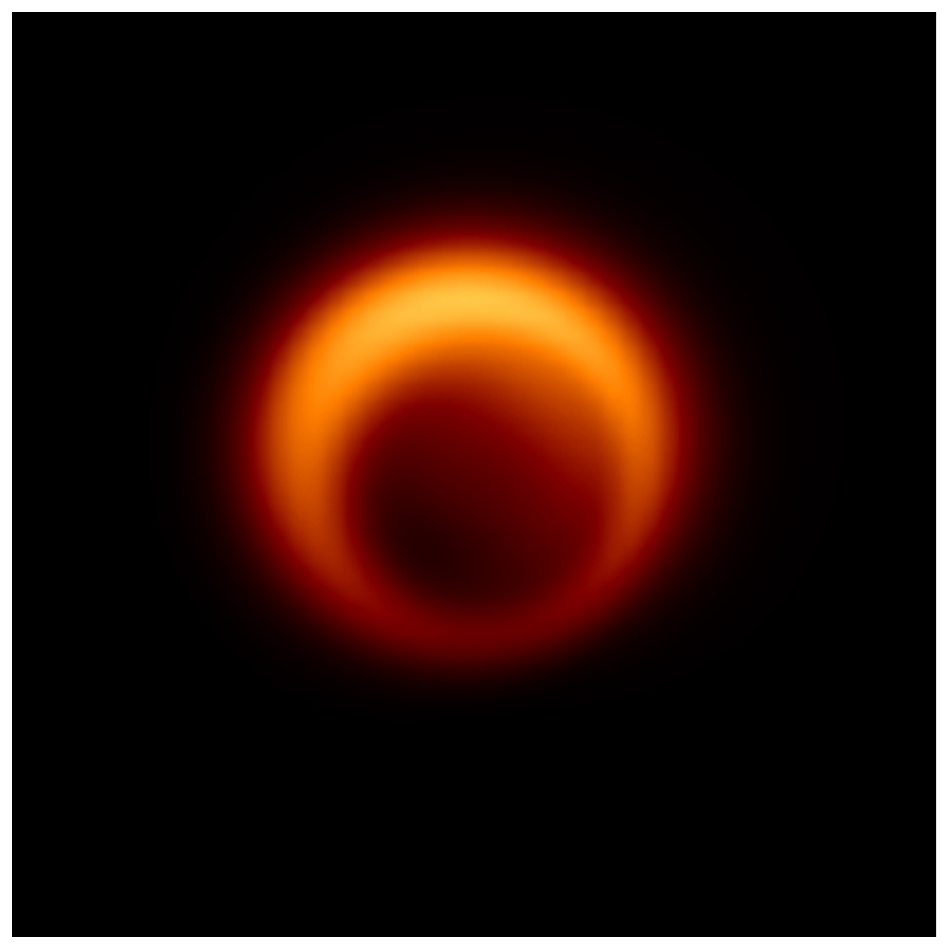}} } \hspace{.55in} &
			\includegraphics[height=.15\linewidth]{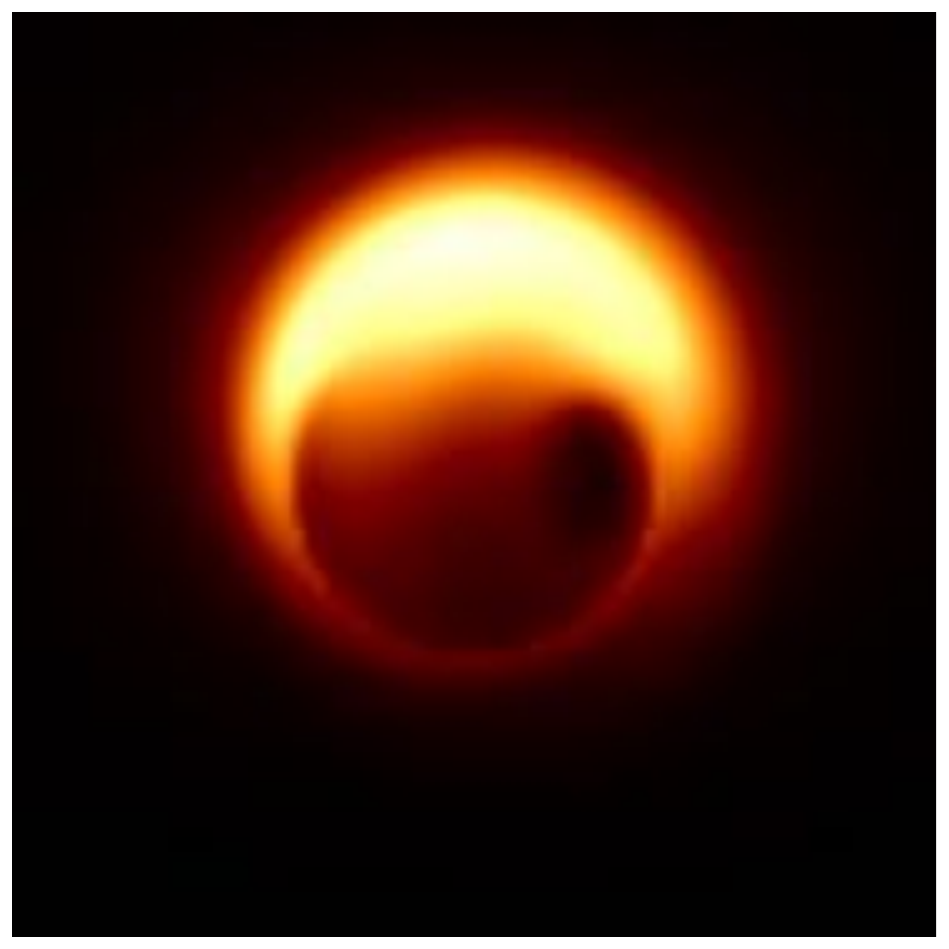} \hspace{.55in} &
			\includegraphics[height=.15\linewidth]{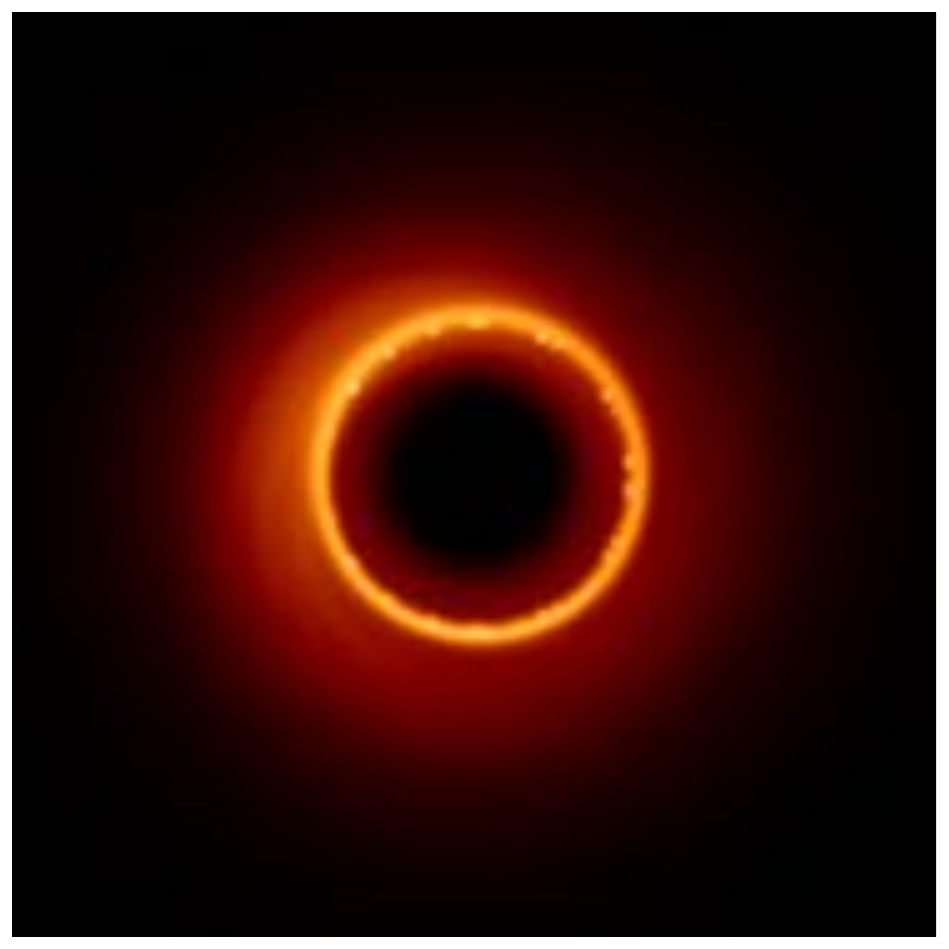} \hspace{.55in} &
			\includegraphics[height=.15\linewidth]{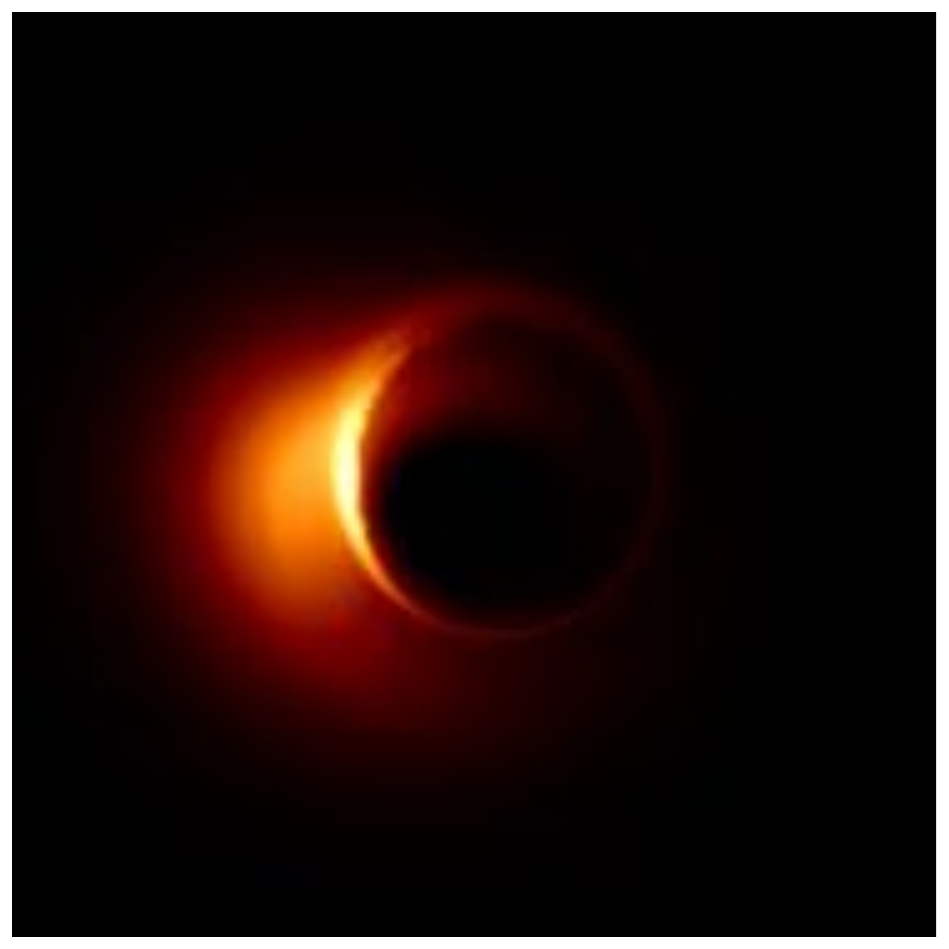} \hspace{.55in}
			\\ \hline
			&\vspace{-.1in}&&&\\
			\multirow{1}{*}[0.9in]{ \rotatebox[origin=t]{90}{\small{\textsf{STD. DEVIATION}} }}
			& \hspace{0.01in}
			{{\includegraphics[height=.147\linewidth]{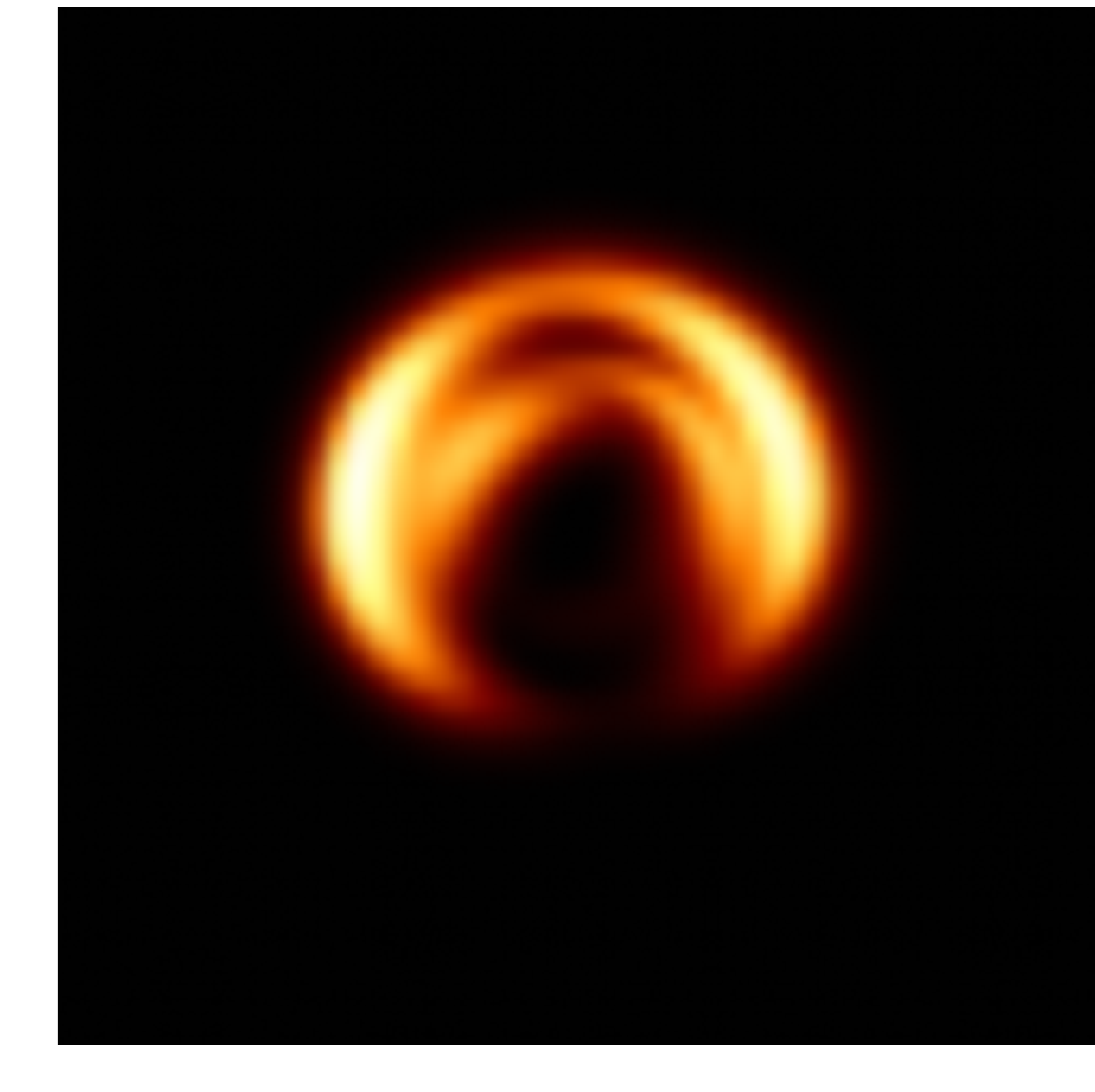}} } \hspace{.59in} & 
			
			\hspace{0.01in}
			 \includegraphics[height=.147\linewidth]{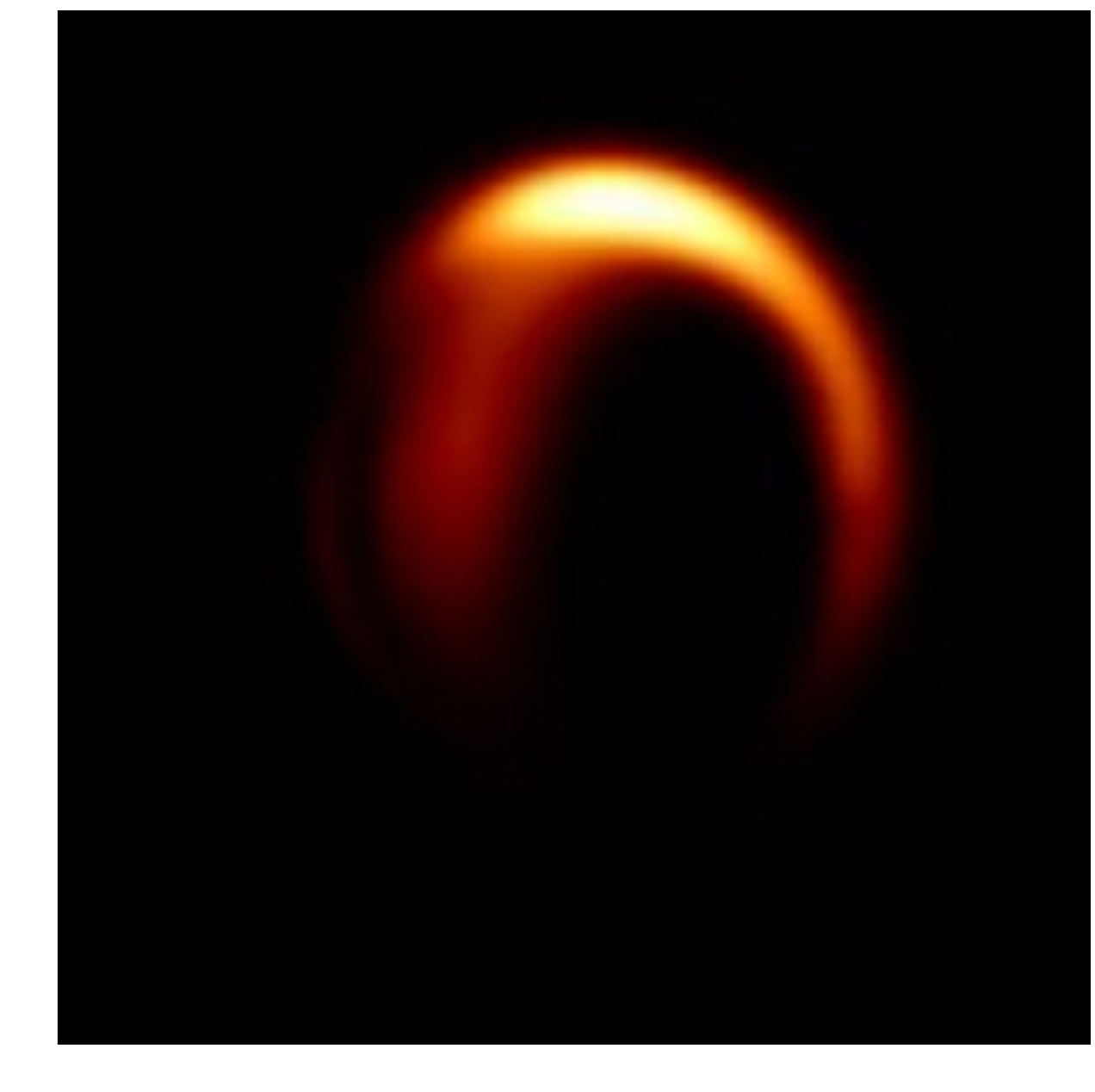}  \hspace{.59in} &
			
			\hspace{-0.01in} \includegraphics[height=.147\linewidth]{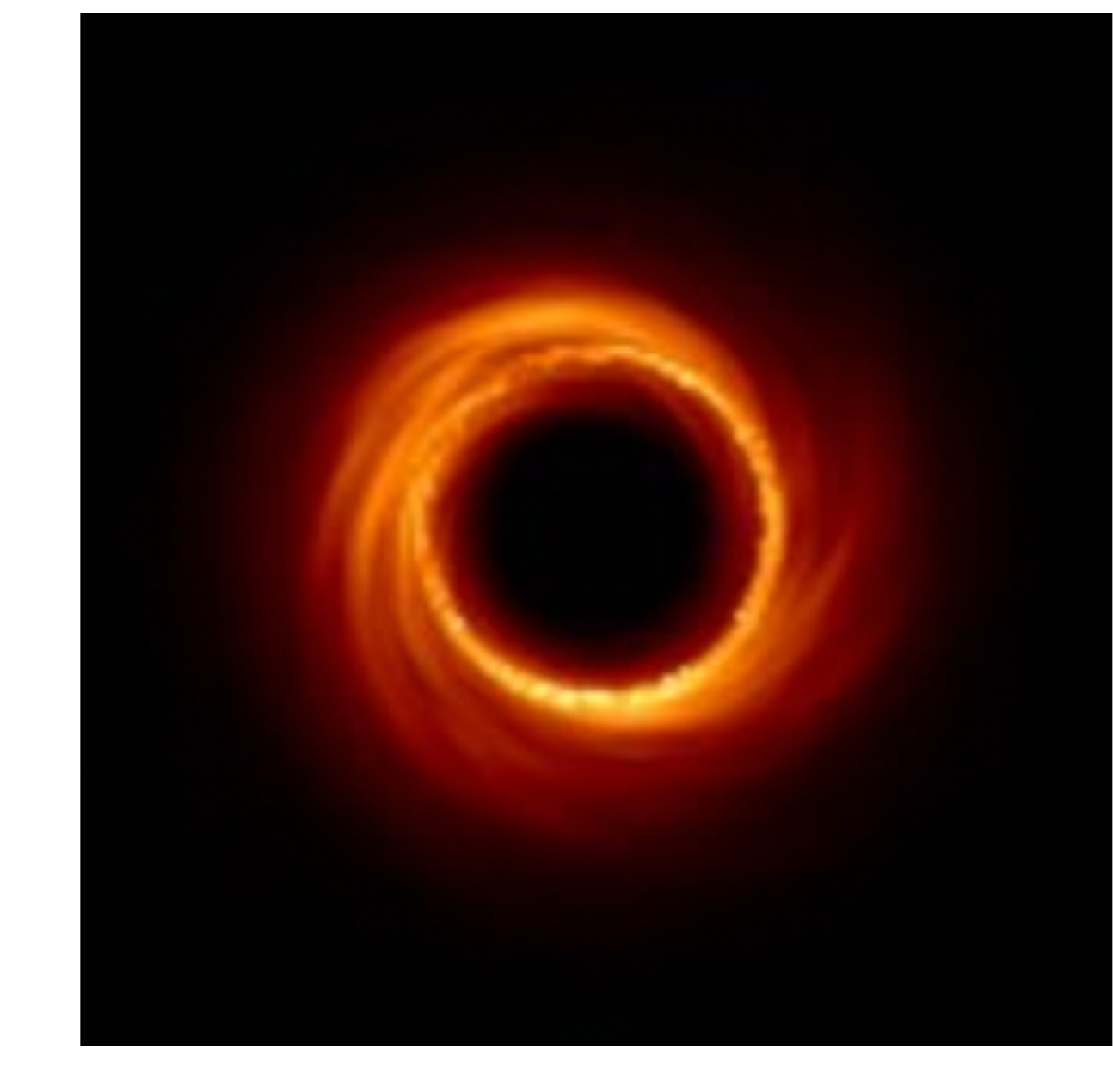}  \hspace{.55in} & 
			\hspace{-0.01in}
			\includegraphics[height=.147\linewidth]{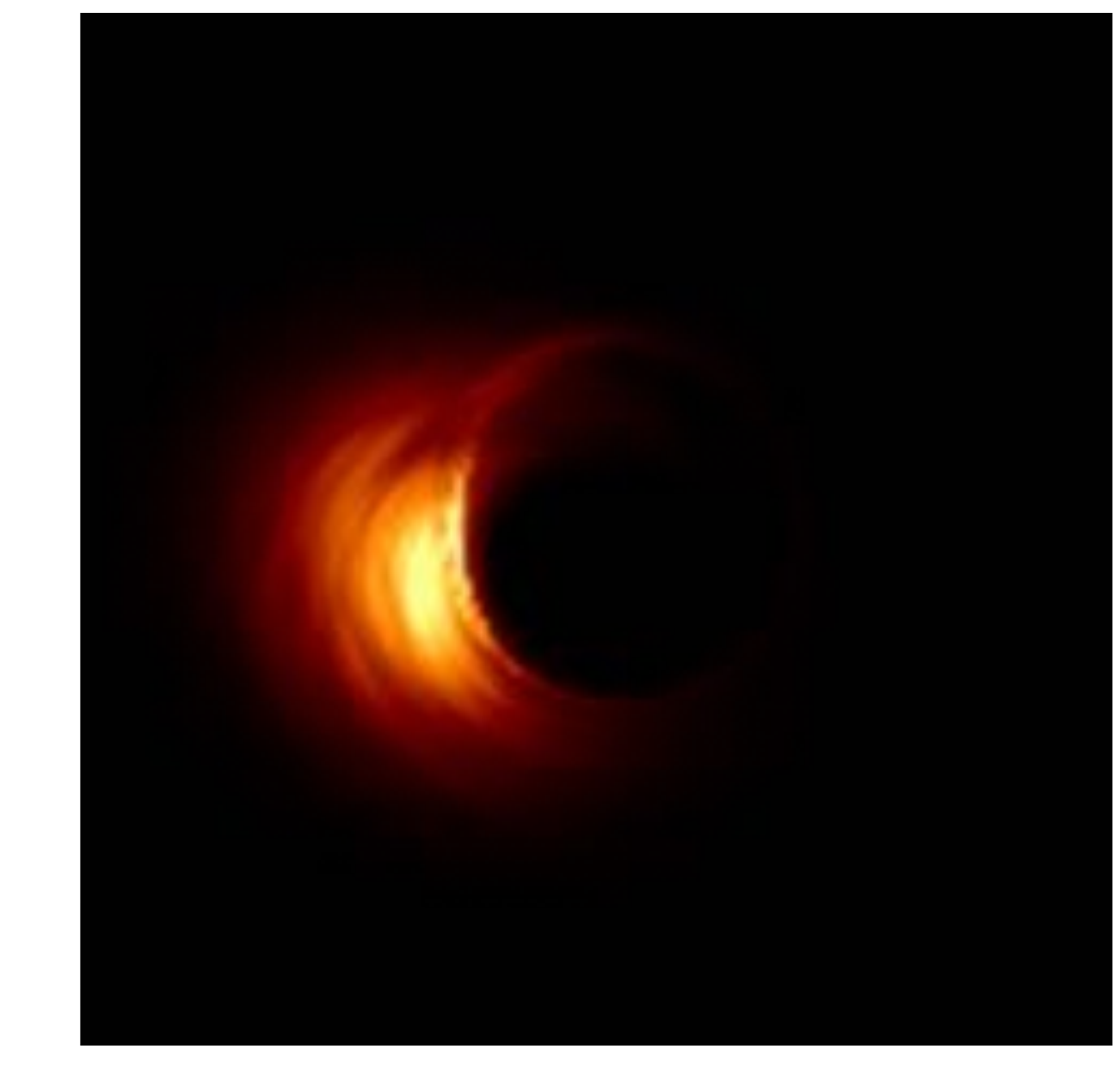} \hspace{.55in} 
			\\
		\end{tabular}
		\vspace{.1in}
		\caption{{\bf Ground truth videos:} The four ground truth sequences used to demonstrate results. We show a single frame from each sequence, the mean frame, and the spatial standard deviation of flux density. Video 1 consists of a $160 \mu$-arcsecond image~\cite{avery} that rotates $180^{\circ}$ over the course of a 12 hour observation (24 hour rotational period). Video 2 is a $120 \mu$-arcsecond view of an edge-on black hole disk with a rotating ``hot spot" predicted by~\cite{Broderick_Loeb_2006} with a rotational period of 2.78 hours. Video 3 and 4 are generated using a model of a black hole observed face on and at a $45^{\circ}$ inclination with a $160 \mu$-arcsecond field of view~\cite{Shiokawa_2013}. They assume a spin of 0.9375 with an Innermost Stable Circular Orbit (ISCO) rotational period of 8.96 minutes. The specified FOV and colormaps for the single frame and mean images are used for each corresponding video throughout the remainder of the paper. }
		\label{fig:groundtruth}
	\end{center}
		\vspace{-.2in}
\end{figure*}

\vspace{-.07in}
\section{Dynamic Imaging Results}
\label{sec:results}

As data from the EHT 2017 campaign is yet to be released, in this section we demonstrate our method on synthetic EHT data and real data from the Very Long Baseline Array (VLBA). Additional results can be seen in the supplemental document and video. The StarWarps algorithm has been implemented as part of the publicly available python \texttt{eht-imaging}\footnote{\url{https://github.com/achael/eht-imaging}} library~\cite{andrew}.

\vspace{-.2in}
\subsection{Synthetic Data Generation} 

We demonstrate our algorithm on synthetic data generated from four different sequences of time-varying sources. These sequences include two realistic fluid simulations of a black hole accretion disk for different observing orientations~\cite{Shiokawa_2013}, a realistic sequence of a ``hot spot" rotating around a black hole~\cite{Broderick_Loeb_2006}, and a toy sequence evolving with pure rotation. The field of view of each sequence ranges from 120 to 160 $\mu$-arcseconds. A still frame from each sequence is shown in Figure~\ref{fig:groundtruth}. To help give a sense of the variation in each sequence, the figure also displays the mean and standard deviation of flux density. We refer to these sequences by their video number, indicated in the figure.

In order to demonstrate the quality of results under various observing conditions, VLBI observations of SgrA* at 1.3 mm (230 GHz) are simulated assuming
 three different telescope arrays.  The first array, EHT2017, consists of the 8 telescopes at 6 distinct locations that were used to collect measurements for the Event Horizon Telescope in the spring of 2017. The uv-coverage for this array can be seen in Figure~\ref{fig:staticimaging}. The second array, EHT2017+, augments the EHT2017 array with 3 potential additions to the EHT: Plateau de Bure (PDB), Haystack (HAY), and Kitt Peak (KP) Observatory. 
Details on telescopes used in the EHT2017 and EHT2017+ array are shown in Table~\ref{tab:telescopes}.  
The third array, FUTURE, consists of 9 additional telescopes. The uv-coverage of these latter two arrays, along with a colorbar indicating the time of each measurement, is shown in Figures~\ref{fig:uvcov2}. 

Visibility measurements are generated using the python \texttt{eht-imaging} library~\cite{andrew}. 
Realistic thermal noise, resulting from a bandwidth ($\Delta \nu$) of 4 GHz and a 100 second integration time ($t_{\rm int}$), is introduced on each visibility. The standard deviation of thermal noise is given by
\begin{align}
\sigma = \frac{1}{0.88} \sqrt{\frac{\mbox{SEFD}_1 \times \mbox{SEFD}_2 }{2 \times \Delta \nu \times t_{\rm int} } },
\label{eq:thermal}
\end{align}

\noindent{for System Equivalent Flux Density (SEFD) of the two telescopes corresponding to each visibility\footnote{The factor of 1/0.88 is due to information loss due to recording 2-bit quantized data-streams at each telescope~\cite{TMS}.}~\cite{taylor1999synthesis}.  
Random station-based atmospheric phases drawn from a uniform distribution at each time step are introduced into measurements using the \texttt{eht-imaging} library. 
In Videos 2-4 a set of measurements is sampled every 5 minutes over a roughly 14 hour duration, resulting in 173 time steps. In Video 1 only 30 time steps are measured over a 12 hour duration. 
}

\begin{figure}[h!]
	\centering
	{\includegraphics[height=.38\linewidth]{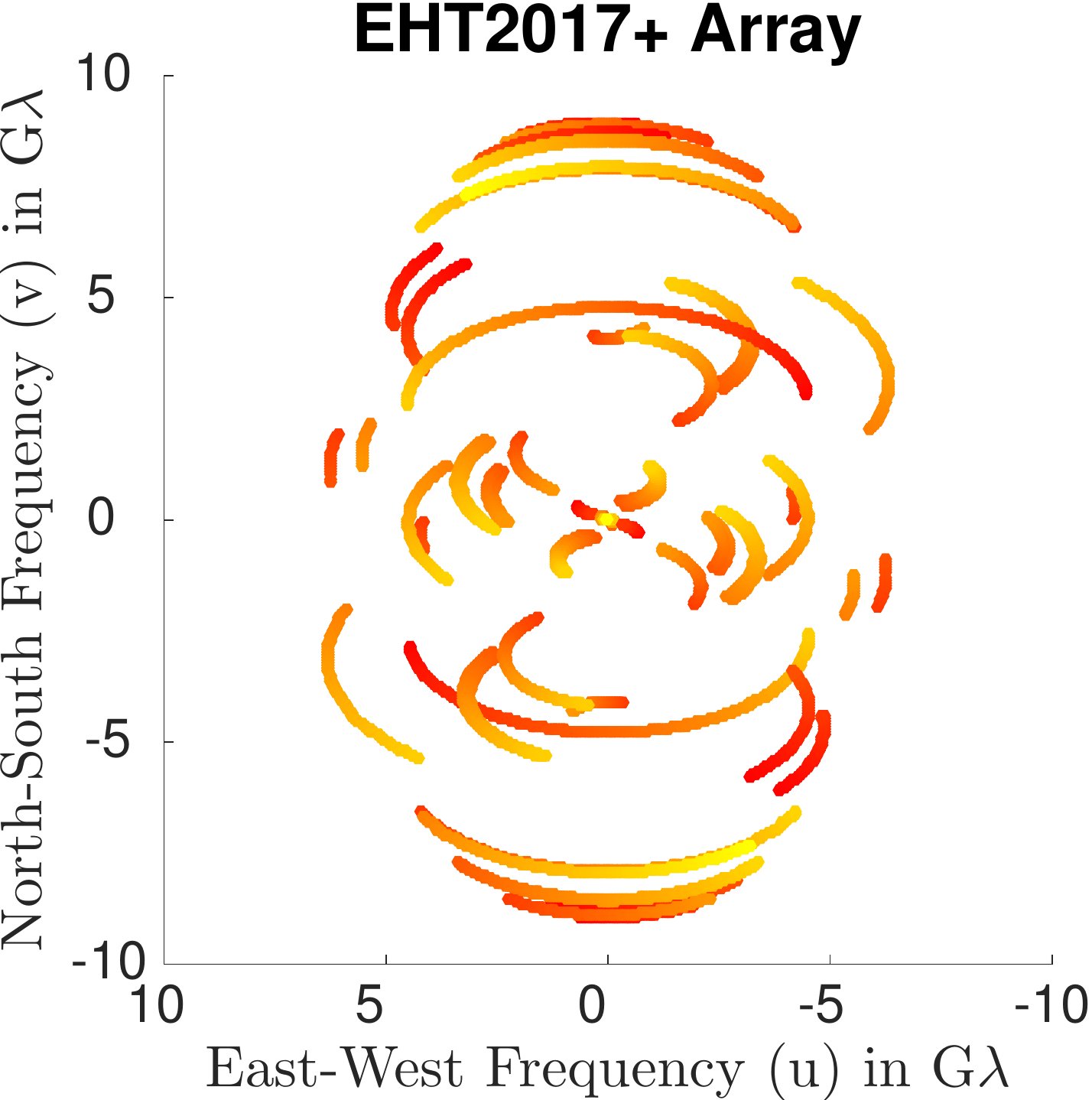}}
	{\includegraphics[height=.38\linewidth]{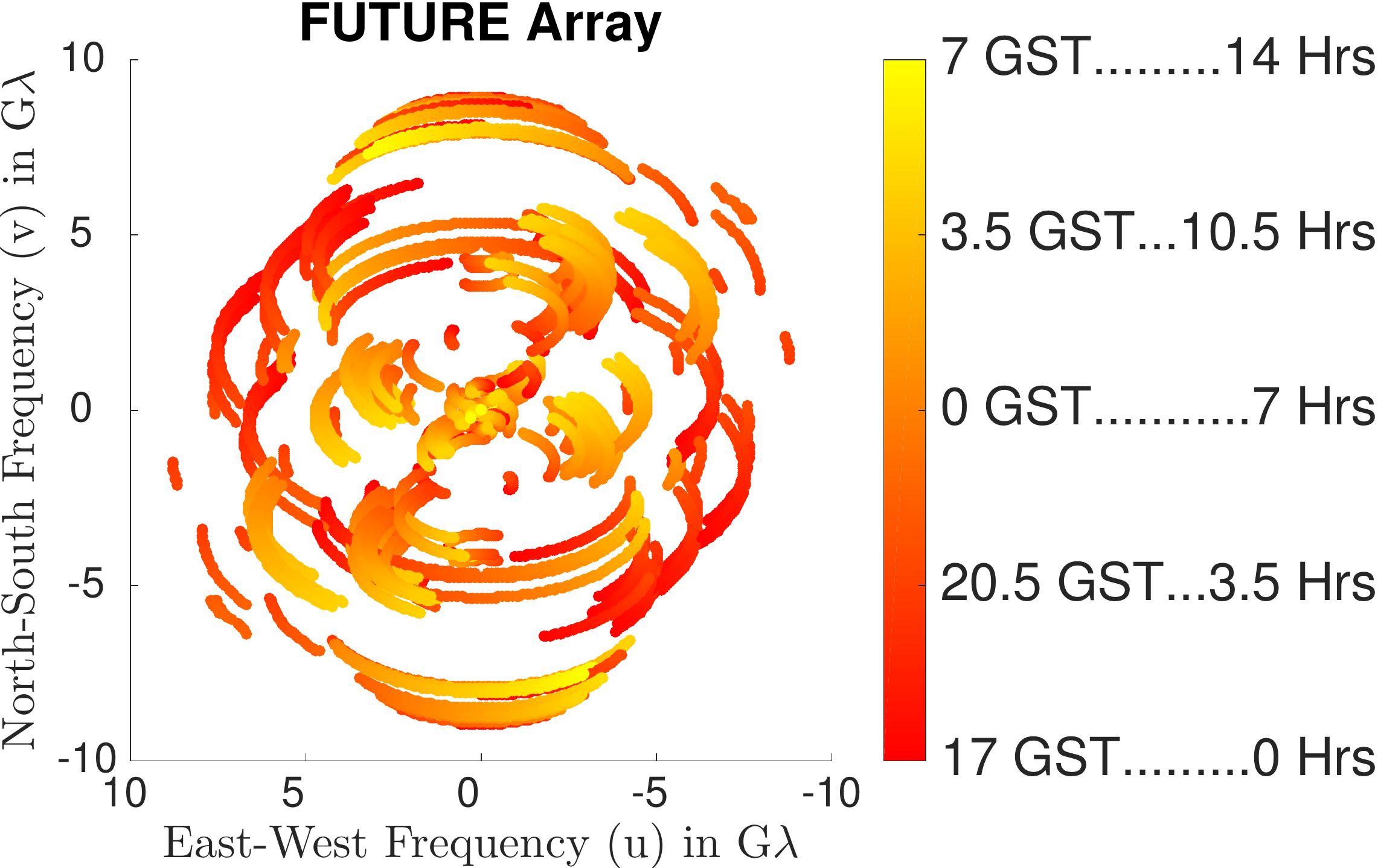}}
	\caption{{\bf Time-varying uv-coverage:} The uv-coverage for EHT2017+ and FUTURE arrays when observing SgrA∗. The uv-coverage for the EHT2017 array can be seen in Figure~\ref{fig:staticimaging}. 
		Baselines are colored by the time of each observation relative to the start time, indicated by the colorbar to the right.}
	\label{fig:uvcov2}

\end{figure}

\begin{table}[!t]
	\begin{center}
		\caption{EHT 2017 Station Parameters}
		\label{tab::eht_station}
		\begin{tabular}{ccc}
 Telescope & Location & SEFD (Jy) \\ \hline
 ALMA & Chile & 110 \\ 
 APEX & Chile & 22000 \\ 
 LMT & Mexico & 560 \\
 SMT & Arizona & 11900 \\ 
 SMA & Hawaii & 4900 \\
 JCMT & Hawaii & 4700 \\ 
 PV & Spain & 2900 \\ 
 SPT & South Pole & 1600 \\ \hline
 PDB* & France & 1600 \\ 
 HAY* & Massachusetts & 2500 \\ 
 KP* & Arizona & 2500 \\ \hline
		\end{tabular}\\
	\end{center}
	\bigskip
	\footnotesize{The location and SEFD of each telescope in the EHT2017 and EHT2017+ arrays. These parameters and locations were used to generate the uv-trajectories in frequency space shown in Figure~\ref{fig:staticimaging} and~\ref{fig:uvcov2}. Telescope names followed by a star (*) were not included in the EHT2017 array.}
	\label{tab:telescopes}
	\vspace{-0.3in}
\end{table}


\vspace{-.1in}
\subsection{Static Evolution Model (No Warp)}
\label{sec:nomotionresults}

We first demonstrate results of our method under a static evolution model. In this case, we fix parameters $\theta$ such that $A=\mathds{1}$. This assumes that there is no global motion under a persistent warp field, but only perturbations around a fairly static scene. Despite this incorrect assumption (especially in Videos 1 and 2), this simple model results in reconstructions that surpass the state-of-the-art methods, and recovers distinctive structures that appear in the underlying source images.

\subsubsection{Synthetic Data Result Comparison}

\begin{figure*}
	\begin{center}
		\setlength{\tabcolsep}{3pt}
		
		\hspace{-0.5in}\normalsize{\textsf{BLURRED TRUE MEAN}}  \hspace{5.5cm} \normalsize{\textsf{NORMALIZED RMSE}} 
		\vspace{0.1in}

						\begin{tabular}{ c " c}
							\hspace{-.06in} \textsf{Video 1} & \hspace{-.06in} \textsf{Video 2} \\
							{{\includegraphics[height=.1\linewidth]{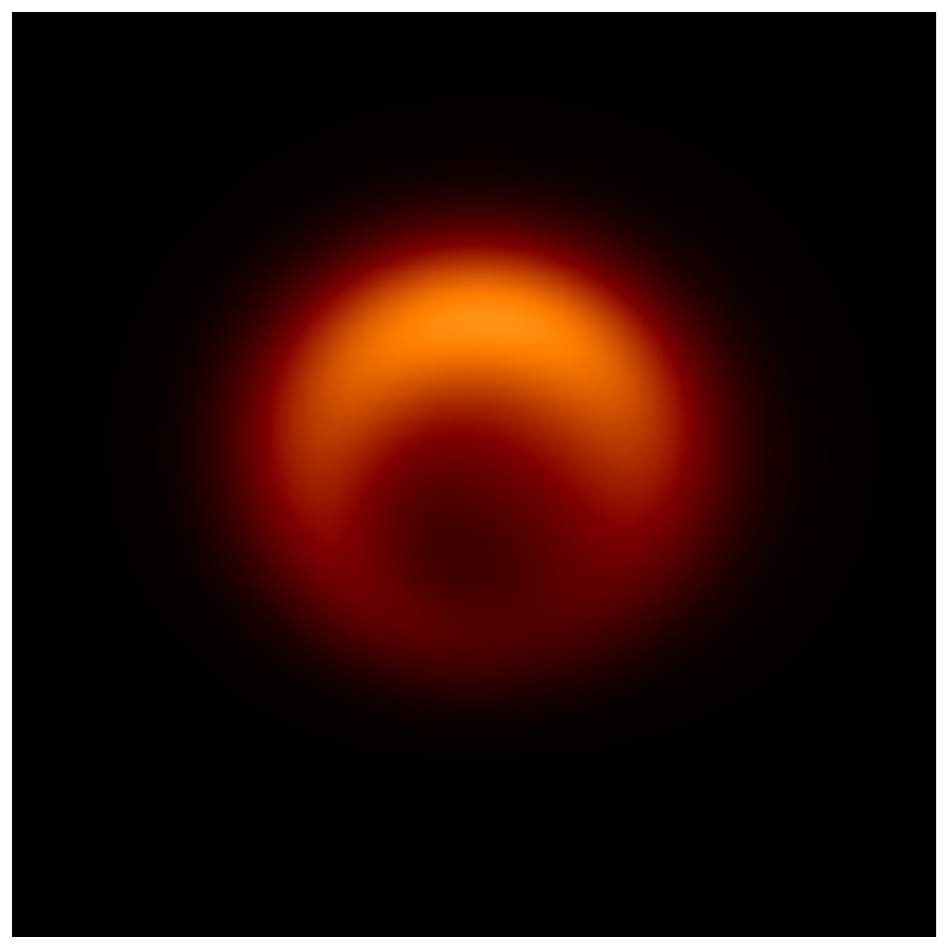}} } & {{\includegraphics[height=.1\linewidth]{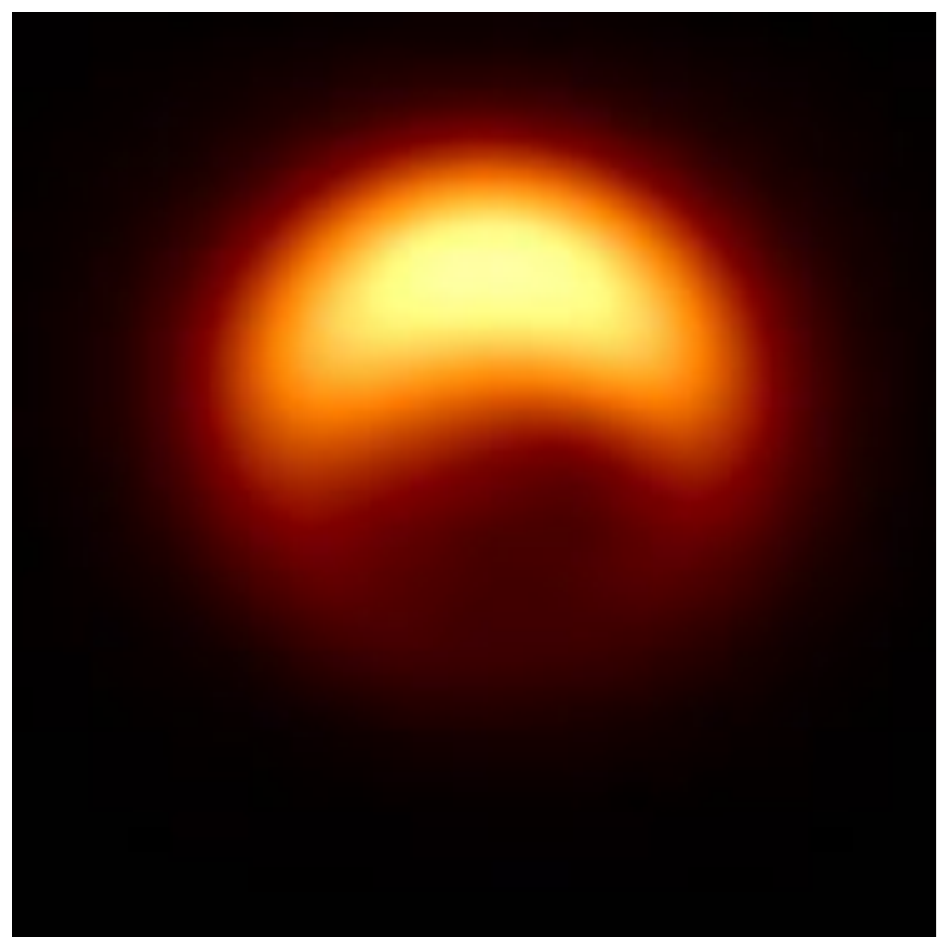}} } \\ \thickhline
							& \vspace{-.1in} \\
							\hspace{-.06in} \textsf{Video 3} & \hspace{-.06in} \textsf{Video 4} \\
							{{\includegraphics[height=.1\linewidth]{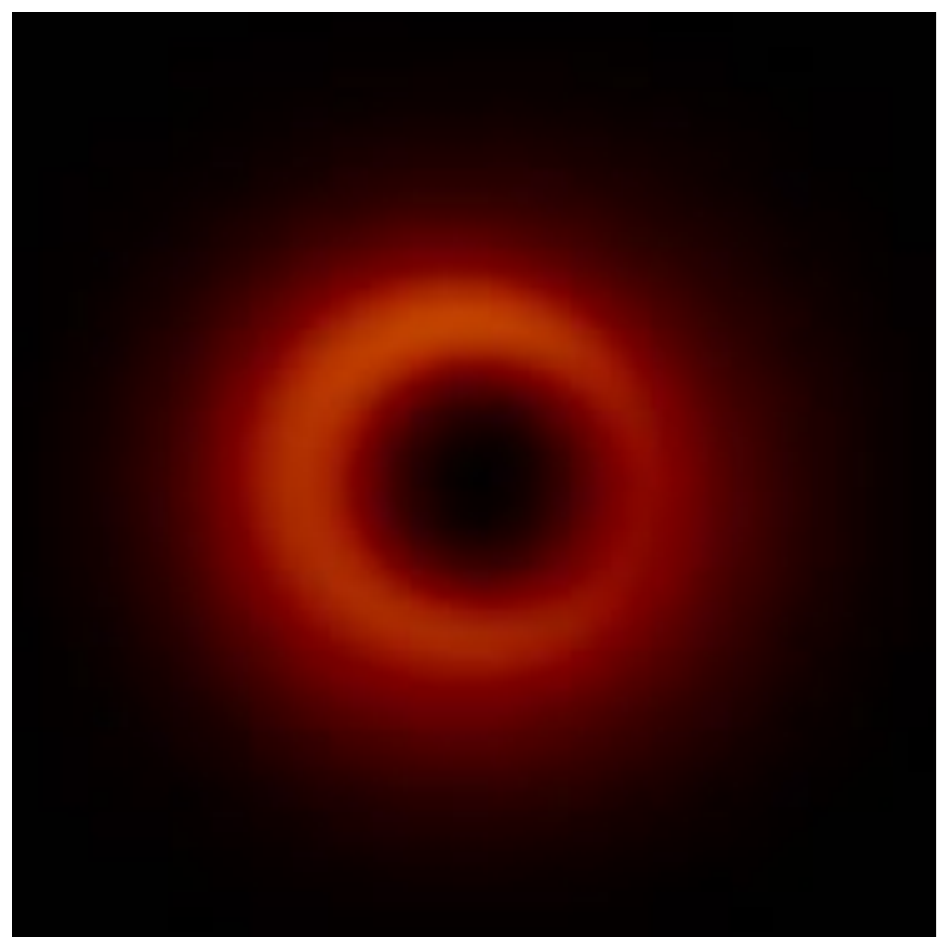}} } & {{\includegraphics[height=.1\linewidth]{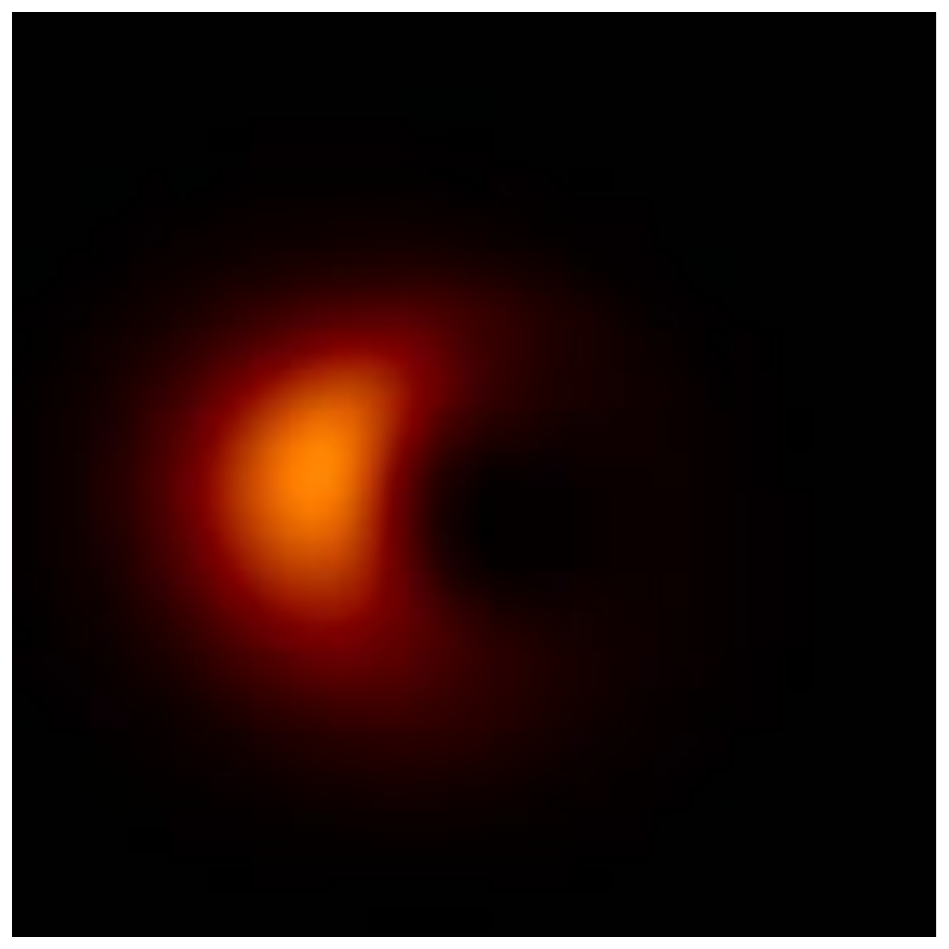}} }
						\end{tabular}	
		\qquad
		\begin{tabular}{ r | c | c | c | c " c | c | c | c}
			& \rotatebox[origin=t]{90}{\small{\textsf{EHT 2017}} } \rotatebox[origin=t]{90}{\small{\textsf{NO ATM.}} }  & \rotatebox[origin=t]{90}{\small{\textsf{EHT 2017}} } \rotatebox[origin=t]{90}{\small{\textsf{ATM.}} } & \rotatebox[origin=t]{90}{\small{\textsf{EHT 2017+}} } \rotatebox[origin=t]{90}{\small{\textsf{ATM.}} } &\rotatebox[origin=t]{90}{\small{\textsf{FUTURE}} } \rotatebox[origin=t]{90}{\small{\textsf{ATM.}} } & \rotatebox[origin=t]{90}{\small{\textsf{EHT 2017}} } \rotatebox[origin=t]{90}{\small{\textsf{NO ATM.}} }  & \rotatebox[origin=t]{90}{\small{\textsf{EHT 2017}} } \rotatebox[origin=t]{90}{\small{\textsf{ATM.}} } & \rotatebox[origin=t]{90}{\small{\textsf{EHT 2017+}} } \rotatebox[origin=t]{90}{\small{\textsf{ATM.}} } &\rotatebox[origin=t]{90}{\small{\textsf{FUTURE}} } \rotatebox[origin=t]{90}{\small{\textsf{ATM.}} }  \\ \hline
			{\small{\textsf{StarWarps Mean} } } & {\bf {0.67} }& {\bf {0.65}} & {\bf {0.55} }&  {\bf {0.55}}& {\bf {0.74}} & {\bf {0.73} }& {\bf {0.73}} &  0.71 \\ 
				\cite{freek} & 1.05  & 0.82 & 0.73  & 0.68 &  0.98 & 1.21 & 0.99 & 0.35 \\ 
				\cite{andrew} & 0.79 & 0.80 & 0.73 & 0.63 & 0.83& 1.05 & 0.81 & {\bf 0.23}  \\  \thickhline
				\small{\textsf{StarWarps Mean} } & {\bf 0.32} & {\bf 0.55} & 0.67 & {\bf 0.36} & {\bf 0.34} & {\bf 0.44} & {\bf 0.23} & {\bf 0.31} \\
				\cite{freek} & 0.93 & 0.90 & {\bf 0.53} & 0.51 & 0.84 & 0.75 & 1.02 & 0.61 \\
				\cite{andrew} &0.84 & 0.85 & 0.71 &  0.39  & 0.60 & 0.50 & 0.60 & 0.47
			\end{tabular}

		\vspace{0.4in}
		
		\hspace*{-1.3cm}
		\begin{tabular}{  c c | c  c  c  c "  c  c  c  c  }
			& \small{\textsf{Array:}} &\small{\textsf{EHT 2017}}   &\small{\textsf{EHT 2017 }} &\small{\textsf{EHT2017+}}    &\small{\textsf{FUTURE}}    &\small{\textsf{EHT 2017}}   &\small{\textsf{EHT 2017 }} &\small{\textsf{EHT2017+}}    &\small{\textsf{FUTURE}}     \\ 
			&\vspace{-.1in} & & & & & & & &\\
			& \small{\textsf{Error:}} &\small{\textsf{NO ATM.}}   &\small{\textsf{ATM.}} &\small{\textsf{ATM.}}    &\small{\textsf{ATM.}}  &\small{\textsf{NO ATM.}}   &\small{\textsf{ATM.}} &\small{\textsf{ATM.}}    &\small{\textsf{ATM.}}   \\ \hline
			&\vspace{-.1in} & & & & & & & &\\
			 \multirow{2}{*}[0.6in]{ \rotatebox[origin=t]{90}{\small{\textsf{StarWarps}} }}   \hspace{-0.3in} &	\multirow{1}{*}[0.45in]{ \rotatebox[origin=t]{90}{\small{\textsf{Mean}} }}
			&
			{{\includegraphics[height=.1\linewidth]{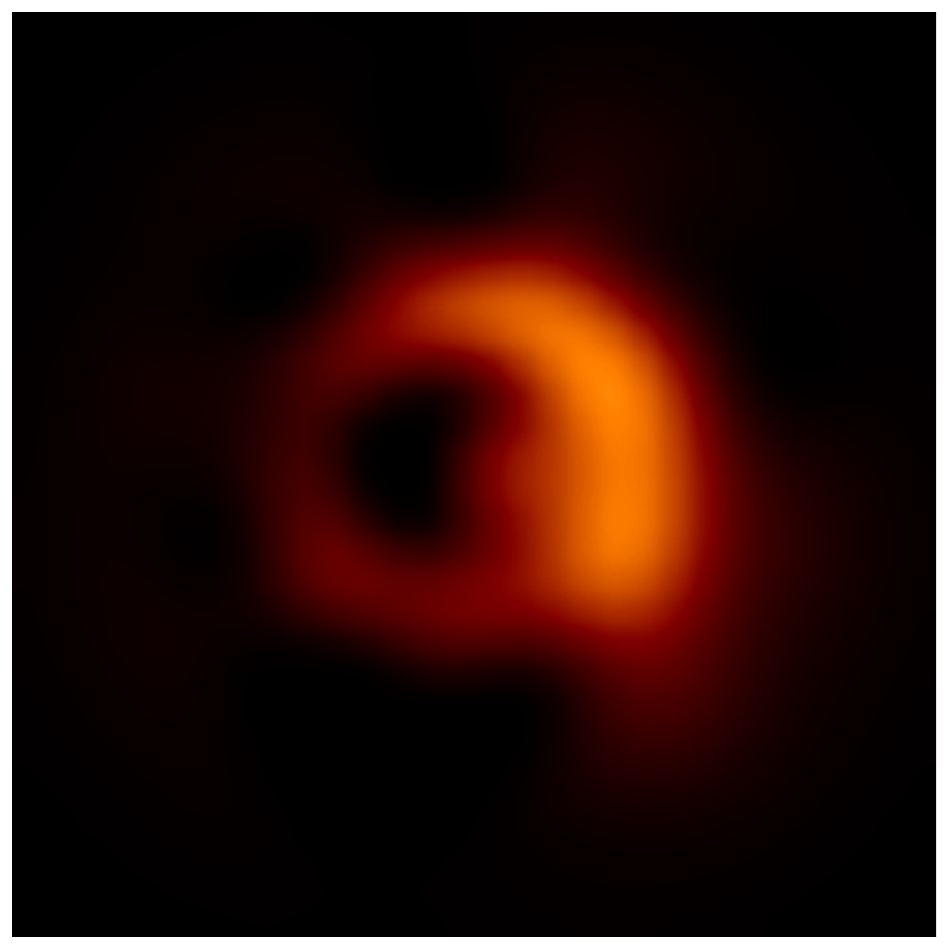}} } &
			\includegraphics[height=.1\linewidth]{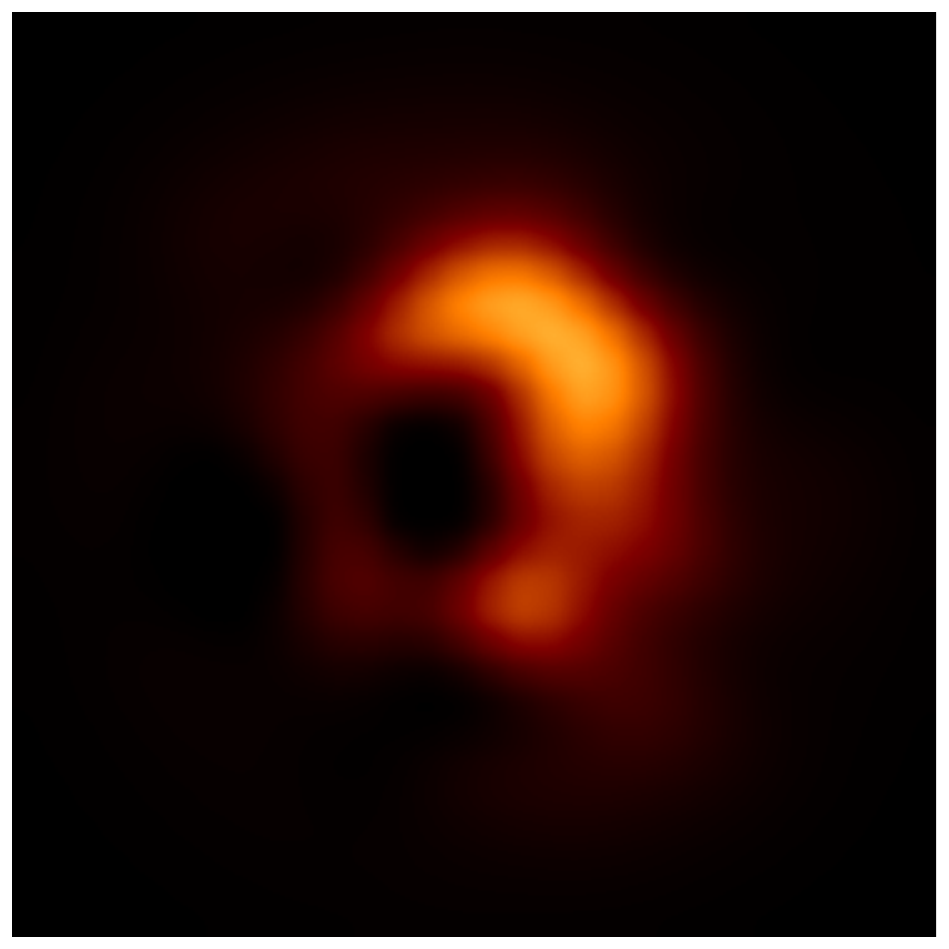} &
			\includegraphics[height=.1\linewidth]{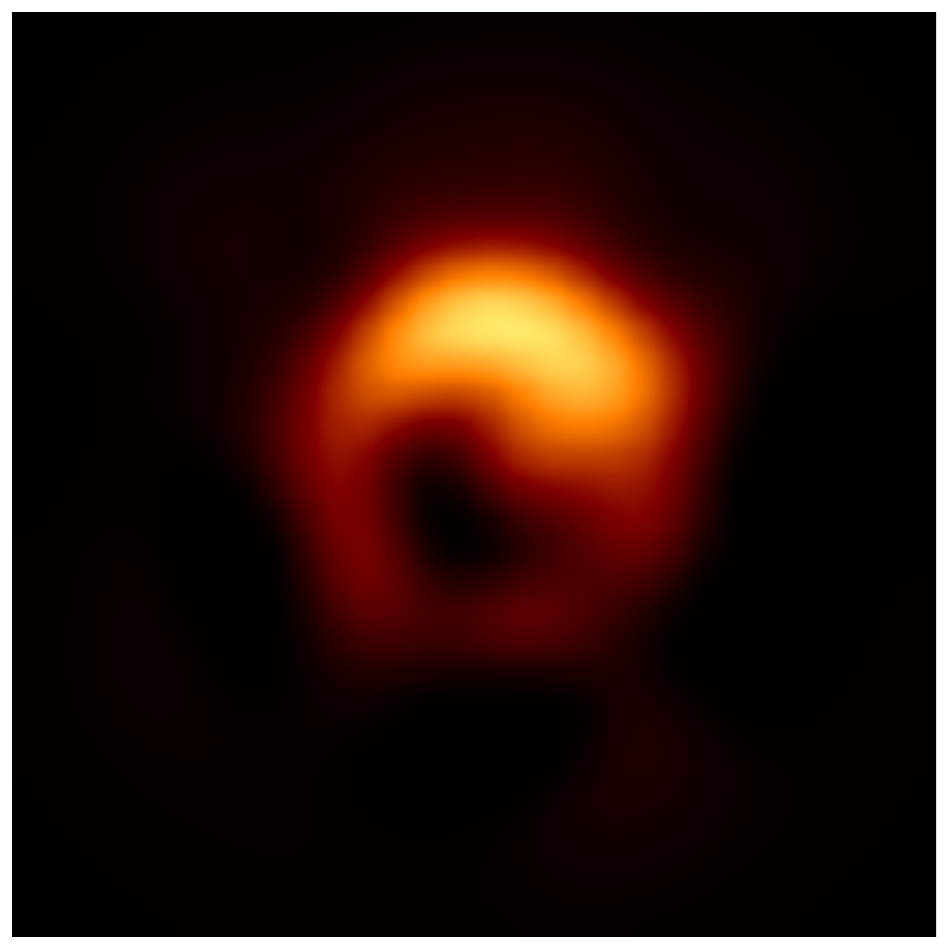} &
			\includegraphics[height=.1\linewidth]{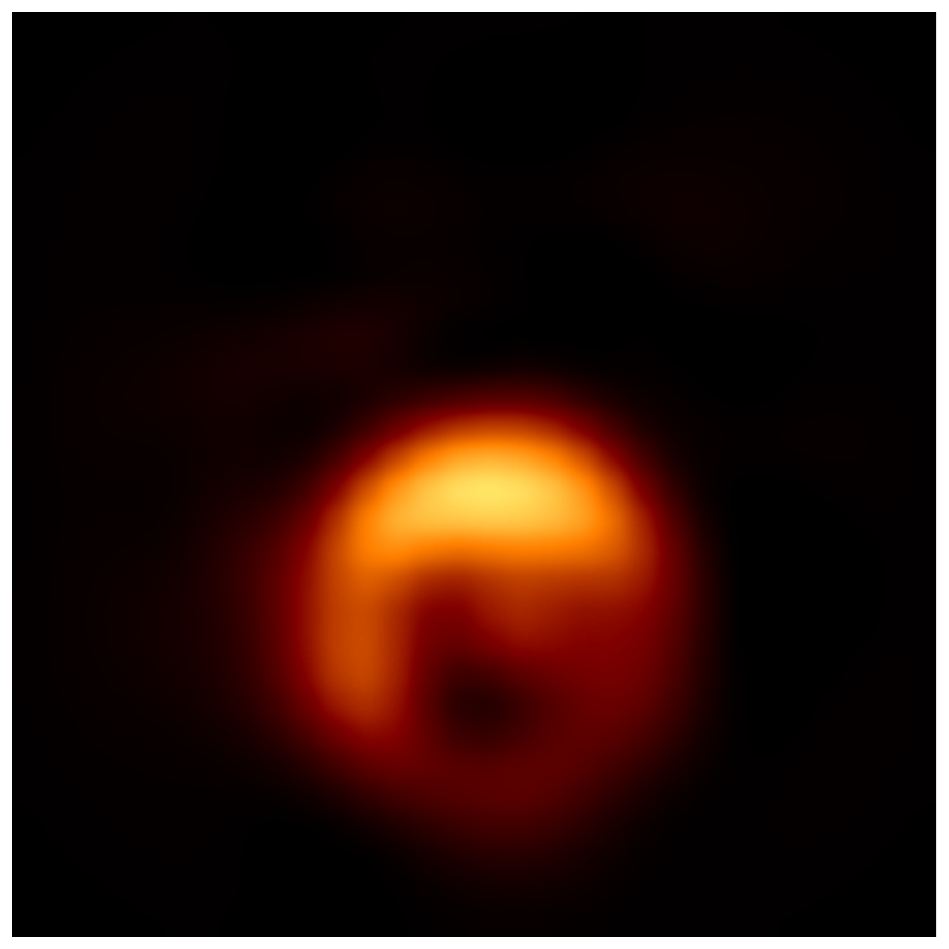} 
			&
			{{\includegraphics[height=.1\linewidth]{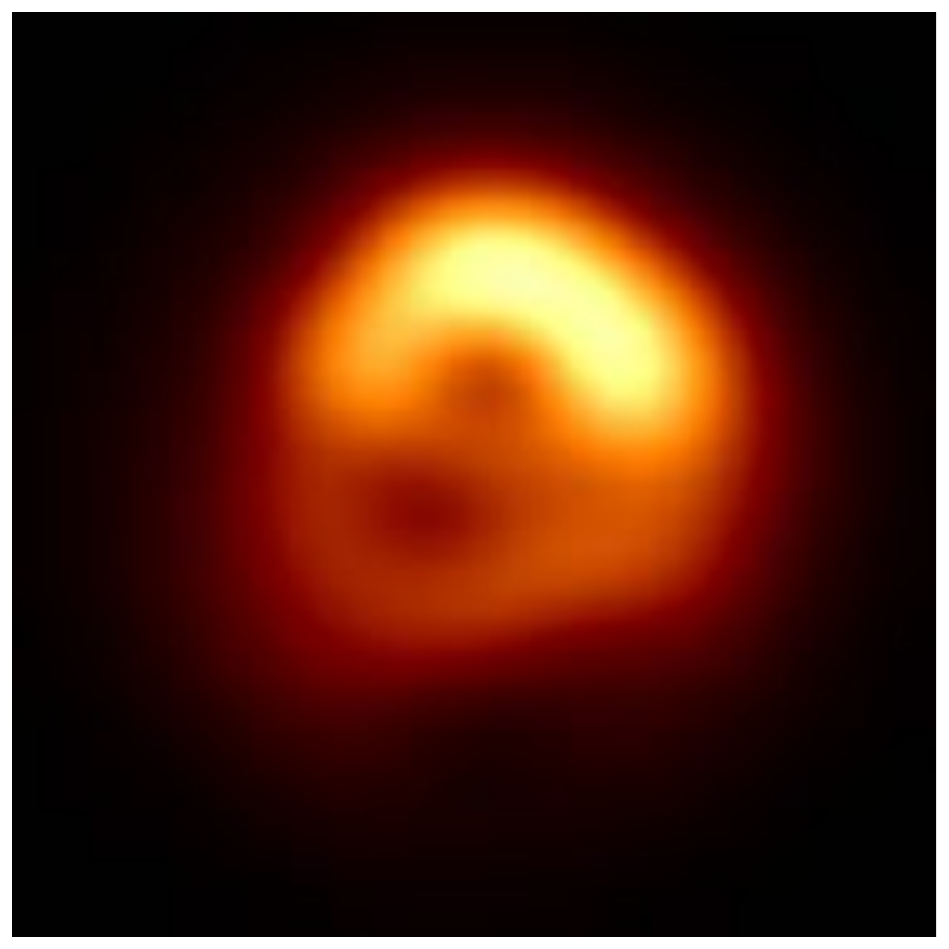}} } &
			\includegraphics[height=.1\linewidth]{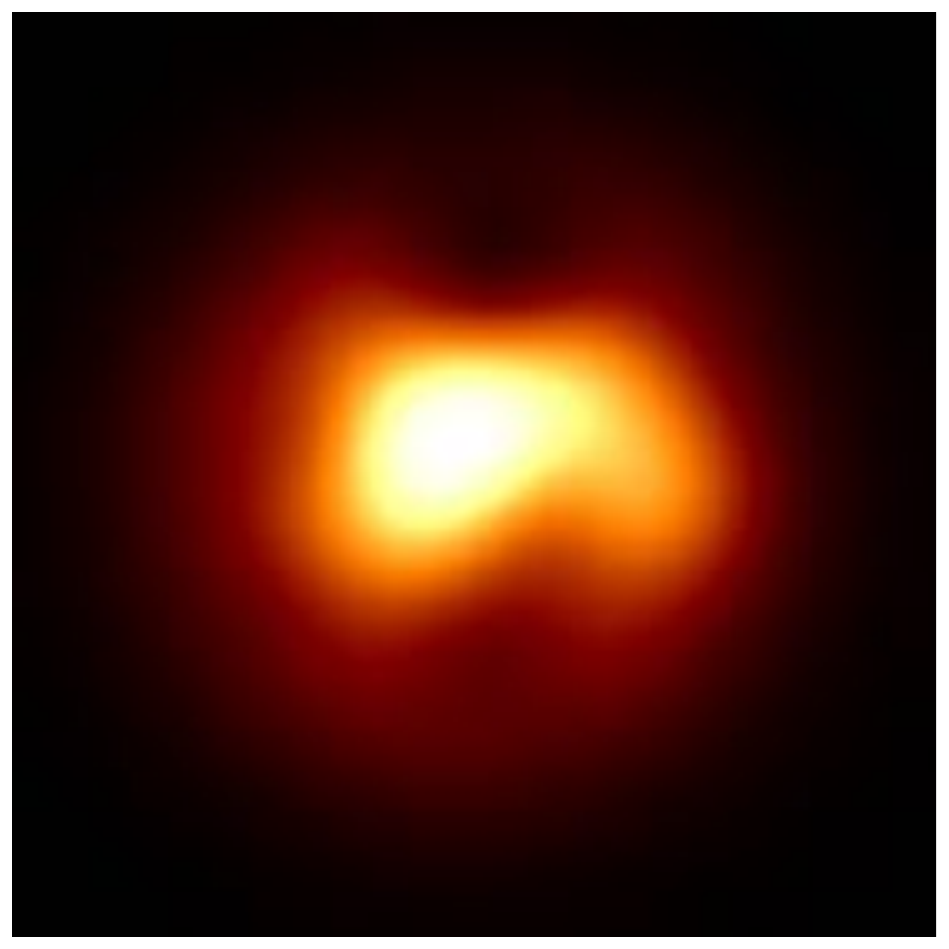} &
			\includegraphics[height=.1\linewidth]{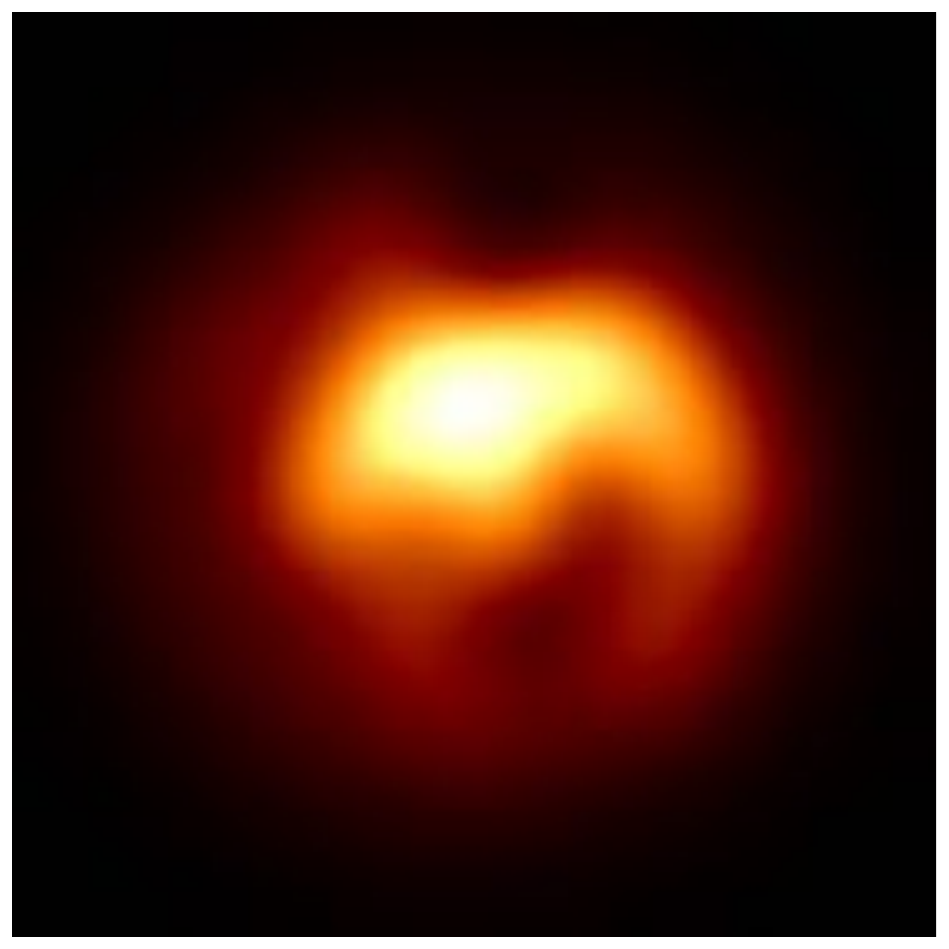} &
			\includegraphics[height=.1\linewidth]{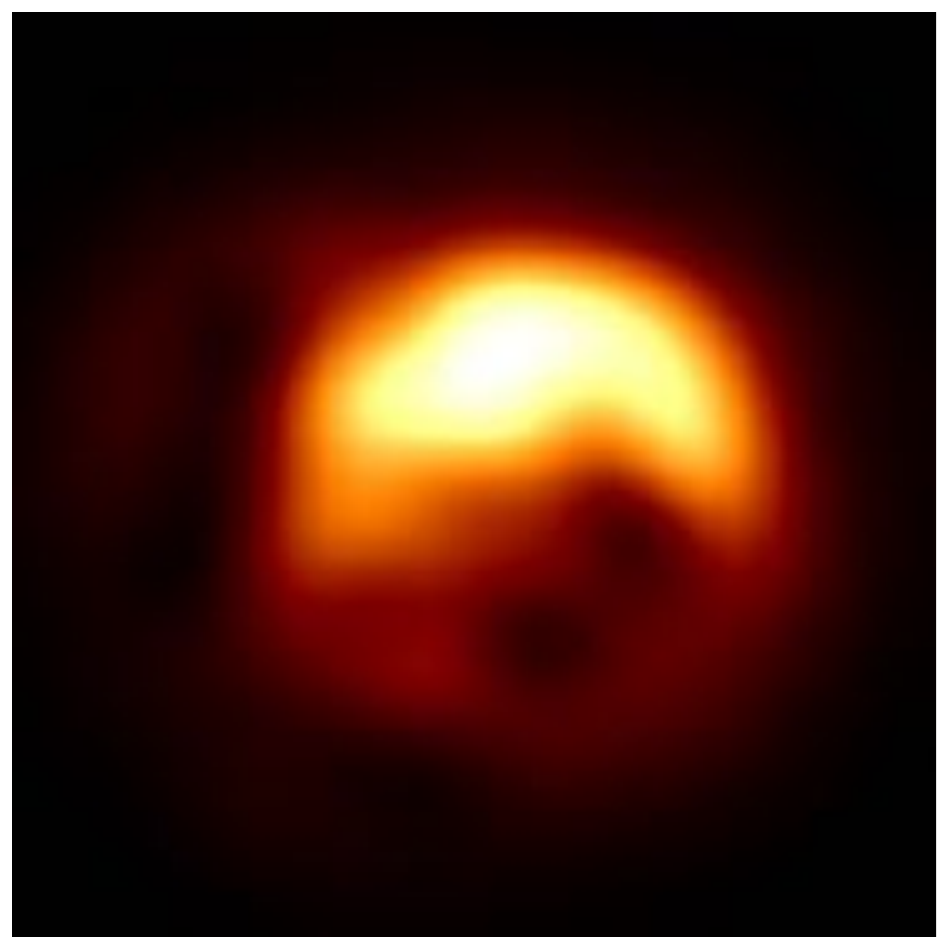} 
			\\   \hline
			&\vspace{-.1in} & & & & & & & &\\
			\multirow{2}{*}[0.6in]{ \rotatebox[origin=t]{90}{\small{\textsf{MEM \& TV Regularization}} }}  \hspace{-0.3in} & \multirow{1}{*}[0.4in]{ \rotatebox[origin=t]{90}{\small{\textsf{\cite{freek}}} }}
			&
			{{\includegraphics[height=.1\linewidth]{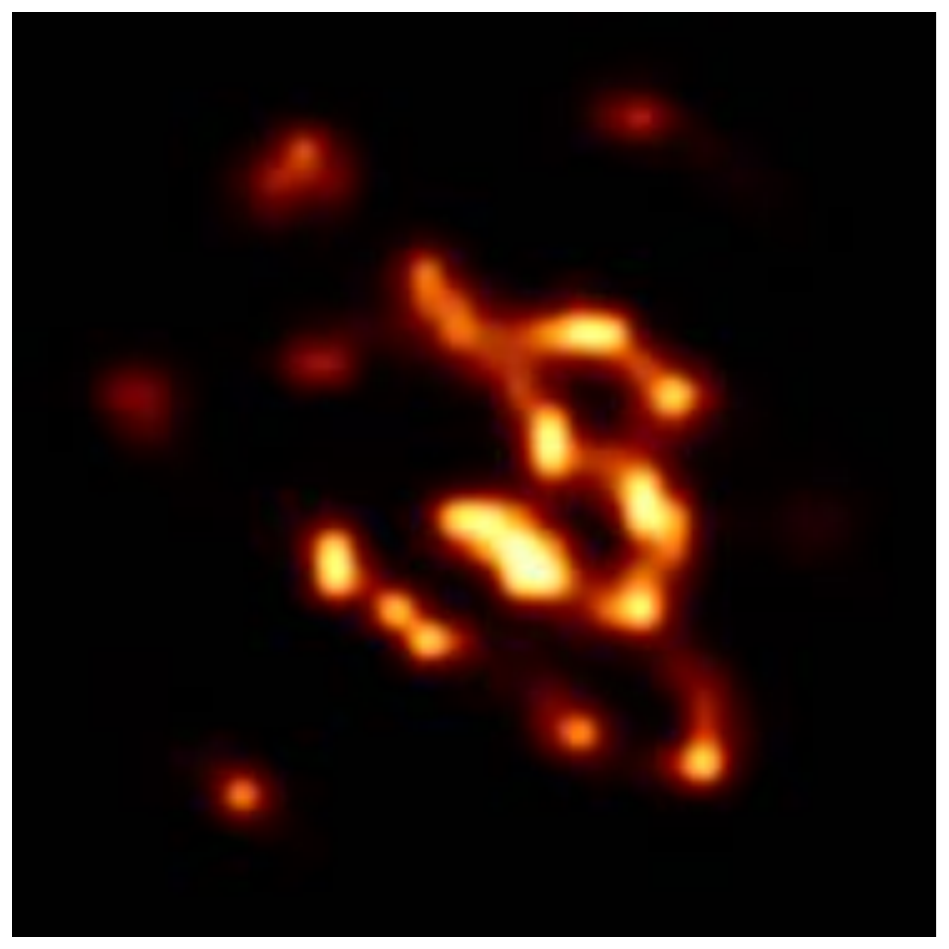}} } &
			\includegraphics[height=.1\linewidth]{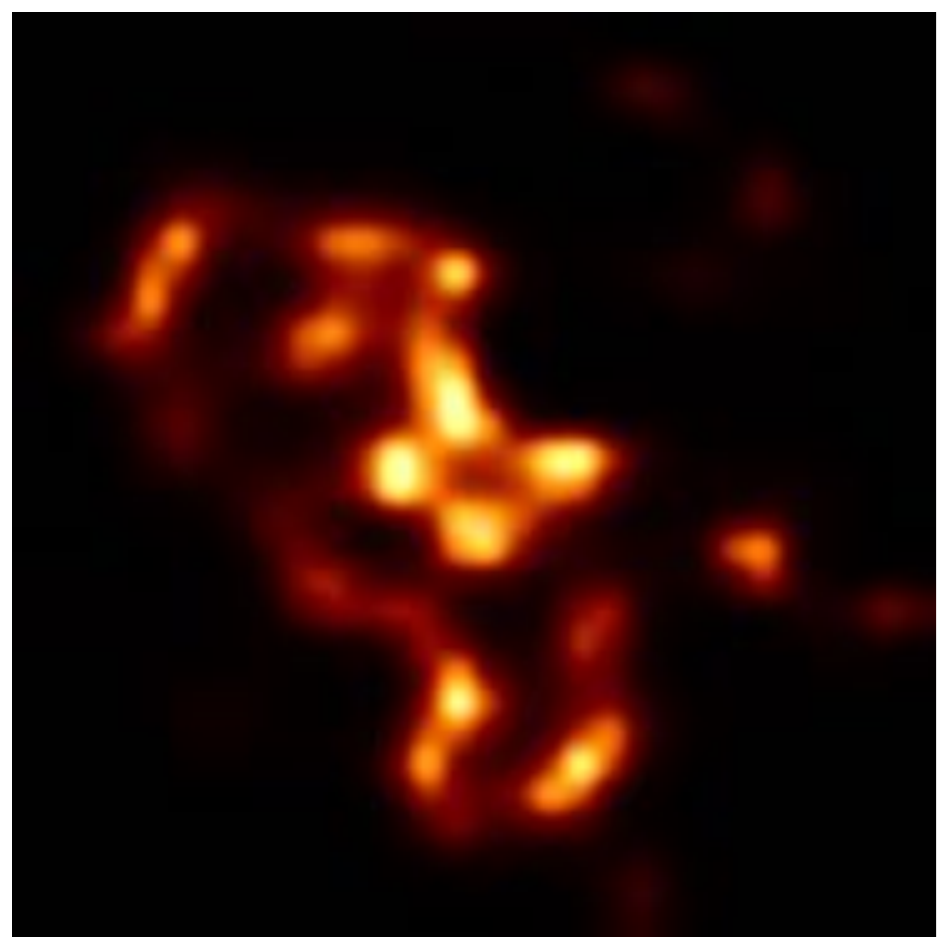} &
			\includegraphics[height=.1\linewidth]{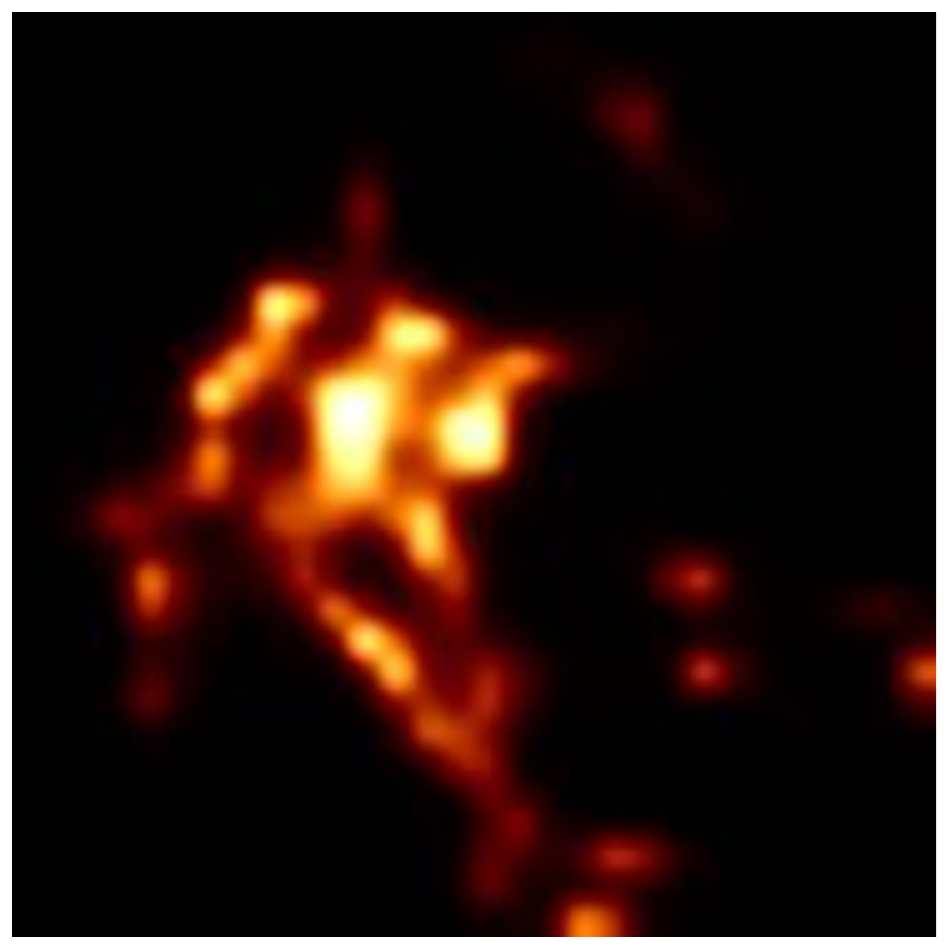} &
			\includegraphics[height=.1\linewidth]{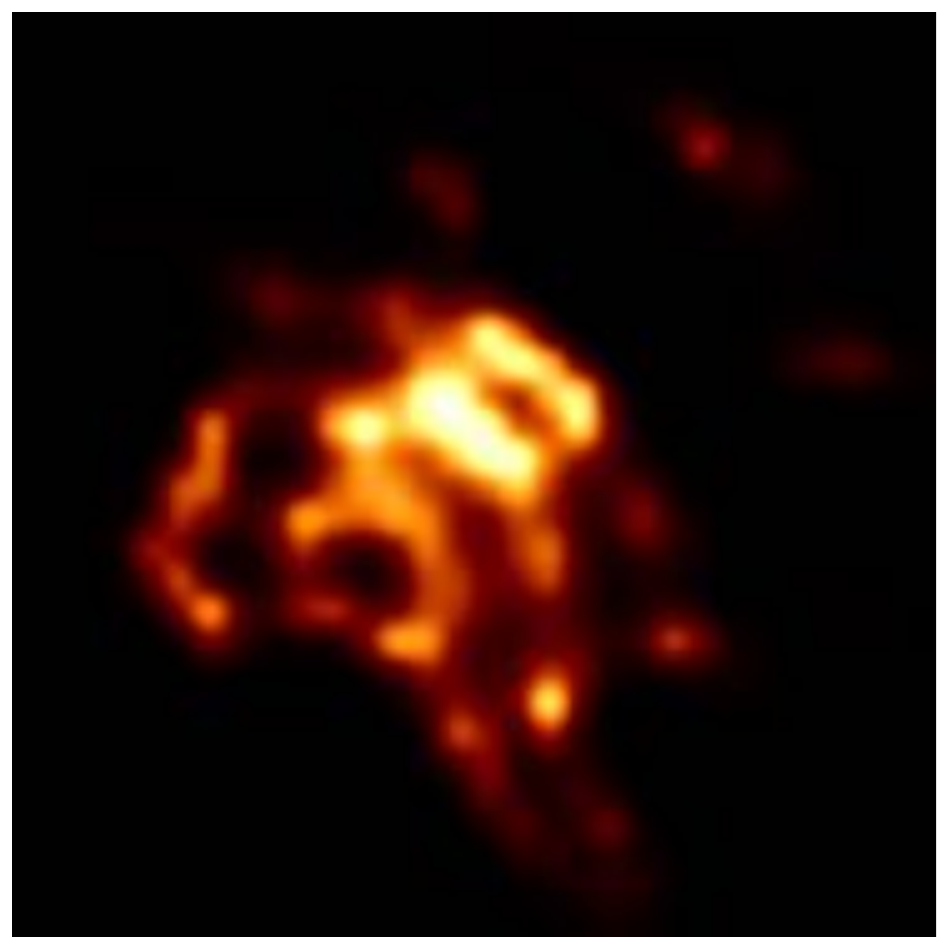} 
			&
			{{\includegraphics[height=.1\linewidth]{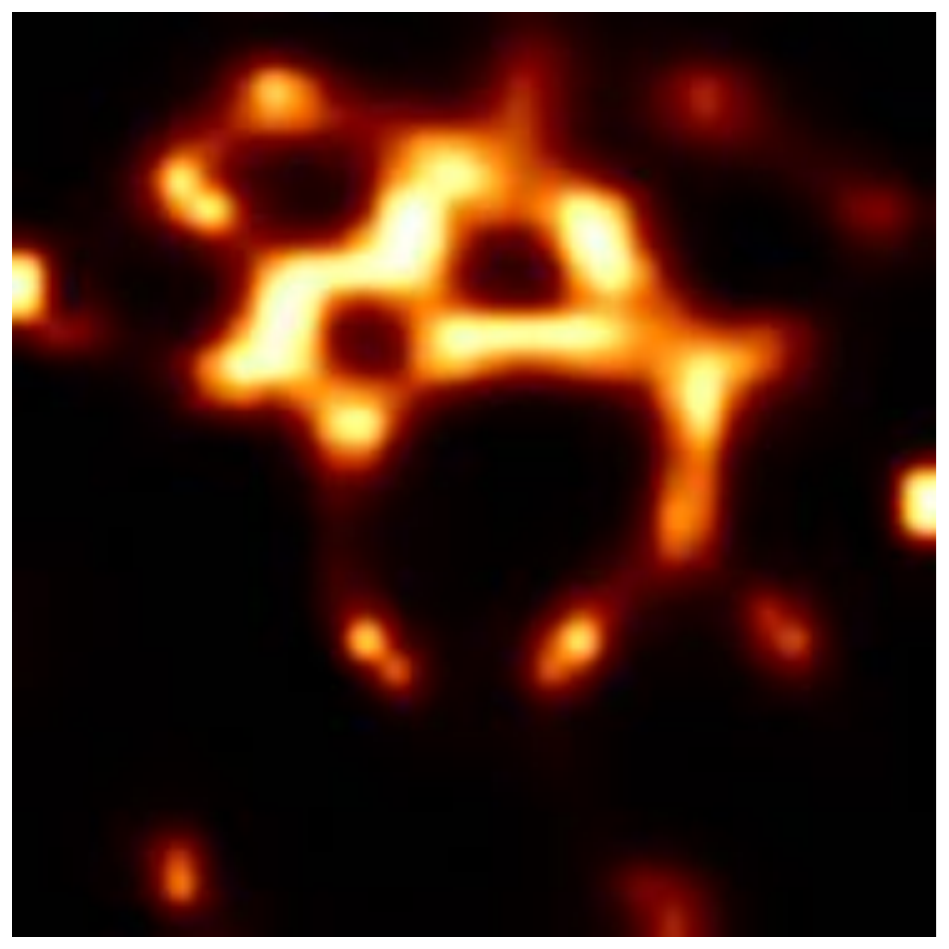}} } &
			\includegraphics[height=.1\linewidth]{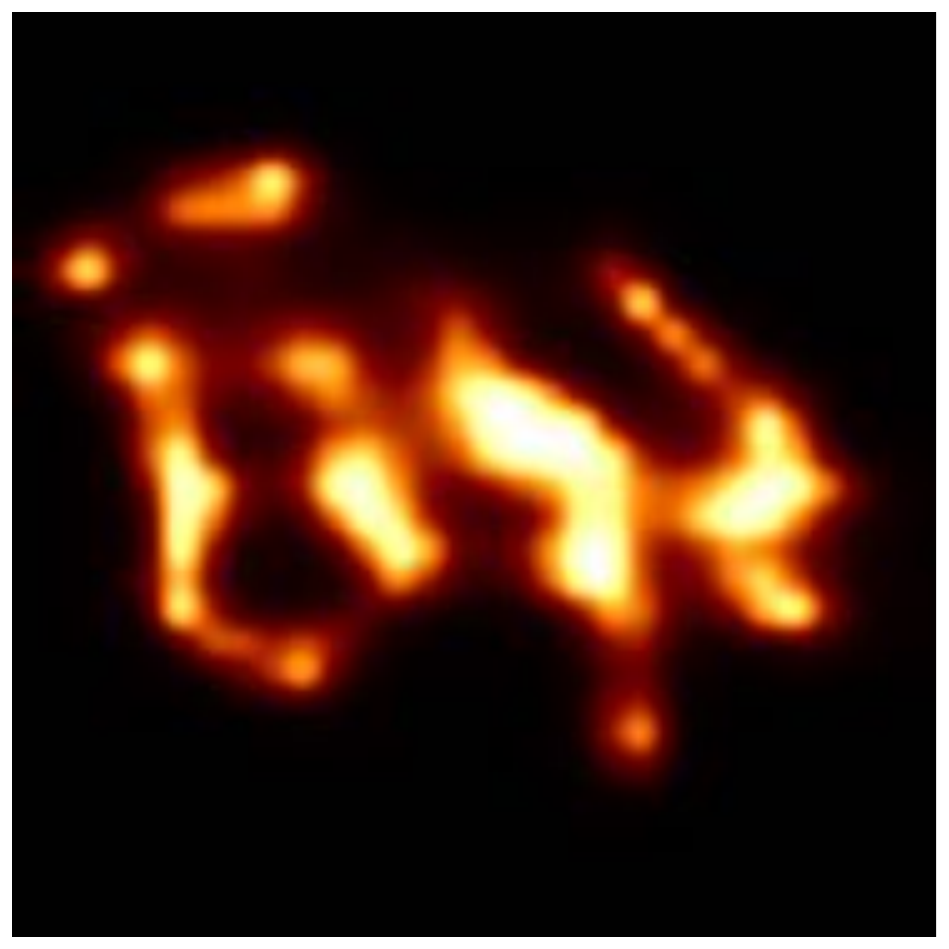} &
			\includegraphics[height=.1\linewidth]{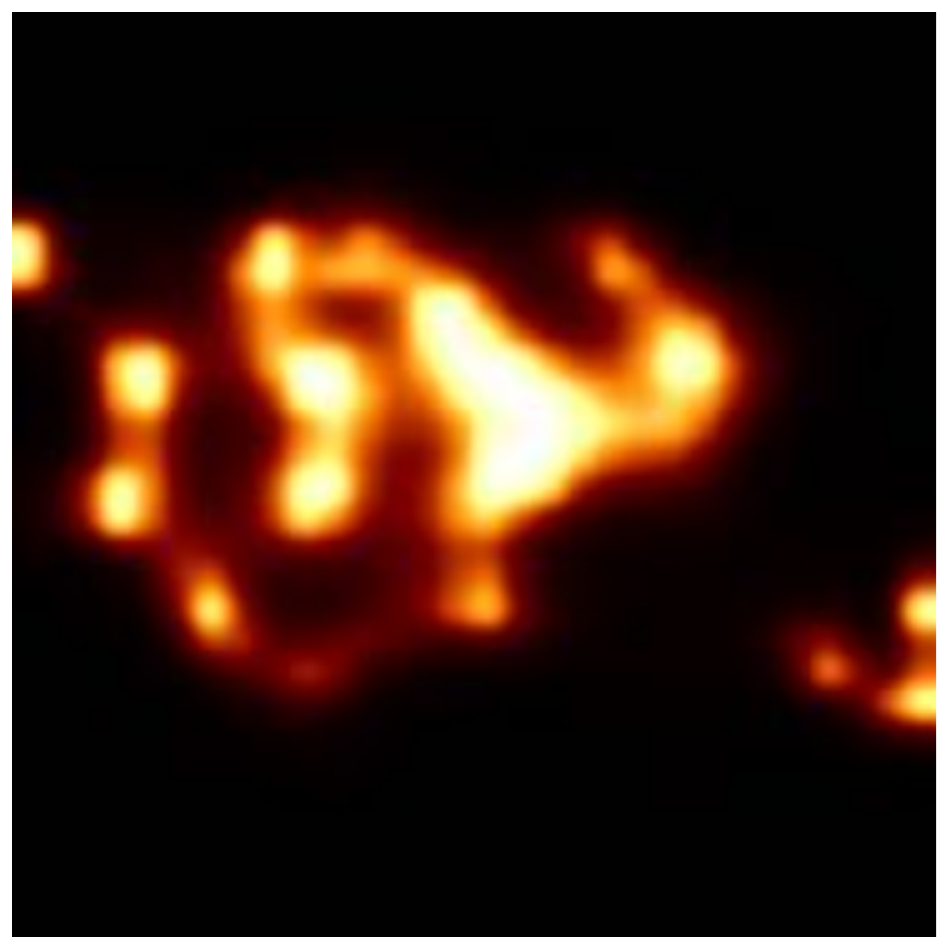} &
			\includegraphics[height=.1\linewidth]{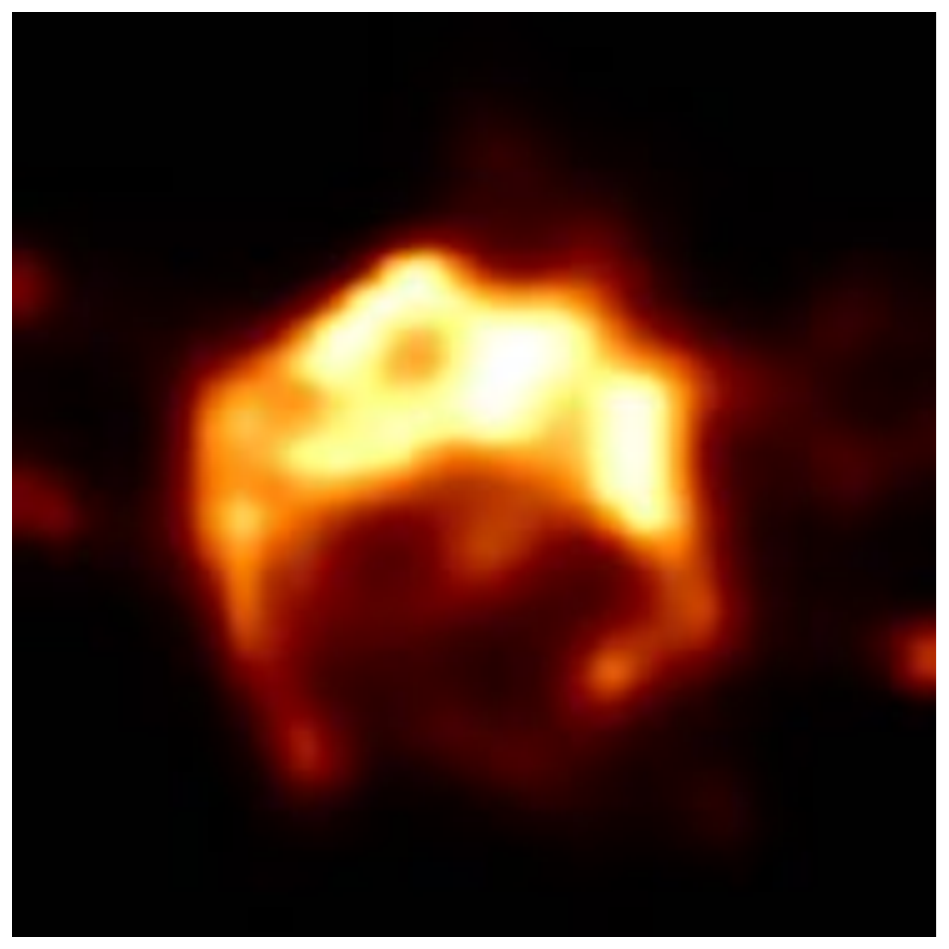} 
			\\
			&\vspace{-.1in} & & & & & & & &\\
			&	\multirow{1}{*}[0.4in]{ \rotatebox[origin=t]{90}{\small{\textsf{\cite{andrew}}} }}
			&
			{{\includegraphics[height=.1\linewidth]{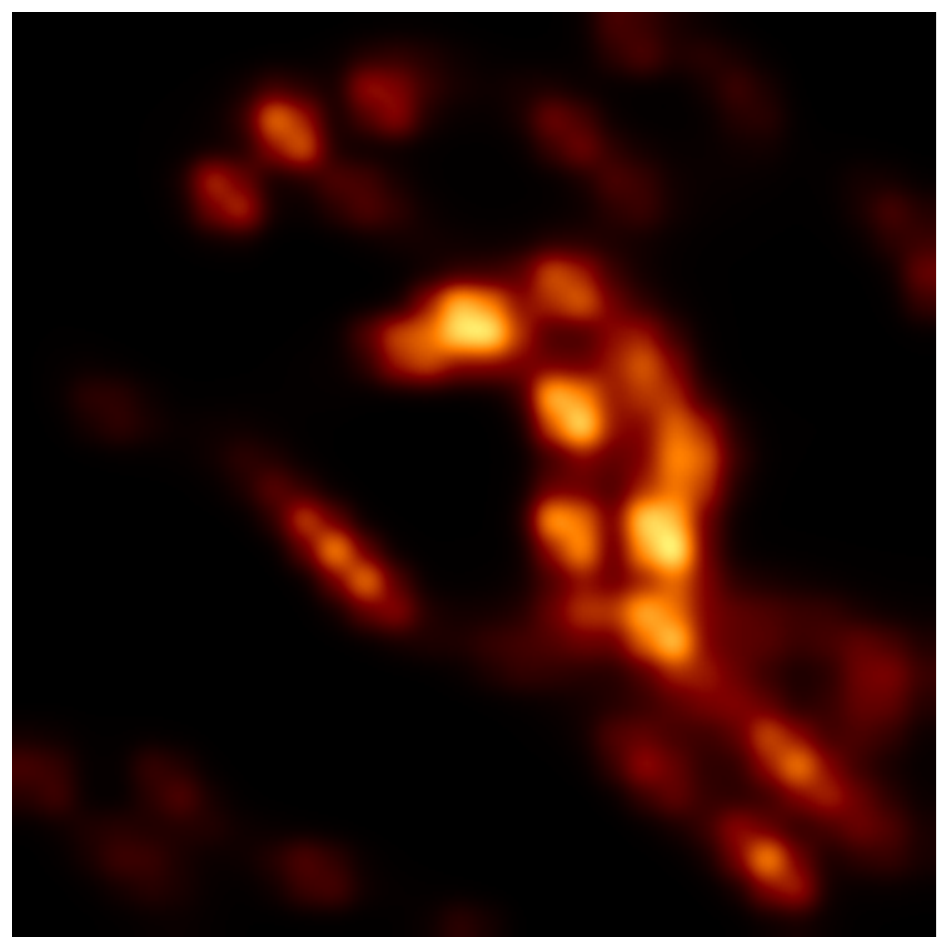}} } &
			\includegraphics[height=.1\linewidth]{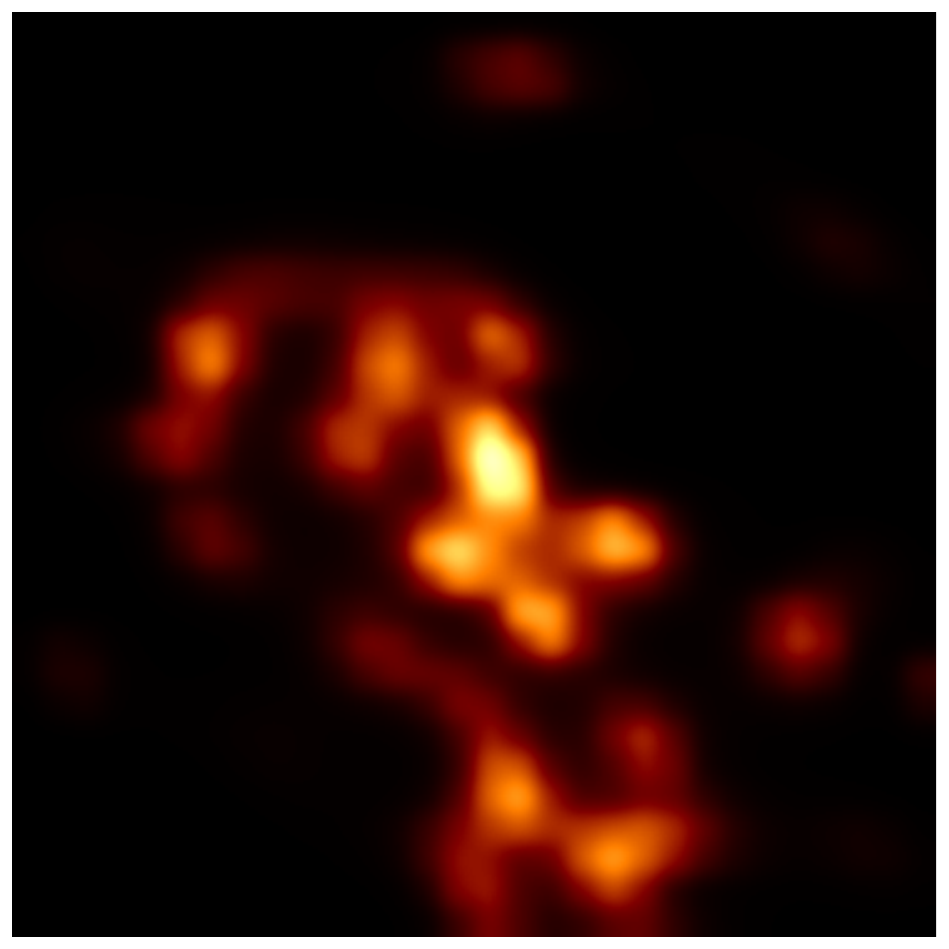} &
			\includegraphics[height=.1\linewidth]{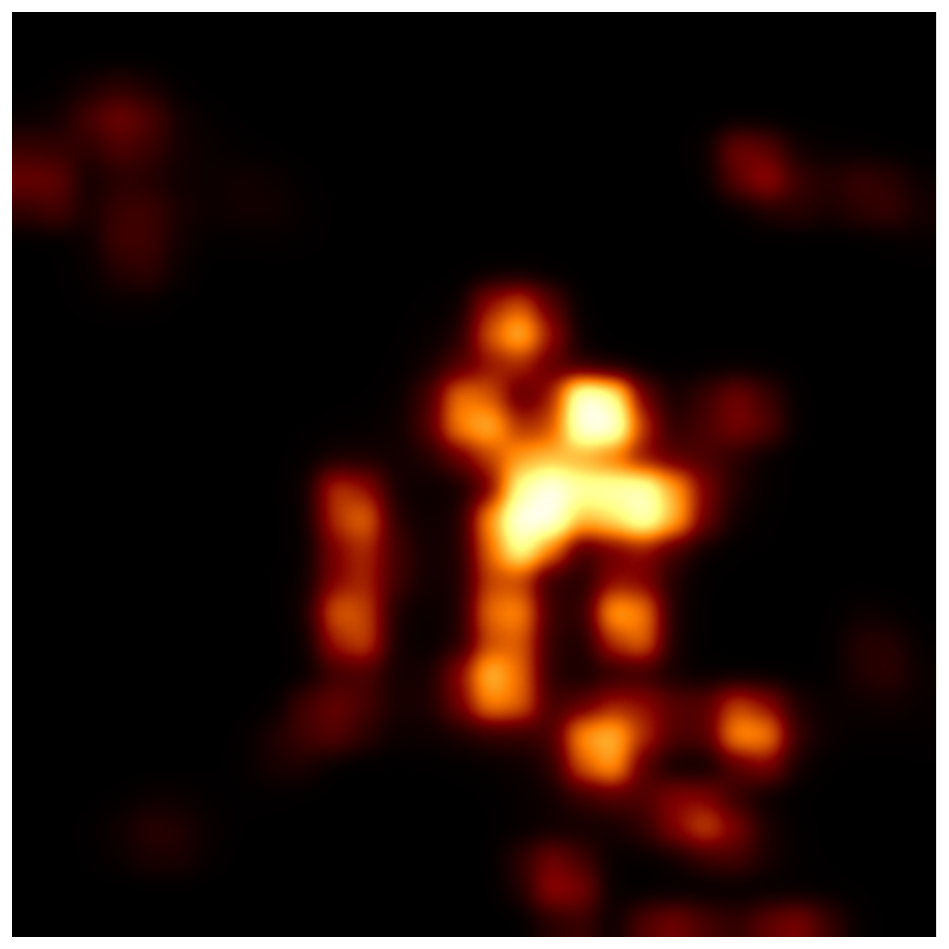} &
			\includegraphics[height=.1\linewidth]{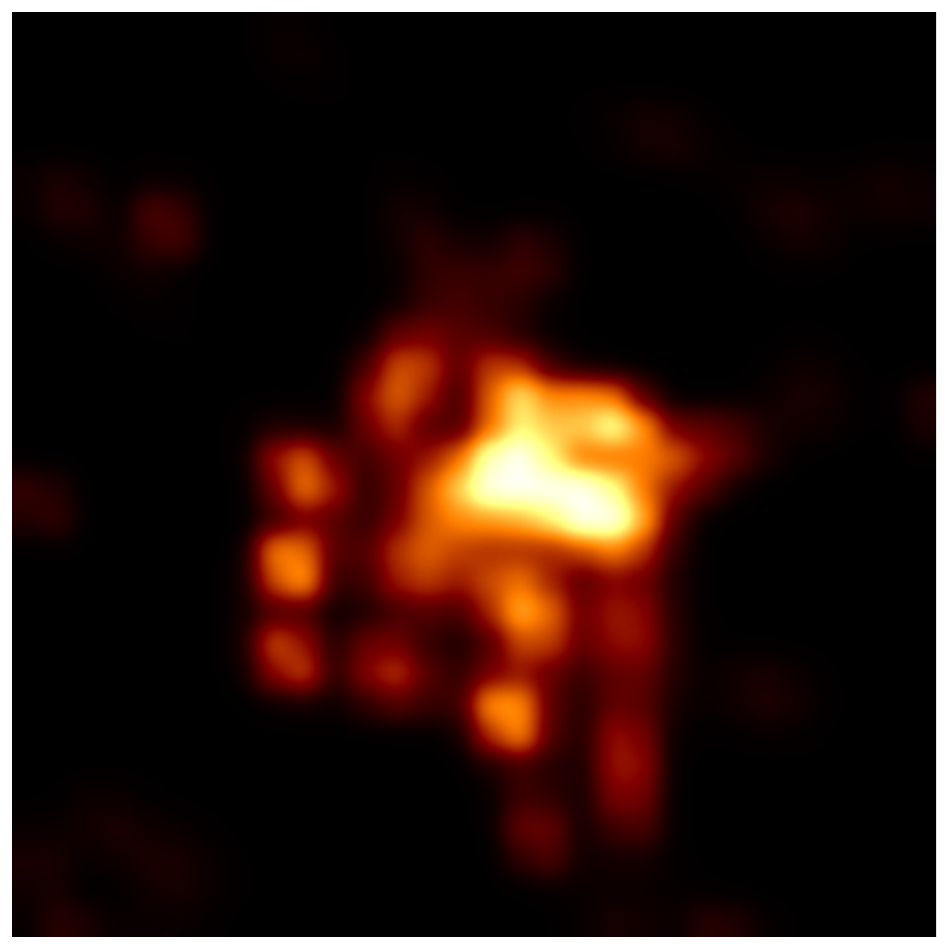} 
			&
			{{\includegraphics[height=.1\linewidth]{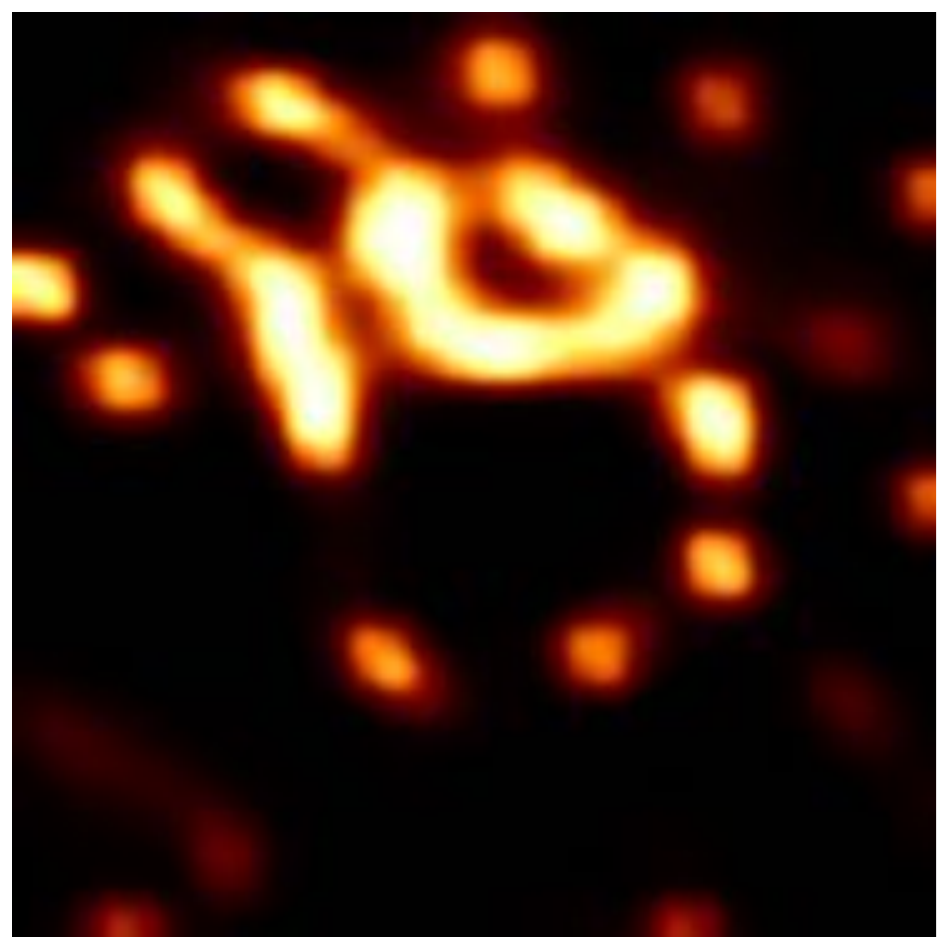}} } &
			\includegraphics[height=.1\linewidth]{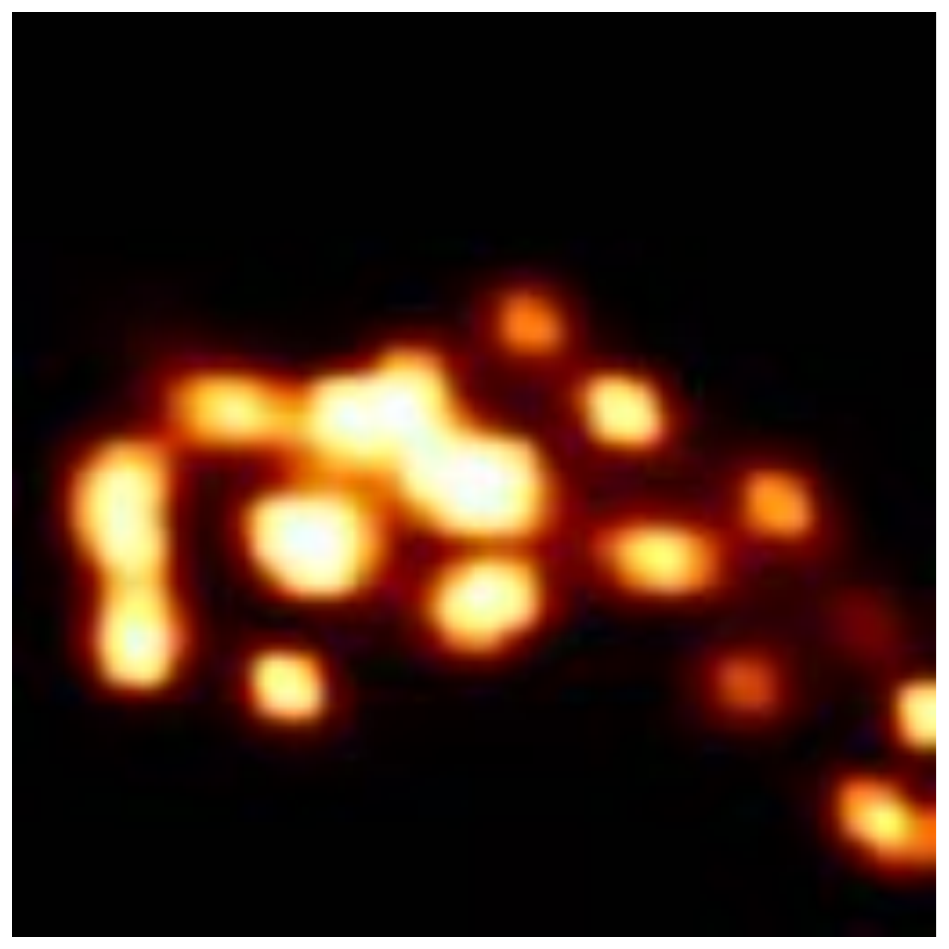} &
			\includegraphics[height=.1\linewidth]{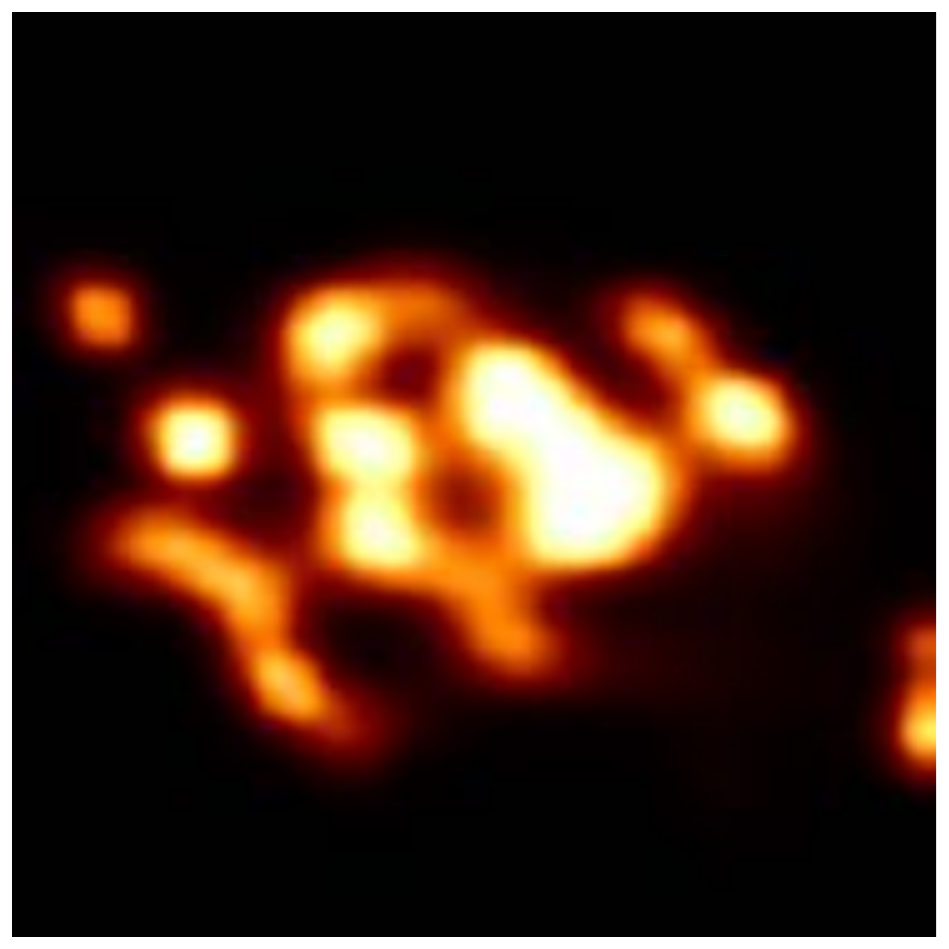} &
			\includegraphics[height=.1\linewidth]{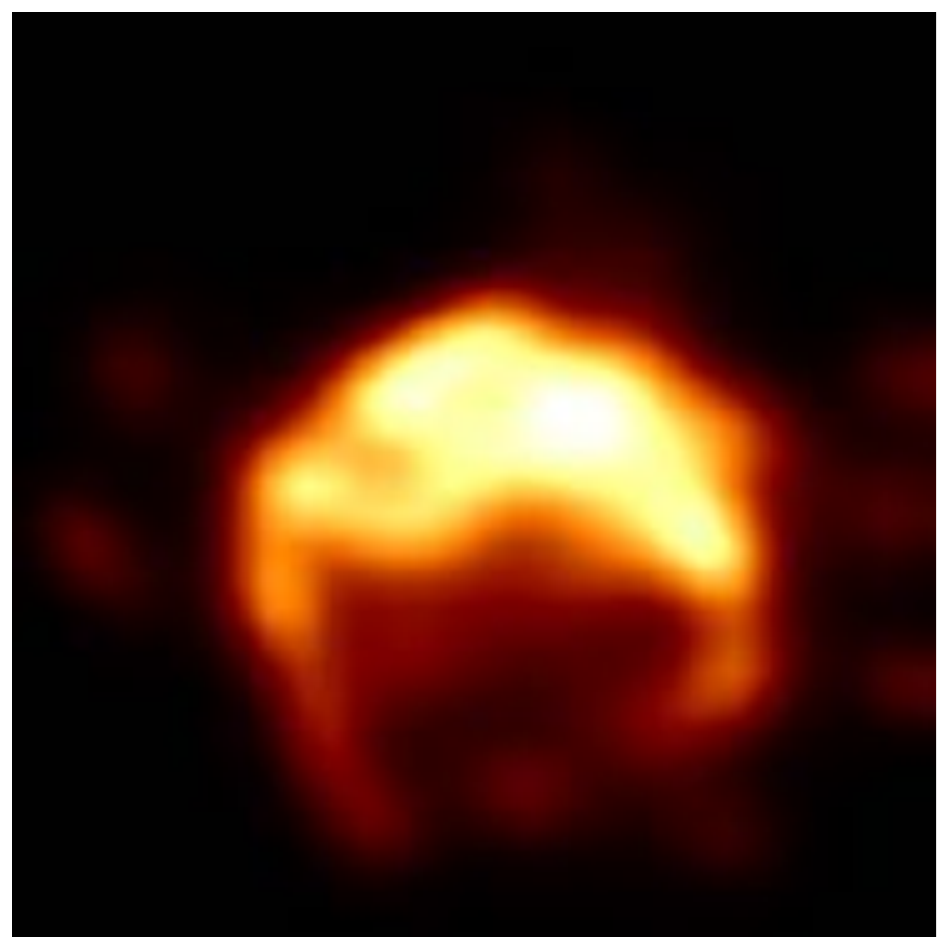} 
			\\   \thickhline
			
			&\vspace{-.1in} & & & & & & & &\\
			 \multirow{2}{*}[0.6in]{ \rotatebox[origin=t]{90}{\small{\textsf{StarWarps}} }}   \hspace{-0.3in} &	\multirow{1}{*}[0.45in]{ \rotatebox[origin=t]{90}{\small{\textsf{Mean}} }}
			&
			{{\includegraphics[height=.1\linewidth]{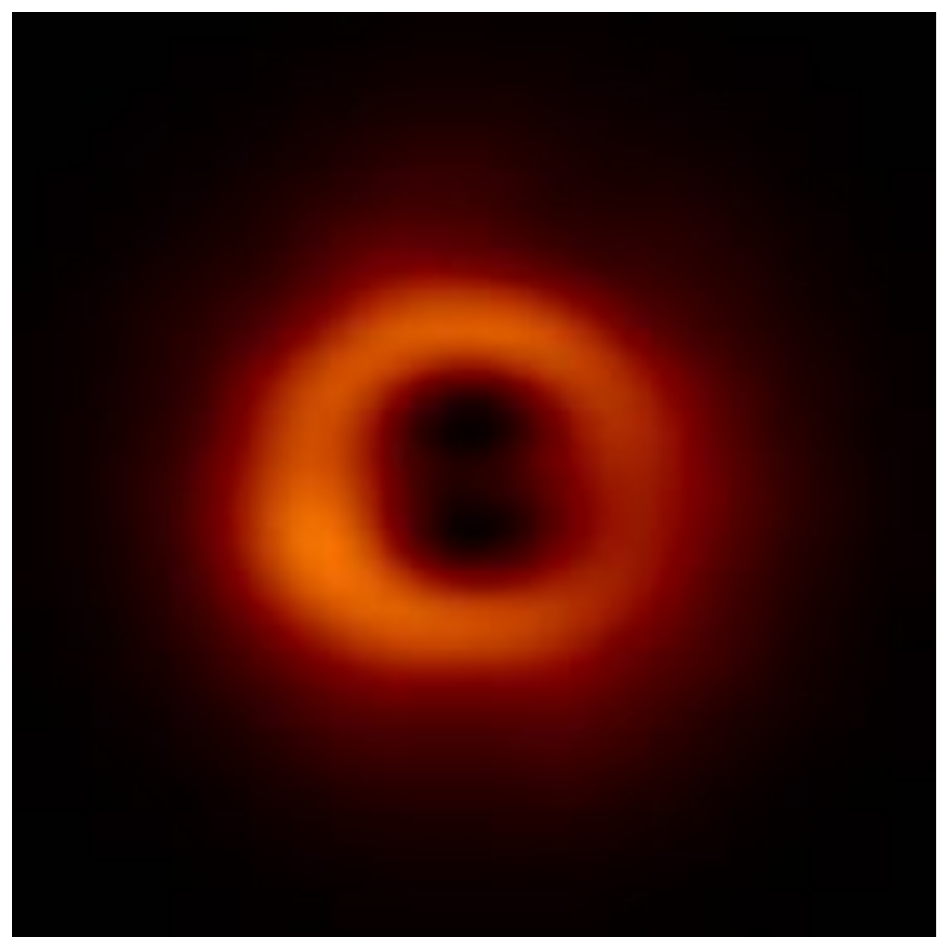}} } 
			&
			\includegraphics[height=.1\linewidth]{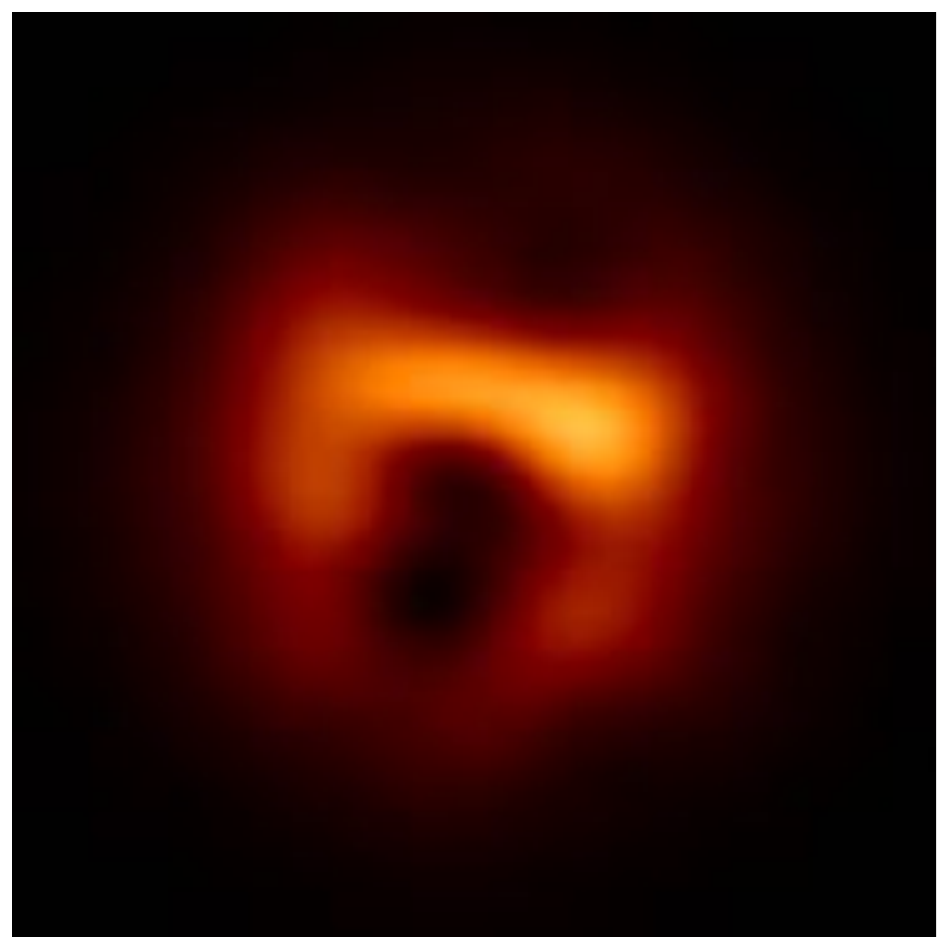} 
			&
			\includegraphics[height=.1\linewidth]{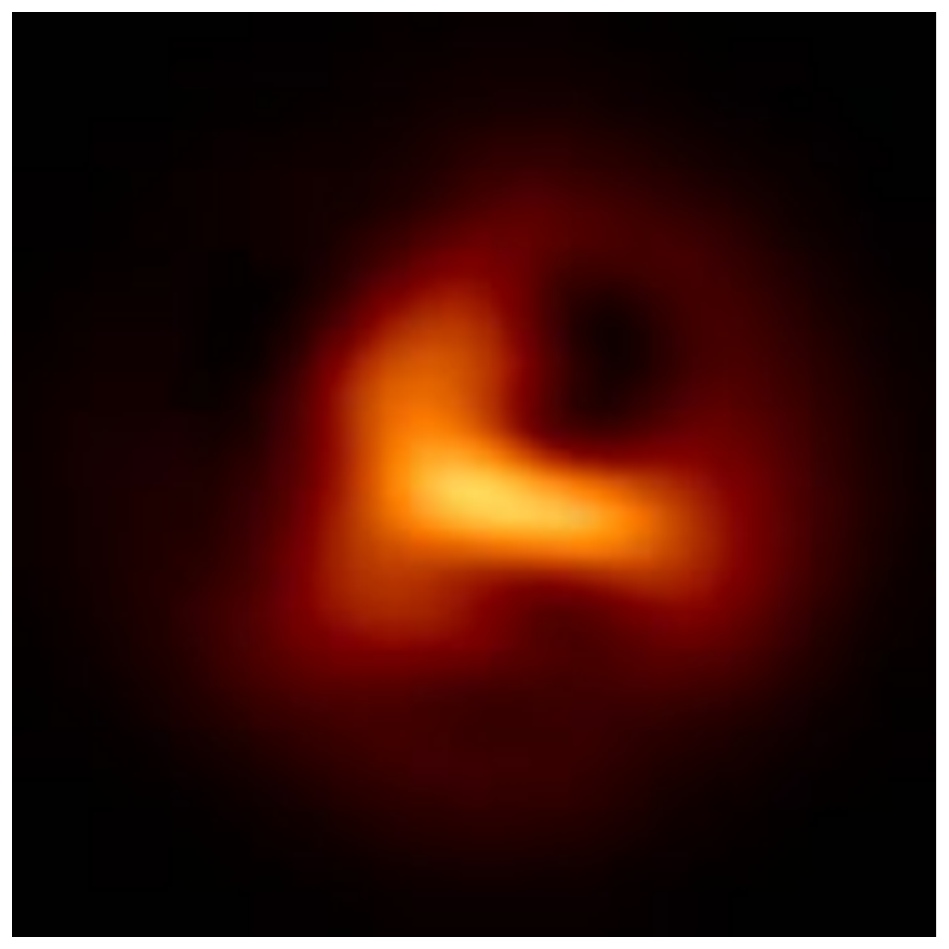} &
			\includegraphics[height=.1\linewidth]{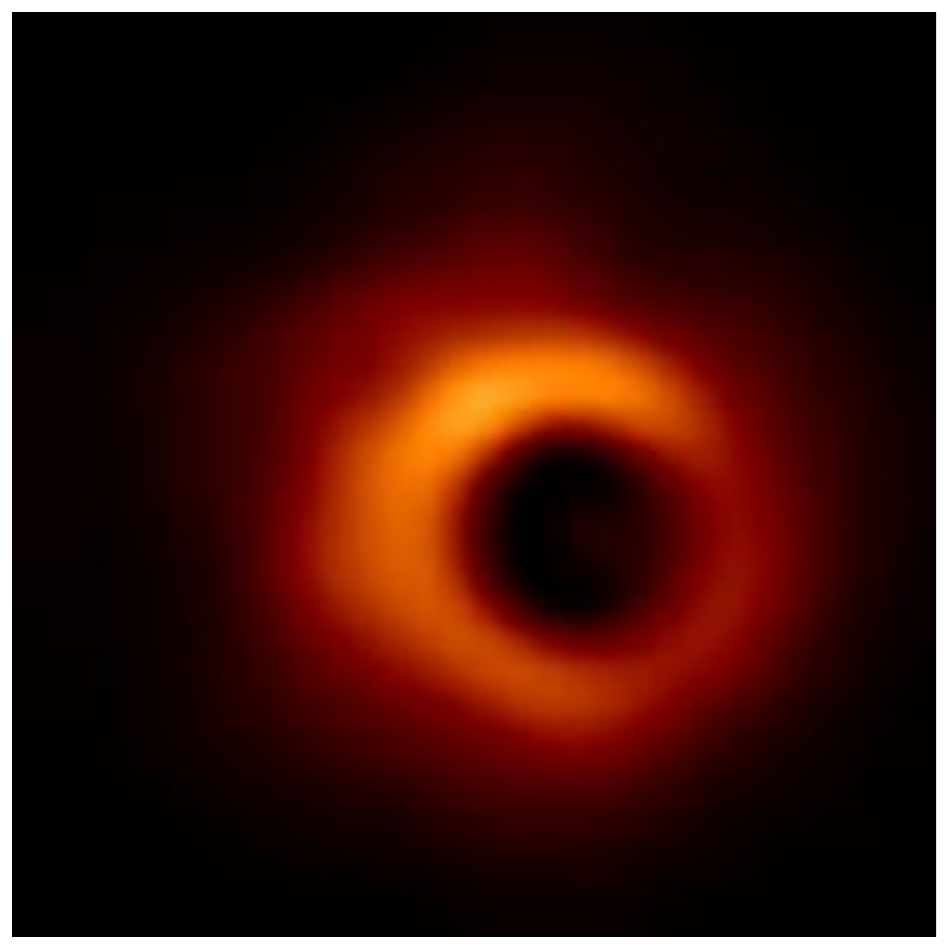} 
			&
			{{\includegraphics[height=.1\linewidth]{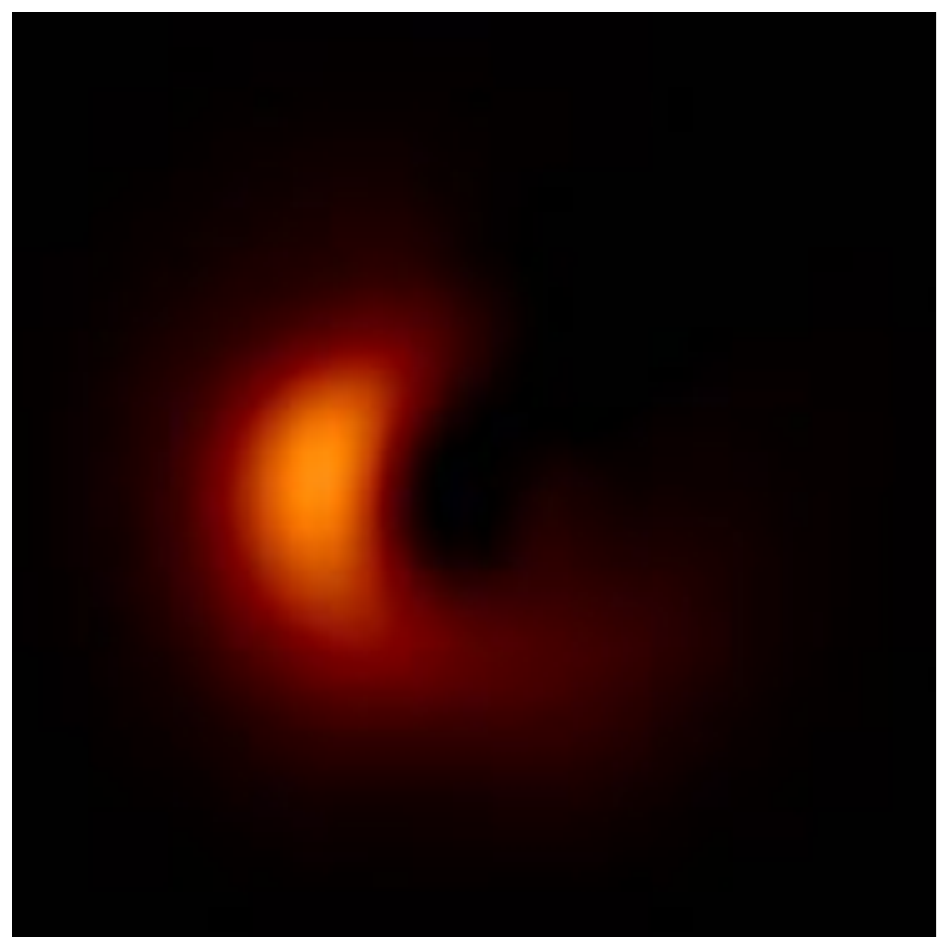}} } &
			\includegraphics[height=.1\linewidth]{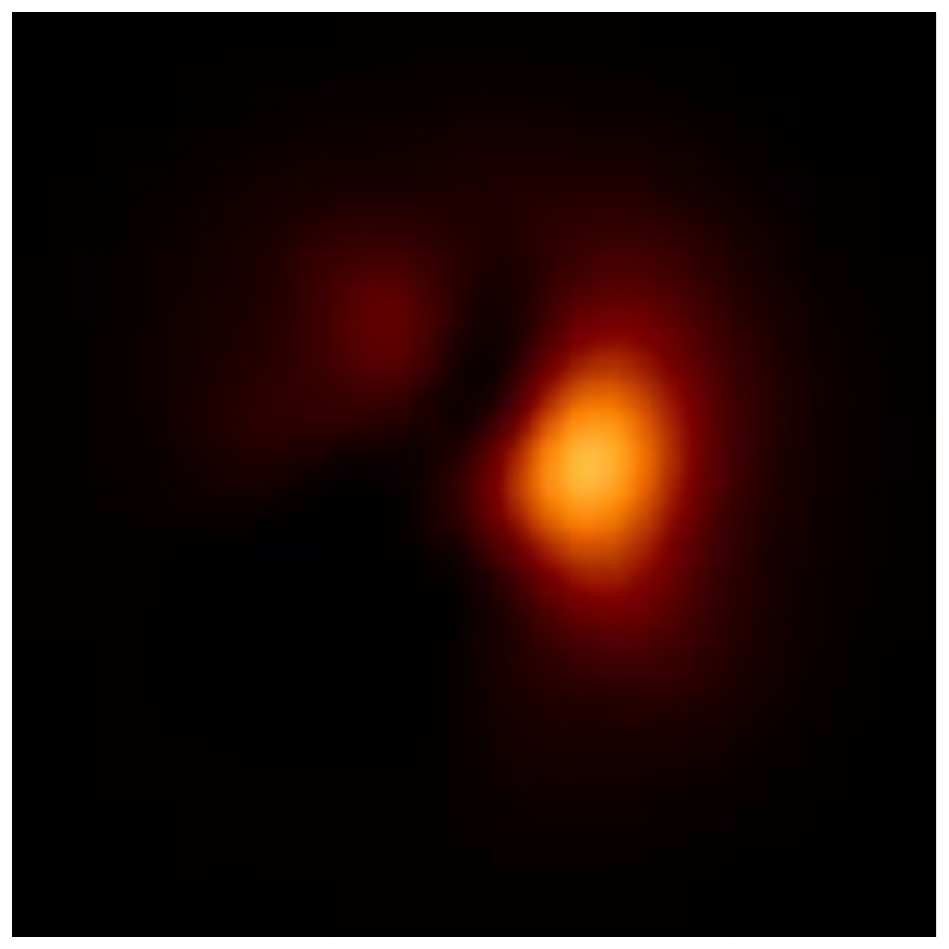} &
			\includegraphics[height=.1\linewidth]{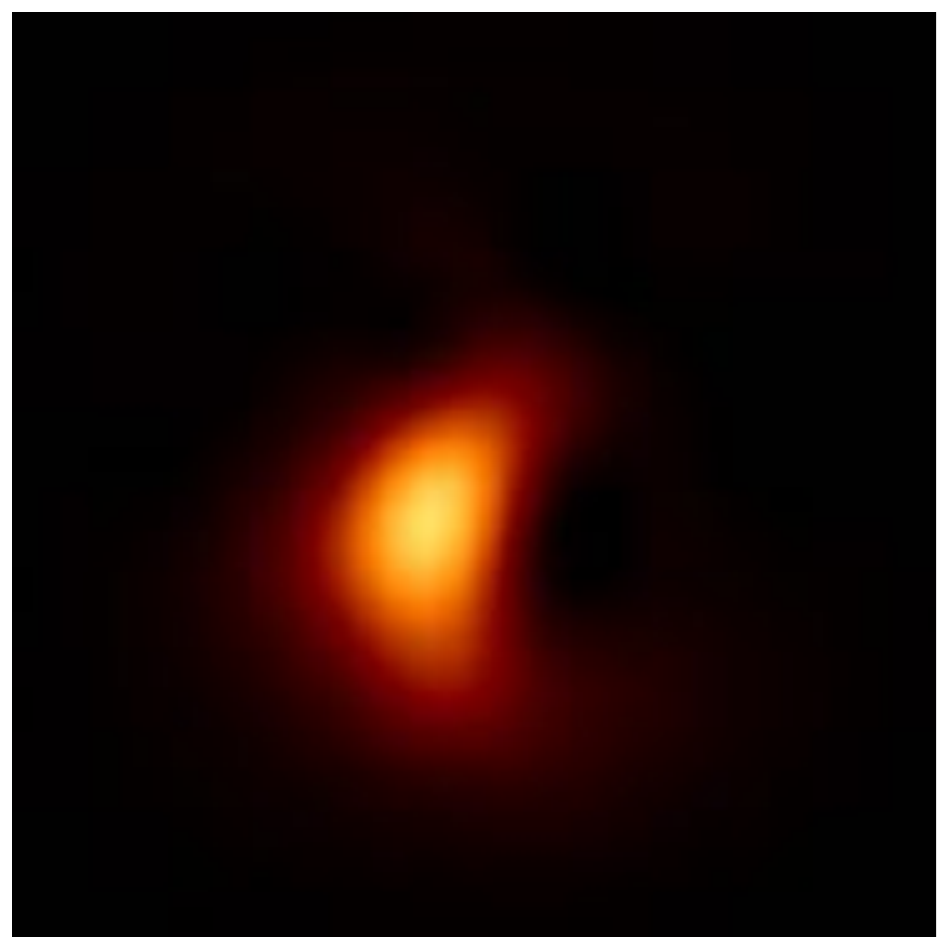} &
			\includegraphics[height=.1\linewidth]{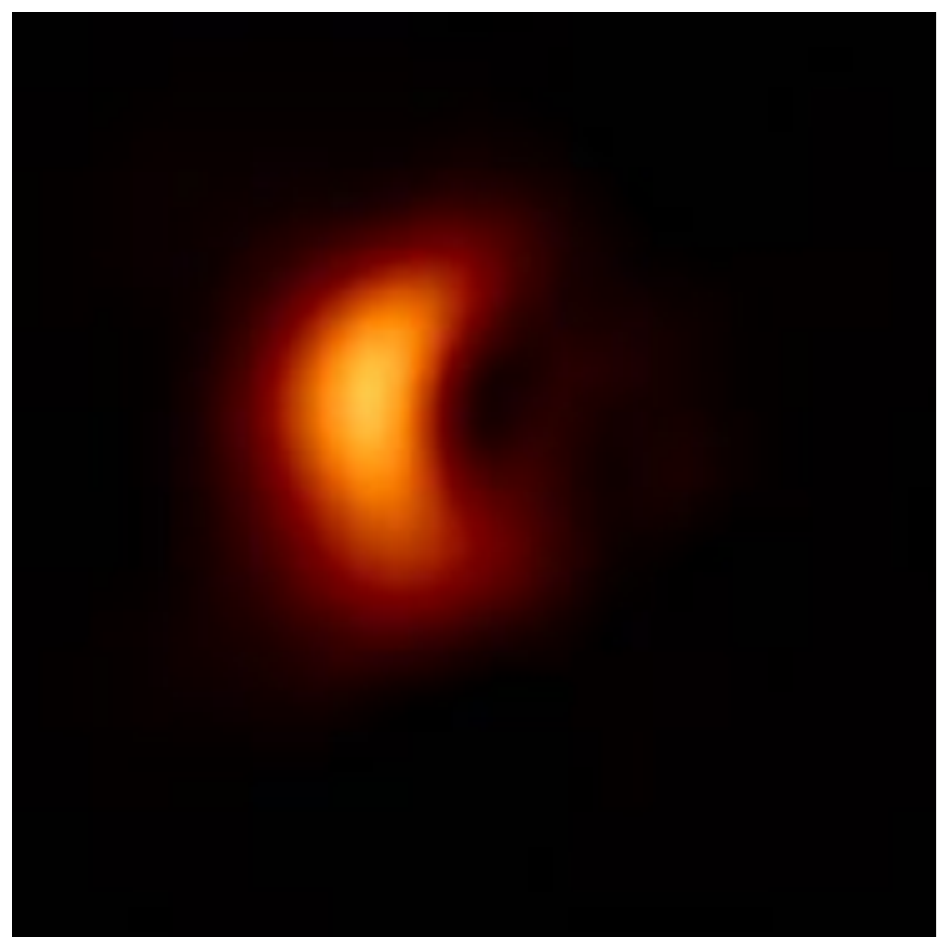} 
			\\   \hline
			&\vspace{-.1in} & & & & & & & &\\
			\multirow{2}{*}[0.6in]{ \rotatebox[origin=t]{90}{\small{\textsf{MEM \& TV Regularization}} }}  \hspace{-0.3in} & 	\multirow{1}{*}[0.4in]{ \rotatebox[origin=t]{90}{\small{\textsf{\cite{freek}}} }}
			&
			{{\includegraphics[height=.1\linewidth]{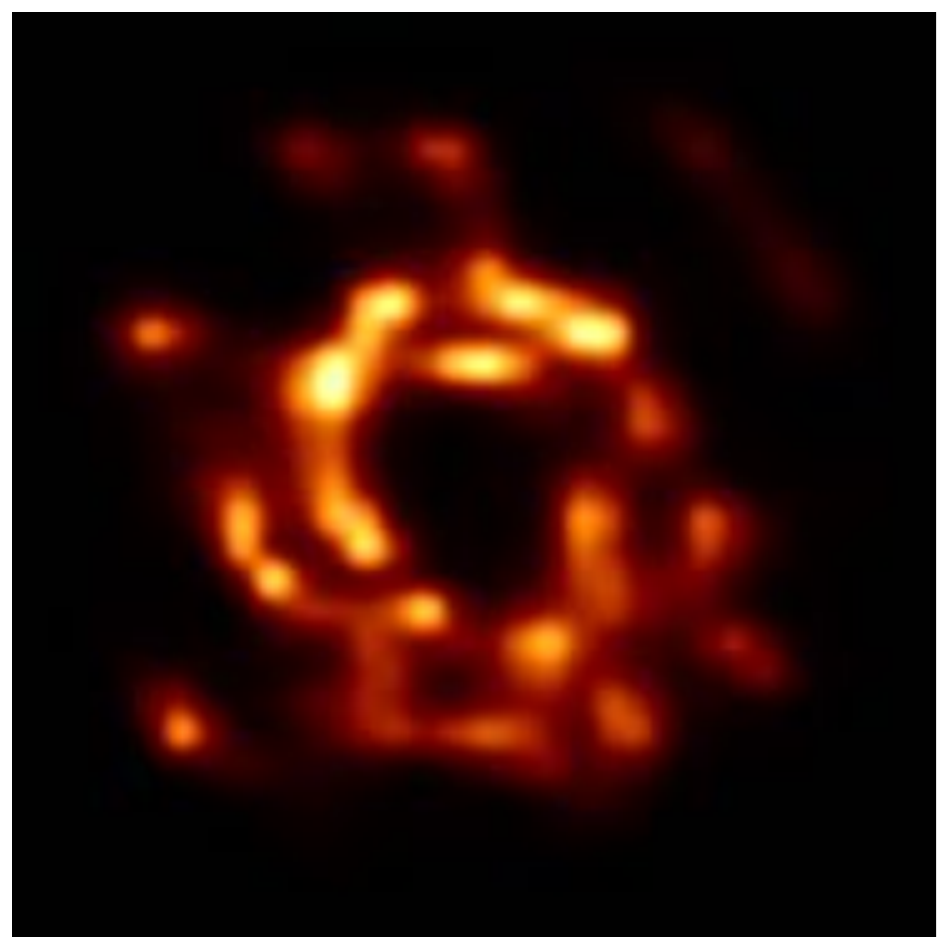}} } &
			\includegraphics[height=.1\linewidth]{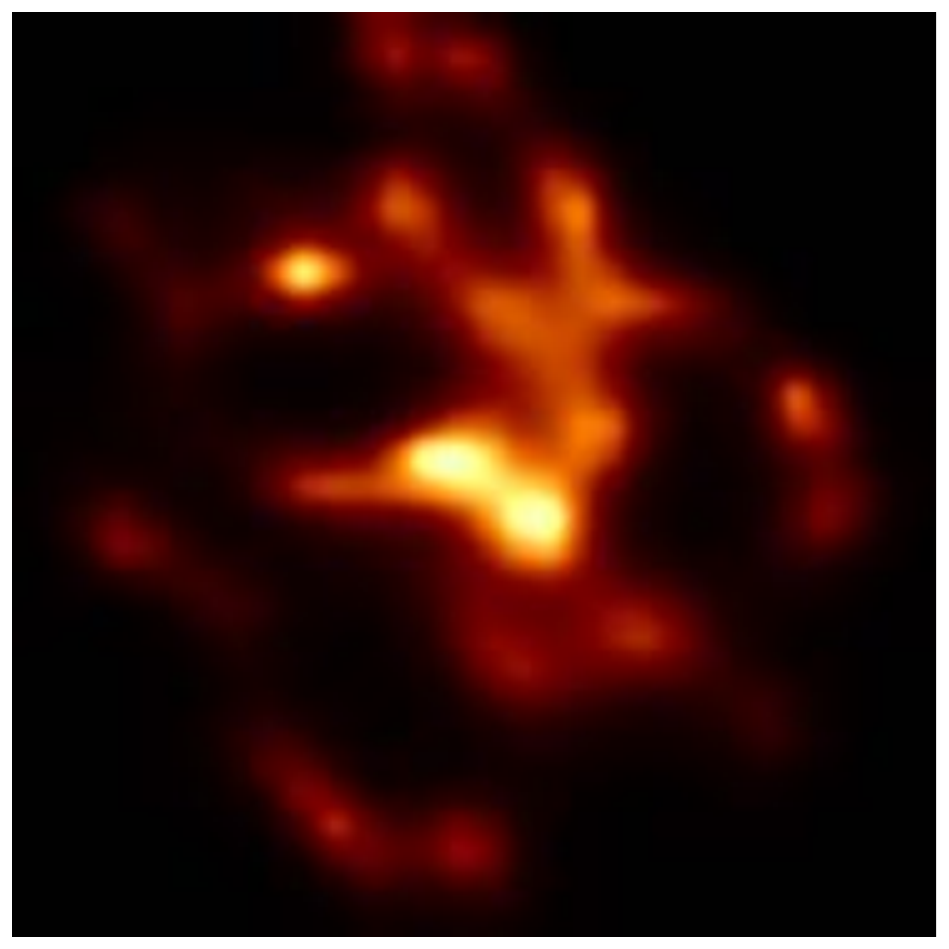} &
			\includegraphics[height=.1\linewidth]{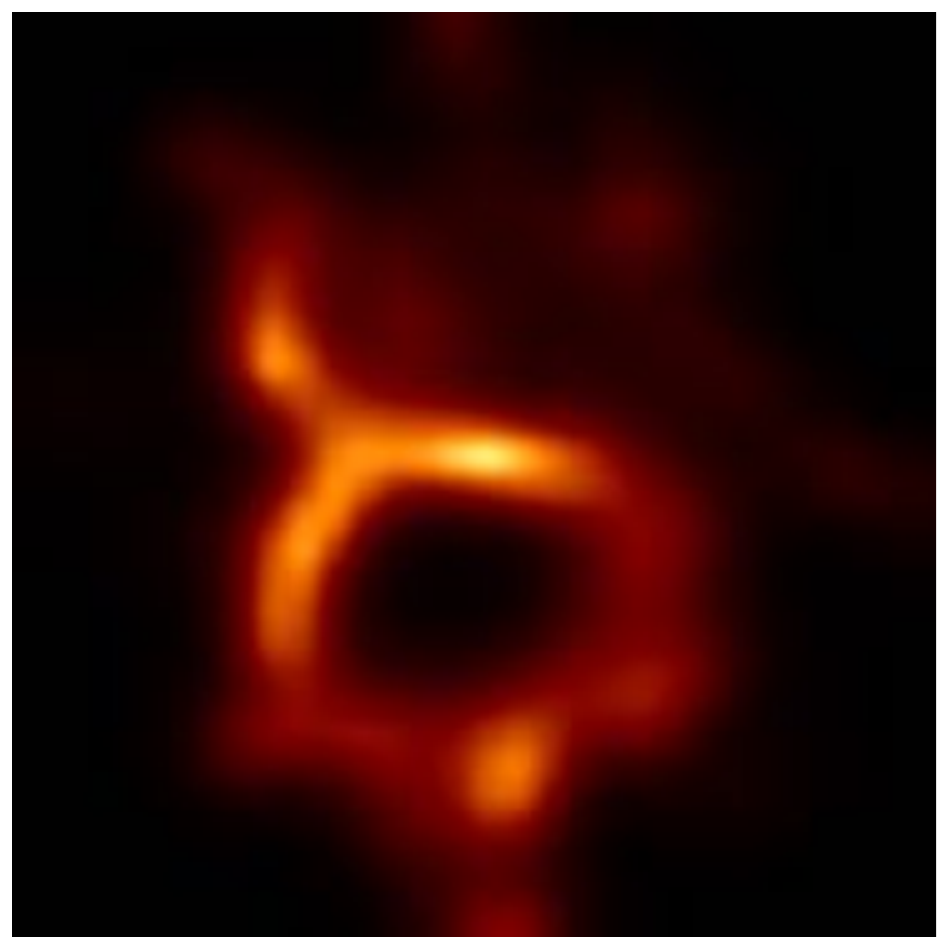} &
			\includegraphics[height=.1\linewidth]{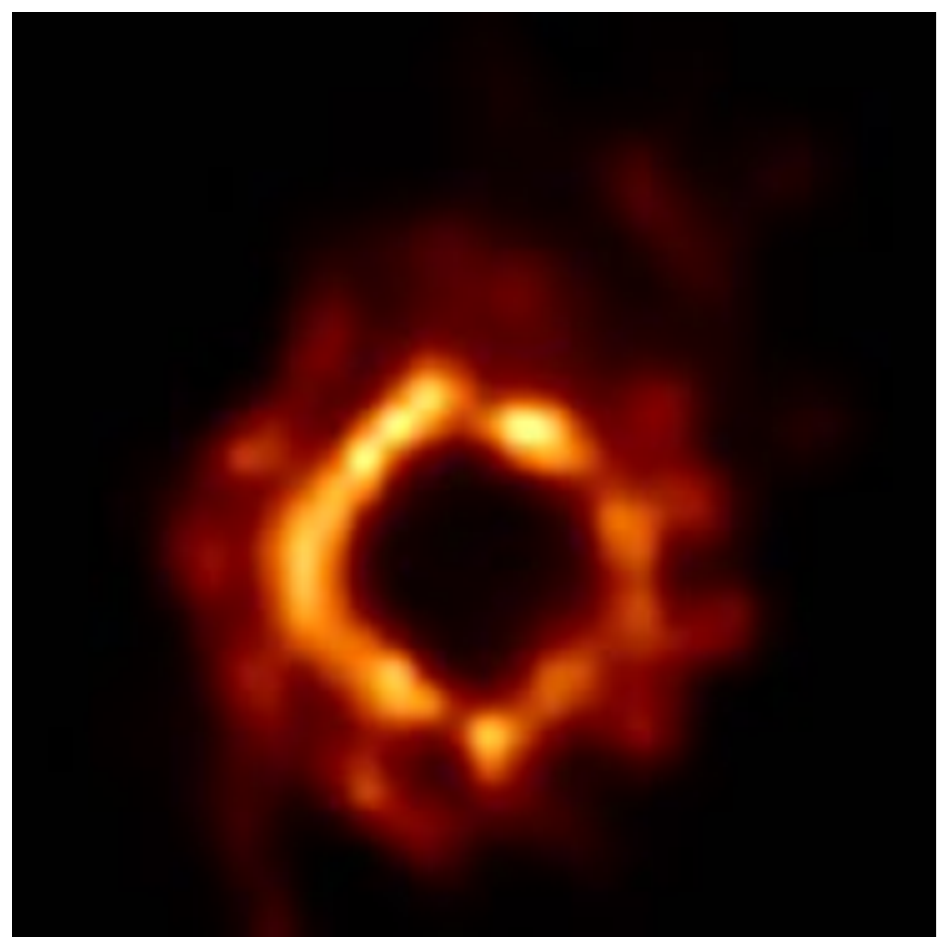} 
			&
			{{\includegraphics[height=.1\linewidth]{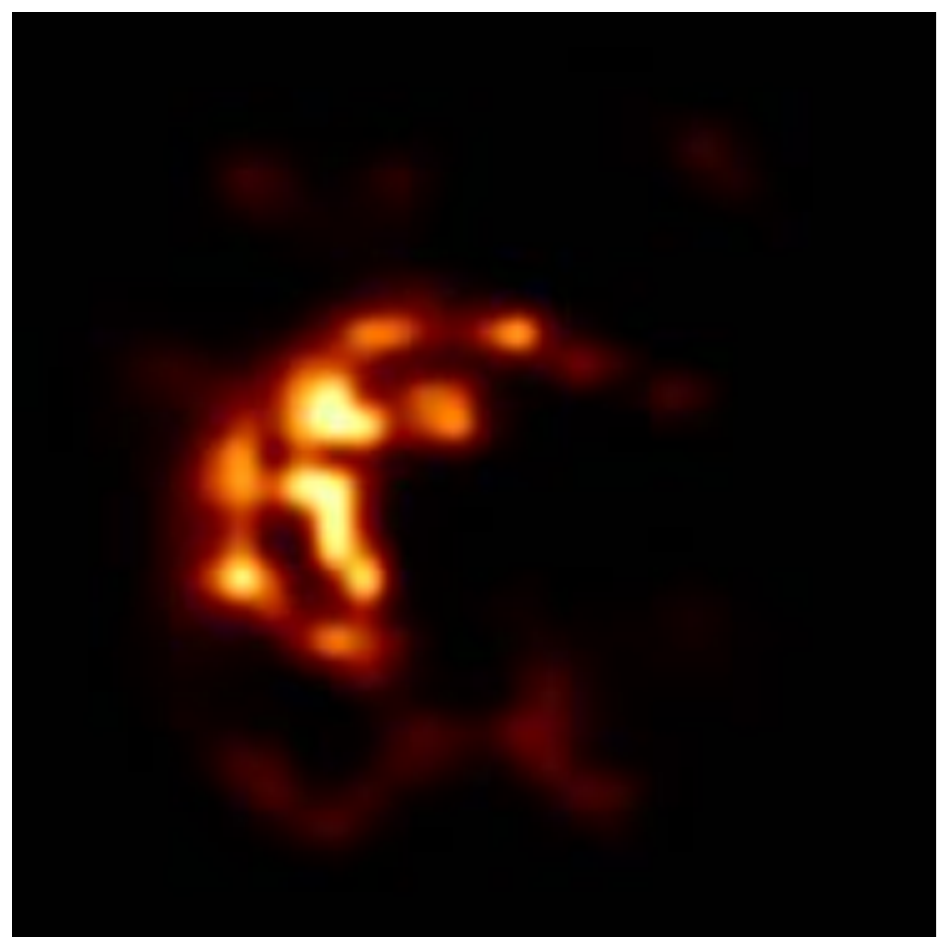}} } &
			\includegraphics[height=.1\linewidth]{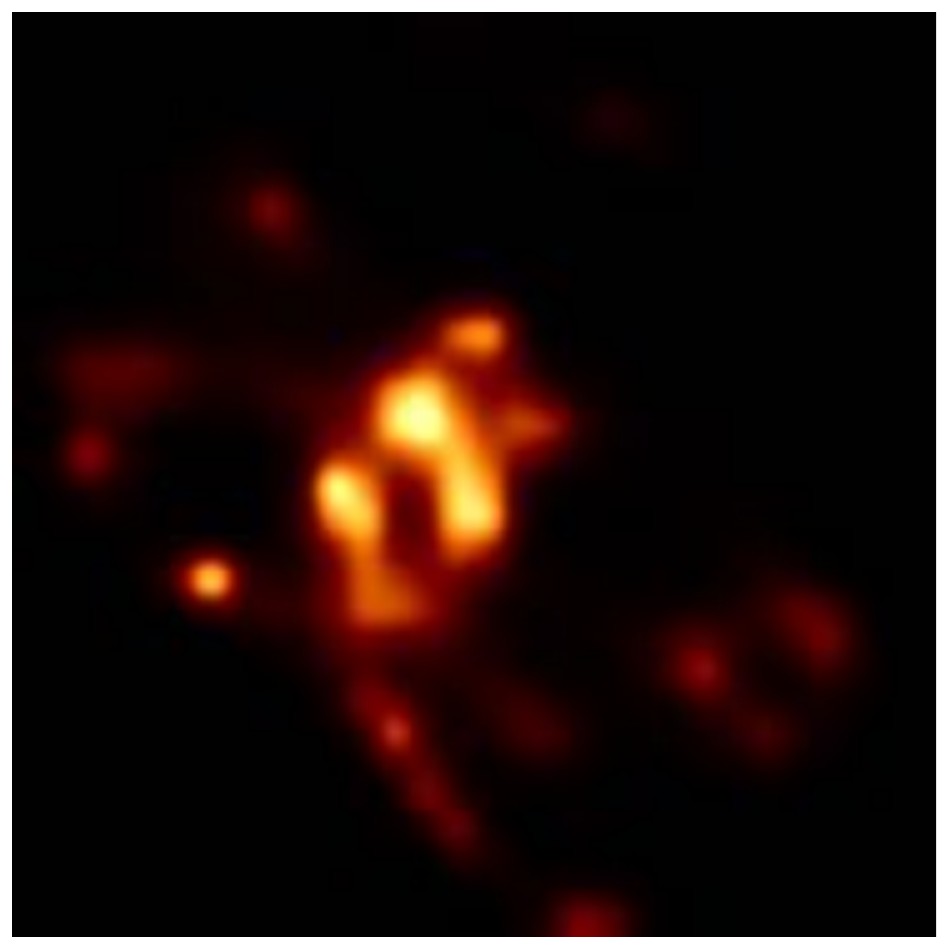} &
			\includegraphics[height=.1\linewidth]{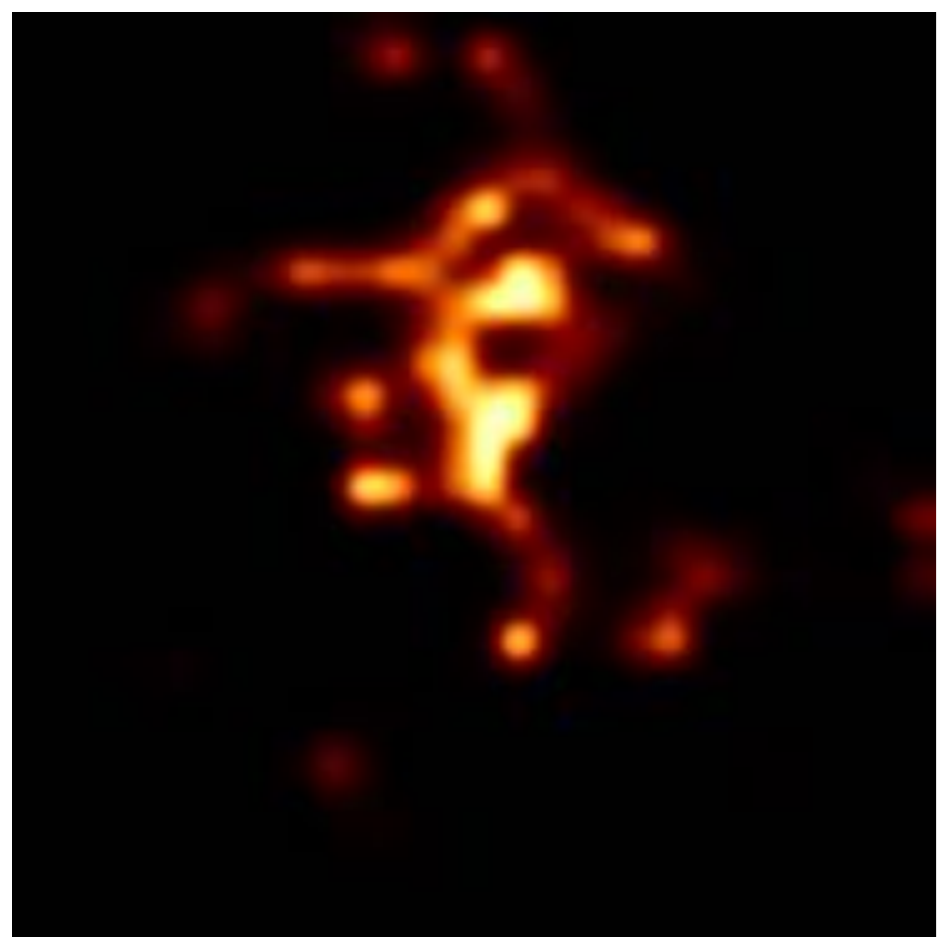} &
			\includegraphics[height=.1\linewidth]{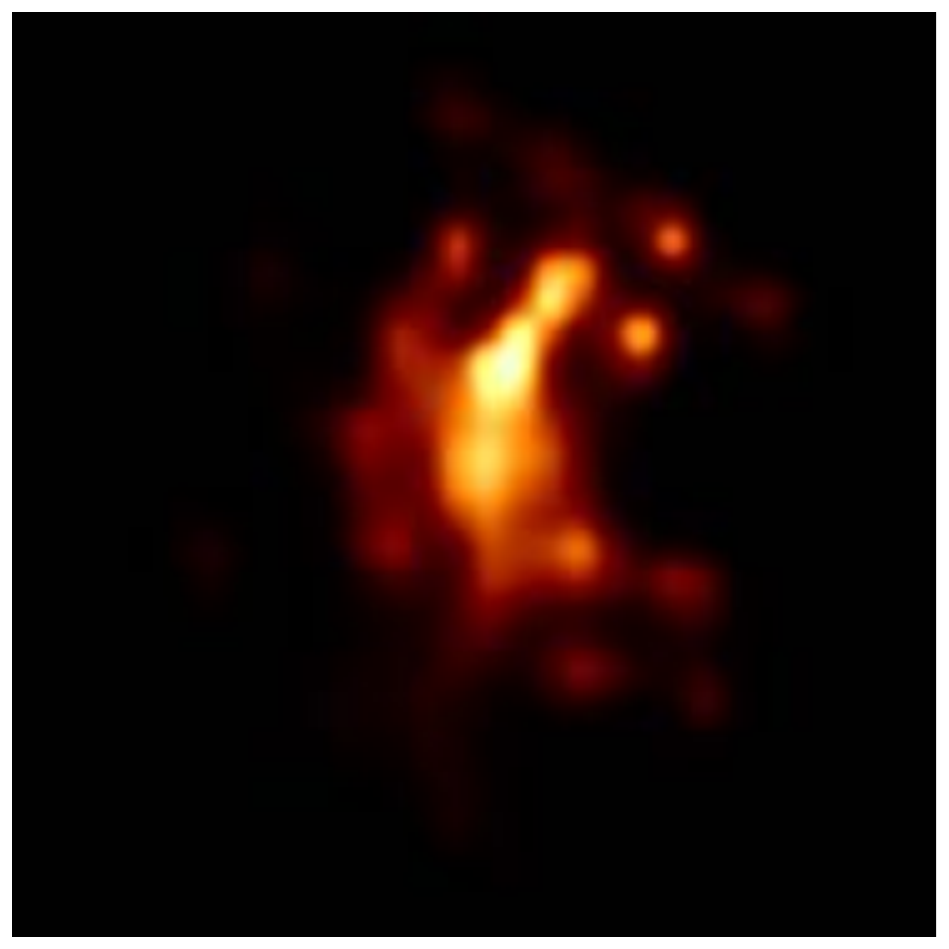} 
			\\
			&\vspace{-.1in} & & & & & & & &\\
			&	\multirow{1}{*}[0.4in]{ \rotatebox[origin=t]{90}{\small{\textsf{\cite{andrew}}} }}
			&
			{{\includegraphics[height=.1\linewidth]{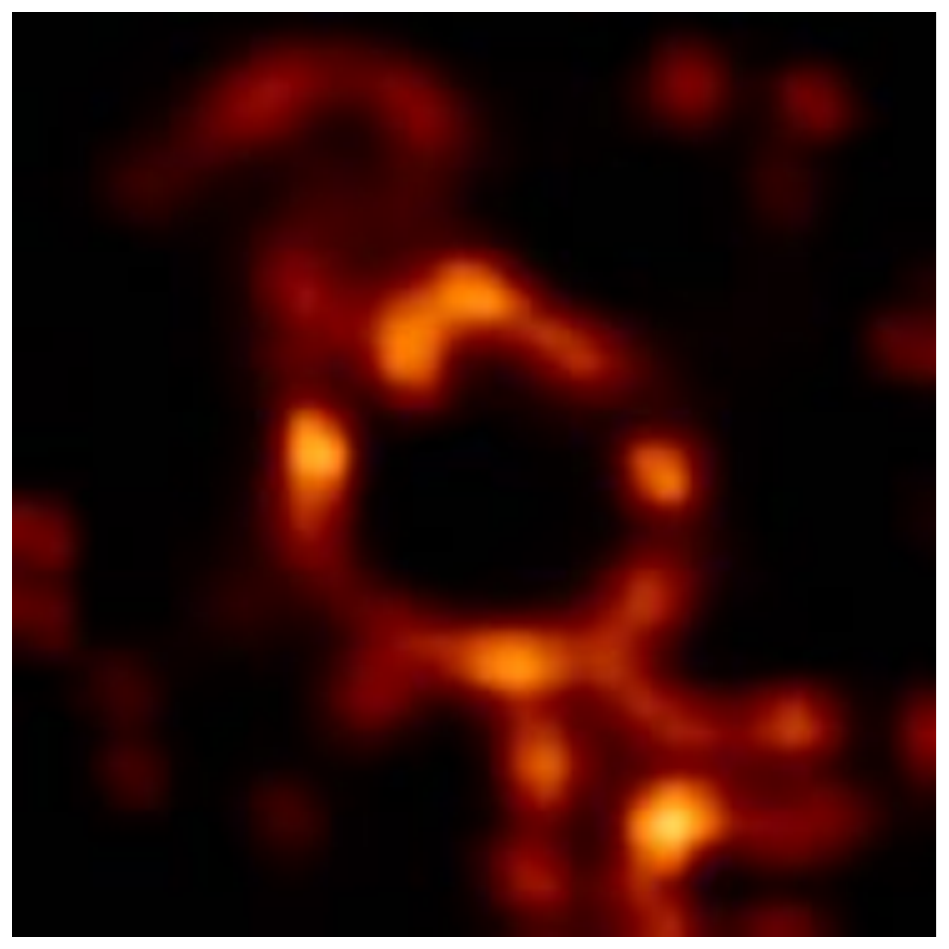}} } &
			\includegraphics[height=.1\linewidth]{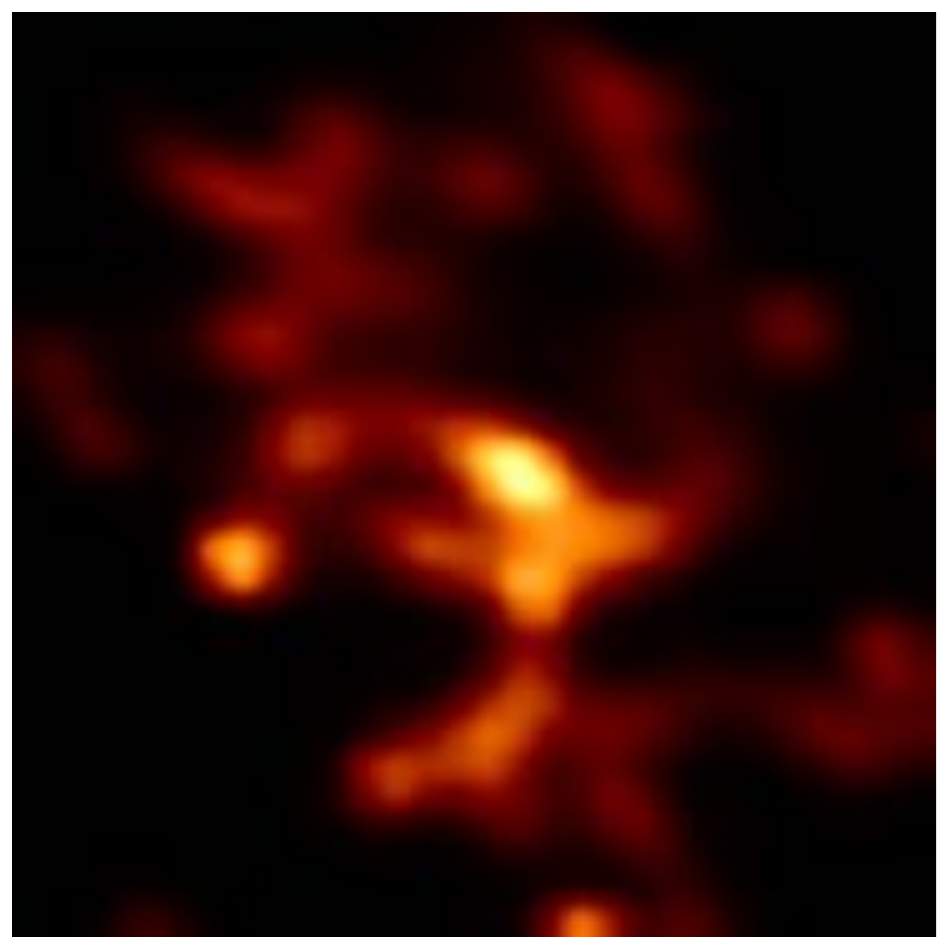} 
			&
			\includegraphics[height=.1\linewidth]{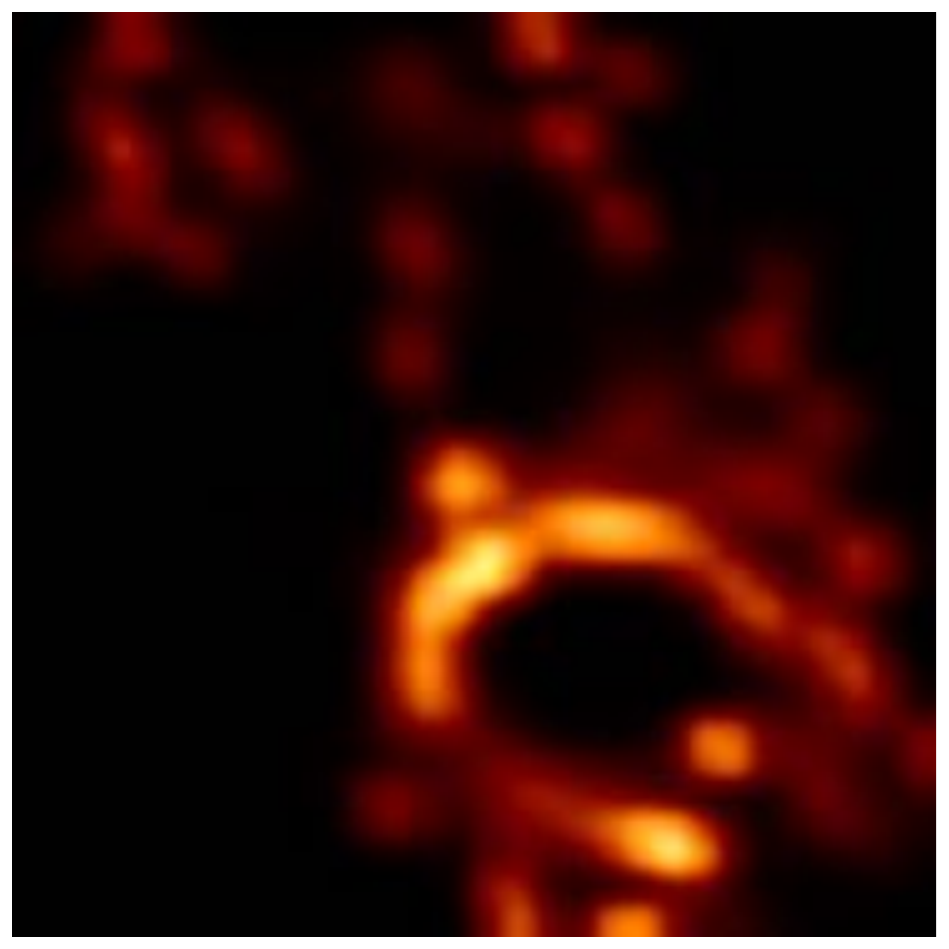} &
			\includegraphics[height=.1\linewidth]{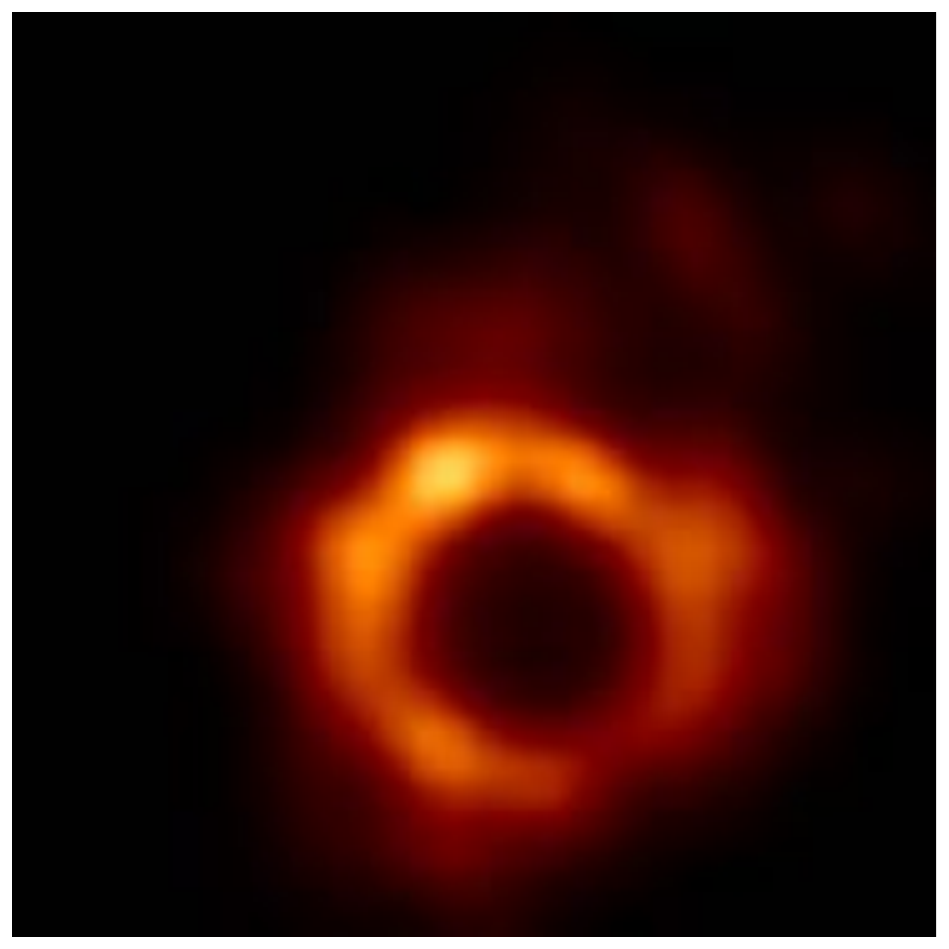} 
			&
			{{\includegraphics[height=.1\linewidth]{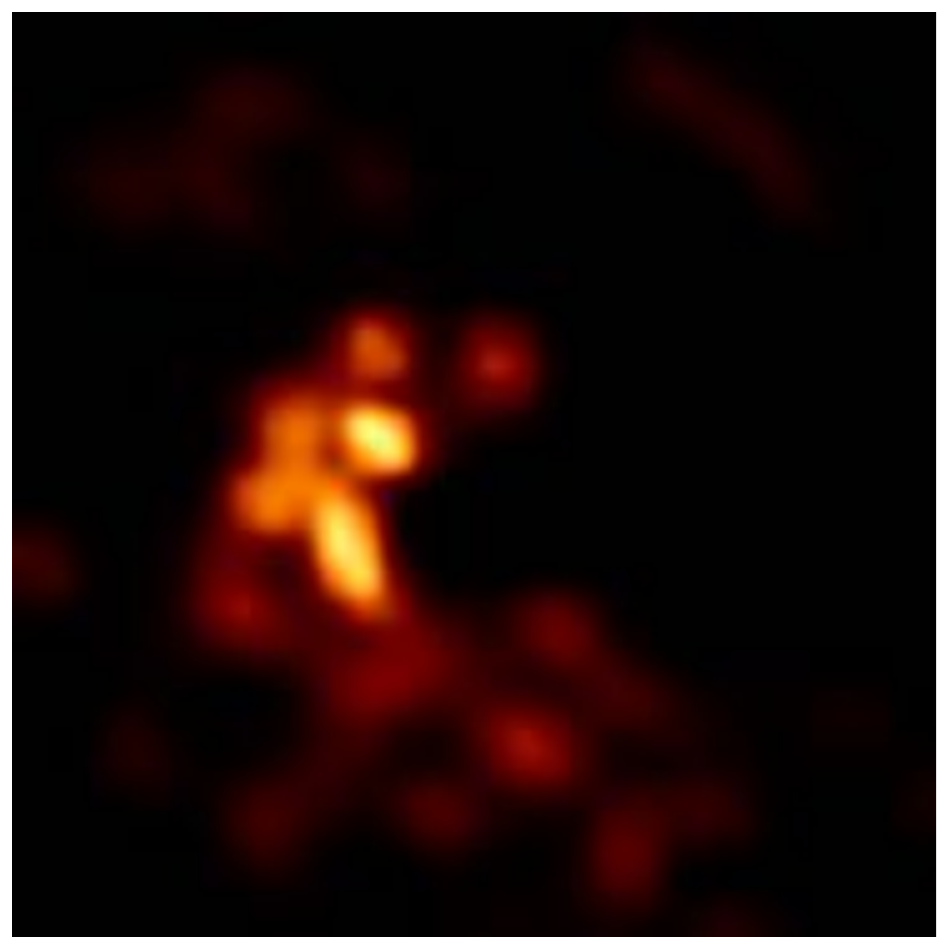}} } &
			\includegraphics[height=.1\linewidth]{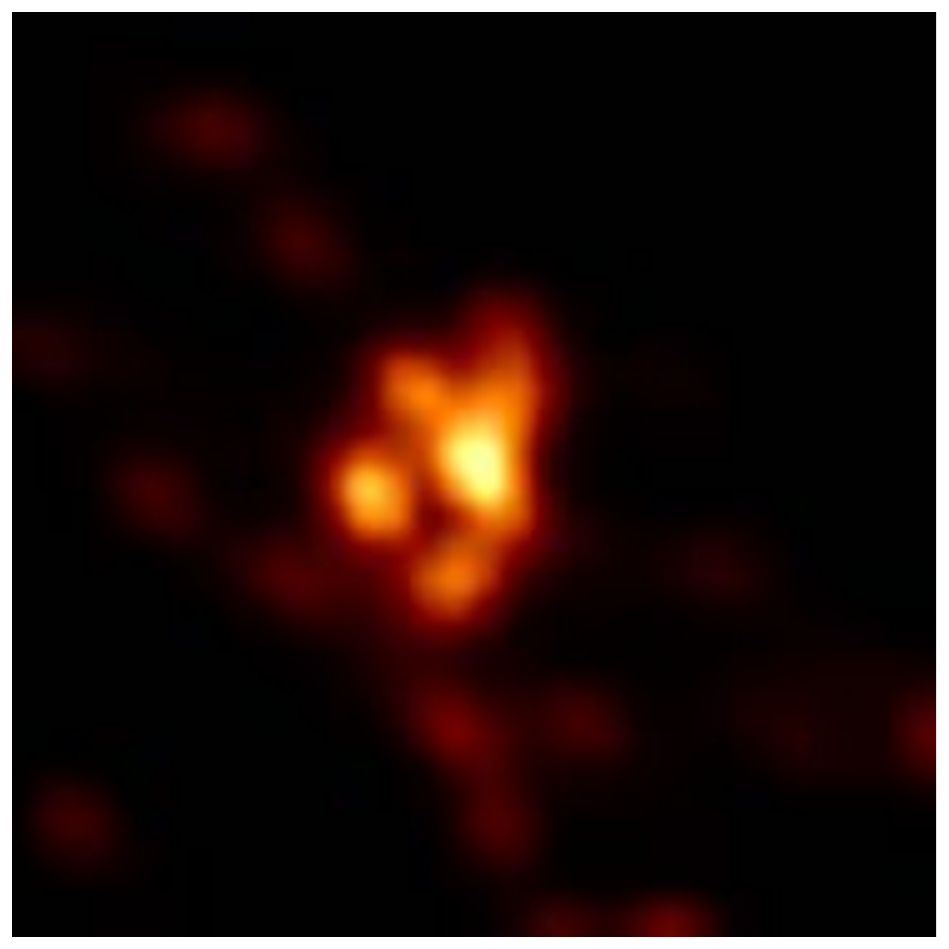} &
			\includegraphics[height=.1\linewidth]{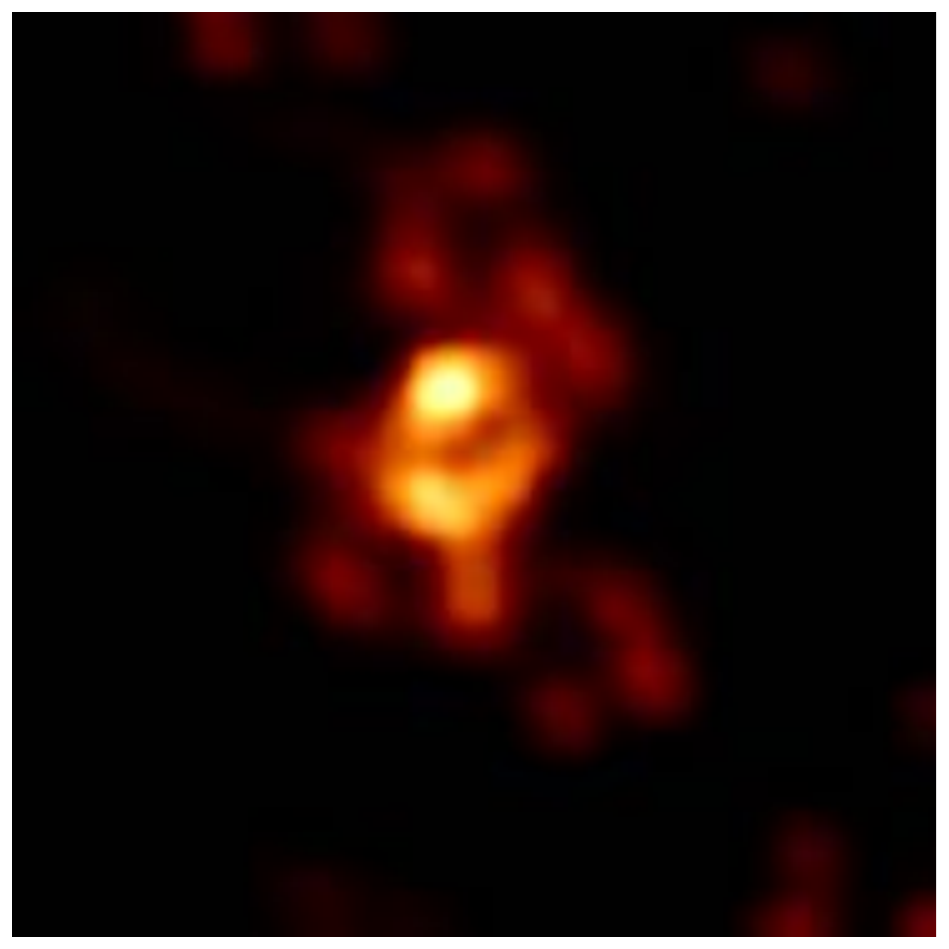} &
			\includegraphics[height=.1\linewidth]{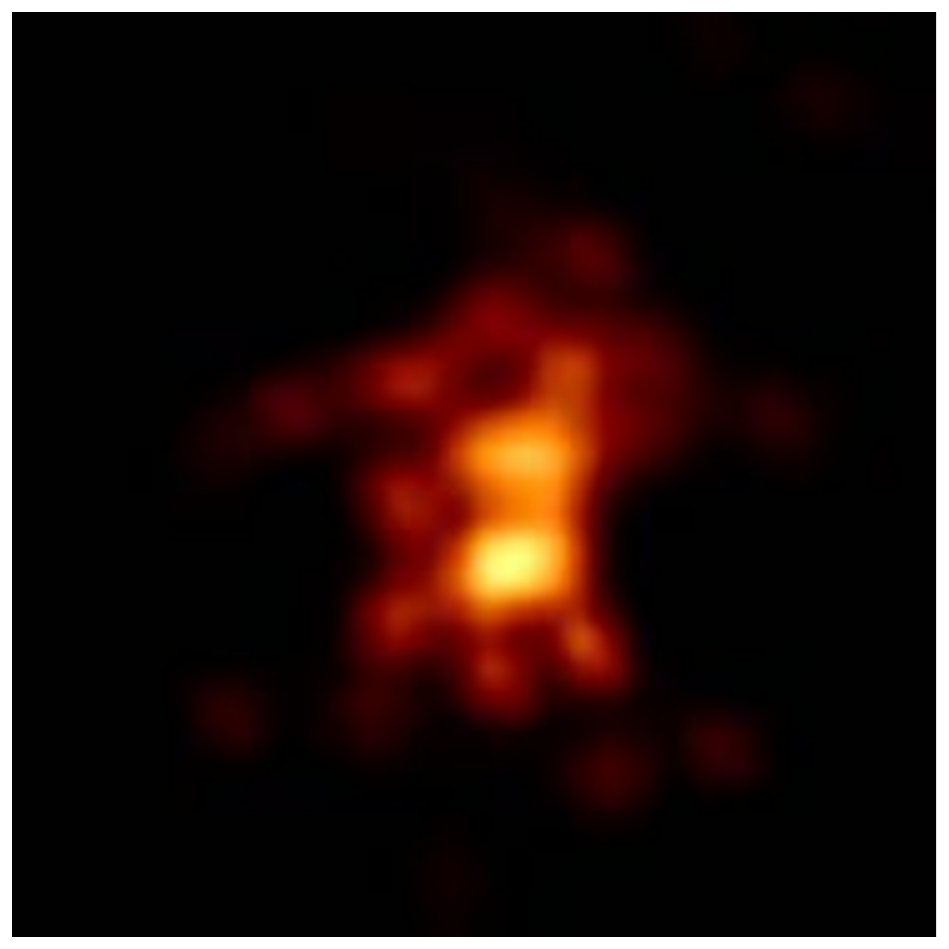} 
			\\ 
		\end{tabular}
		\caption{{\bf Static evolution model:} Results obtained using data simulated from each of the 4 video sequences (see Figure~\ref{fig:groundtruth}) under different telescope arrays (see Figures~\ref{fig:staticimaging} and~\ref{fig:uvcov2}) and noise conditions. The main portion of the figure is broken up into 4 quadrants corresponding to Videos 1-4 when moving from left to right, top to bottom. The true mean image from the ground truth videos, blurred to 3/4 the nominal resolution of the array, is shown on the top. We compare results of our proposed method, StarWarps, to that of the single imaging methods presented in~\cite{freek} and~\cite{andrew}. In particular, we compare the mean image obtained using StarWarps video reconstruction. The error type NO ATM. indicates reconstructing using visibilities on data with no atmospheric error, while the error type ATM. indicates using the visibility amplitudes and bispectrum on data where atmospheric phase errors have been introduced. The quality of each result, compared to the ground truth mean image, is indicated in the table of normalized root mean squared errors (Normalized RMSE). To account for the loss of absolute position in the presence of atmospheric phase error, images were rigidly aligned to the true mean before computing the error. The FOV and colorbar used for each reconstruction can be seen in Figure~\ref{fig:groundtruth}. }
		\label{fig:staticevolutionresults}
	\end{center}
\end{figure*}

Figure~\ref{fig:staticevolutionresults} shows example reconstructions, and corresponding measured error (NRMSE), for combinations of the 4 source videos observed under the 3 telescope arrays. For these results we have set $a=2$, $c=1/2$, and $\bQ = 10^{\text{-}7} \mathds{1}$. The main portion of this figure is broken up into 4 quadrants, each containing results for one video. From left to right, up to down, each quadrant corresponds to Video 1-4 respectively. The ground truth mean image for each video is shown in the upper table. These images correspond to those shown in Figure~\ref{fig:groundtruth}, but are smoothed to 3/4 the nominal resolution of the interferometer to help illustrate the level of resolution we aim to recover. 

Horizontally within each quadrant we present results obtained using data with varying degrees of difficulty. As the number of telescopes in the array increases, so does the spatial frequency coverage. Therefore, reconstructing an accurate video with the FUTURE array is a much easier task than with the EHT2017 array.
Additionally, using complex visibilities that are not subject to atmospheric errors is much easier than having to recover images from phase corrupted measurements. 
In the case where there are atmospheric phase errors (ATM.), we constrain the reconstruction problem using a combination of visibility amplitude and bispectrum data products. This results in a non-convex problem (that we approximate with series of linearizations) that is much more difficult to solve than when using complex visibilities when there is no atmospheric phase error (NO ATM.). We demonstrate results on the EHT2017 array for both cases, and the EHT2017+ and FUTURE arrays in the case of atmospheric error.

Vertically within each quadrant we illustrate the results of our method, StarWarps, by displaying the average frame reconstructed. 
We compare our method to two state-of-the-art Bayesian-style methods.~\cite{andrew} solves for a single image by imposing a combination of MEM and TV priors. This method performs well in the case of a static source (see Figure~\ref{fig:staticimaging}), however, in the case of an evolving source it often results in artifact-heavy reconstructions that are difficult to interpret. In~\cite{freek} the authors attempt to mitigate this problem by first smoothing the time-varying data products before imaging.
This approach was originally designed to work on mutli-epoch data; we find it is unable to accurately recover the source structure from a single day (epoch) observation. 
Results of~\cite{freek} are reconstructed by an author of the method.

\begin{figure}[tb]
	\begin{center}
		\vspace{-.2in}
		
		\begin{tabular}{   c c | c  c  c   }
			& & \large{\textsf{23 GST}} &\large{\textsf{2 GST}}   &\large{\textsf{6 GST}}    \\ 
			&\vspace{-.1in} & & & \\ \hline
			&\vspace{-.1in} & & & \\
			& \multirow{1}{*}[0.45in]{ \rotatebox[origin=t]{90}{\small{\textsf{Truth}} }} & 
			{{\includegraphics[height=.2\linewidth]{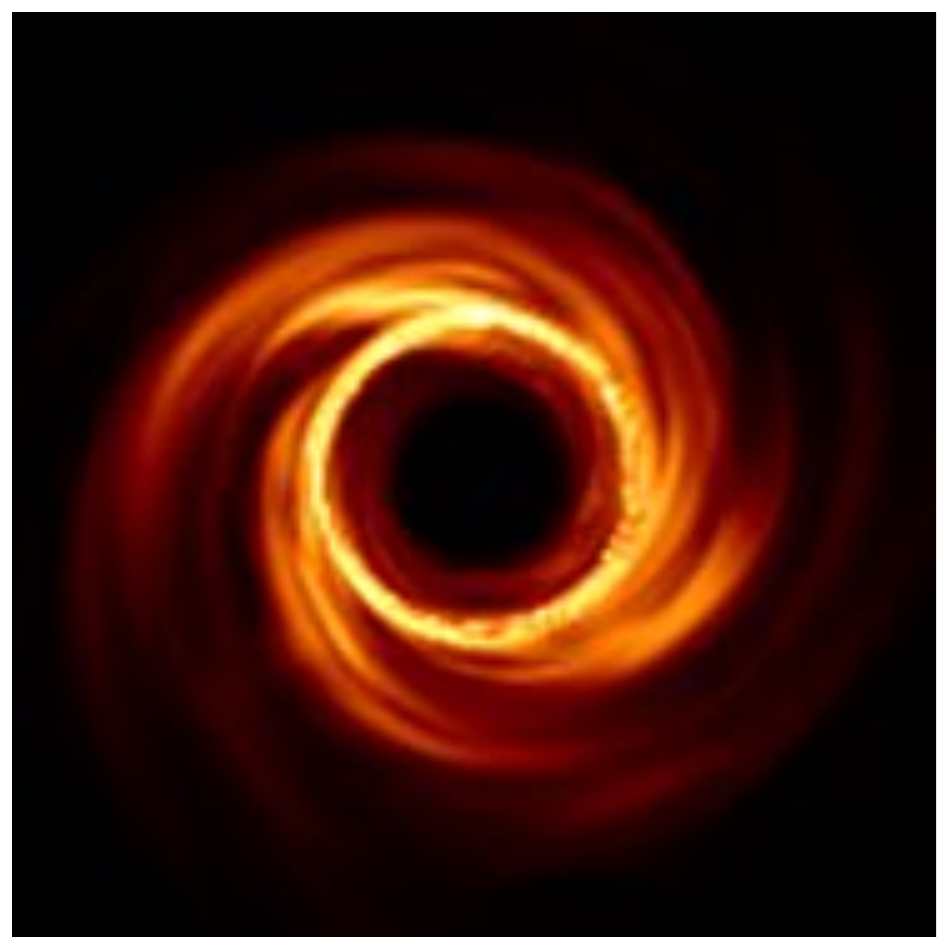}} } &
			\includegraphics[height=.2\linewidth]{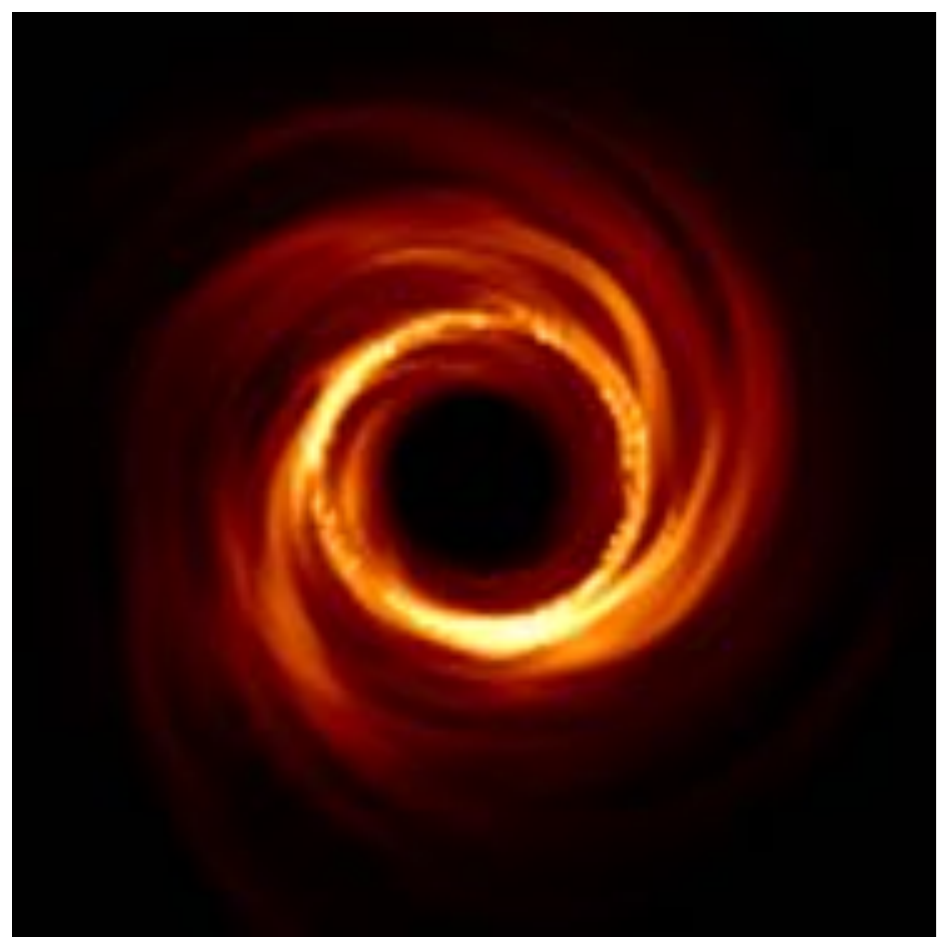} &
			\includegraphics[height=.2\linewidth]{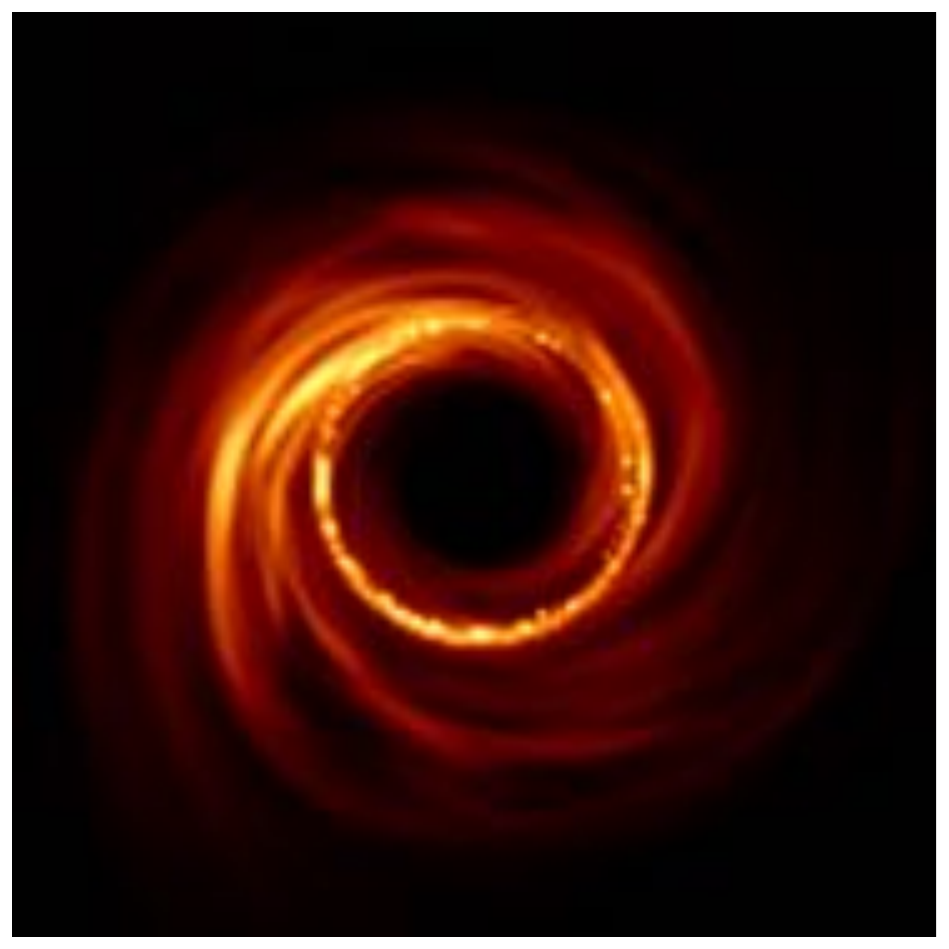}  
			\\ \hline
			&\vspace{-.1in} & & & \\
			\multirow{1}{*}[0.6in]{ \rotatebox[origin=t]{90}{\small{\textsf{NO ATM. \& }} }} \hspace{-0.25in} &
			\multirow{1}{*}[0.55in]{ \rotatebox[origin=t]{90}{\small{\textsf{ NO PROP.}} }} &
			{{\includegraphics[height=.2\linewidth]{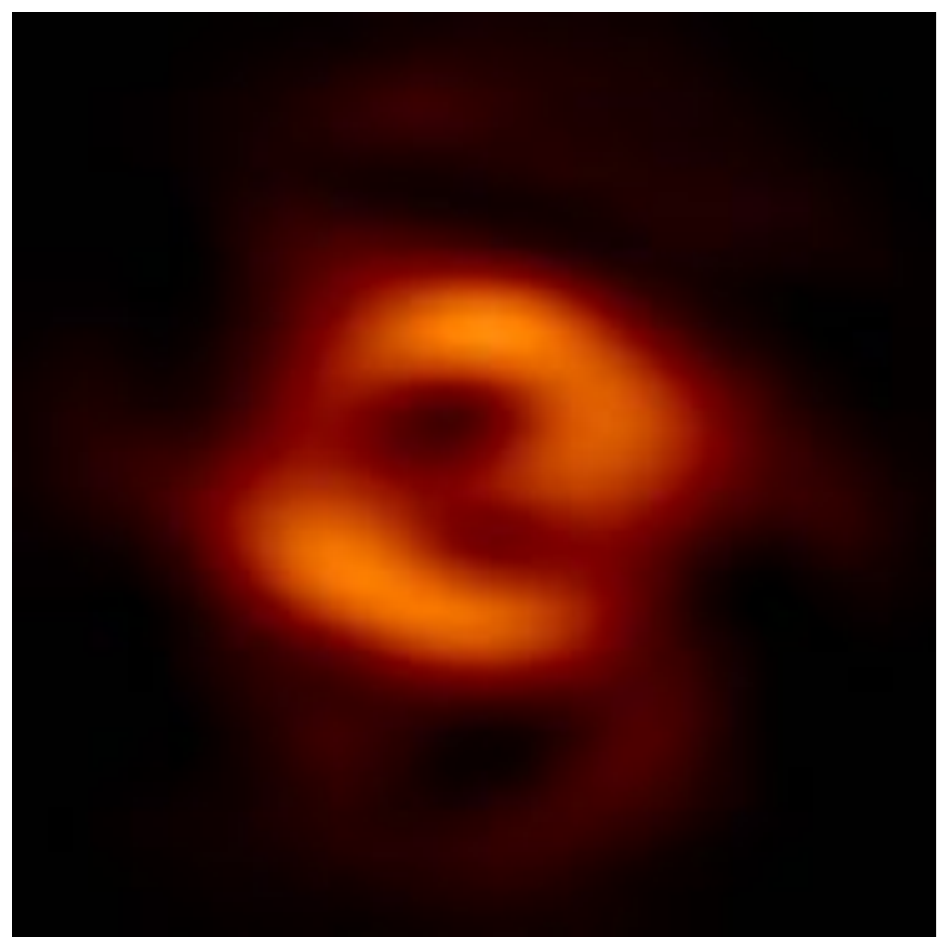}} } &
			\includegraphics[height=.2\linewidth]{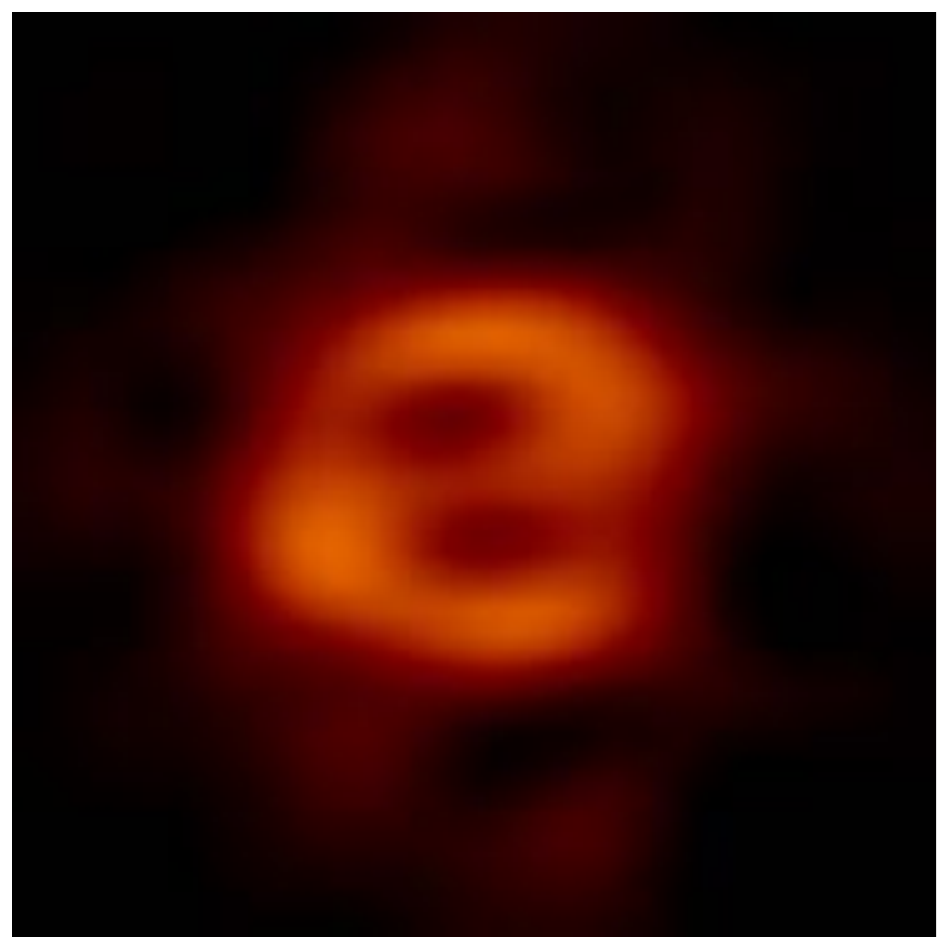} &
			\includegraphics[height=.2\linewidth]{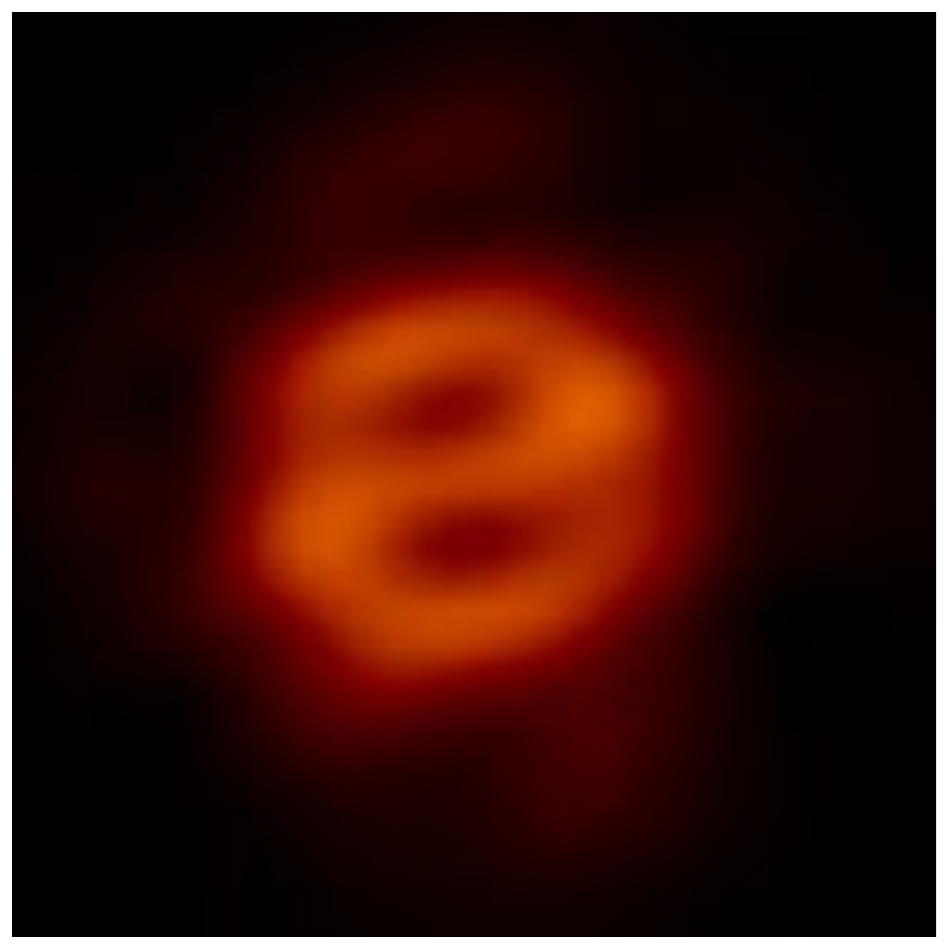} 
			\\
			\multirow{1}{*}[0.55in]{ \rotatebox[origin=t]{90}{\small{\textsf{NO ATM. }} }} \hspace{-0.25in} &
			\multirow{1}{*}[0.5in]{ \rotatebox[origin=t]{90}{\small{\textsf{ \& PROP.}} }} &
			{{\includegraphics[height=.2\linewidth]{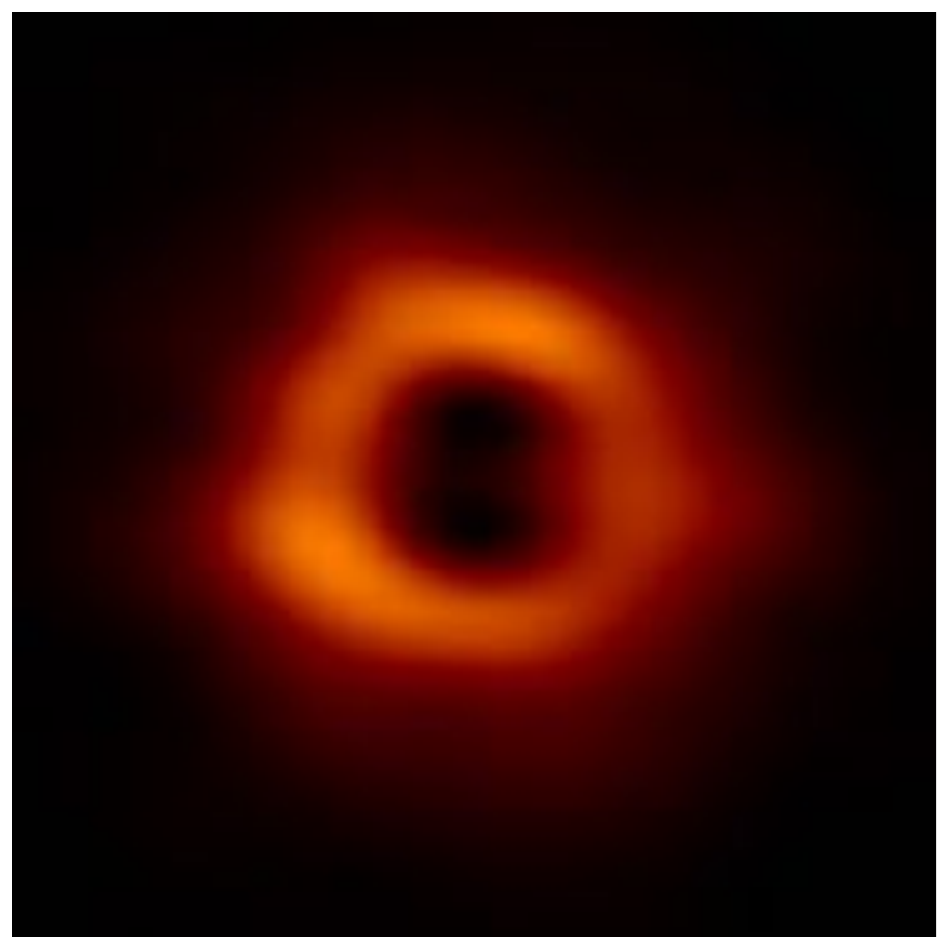}} } &
			\includegraphics[height=.2\linewidth]{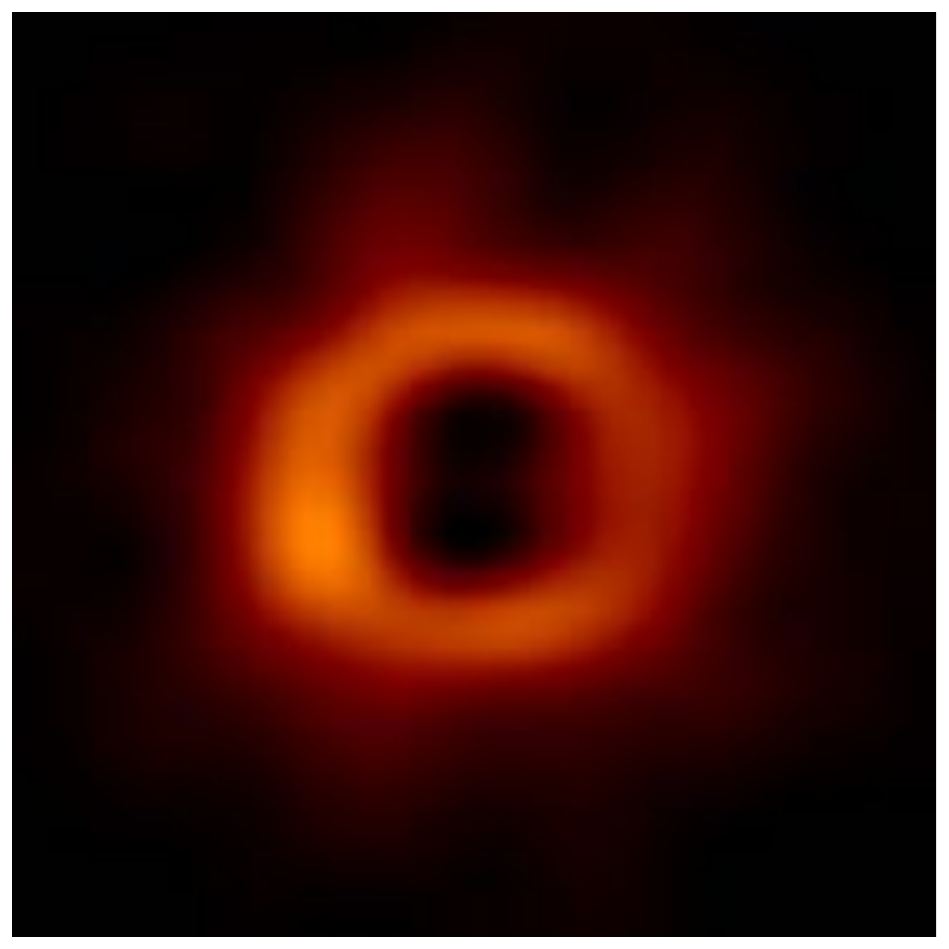} &
			\includegraphics[height=.2\linewidth]{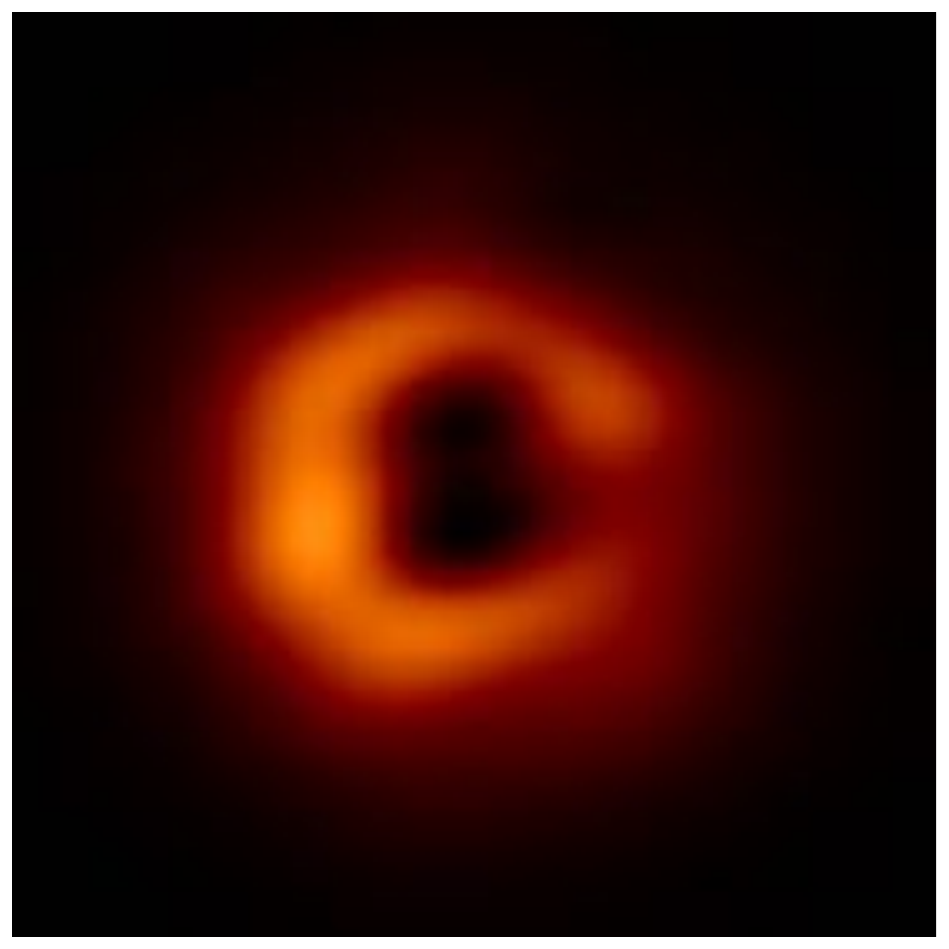} 
			\\ \hline
			&\vspace{-.1in} & & & \\
			\multirow{1}{*}[0.45in]{ \rotatebox[origin=t]{90}{\small{\textsf{ATM. \&}} }} \hspace{-0.25in} &
			\multirow{1}{*}[0.55in]{ \rotatebox[origin=t]{90}{\small{\textsf{NO PROP.}} }}
			&
			{{\includegraphics[height=.2\linewidth]{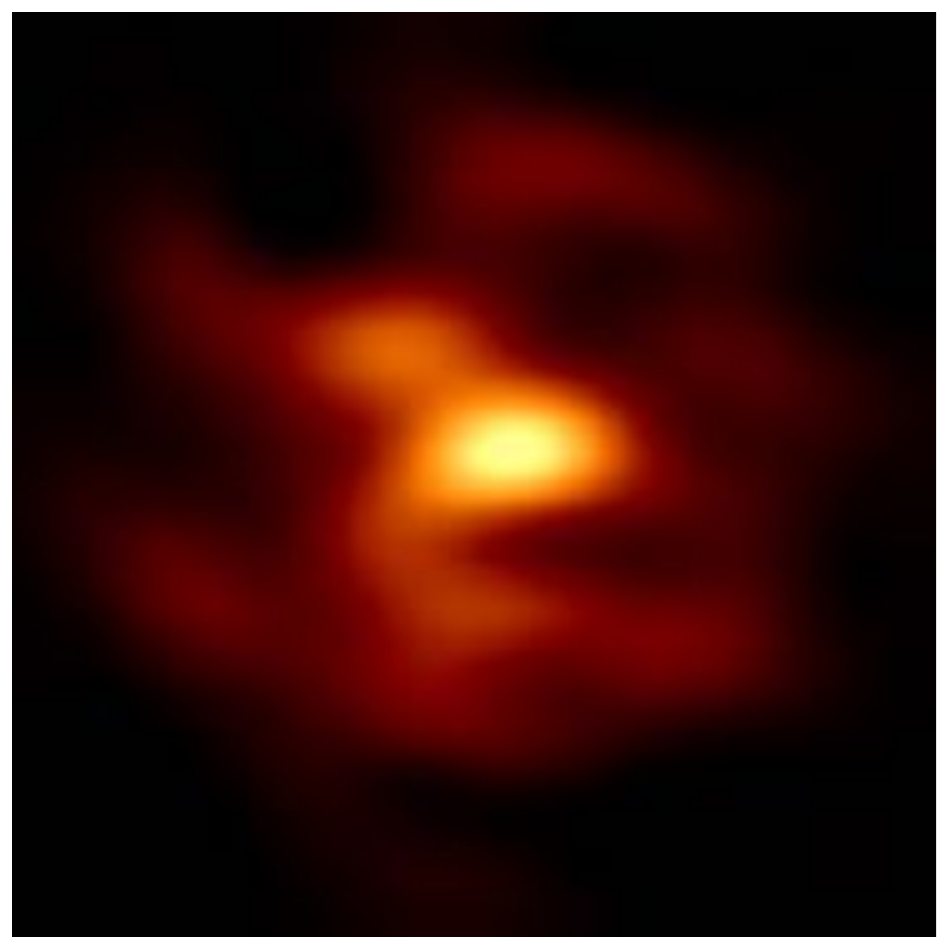}} } &
			\includegraphics[height=.2\linewidth]{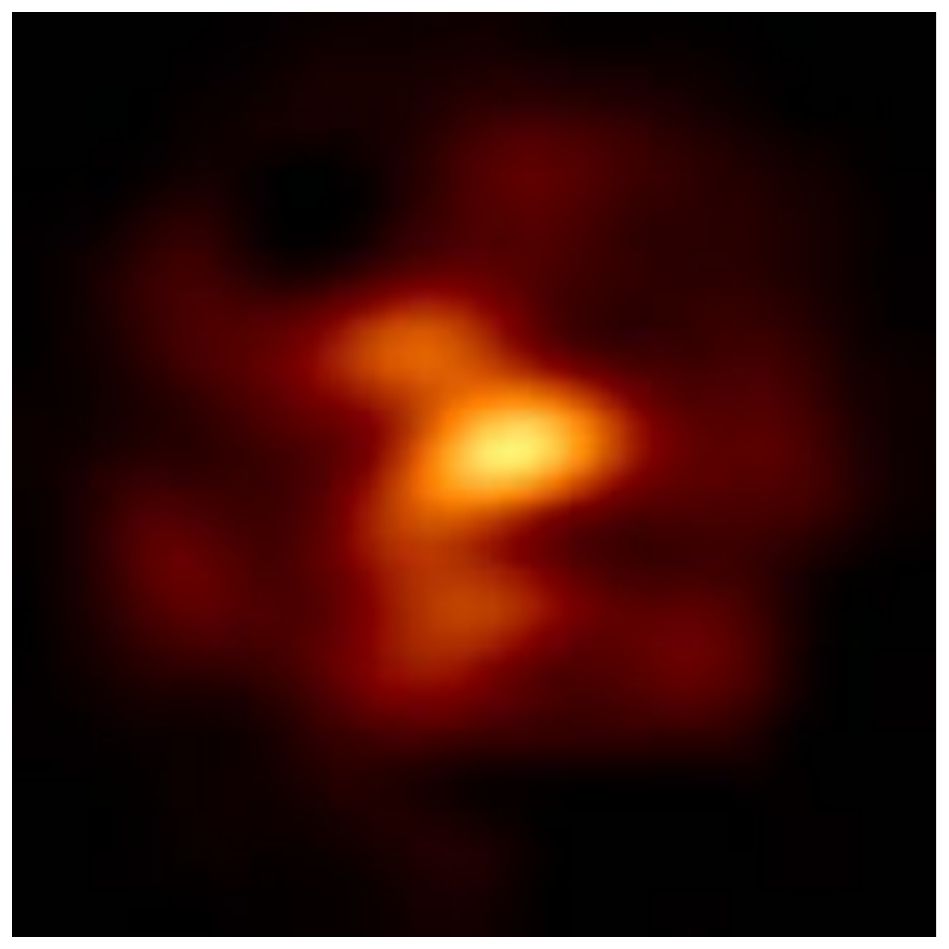} &
			\includegraphics[height=.2\linewidth]{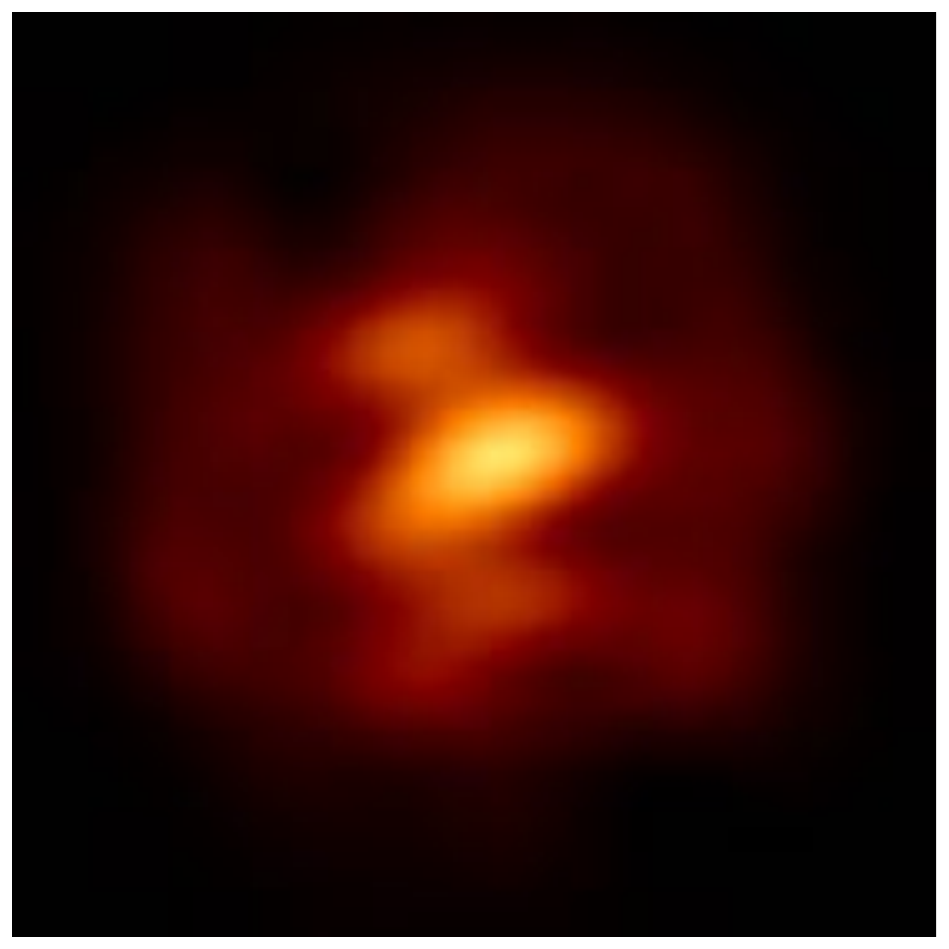} 
			\\ 
			\multirow{1}{*}[0.48in]{ \rotatebox[origin=t]{90}{\small{\textsf{ATM. \&}} }} \hspace{-0.25in} &
			\multirow{1}{*}[0.45in]{ \rotatebox[origin=t]{90}{\small{\textsf{PROP.}} }} &
			{{\includegraphics[height=.2\linewidth]{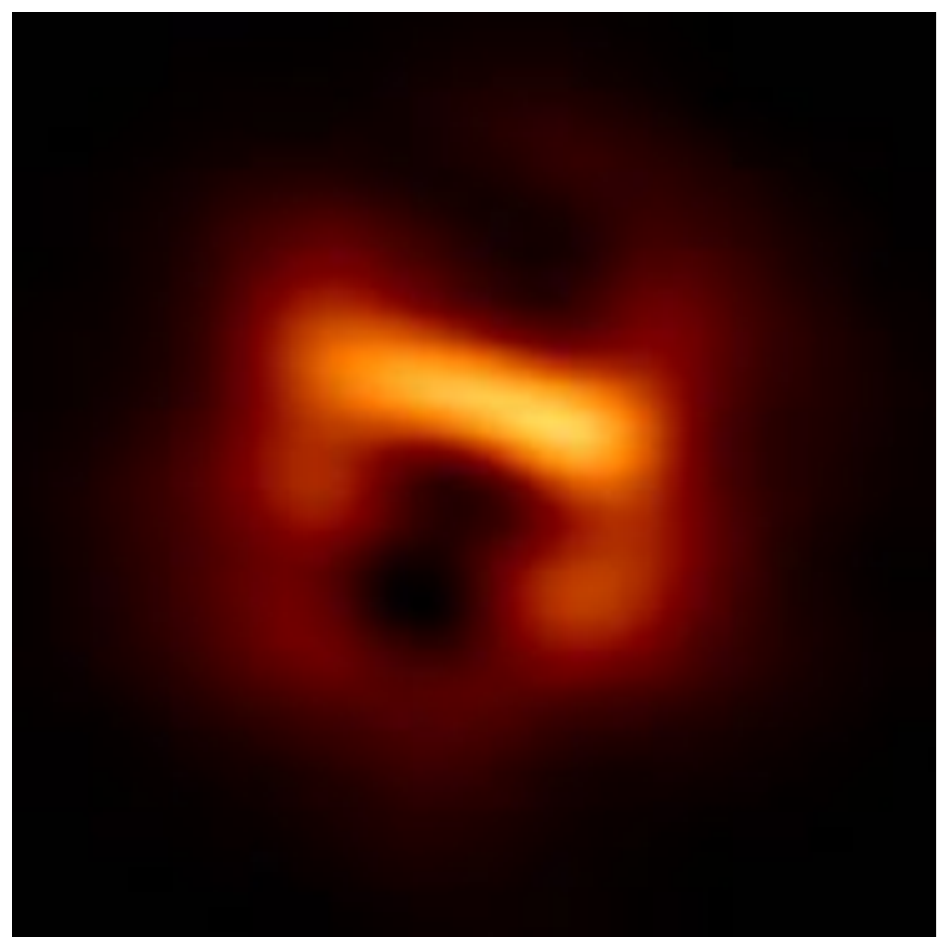}} } &
			\includegraphics[height=.2\linewidth]{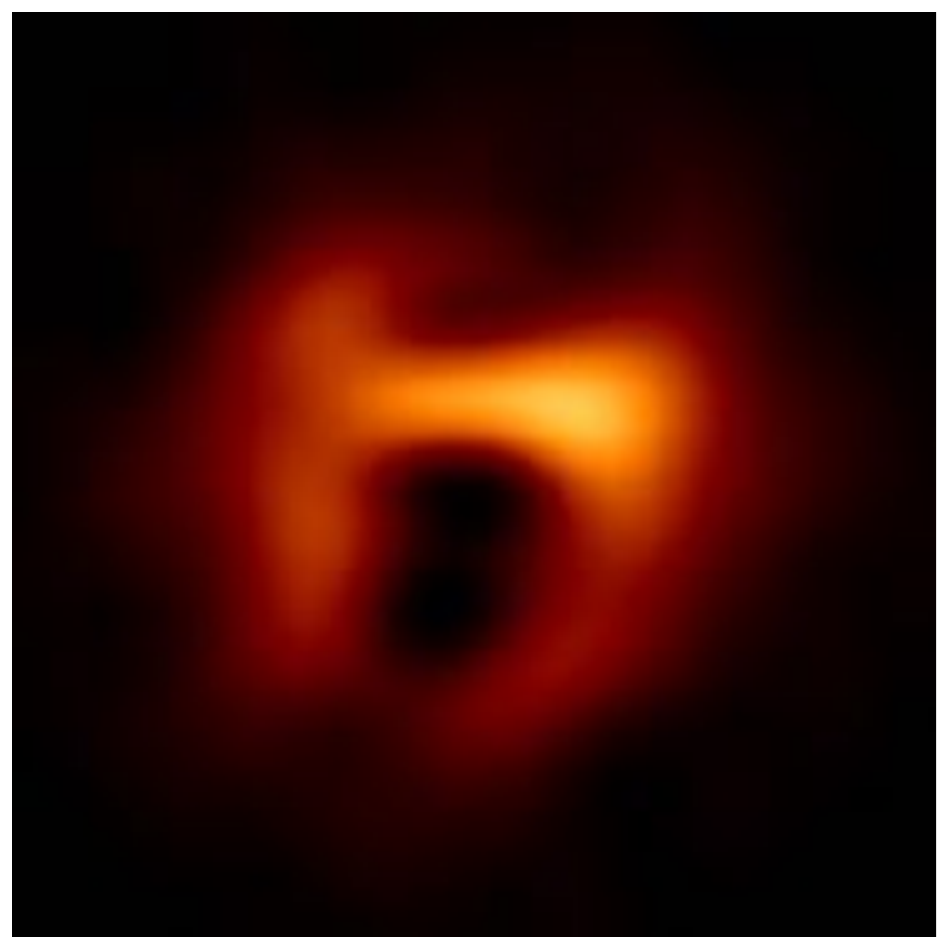} &
			\includegraphics[height=.2\linewidth]{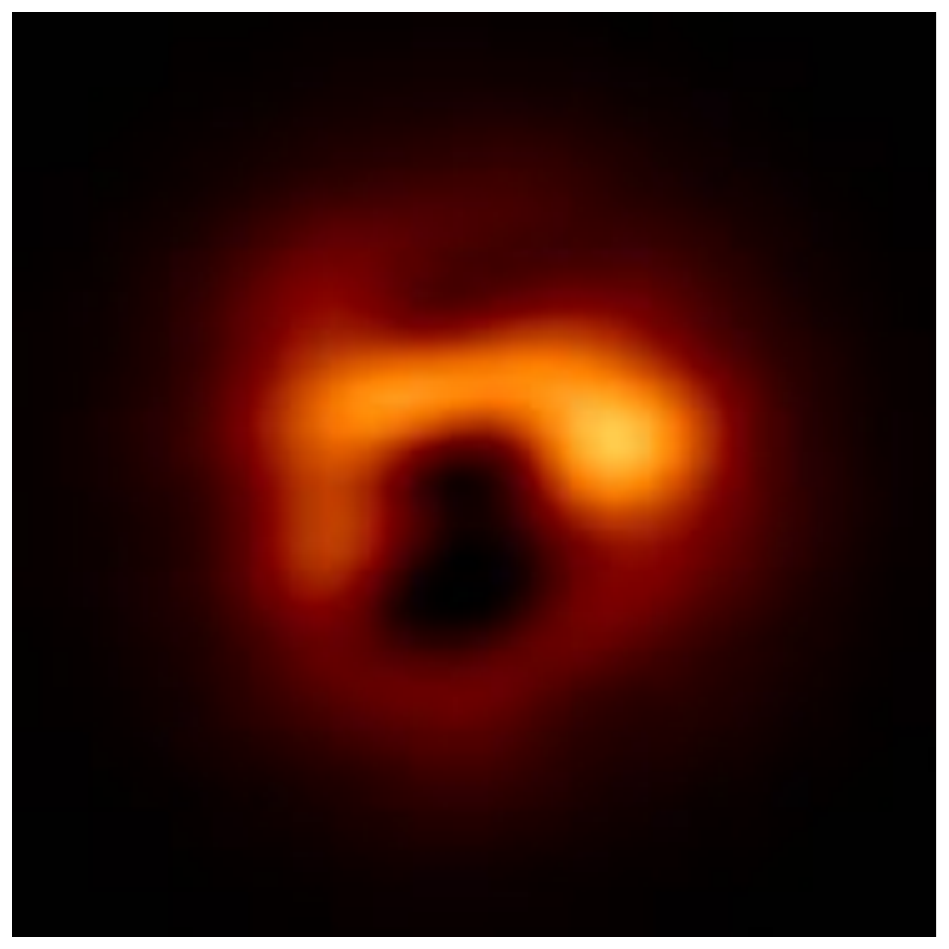} 
			\\ \hline
		\end{tabular}
		\caption{{\bf Propagating Uncertainty:} During inference, StarWarps approximates each image's covariance matrix in order to propagate its uncertainty to neighboring frames in time. Propagating this information is crucial when using very few measurements. We show frames resulting from the same EHT2017 data of Video 3 when the covariance is propagated (PROP.) as described in this paper, versus not propagated (NO PROP.). Note that propagating the covariance results in significantly improved results. This is true even in the case of using non-linear measurements when atmospheric error is present (ATM.). In this non-linear case the covariance matrix is simply a crude approximation of the uncertainty, but proves critical in obtaining a result that captures the ring structure of the underlying source.   }
		\label{fig:propinfo}
	\end{center}
	\vspace{-.3in}
\end{figure}

\begin{figure}
	\begin{center}
			\vspace{-0.2in}
		\setlength{\tabcolsep}{3pt}
		
		\hspace*{-.3cm}
		\begin{tabular}{  c c | ccccc  }
			& \small{\textsf{GST:}} &\small{\textsf{1:13}}   &\small{\textsf{1:33}} &\small{\textsf{1:53}}    &\small{\textsf{2:13}}  &\small{\textsf{2:33}}   \\ \hline
		&	&\vspace{-.1in} & & & &\\
			
	&	\multirow{1}{*}[0.33in]{ \rotatebox[origin=t]{90}{\small{\textsf{Truth}} }}
			&
			{{\includegraphics[height=.15\linewidth]{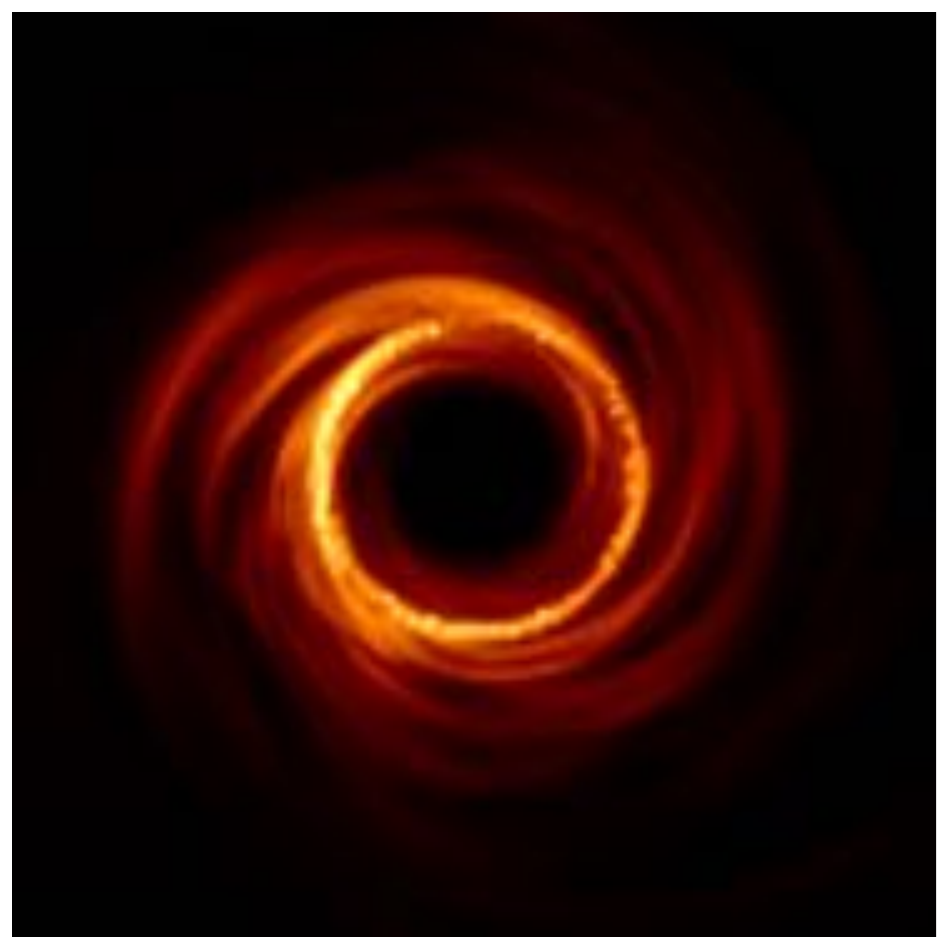}} } &
			{{\includegraphics[height=.15\linewidth]{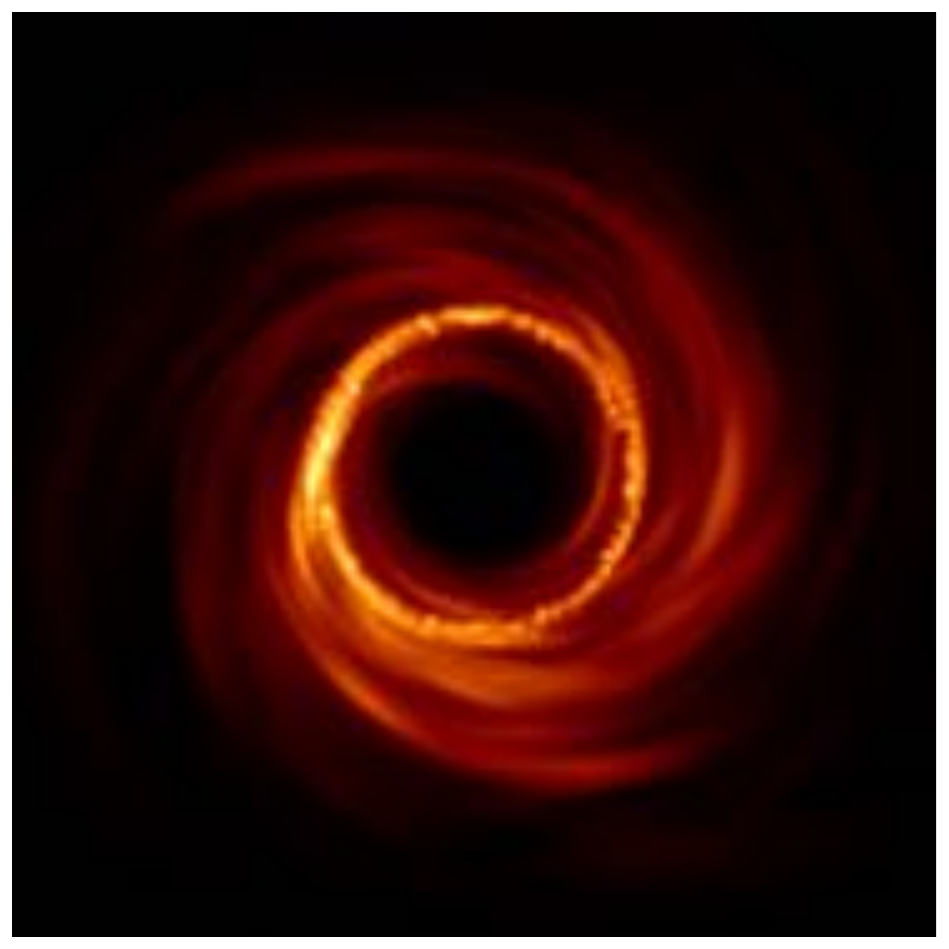}} } &
			\includegraphics[height=.15\linewidth]{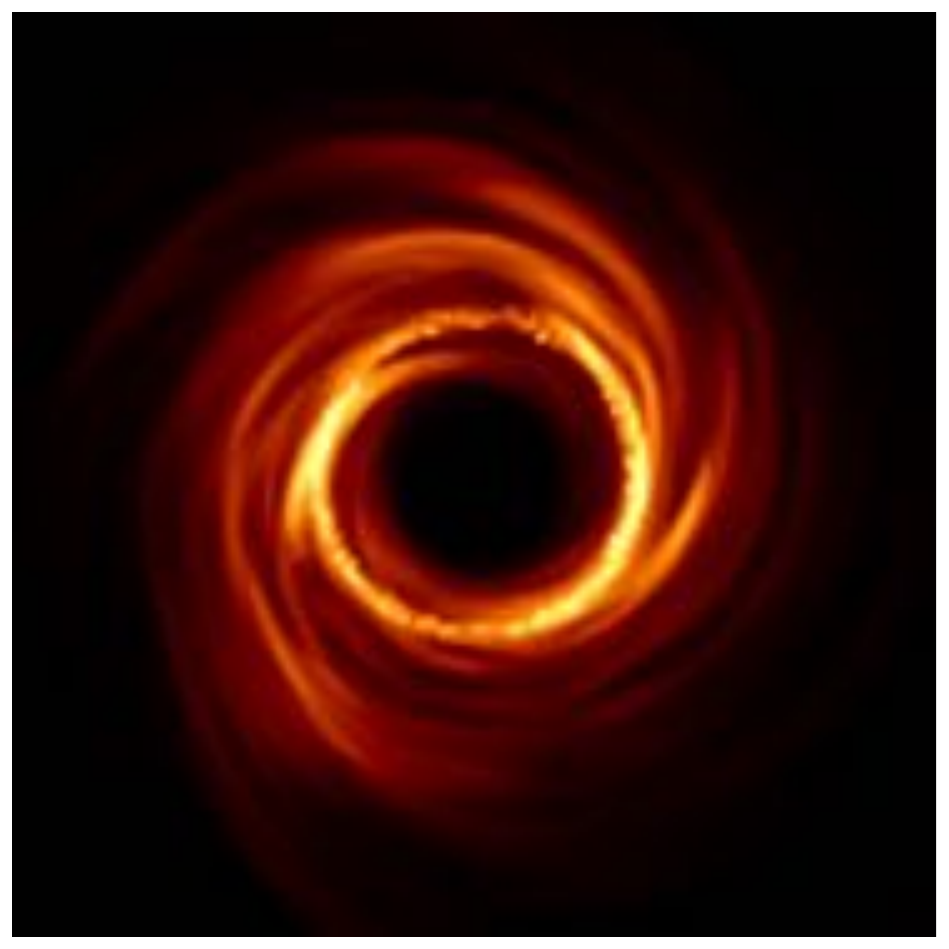} &
			\includegraphics[height=.15\linewidth]{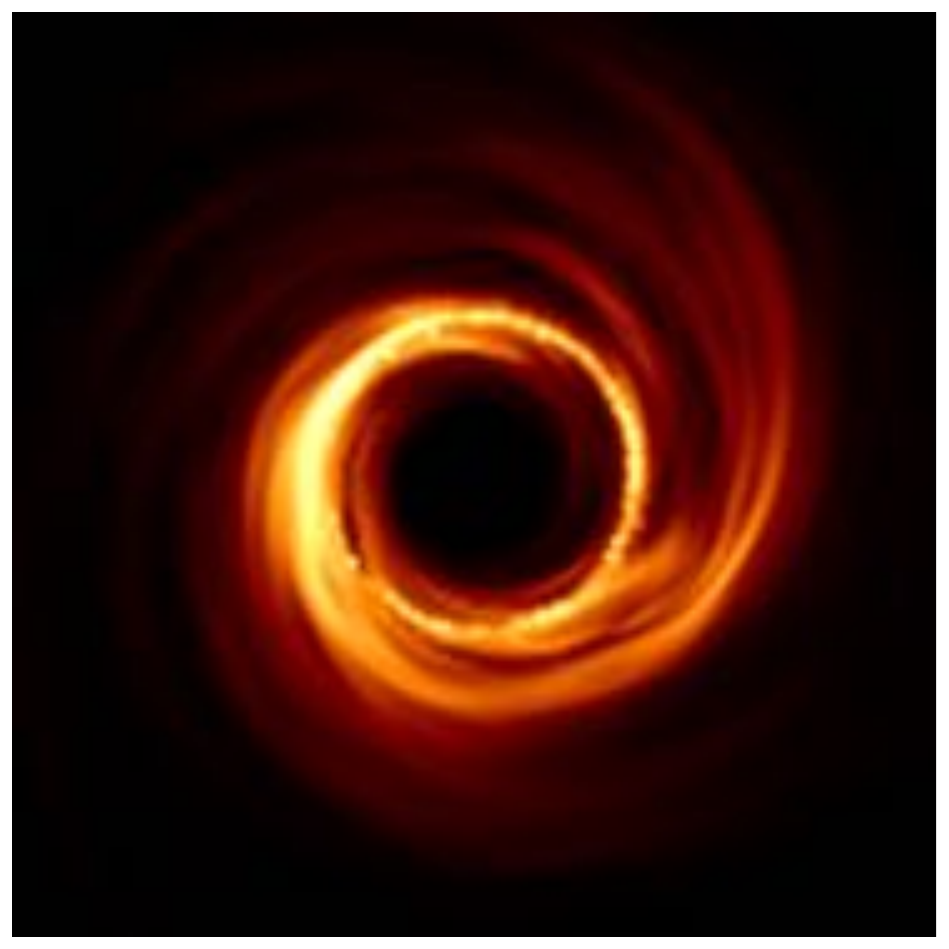} &
			\includegraphics[height=.15\linewidth]{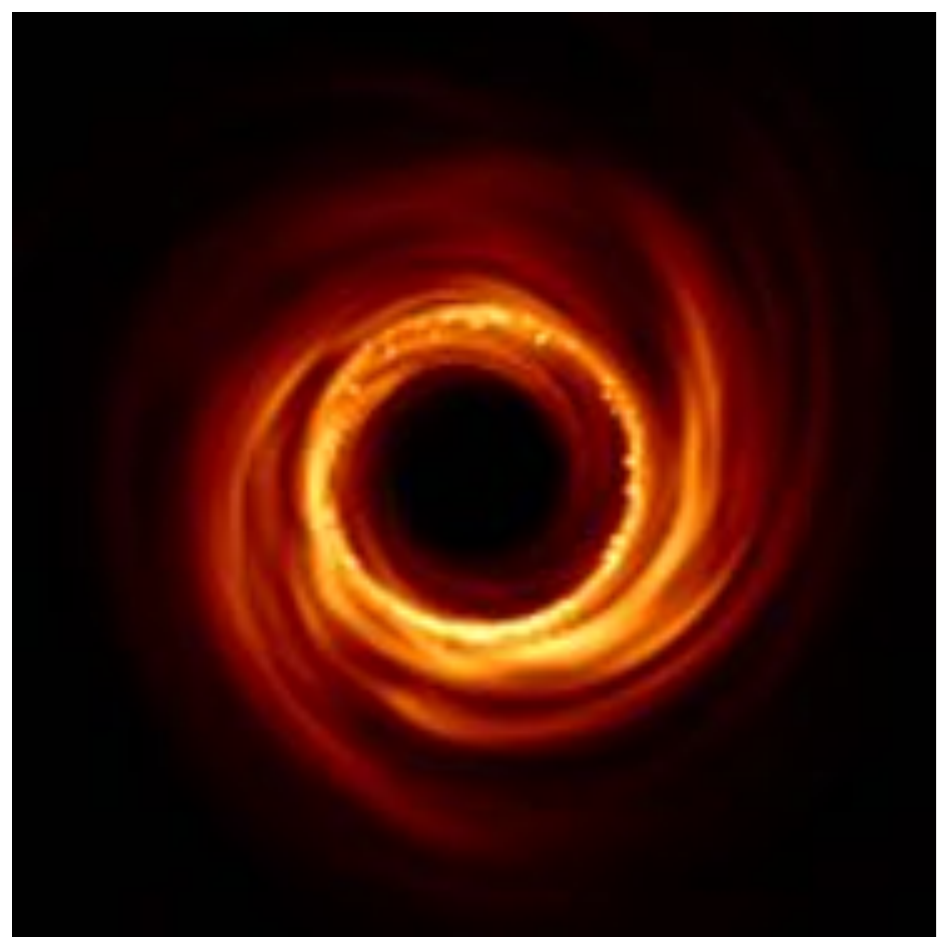} 
			\\   \hline
		&	&\vspace{-.1in} & & & &\\
			 & 	\multirow{1}{*}[0.45in]{ \rotatebox[origin=t]{90}{\small{\textsf{Snapshot}} }}
			&
			{{\includegraphics[height=.15\linewidth]{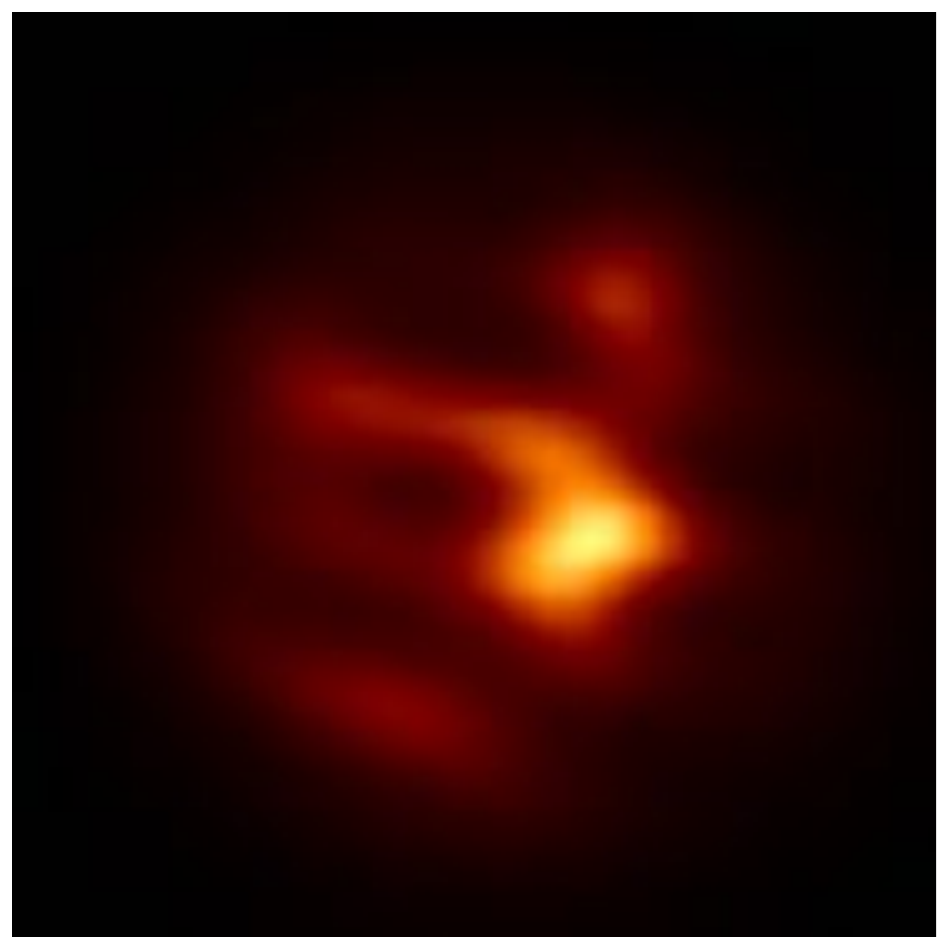}} } &
			{{\includegraphics[height=.15\linewidth]{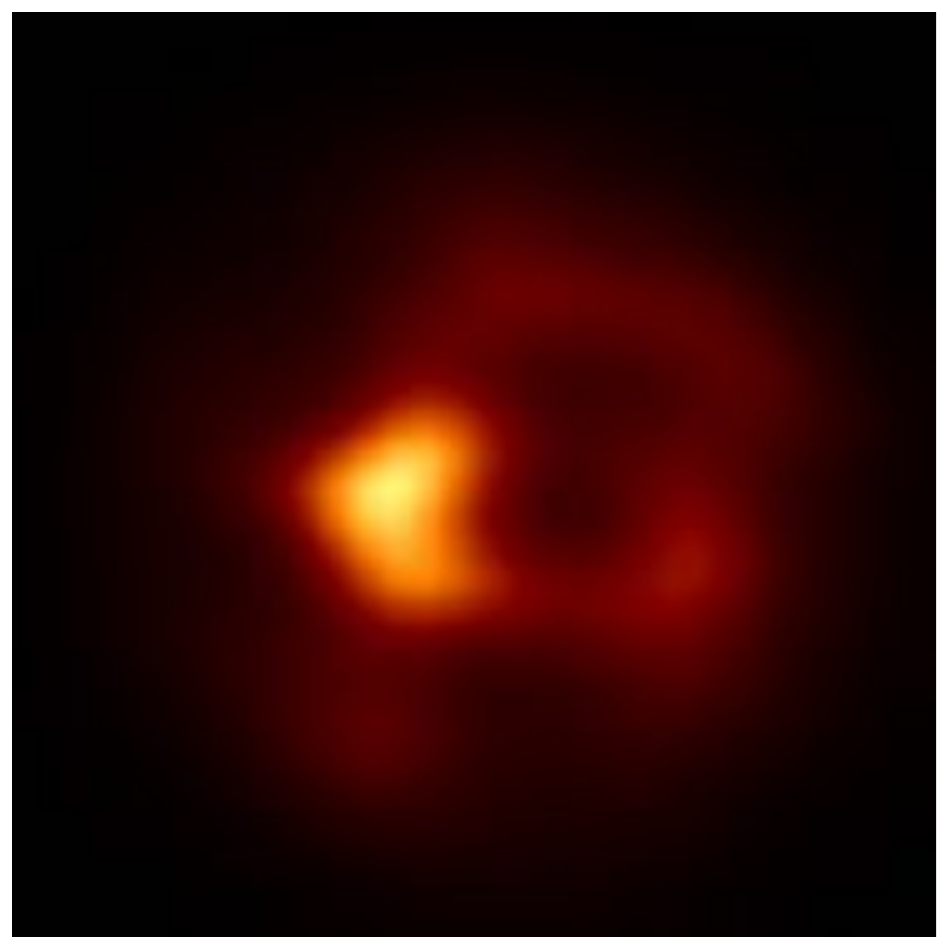}} } &
			\includegraphics[height=.15\linewidth]{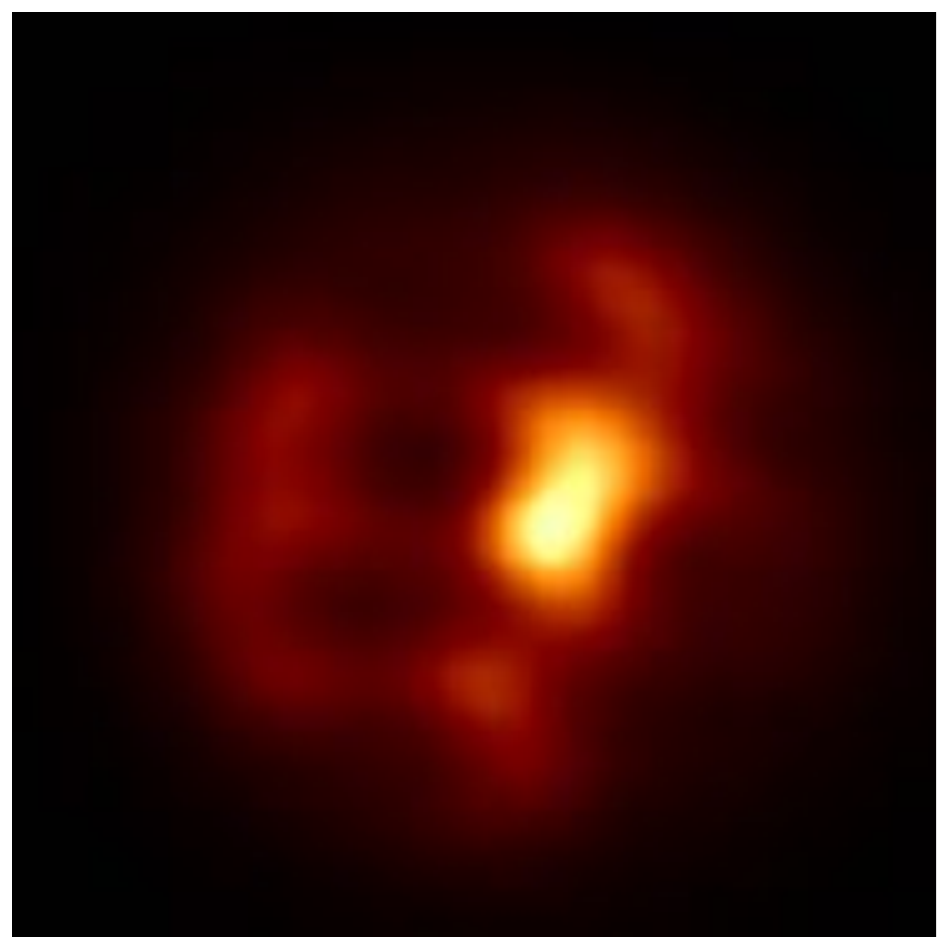} &
			\includegraphics[height=.15\linewidth]{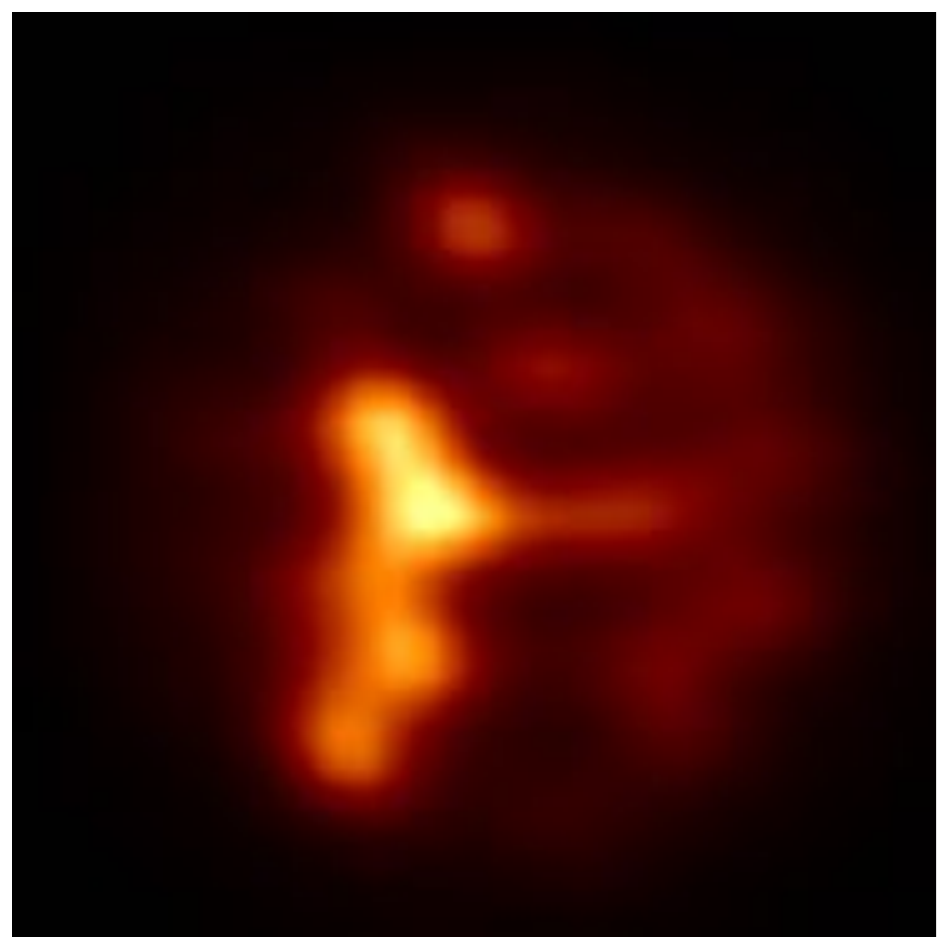} &
			\includegraphics[height=.15\linewidth]{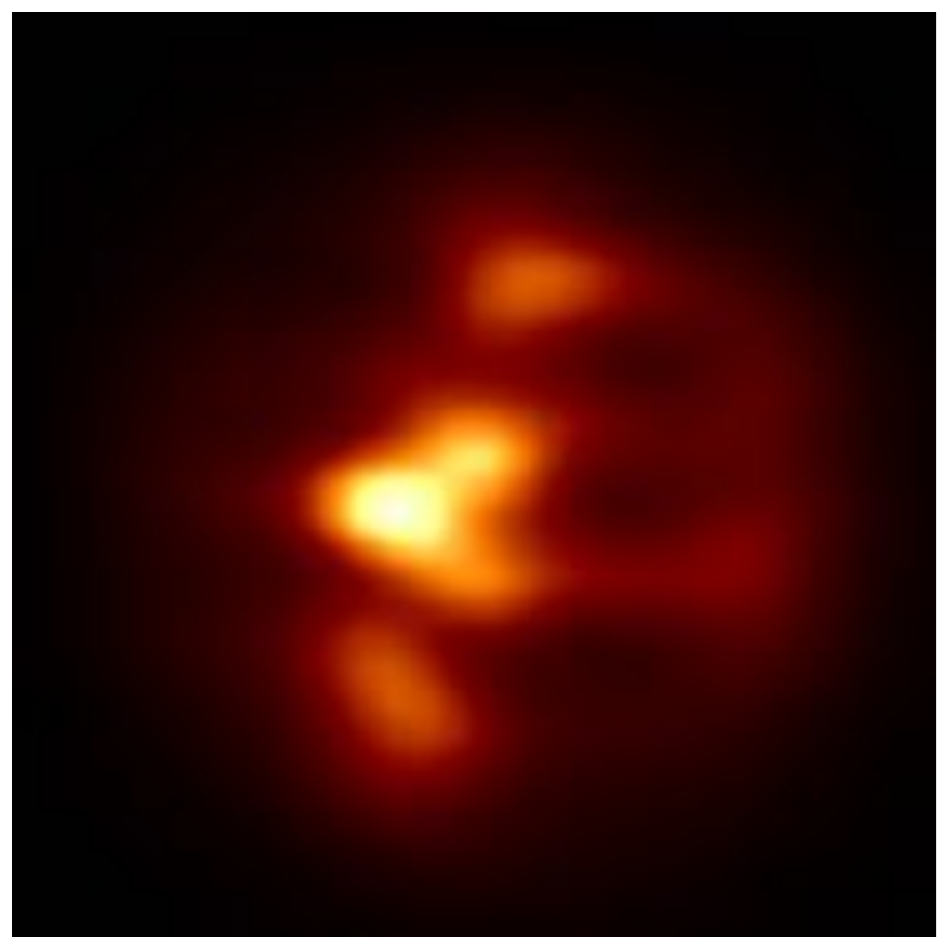}  
			\\   \hline
		&	&\vspace{-.1in} & & & &\\
		&	 	\multirow{1}{*}[0.3in]{ \rotatebox[origin=t]{90}{\small{\textsf{~\cite{Johnson_dynamical}}} }}
			&
			{{\includegraphics[height=.15\linewidth]{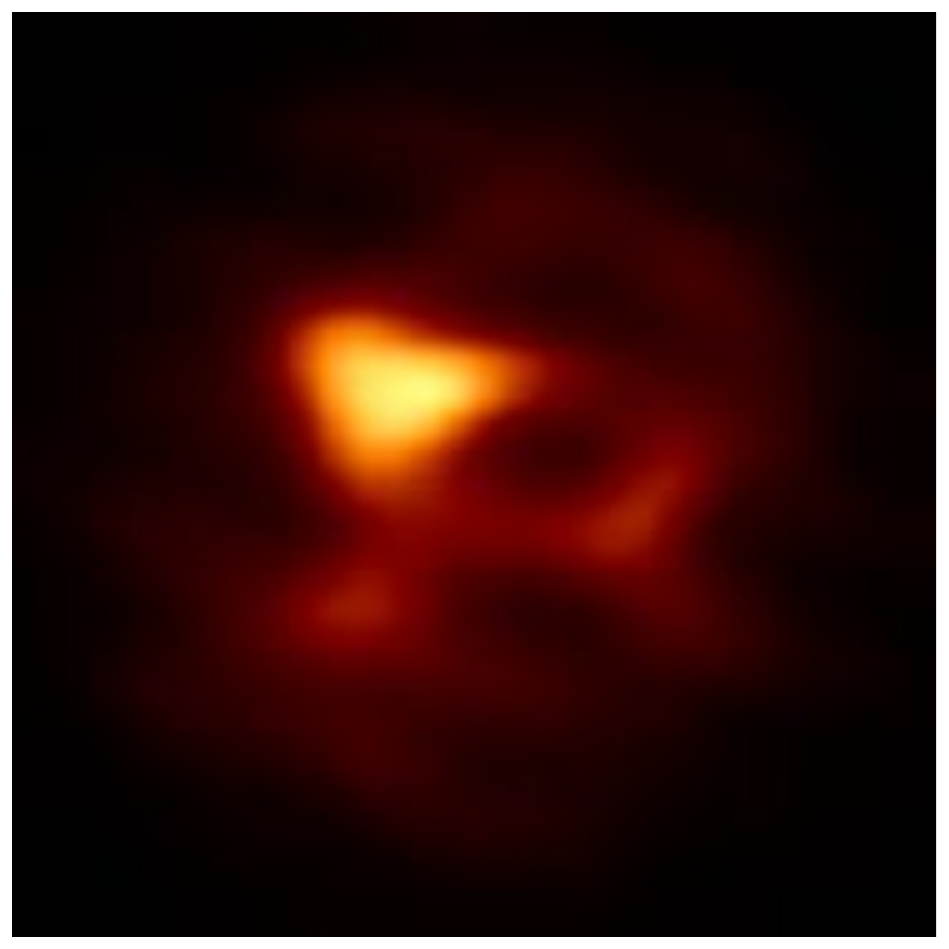}} } &
			{{\includegraphics[height=.15\linewidth]{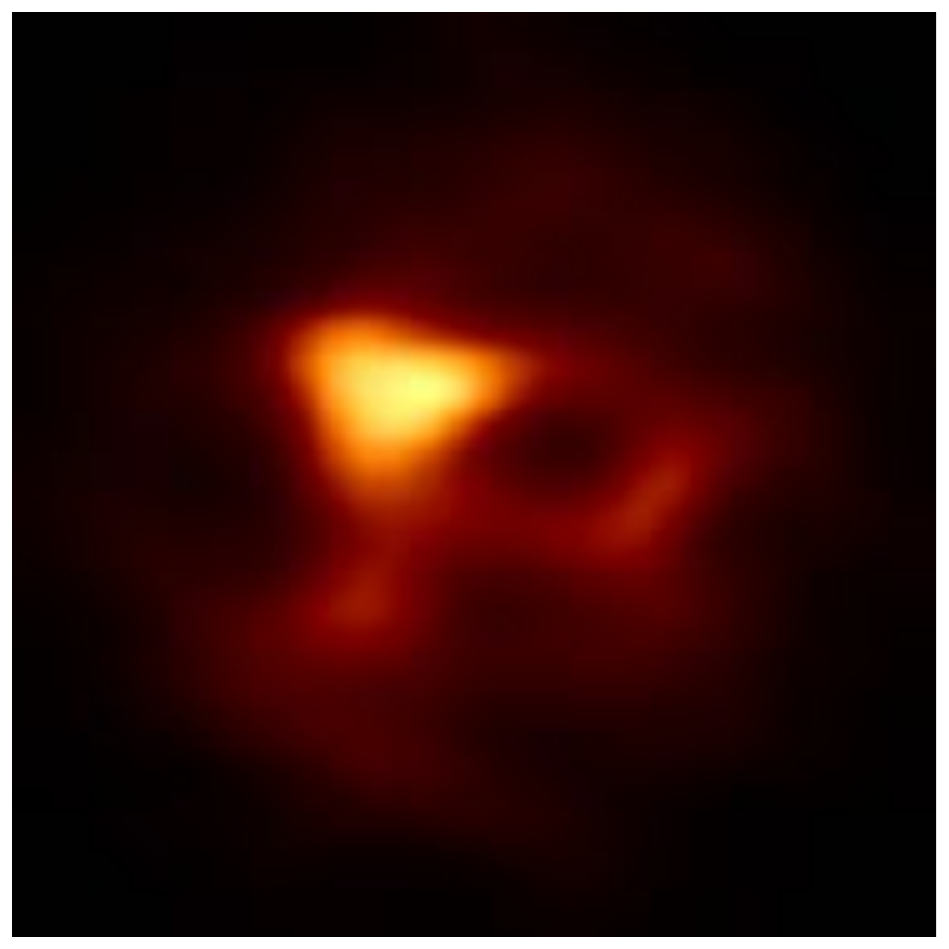}} } &
			\includegraphics[height=.15\linewidth]{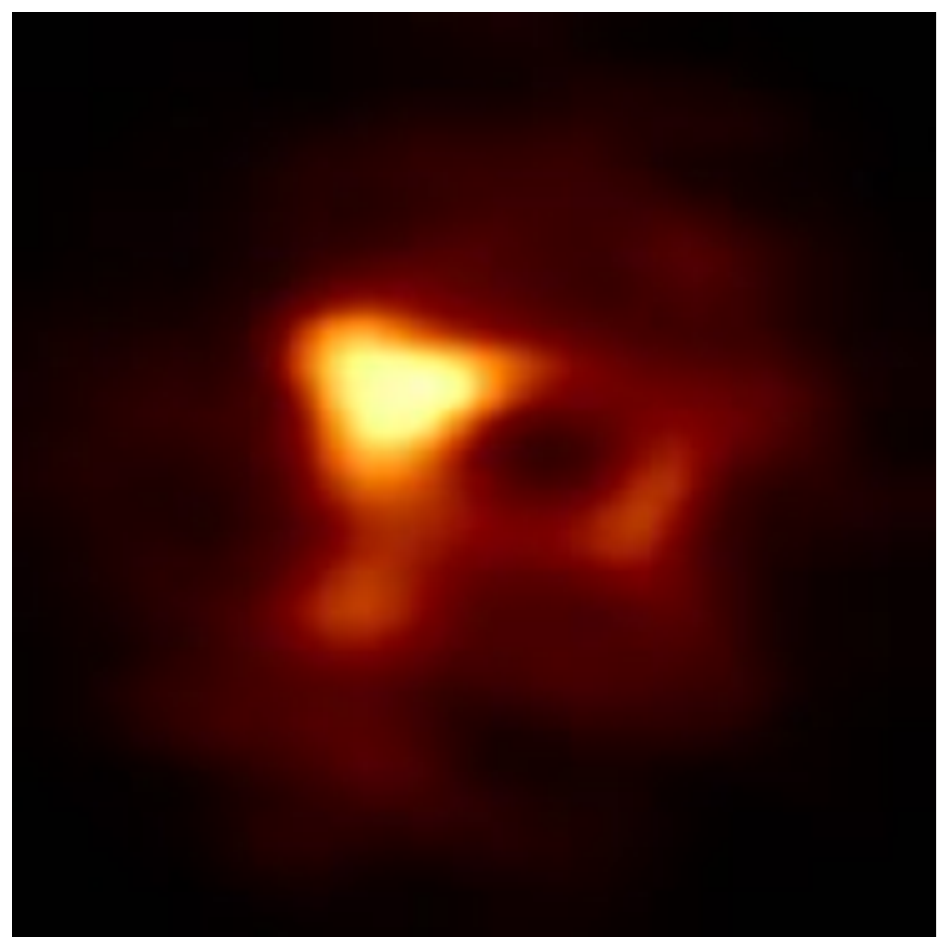} &
			\includegraphics[height=.15\linewidth]{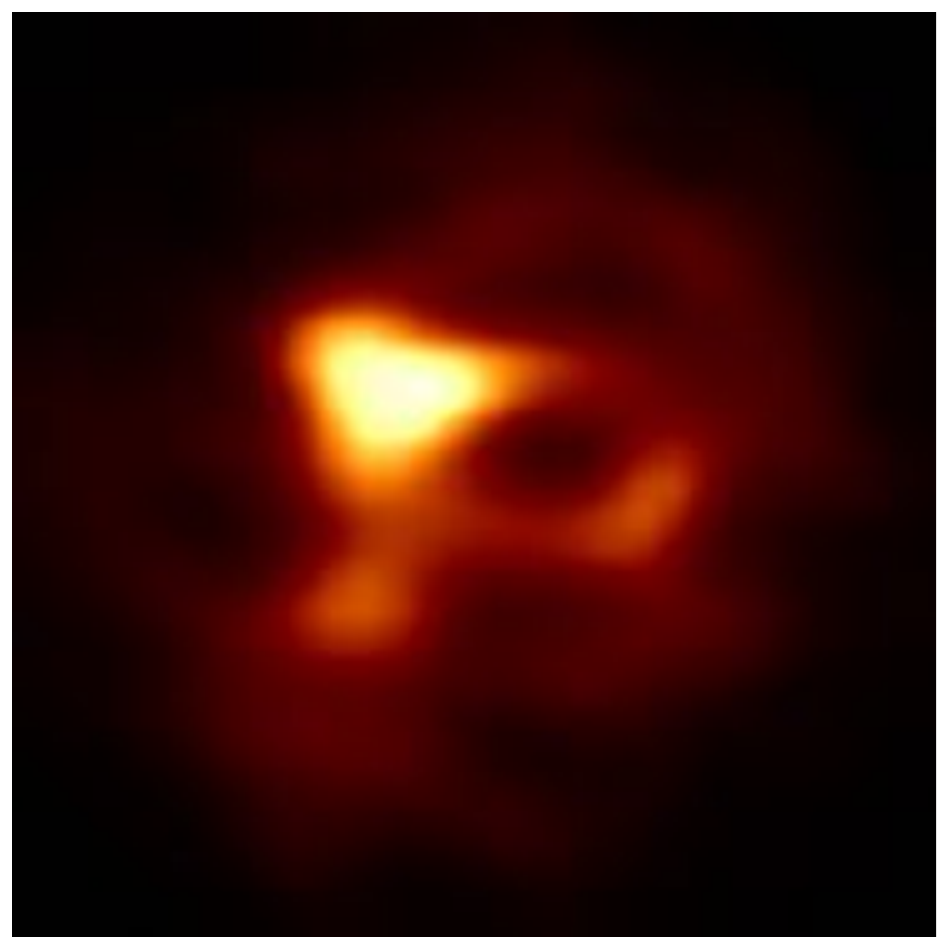} &
			\includegraphics[height=.15\linewidth]{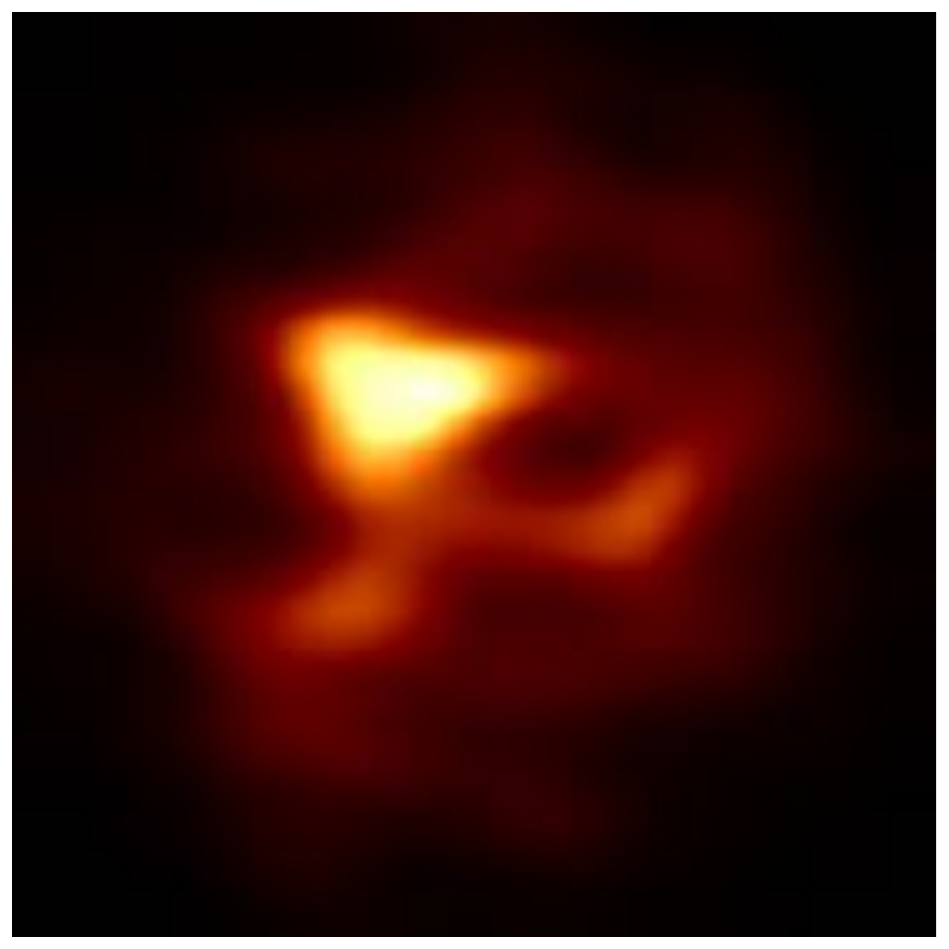} 
			\\   \hline
			& &\vspace{-.1in} & & & &\\
		 	& \multirow{1}{*}[0.48in]{ \rotatebox[origin=t]{90}{\small{\textsf{StarWarps}} }}
			&
			{{\includegraphics[height=.15\linewidth]{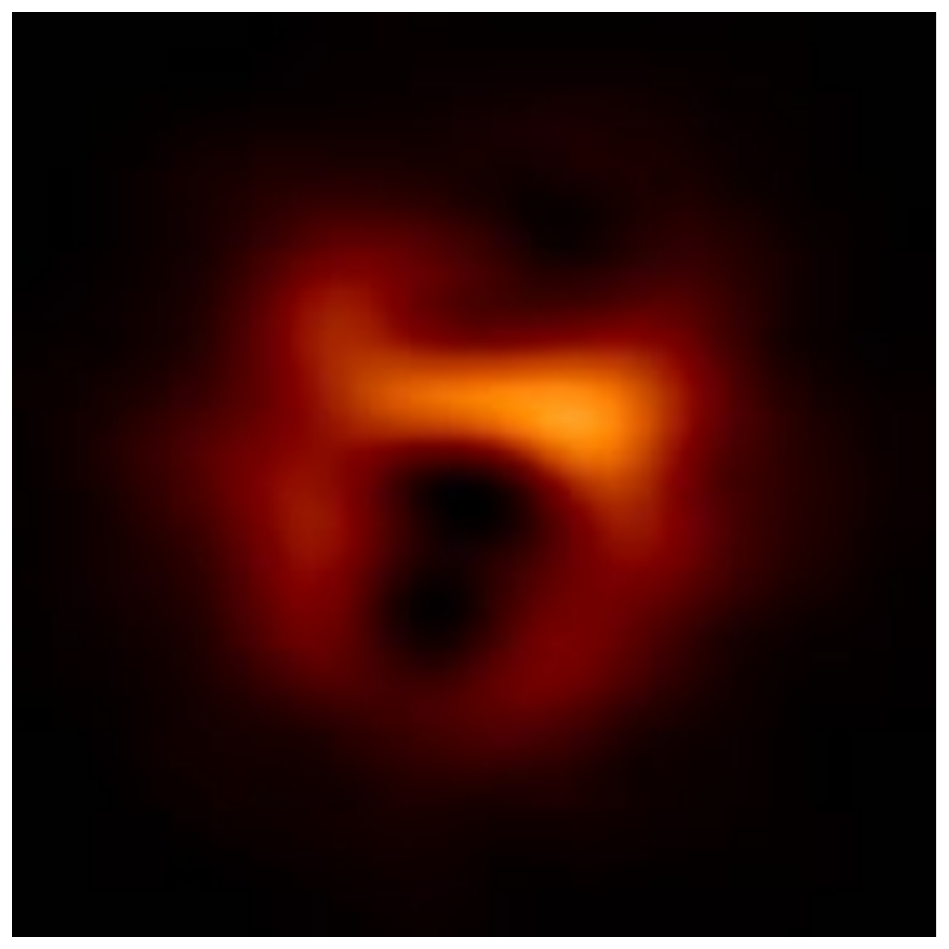}} } &
			{{\includegraphics[height=.15\linewidth]{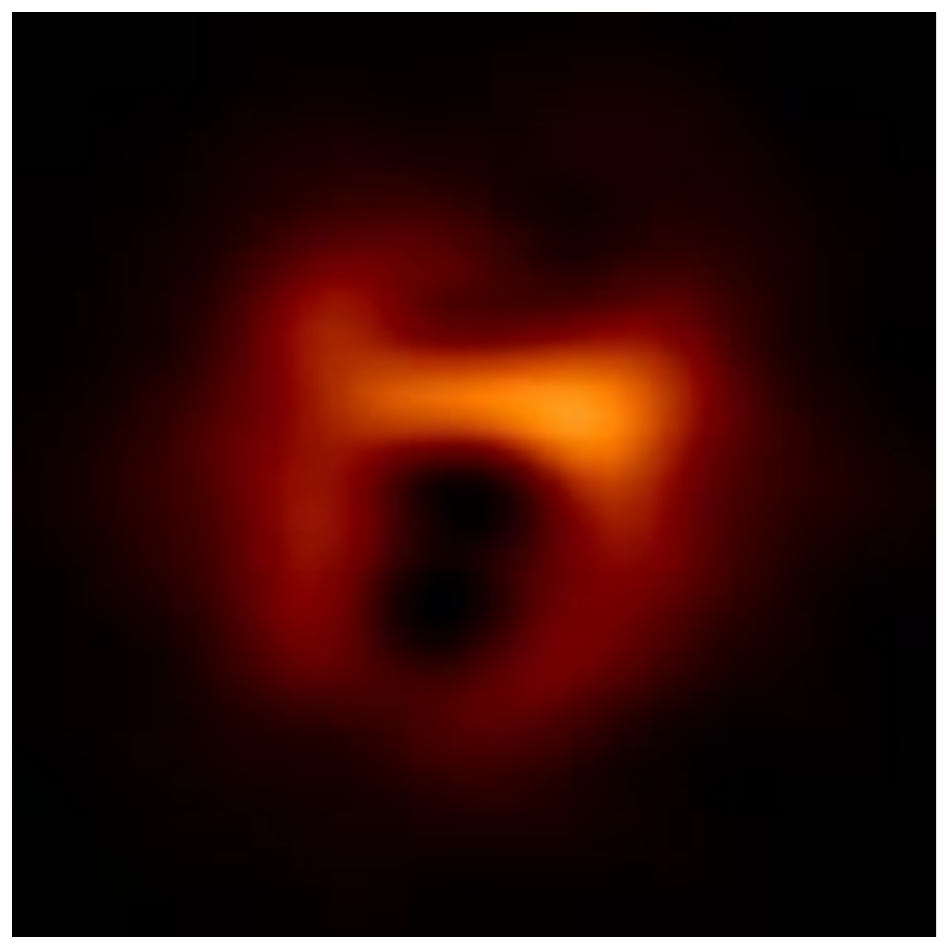}} } &
			\includegraphics[height=.15\linewidth]{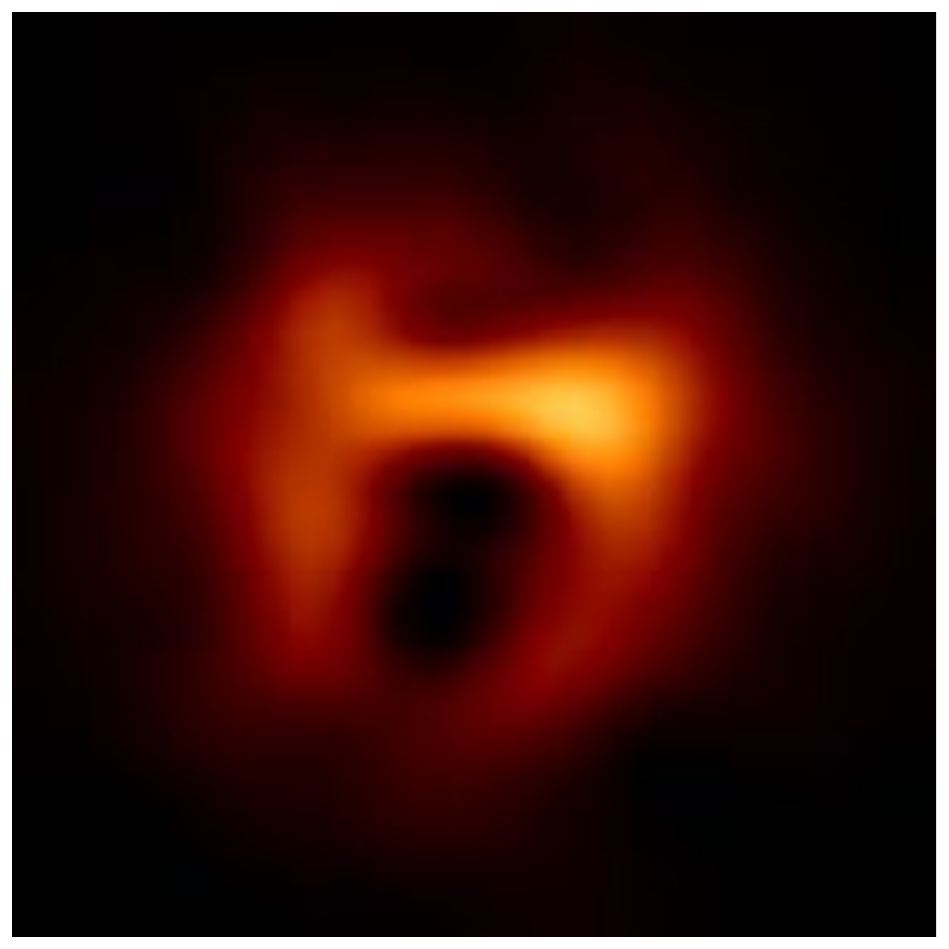} &
			\includegraphics[height=.15\linewidth]{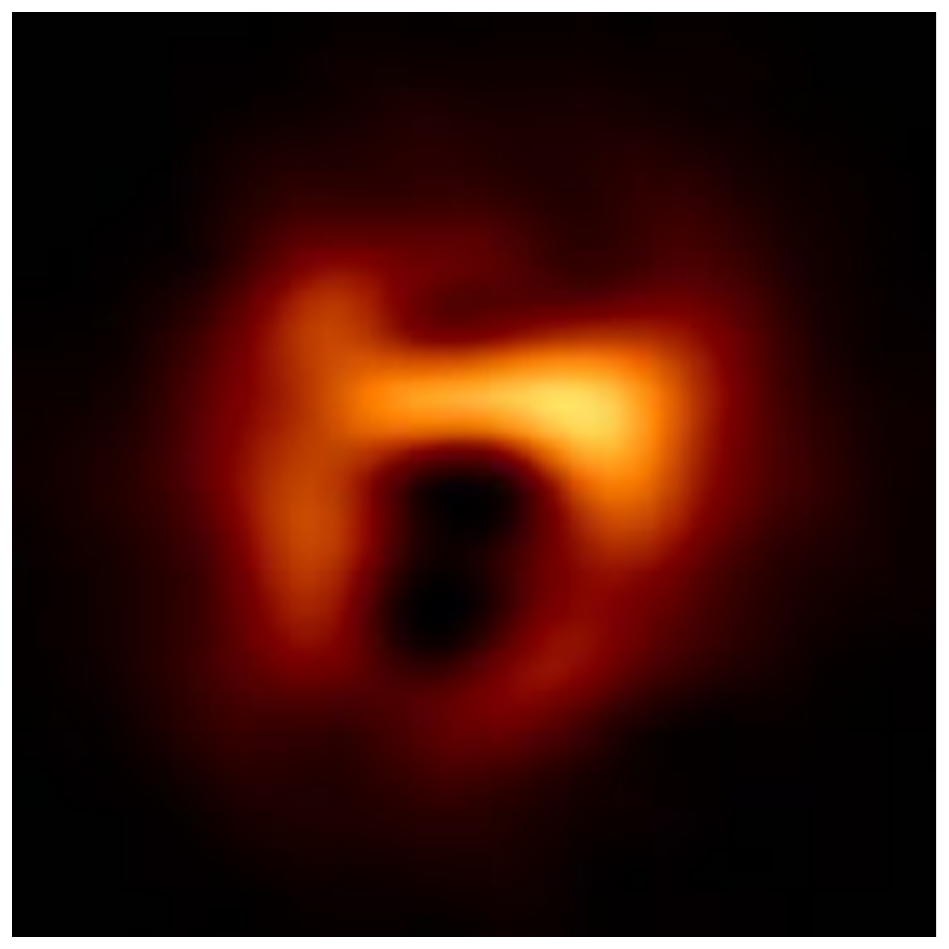} & 
			\includegraphics[height=.15\linewidth]{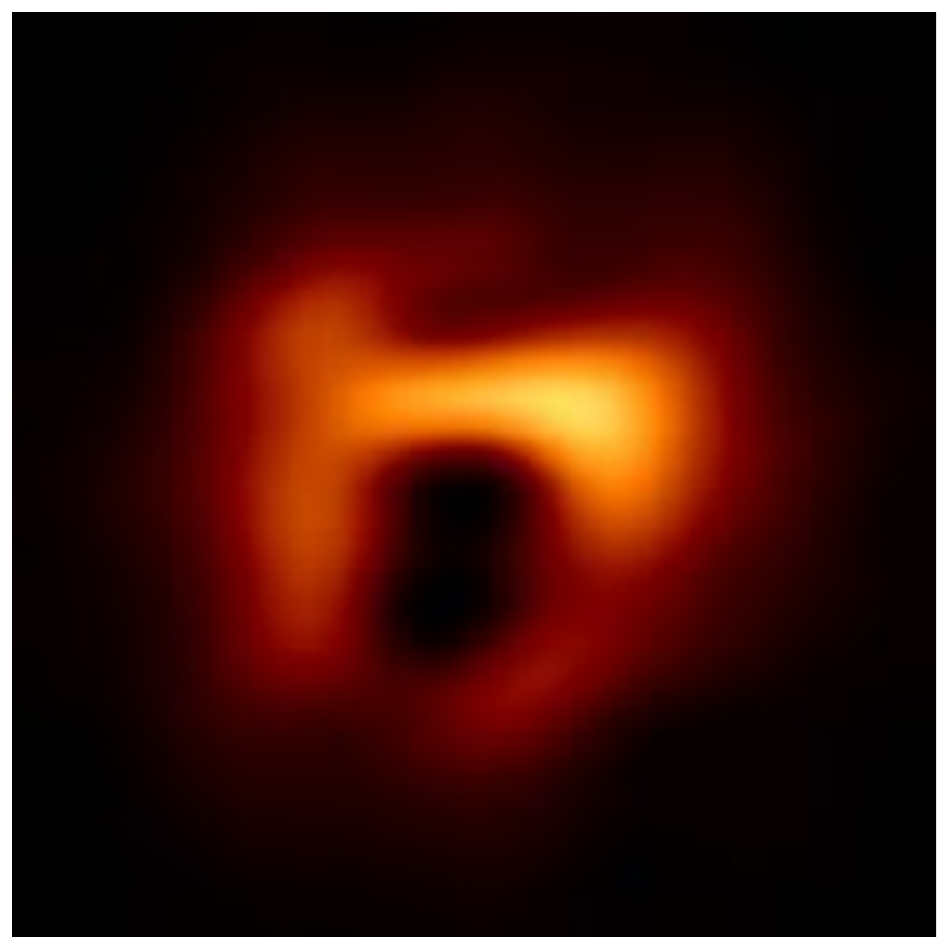} 
			\\   \hline
			& &\vspace{-.1in} & & & &\\
		 		\multirow{2}{*}[0.52in]{ \rotatebox[origin=t]{90}{\small{\textsf{StarWarps }} }}   \hspace{-0.25in} &\multirow{1}{*}[0.33in]{ \rotatebox[origin=t]{90}{\small{\textsf{+~\cite{Johnson_dynamical}}} }}
			&
			{{\includegraphics[height=.15\linewidth]{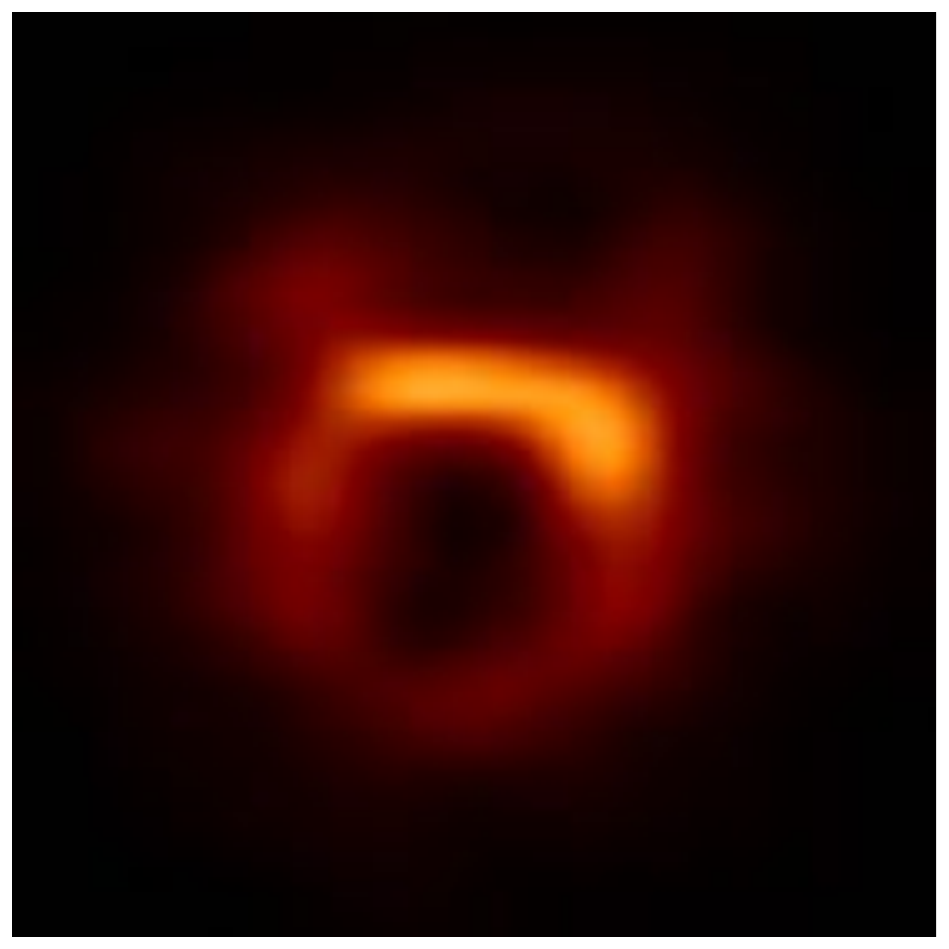}} } &
			{{\includegraphics[height=.15\linewidth]{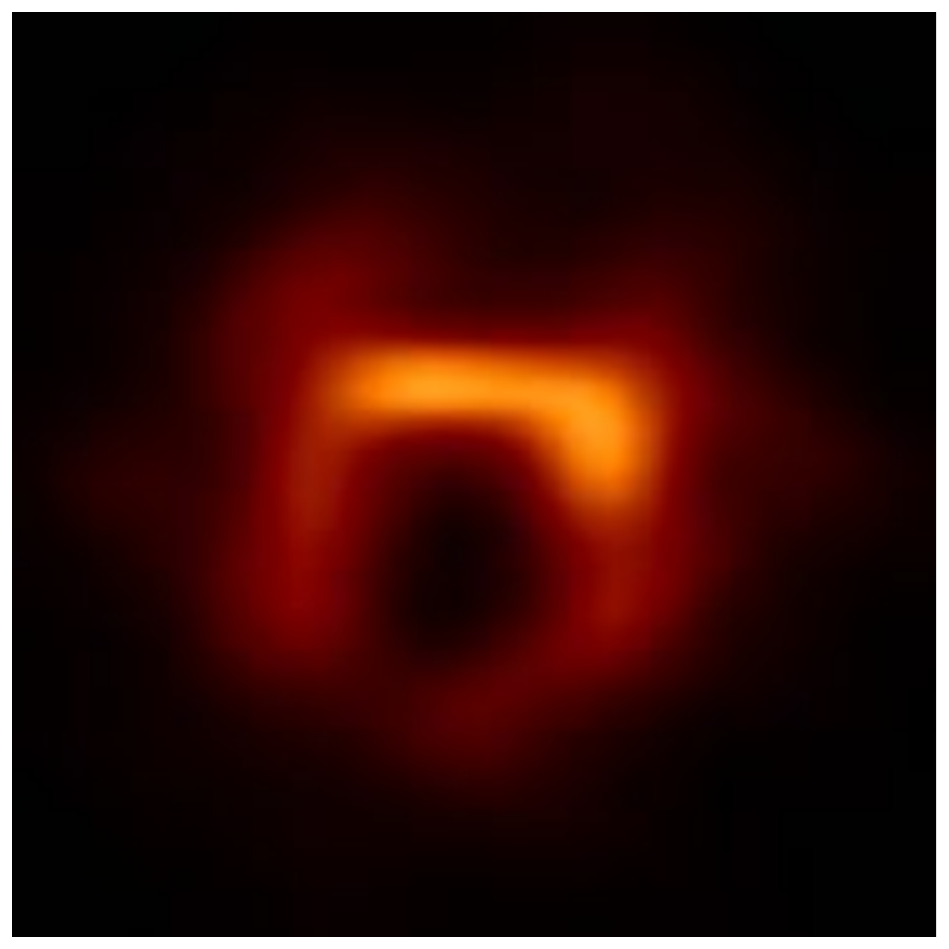}} } &
			\includegraphics[height=.15\linewidth]{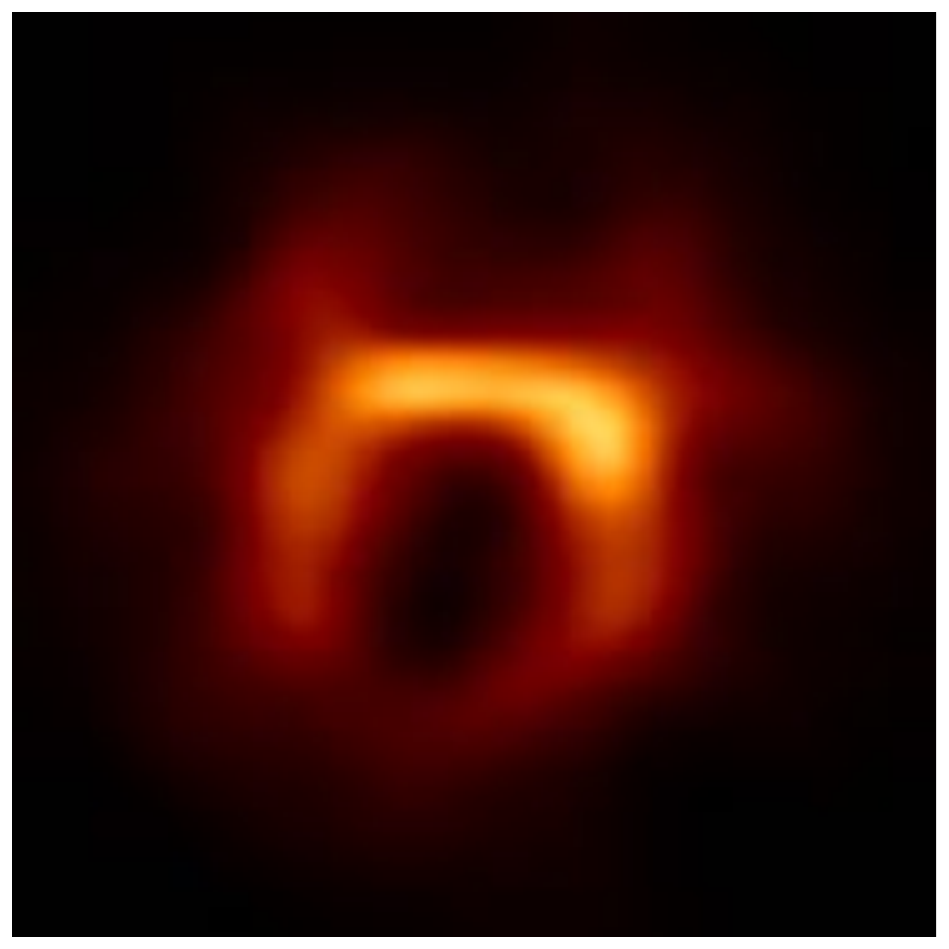} &
			\includegraphics[height=.15\linewidth]{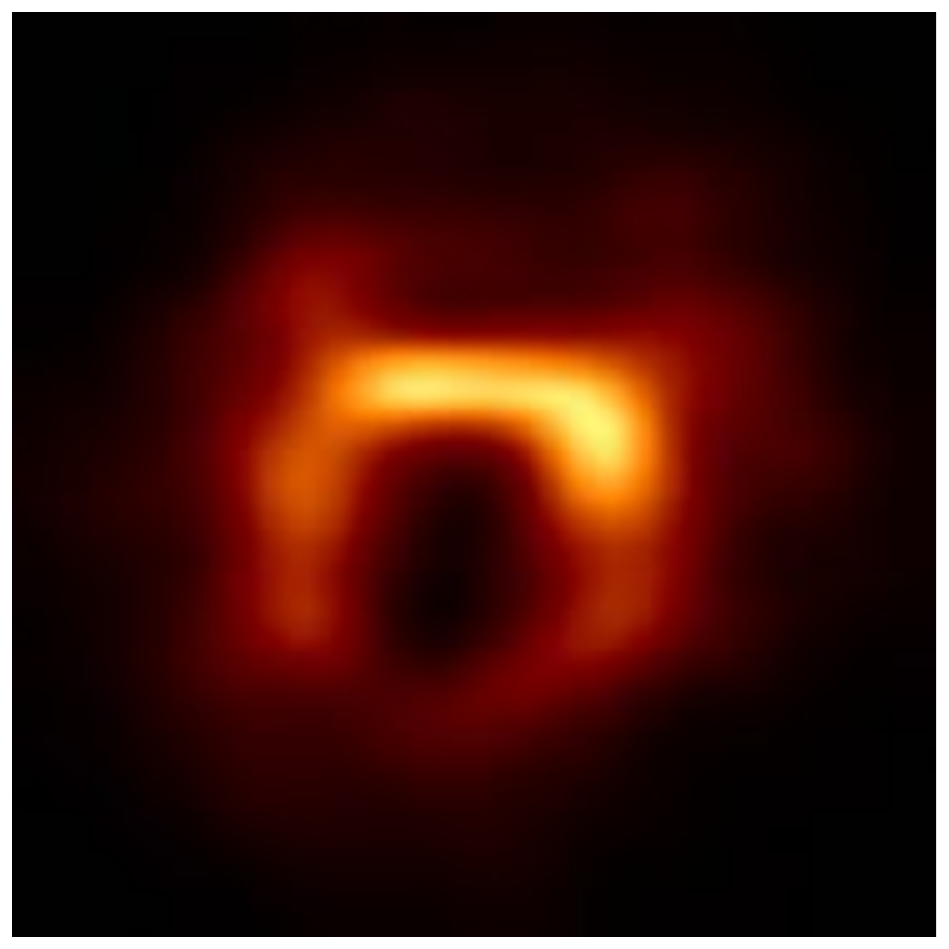} &
			\includegraphics[height=.15\linewidth]{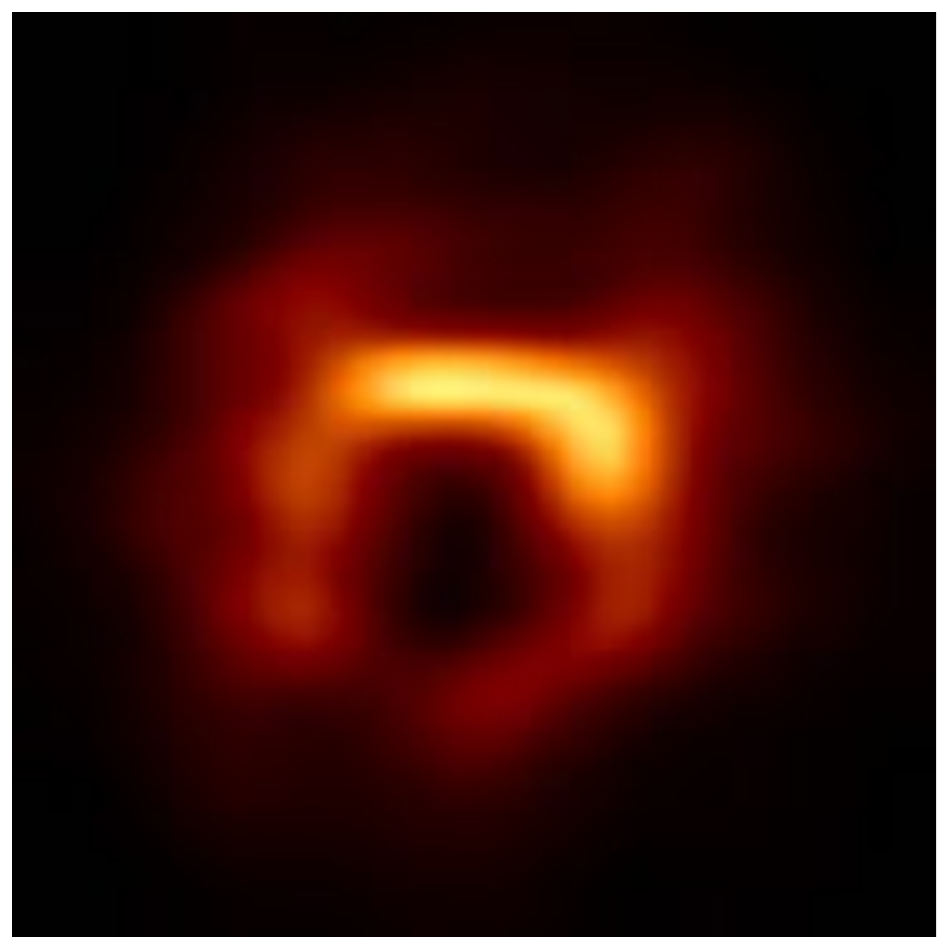} 
			\\   \hline
		\end{tabular}
		\caption{{\bf Comparison to~\cite{Johnson_dynamical} and Snapshot Imaging:} A comparison of results obtained using the proposed StarWarps method to Snapshot imaging and a method presented in~\cite{Johnson_dynamical}. The same simulated EHT2017 data of Video 3 was used for each result, and contained atmospheric noise. Snapshot imaging, which independently reconstructs each frame, is unable to produce reasonable results, and has no continuity through time due to the loss of absolute location information when using atmosphere corrupted measurements. The more flexible framework of~\cite{Johnson_dynamical} often makes it possible to obtain sharper and cleaner images, however struggles when working with very few measurements, as is the case for the EHT2017 array. Although results are consistent through time,~\cite{Johnson_dynamical} fails to recover the true ring structure of the source. StarWarps is able to begin recovering this ring structure, but contains a number of artifacts spurring from the main ring structure. Initializing~\cite{Johnson_dynamical} with the result of StarWarps produces a cleaner and sharper result.  
			 }
		\label{fig:dynamicimagingcmp}
	\end{center}
\end{figure}

\begin{figure}[h!]
	\vspace*{-.3in}
	\centering
	\subfigure[Fig.~\ref{fig:dynamicimagingcmp} frame uv-coverage  ]{\includegraphics[height=.43\linewidth]{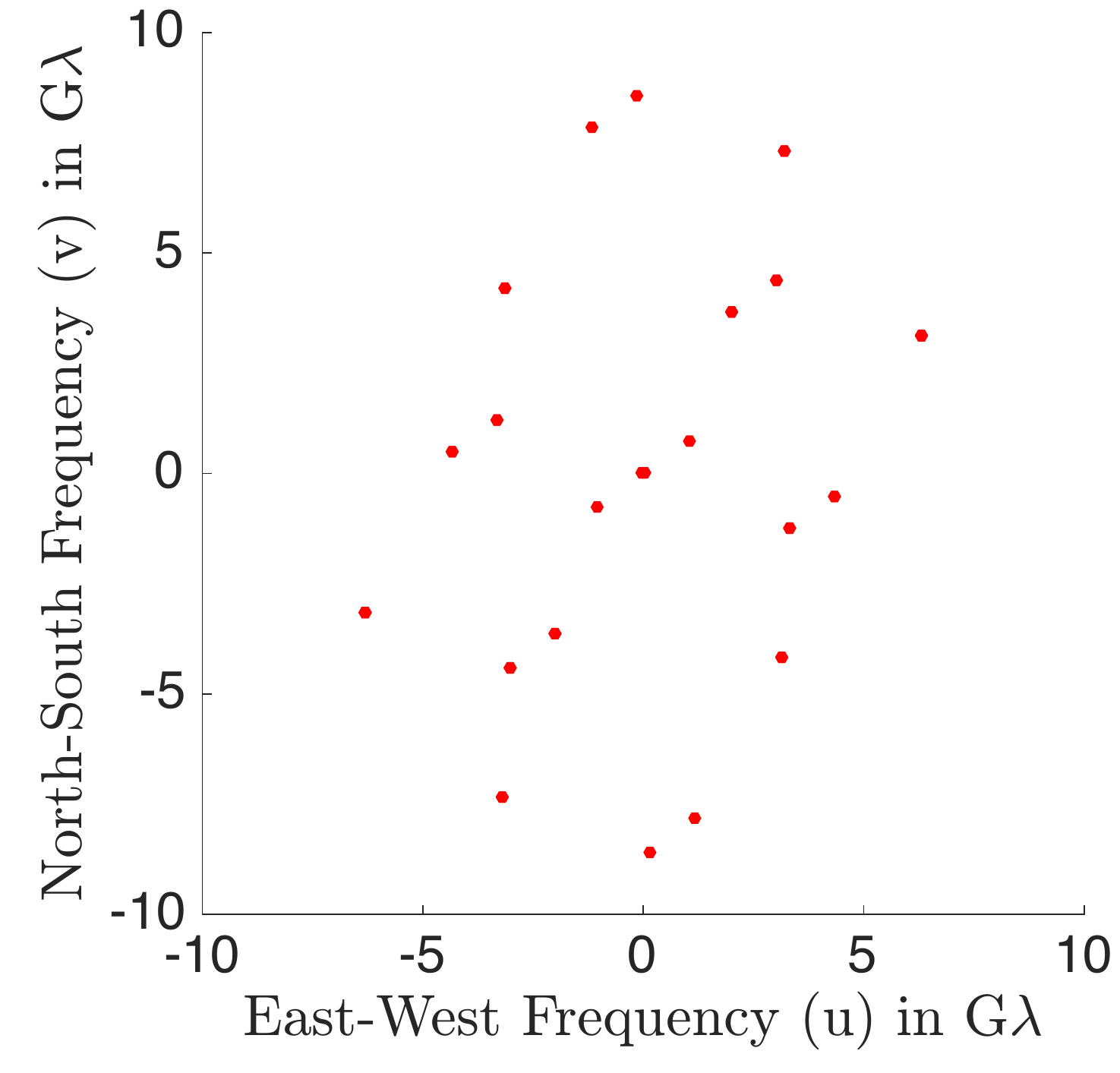}}
	\subfigure[Fig.~\ref{fig:m87} frame uv-coverage]{\includegraphics[height=.43\linewidth]{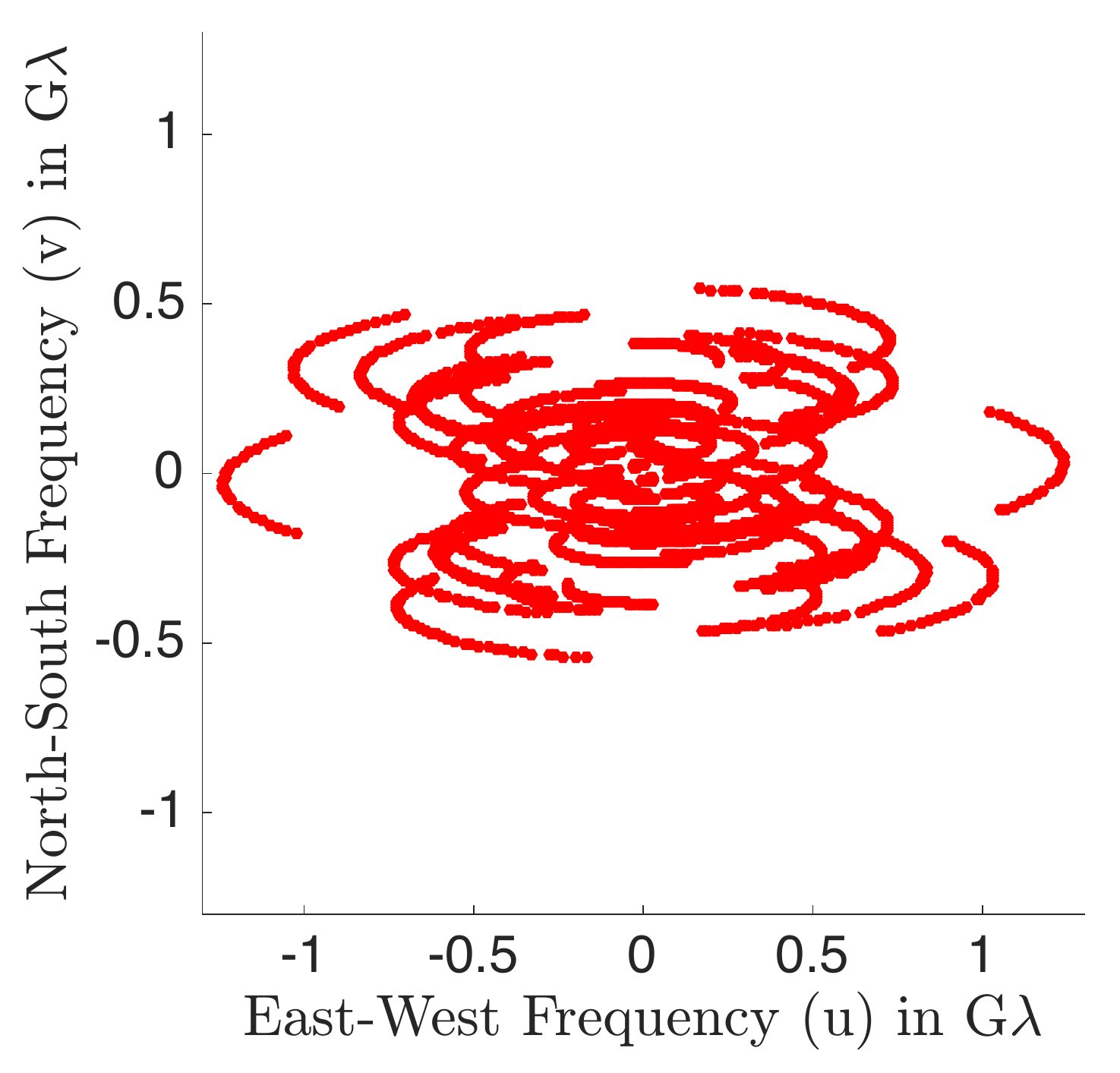}}
	\vspace{-.1in}
	\caption{{\bf Single frame uv-coverage:} (a) The uv-coverage for the first frame shown in Figure~\ref{fig:dynamicimagingcmp} contains 21 measurements while (b) the uv-coverage for the first frame shown in Figure~\ref{fig:m87} contains 1736. When the measurements provided are very sparse, as in (a), StarWarps significantly outperforms~\cite{Johnson_dynamical}. However, in the case of many measurements, as in (b),~\cite{Johnson_dynamical} achieves better results with a higher dynamic range.  }
	\label{fig:uvcov3}
	\vspace{-.25in}
	
\end{figure}

\begin{figure*}
	\begin{center}

			\vspace*{-.35in}

	\begin{tabular}{  c c c  }
					\multicolumn{3}{c}{{\includegraphics[width=1\linewidth]{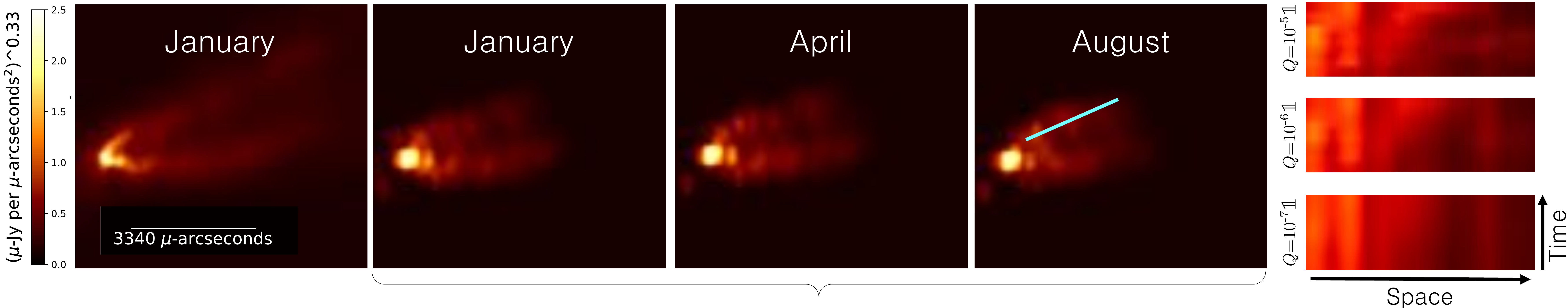}} } 
					\\
					\vspace{-.25in} && \\
	\hspace{.8in}	\normalsize{\textsf{\cite{Johnson_dynamical} }}& \hspace{2.1in} \normalsize{\textsf{StarWarps }}&\hspace{1.5in} 	 \\
		\vspace{.1in} && 
\end{tabular}

\vspace*{-.27in}
		\caption{{\bf Video Reconstruction of Real Observations:} A StarWarps movie reconstruction obtained using real VLBI data taken of the M87 jet over the course of a year. Frames are shown with a gamma correction of $\gamma={1}/{3}$ to highlight weak emission. Data was collected in 2007 using the Very Long Baseline Array (VLBA) at 43 GHz~\cite{walker2016observations}. As the source structure does not evolve over the course of a night, traditional imaging approaches can be used to reconstruct `snapshot' images from this data. 
			The forked structure appearing in the StarWarps reconstructions also appears in images reconstructed using the CLEAN static imaging approach~\cite{walker2016observations} and the dynamic imaging approach presented in~\cite{Johnson_dynamical} (see left-most image). 
StarWarps produces a video that allows us to easily visualize the moving arms of the jet; the reconstructed video appears to contain outward motion, with a brighter region propagating down the arms. By visualizing the same slice of each frame (indicated by the cyan line) it becomes easier to see this motion as a static image (see the 3 Space $\times$ Time images on the far right). Note the diagonal `line' shown in the top 2 of these Space $\times$ Time images indicates a bright region moves down the arm, towards the right of the image. The intensity of these slices has been increased by $70 \%$ to highlight the evolving, weak emission. By controlling the amount of temporal regularization, $\bQ$, we control the amount of motion that appears in the reconstructed video. Increasing the temporal regularization, by decreasing $\bQ$, results in a Space $\times$ Time slice that varies less with Time. Individual frames shown were generated using $\bQ = 10^{\text{-}6} \mathds{1}$. }
\vspace{-.3in}
		\label{fig:m87}
	\end{center}
\end{figure*}

\subsubsection{The Importance in Propagating Uncertainty}

StarWarps uses a multivariate Gaussian regularizer for imaging, which leads to a straightforward optimization method that propagates information through time. The uncertainty of each reconstructed image is encompassed in its approximated covariance matrix (${\bf P}_{t|t}$), which informs the reconstruction of each neighboring latent image. Although this covariance matrix is sometimes a crude estimate of the true uncertainty, it is still crucial in reconstructing faithful images when measurements are especially sparse.

The importance of propagating uncertainty through the covariance matrix is demonstrated in Figure~\ref{fig:propinfo}. This figure shows the effect of turning on and off the covariance propagation. Covariance propagation can be easily turned off by setting ${\bf P}_{t|t}=0$ at each forward and backward update. Results in the figure are shown on simulated data from the EHT2017 array on Video 3, with and without atmospheric error. Note that in both cases, propagating the covariance matrix helps to substantially improve results. This is true even even in the case of atmospheric error, when the measurement function $f(\im)$ is non-linear and the covariance matrix is only a rough approximation of the true uncertainty.

\subsubsection{Dynamical Imaging Comparison}


As discussed in Section~\ref{sec:setup}, the Dynamical Imaging method presented in~\cite{Johnson_dynamical} was developed simultaneously, and shares many similarities to the work presented in this paper: they both aim to solve for a video rather than a static image. However, they have significant differences, leading to different strengths and weaknesses. The framework of~\cite{Johnson_dynamical} allows for more sophisticated image and temporal regularization, at the cost of a difficult optimization problem that does not propagate uncertainty. This results in sharper and cleaner videos when there is sufficient data, but can lead to poor results when there are very few measurements. 
Conversely, StarWarps' use of very simple Gaussian image and temporal regularization results in blurry results, but allows us to propagate an approximation of uncertainty (through the covariance matrix) and produce better results when very few measurements are available.

A comparison of results from~\cite{Johnson_dynamical} and StarWarps on EHT2017 simulated data can be seen in Figure~\ref{fig:dynamicimagingcmp}. Results of~\cite{Johnson_dynamical} were produced using $\mathcal{R}_{\Delta I}$ and KL $\mathcal{R}_{\Delta t}$ temporal regularization, and Maximum Entropy and Total Variation Squared image regularization. Note that for this especially sparse data,~\cite{Johnson_dynamical} on its own does not faithfully reconstruct the ring structure of the underlying source. StarWarps is able to produce a ring, but with a number of blurry artifacts. Initializing~\cite{Johnson_dynamical} with the output of StarWarps produces the cleanest result. 
Although the StarWarps method runs faster on this example than~\cite{Johnson_dynamical} (84 seconds vs 204 seconds in Python on a 2.8 GHz Intel Core i7), StarWarps is memory intensive and its computational complexity scales poorly with increasing image size compared to~\cite{Johnson_dynamical}.  
To help solve these issues, in the future ideas from Ensemble Kalman Filters could be adapted in order to avoid StarWarp's costly matrix inversions and reduce the method's memory footprint~\cite{evensen2003ensemble}.  

An additional result comparing the two methods can be seen in Figure~\ref{fig:m87}, which is discussed in the next section. In this example there is sufficient data to reconstruct each frame independently, and~\cite{Johnson_dynamical} is able to produce a cleaner image with a higher dynamic range than StarWarps. 

Figure~\ref{fig:uvcov3} compares the uv-coverage of a single frame for Figures~\ref{fig:dynamicimagingcmp} and Figure~\ref{fig:m87}, highlighting that StarWarps is comparatively strongest  in the case of sparse data, as will be available for the EHT. These examples demonstrate that StarWarps and~\cite{Johnson_dynamical} are complementary methods, and may ultimately lead to hybrid approaches for video reconstruction that produce higher quality results.




\subsubsection{Application to Real VLBI Data}

Although StarWarps was developed with the considerations of the EHT in mind, it can be applied to VLBI data taken from other sources and telescope arrays. For instance, galactic relativistic jet sources (``microquasars") often show variability over the course of a single observation~\cite{timedeprecon}. 
However, due to physical constraints, most VLBI telescope networks observe sources that do not evolve this quickly, such as distant jets from the cores of Active Galactic Nuclei. In these cases, traditional static imaging approaches can be applied to each night of data to produce faithful reconstructions. Yet, by jointly processing the data taken over a larger span of time, we are able to make movies of long-term source evolution that preserve continuity of features through time, thus reducing the flickering that occurs when independently reconstructing each frame. 

In Figure~\ref{fig:m87} we demonstrate StarWarps on archival data taken of the M87 jet. This data was taken using the Very Long Baseline Array (VLBA) as part of the M87 Movie Project~\cite{walker2016observations}. Ten epochs of data between the beginning of January and end of August in 2007 were processed simultaneously. Images were reconstructed with a 10 m-arcsecond field of view with $\npix=70$ pixels. 

Unlike as expected in EHT observations, the dynamic range of the M87 Jet is very high. In order to faithfully reconstruct a high dynamic range image using the simple Gaussian prior, we have incorporated gamma correction into our measurement function. Rather than reconstruct a video containing linear-scale images, we instead reconstruct gamma-corrected images. To do this we replace the measurement function $f(\im)$ with $f(\im^{\frac{1}{\gamma}})$. During reconstruction of this M87 Jet video we have used $\gamma=1/2$. Although these images still do not have the same dynamic range that is achieved through other imaging methods~\cite{walker2016observations,Johnson_dynamical}, StarWarps is still able to recover the faint arms of the jet. 

The reconstructed movie produced by StarWarps shows outward motion along the jet. While this motion is hard to see in Figure~\ref{fig:m87}'s static frames, by visualizing a slice of each image (indicated by the cyan line) through time the motion becomes more apparent. The resulting Space $\times$ Time image shows a brighter region of emission moving along the arm of the jet. We show the same Space $\times$ Time reconstruction for different weightings of temporal regularization, $\bQ$. Note that as temporal regularization increases, by decreasing $\bQ$, the Space $\times$ Time image becomes more uniform in time. 
 
\vspace{-.2in}
\subsection{Unknown Evolution  Model (Learn Warp)}

\begin{figure}[tb]
\vspace{-.35in}
\hspace*{-.5in}
\centering
	\begin{center}
		\setlength{\tabcolsep}{1pt}
		\begin{tabular}{  c | c | c }
\hspace*{-.1in} & \large{\textsf{NO ATM. ERROR}}   &\large{\textsf{ATM. ERROR}}      \\  \hline

& \vspace{-0.05in} & \\

\hspace*{-.1in} \multirow{1}{*}[0.9in]{ \rotatebox[origin=t]{90}{  \large{\textsf{Video 1}} }} & \hspace{0.05in} {{\includegraphics[height=0.35\linewidth]{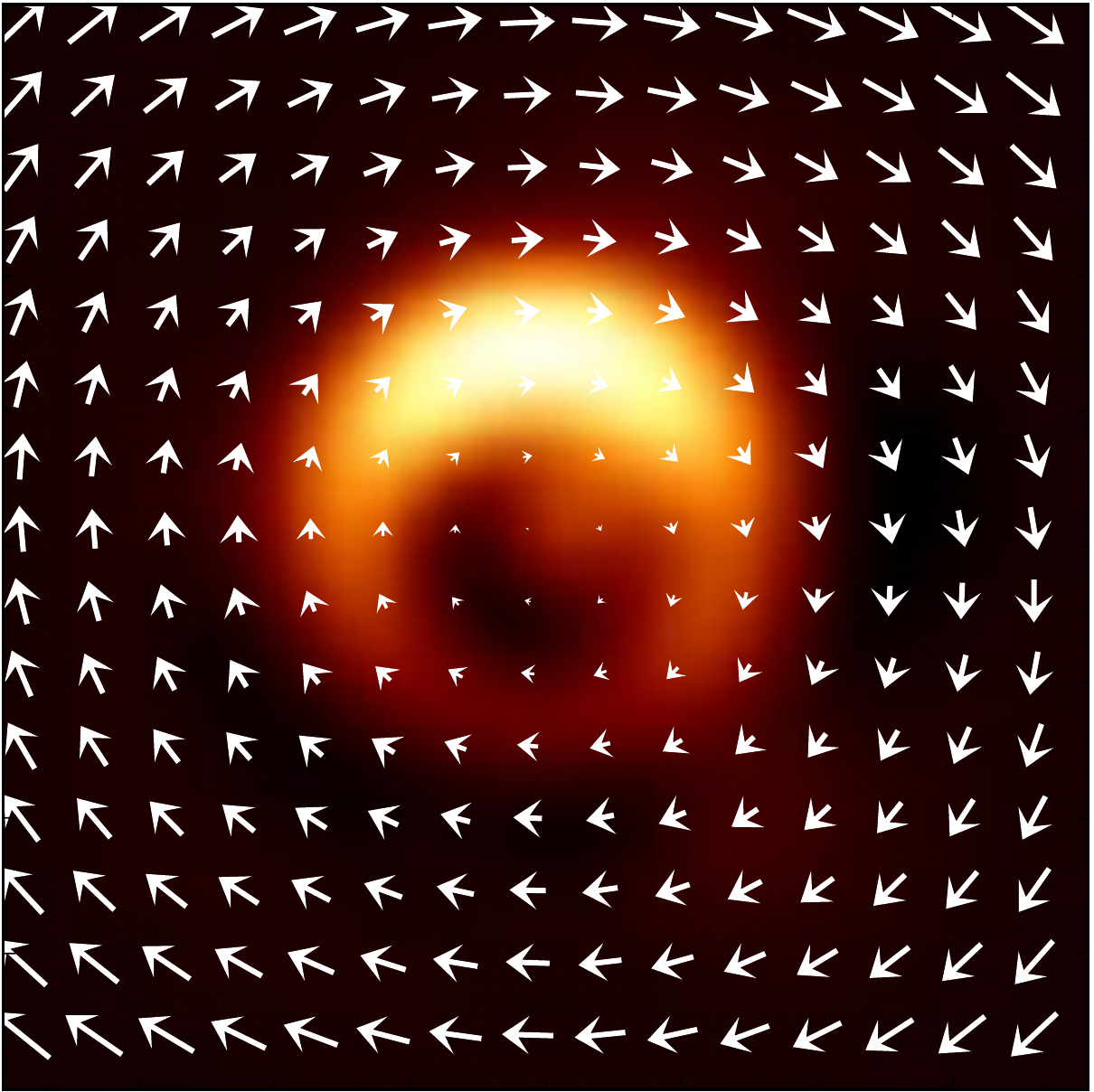}} } \hspace{0.005in}  & \hspace{0.05in}
{{\includegraphics[height=0.35\linewidth]{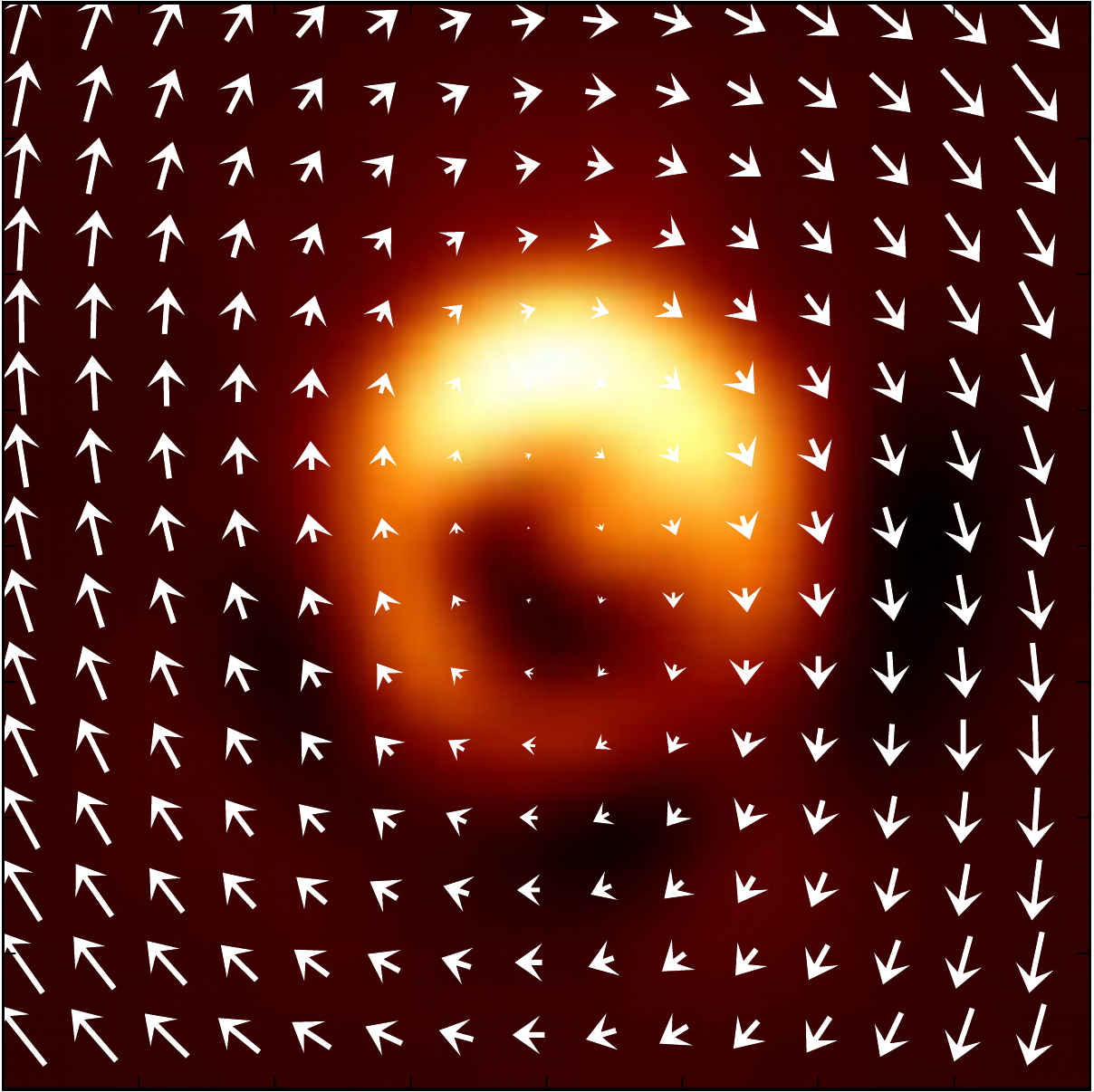}} } \\ 

& \vspace{-0.09in} & \\

\hline

& \vspace{-0.05in} & \\

\hspace*{-.1in} \multirow{1}{*}[0.9in]{ \rotatebox[origin=t]{90}{ \large{\textsf{Video 2}} }} \hspace{0.06in}  &  \hspace{0.05in} {{\includegraphics[height=0.35\linewidth]{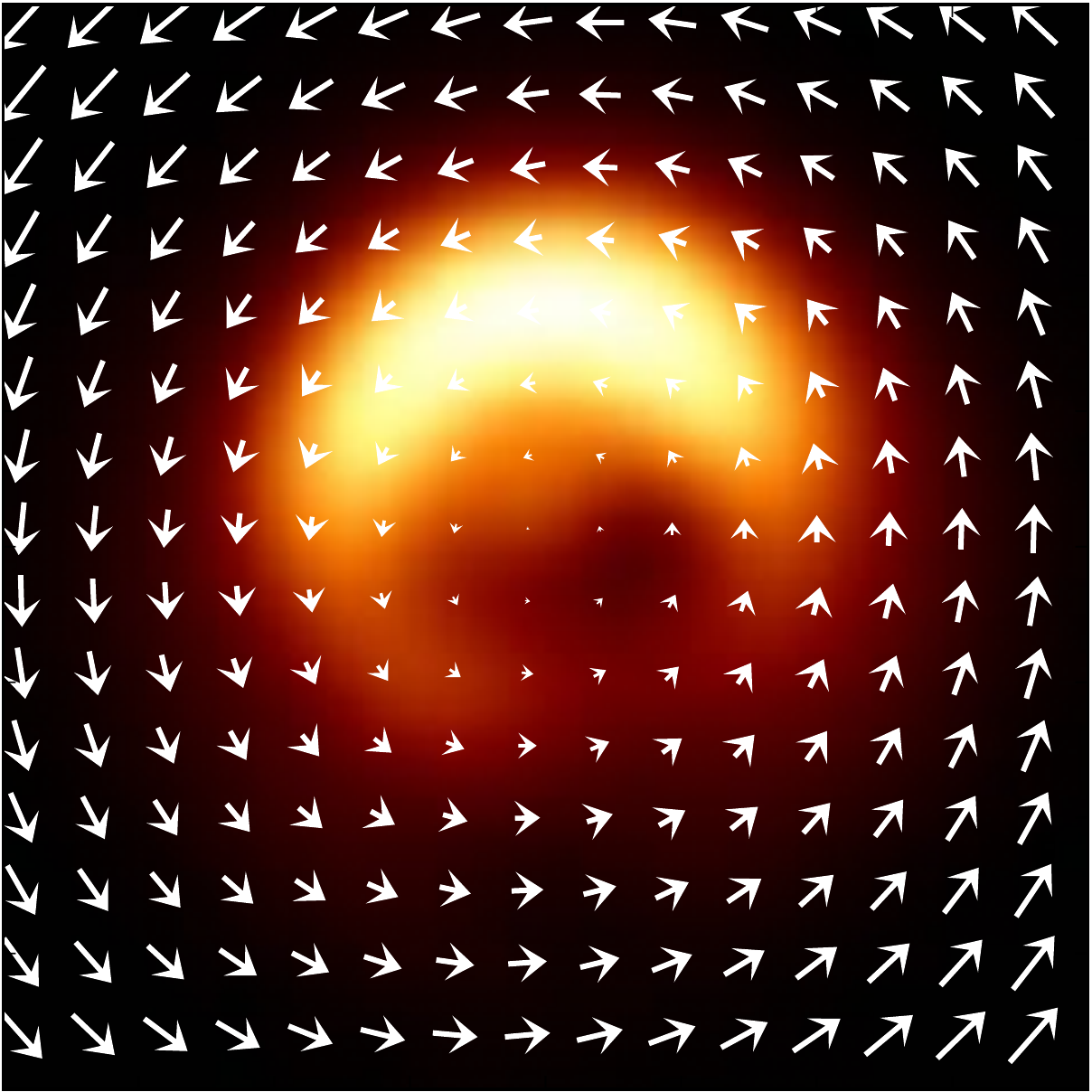}} }  \hspace{0.005in}& \hspace{0.05in}
{{\includegraphics[height=0.35\linewidth]{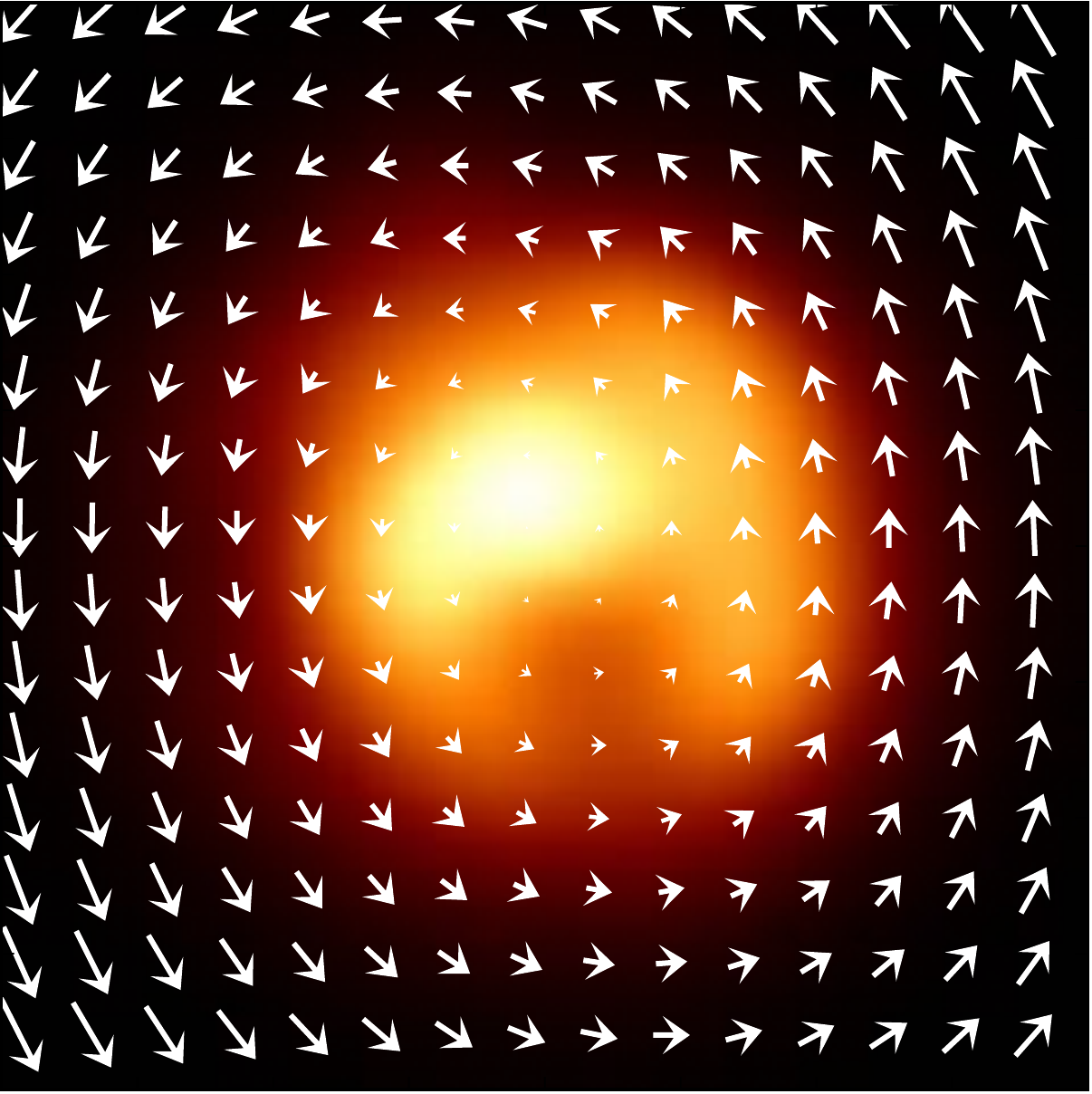}} }
\end{tabular}
\end{center}
\caption{{\bf Recovering Warp Field:} By solving for the parameters of a persistent warp field using the proposed EM algorithm, we are able to recover a low-dimensional representation of the source dynamics. Results are shown using the EHT2017+ array with and without atmospheric error (ATM. and NO ATM. ERROR, respectively). Arrows showing the direction of recovered motion are overlaid on the mean image for a recovered video. Refer to the supplemental video for a visualization of the true underlying and recovered videos. In Video 1 the true underlying motion can be described by a clockwise rotation. The proposed method is able to recover Video 1's motion from the observed data. Video 2 contains a `hot spot' rotating counter-clockwise around a static emission. Video 2 cannot be described using a single persistent flow field. Yet, despite this, the proposed method is still able to recover the general direction of counter-clockwise motion. 
}
\label{fig:warpfield}
\vspace{-.15in}
\end{figure}
\begin{figure}[tb]
	\begin{center}
		\vspace{-.35in}
		\hspace*{-0.5cm}
		\begin{tabular}{   c | c  c   }
			&\vspace{-.1in}&\\
			& \small{\textsf{Single Frame}} &\small{\textsf{ Unwrapped }}       \\ 
			&\vspace{-.1in}&\\
			& \small{\textsf{with Overlaid Circle}} &\small{\textsf{Space $\times$ Time Image }}       \\ \hline
			&\vspace{-.1in}&\\
			\multirow{1}{*}[0.5in]{ \rotatebox[origin=t]{90}{\small{\textsf{ Truth }} }}
				&
				{{\includegraphics[height=.2\linewidth]{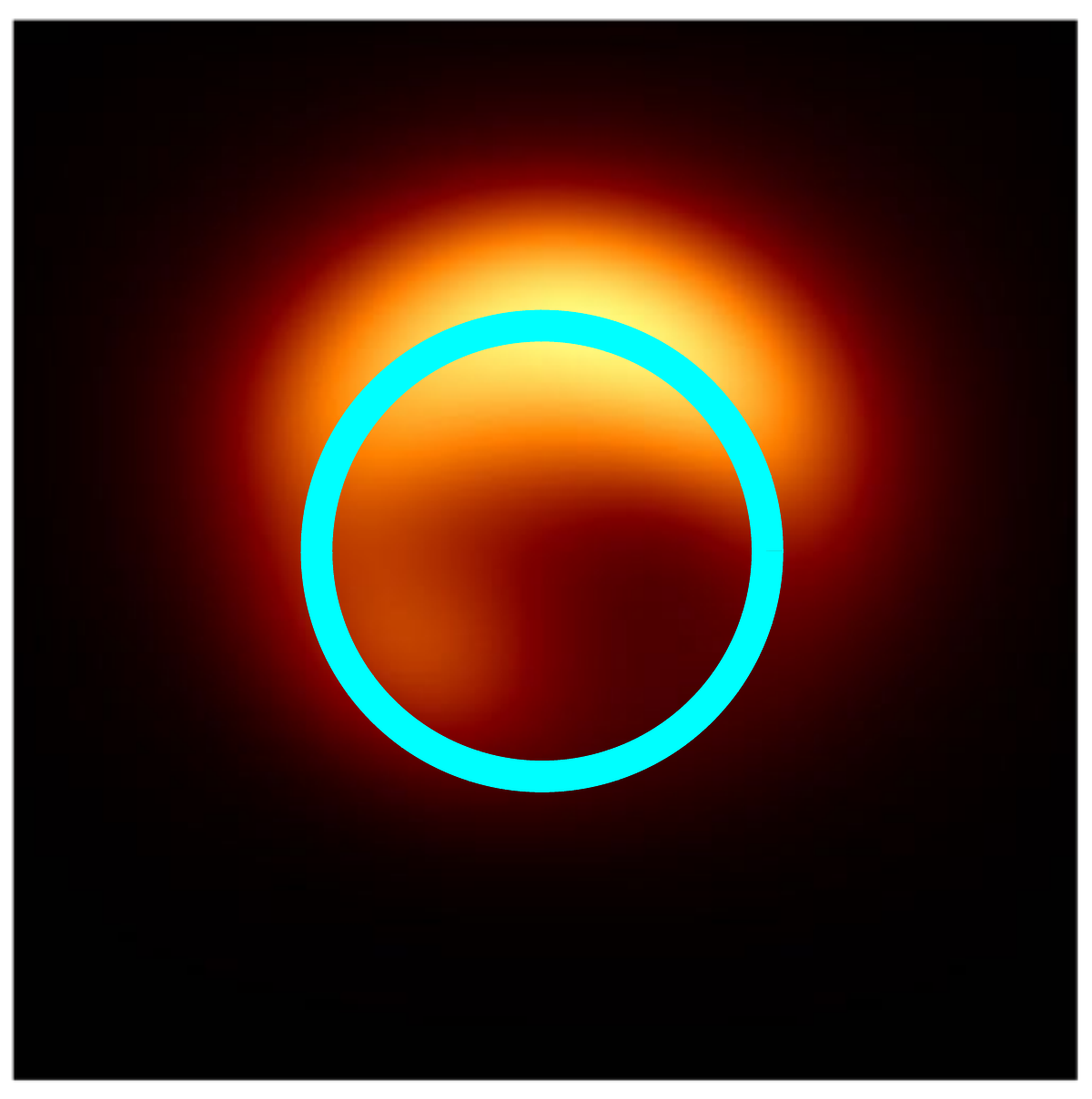}} } &
				\includegraphics[height=.2\linewidth]{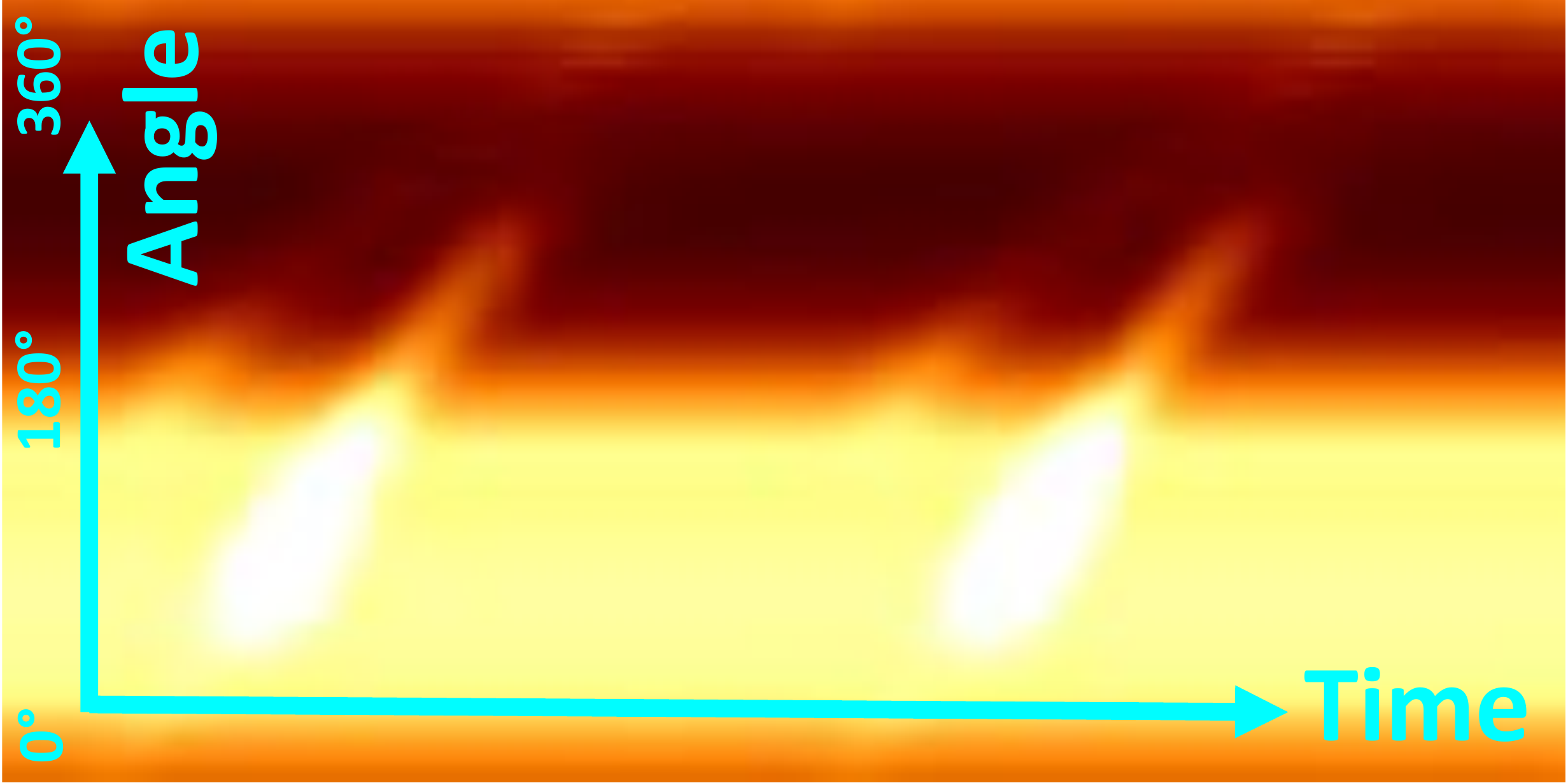} 
				\\
				&\vspace{-.1in}&\\
				\multirow{1}{*}[0.65in]{ \rotatebox[origin=t]{90}{  \specialcell{ \small{\textsf{Blurred Truth}} \\  \tiny{\textsf{ $\frac{3}{4}$ Nominal Beam}}}  }}
			&
			{{\includegraphics[height=.2\linewidth]{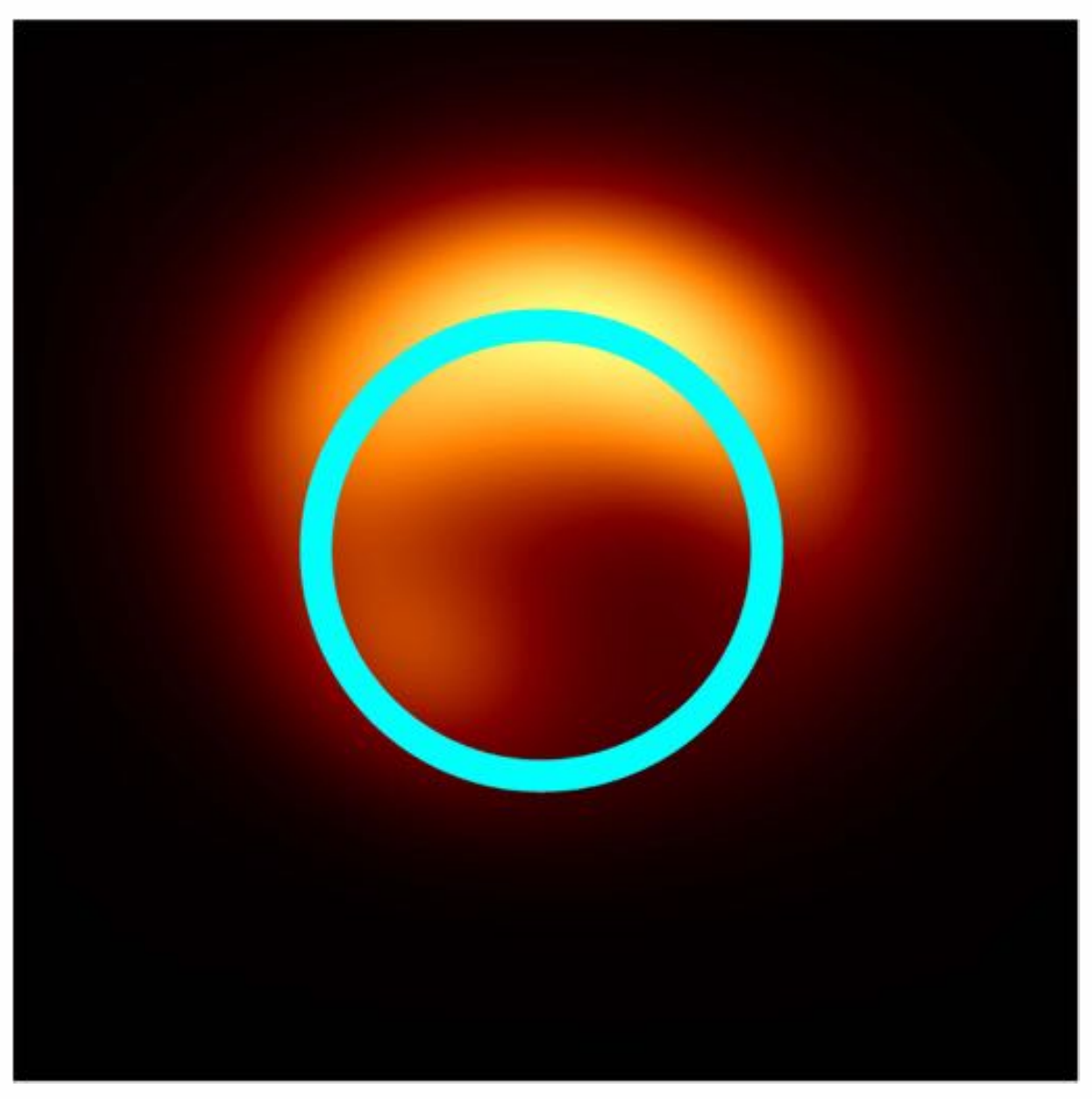}} } &
			\includegraphics[height=.2\linewidth]{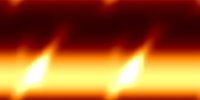} 
			\\
			&\vspace{-.1in}&\\
			\multirow{1}{*}[0.6in]{ \rotatebox[origin=t]{90}{\small{\textsf{ Snapshot }} }}
			&
			{{\includegraphics[height=.2\linewidth]{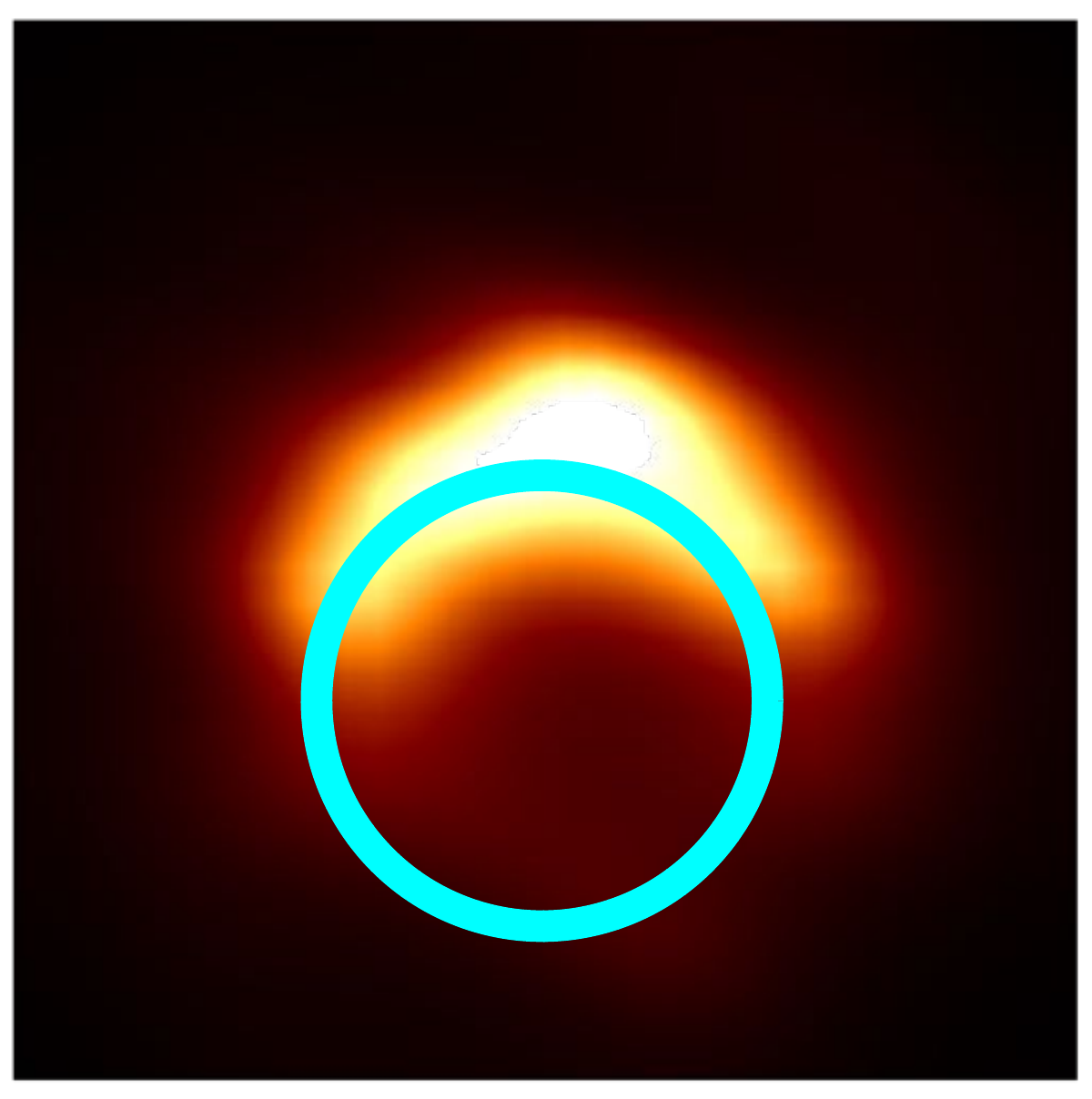}} } &
			\includegraphics[height=.2\linewidth]{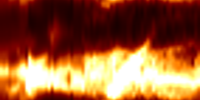} 
			\\
				&\vspace{-.1in}&\\
					\multirow{1}{*}[0.65in]{ \rotatebox[origin=t]{90}{  \specialcell{ \small{\textsf{StarWarps:}} \\  \small{\textsf{No Warp}}}  }}
				&
				{{\includegraphics[height=.2\linewidth]{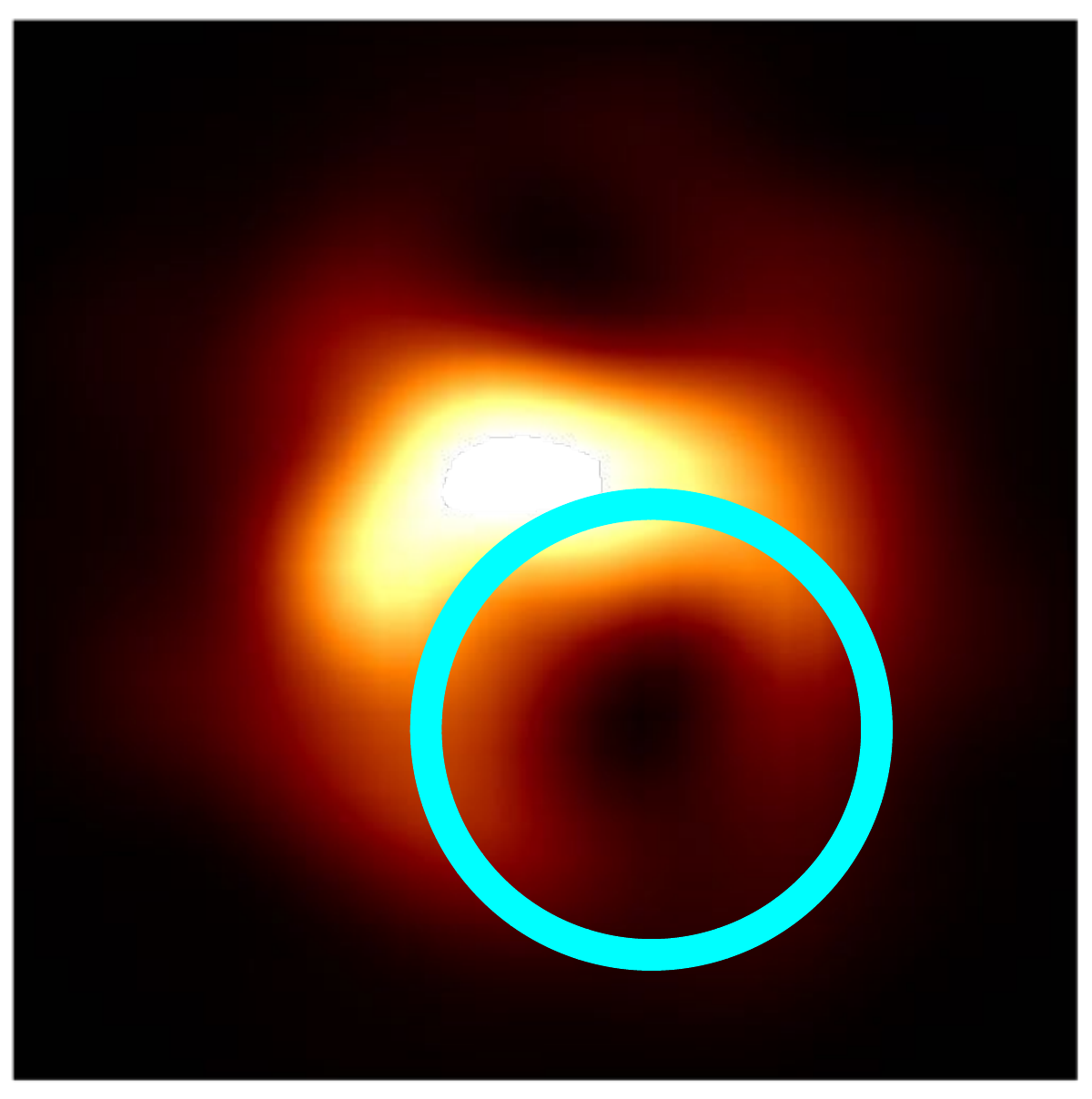}} } &
				\includegraphics[height=.2\linewidth]{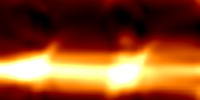} 
				\\
				&\vspace{-.1in}&\\
				\multirow{1}{*}[0.65in]{ \rotatebox[origin=t]{90}{  \specialcell{ \small{\textsf{StarWarps:}} \\  \small{\textsf{Learn Warp}}}  }}
				&
				{{\includegraphics[height=.2\linewidth]{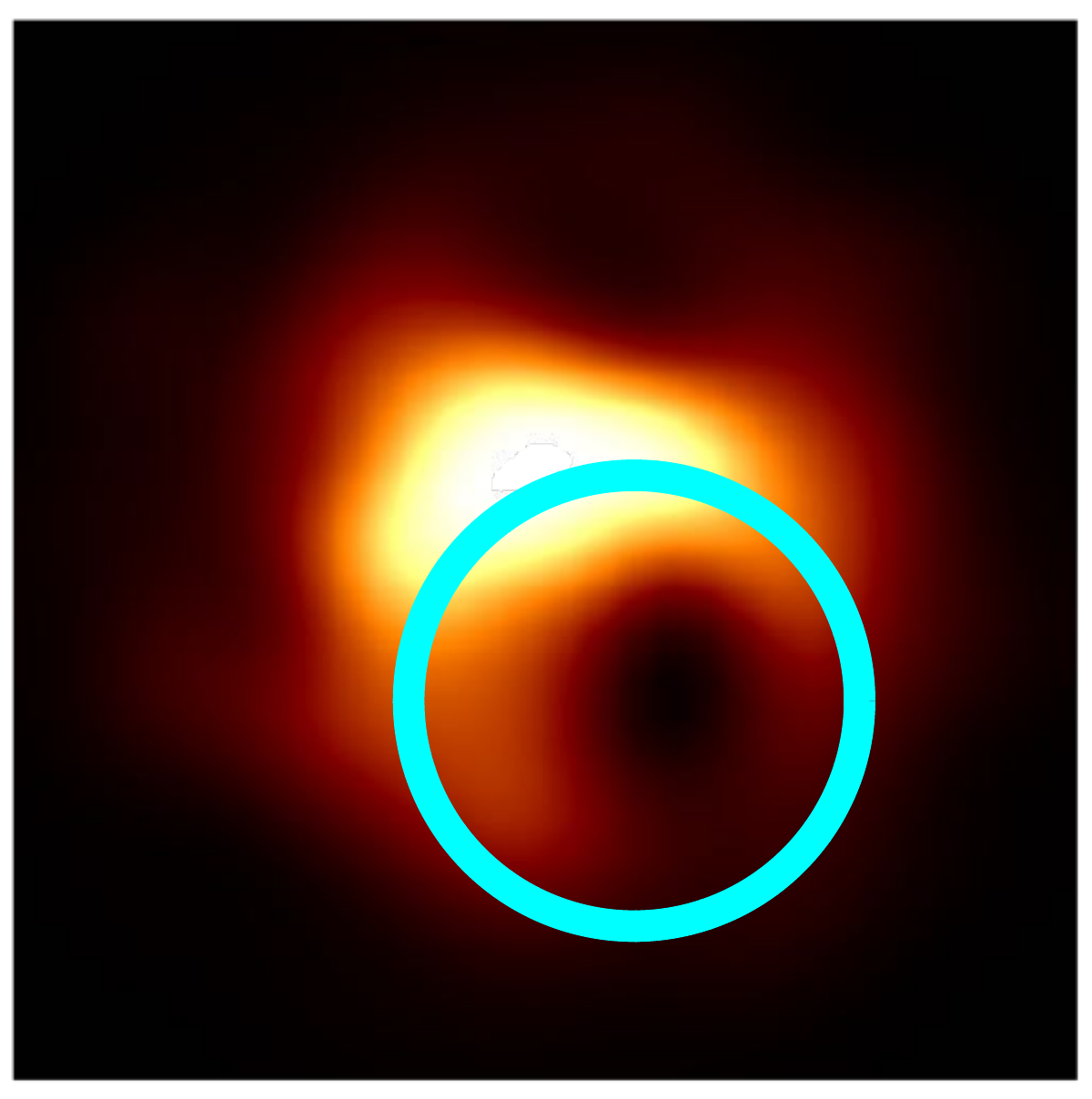}} } &
				\includegraphics[height=.2\linewidth]{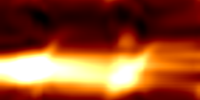} 
				\\
		\end{tabular}
		\caption{{\bf Visualizing Recovered Motion:} We visualize the recovered motion in Video 2 by displaying the change in intensity around a circle in the image over time. After fitting a circle of constant radius to each video,  the intensities around the circle in each image are unwrapped and placed in a single column in the unwrapped space $\times$ time image. As the hot spot rotates around the black hole a distinctive line appears in the true angle $\times$ time image. These lines also appear in the StarWarps angle $\times$ time images, but are harder to discern among the other artifacts in the snapshot imaging result. Results were obtained using the EHT2017+ array with added atmospheric noise, and correspond to results shown in Figure 2 of the supplemental document. As the absolute position of the source is lost when using the closure phase or bispectrum, the position of the recovered black hole moves slightly over the course of the video. This causes the fluctuation in the intensity of the bright horizontal line in the StarWarps recovered angle $\times$ time images, as we do not shift the position of the fitted circle.  }
		\label{fig:motion}
	\end{center}
	\vspace{-.35in}
\end{figure}

In Section~\ref{sec:nomotionresults} we showed that a static model can often substantially improve results over the state-of-the-art methods, even when there is significant global motion. However, when a source's emission region evolves in a similar way over time, we are able to further improve results by simultaneously estimating a persistent warp field along with the video frames.
We demonstrate the StarWarps EM approach proposed in Section~\ref{sec:dynamic_inference_unknown}, on Videos 1 and 2. In results presented, we have assumed an affine motion basis with no translation ($\theta$ consists of 4 parameters), and have allowed the method to converge over 30 EM iterations. 

Figure~\ref{fig:warpfield} shows the warp field recovered by our EM algorithm. 
Results were obtained from data with and without atmospheric error. In Video 1 the true underlying motion of the emission region can be perfectly captured by the affine model we assume. This allows us to freely recover a very similar warp field. 
However, in the ``hot spot" video (Video 2), there does not exist a persistent warp field that fits the data, let alone an affine warp field. Although the true motion cannot be described by our model, we still recover an accurate estimate indicating the direction of motion. 

Figure~\ref{fig:motion} helps to further visualize the recovered motion in the ``hot spot" video by showing how the intensities of a region evolve over time. Results of our method are compared to that of a simple baseline method that we refer to as `snapshot imaging'. In snapshot imaging each frame is independently reconstructed using only the small number of measurements taken at that time step.  In particular, we use the MEM \& TV method shown in Figure~\ref{fig:staticimaging} to reconstruct each snapshot. 
Our results using StarWarps show substantial improvement over snapshot imaging, especially in the case of data containing atmospheric phase error.

We expand upon these results in the supplemental material's video and document. 
Figures 1 and 2 in the supplemental material document compare results obtained when we assume no global motion ($A=\mathds{1}$) to those when we allow the method to search for a persistent warp field. 
In the case of large global motion, most of the reconstructed motion is suppressed when we assume $A=\mathds{1}$. However, by solving for the low dimensional parameters of the warp field, $\theta$, we can learn about the underlying dynamics and sometimes produce higher quality videos. 



\vspace{-.15in}
\section{Conclusion}
\label{sec:conclusion}

Traditional interferometric imaging methods are designed under the assumption that the target source is static over the course of an observation~\cite{TMS}. However, as we continue to push instruments to recover finer angular resolution, this assumption may no longer be valid. For instance, the innermost orbital periods around the Milky Way's supermassive black hole, Sgr A*, are just minutes~\cite{orbitalperiod}. In these cases, we have demonstrated that traditional imaging methods often break down. 

In this work, we propose a way to model VLBI measurements that allows us to recover both the appearance and dynamics of a rapidly evolving source. Our proposed approach, StarWarps, reconstructs a video rather than a static image. By propagating information across time, it produces significant improvements over conventional approaches to create static images or a series of snapshot images in time.

Our technique will hopefully soon allow for video reconstruction of sources that change on timescales of minutes, allowing a real-time view of the most energetic and explosive events in the universe. 

\vspace{-.2in}

\begingroup
\setstretch{0.79}
\bibliographystyle{splncs}
\bibliography{egbib}
\endgroup


\end{document}